\documentclass{aa}
\usepackage{graphicx}
\usepackage{longtable,lscape}
\usepackage{natbib}
\bibpunct{(}{)}{;}{a}{}{,} % to follow the A&A style
\usepackage{color}

\newcommand{\geqsim}{\,\raisebox{-0.6ex}{$\buildrel > \over \sim$}\,}

\newcommand{\teff}{$T_{\rm eff}$}
\newcommand{\logg}{$\log g$}
\newcommand{\vsini}{$v\sin i$}

\newcommand{\feh}{[Fe/H]}

\newcommand{\kms}{km\,s$^{-1}$}

\newcommand{\Msun}{{$M_{\odot}$}}

\begin{document}
\title{A Spectroscopic Survey of the Youngest Field Stars in the Solar Neighborhood}
\subtitle{II. The optically faint sample \thanks{Based on observations collected at the 
Italian \textit{Telescopio Nazionale Galileo} (TNG) operated by the \textit{Fundaci\'on 
Galileo Galilei -- INAF} (Canary Islands, Spain),  at the \textit{Observatoire de Haute Provence} (OHP, France), 
and the \textit{Osservatorio Astrofisico di Catania} (OAC, Italy)}~\fnmsep\thanks{Tables~\ref{Tab:RV}, \ref{Tab:RV_SB2}, \ref{Tab:RV_SB3},
and \ref{Tab:APs} are available at the CDS via anonymous ftp to {\tt cdsarc.u-strasbg.fr (130.79.128.5)} or via 
{\tt http://cdsarc.u-strasbg.fr/viz-bin/qcat?J/A+A/?/?}. }}

\author{A. Frasca\inst{1}\and 
	P. Guillout \inst{2}\and
	A. Klutsch \inst{1}\and 
	R. Freire Ferrero \inst{2}\thanks{Rubens Freire Ferrero passed away on September 10, 2015.}\and
	E. Marilli\inst{1}\and 
	K. Biazzo\inst{1}\and
	D. Gandolfi\inst{3}\and 
	D. Montes\inst{4}	
	} 

\offprints{A. Frasca\\ \email{antonio.frasca@oact.inaf.it}}

\institute{
INAF - Osservatorio Astrofisico di Catania, via S. Sofia, 78, 95123 Catania, Italy
\and
Universit\'e de Strasbourg, CNRS, Observatoire astronomique de Strasbourg, UMR 7550, F-67000 Strasbourg, France 
\and
Dipartimento di Fisica, Universit\`a di Torino, via P. Giuria, 1, 10125 Torino, Italy
\and
Departamento de Astrof\'{\i}sica y Ciencias de la Atm\'osfera, Universidad Complutense de Madrid, 28040 Madrid, Spain
}

\date{Received  / accepted}
 
\abstract 
  % context heading (optional)
{Star formation in the solar neighborhood is mainly traced by young stars in open clusters, associations and in the field, which 
can be identified, e.g., by their X-ray emission. The determination of stellar parameters for the optical counterparts of X-ray sources is crucial 
for a full characterization of these stars.  }
 %leave it empty if necessary  
  % aims heading (mandatory)
{This work extends the spectroscopic study of the \textit{RasTyc} sample, obtained by the cross-correlation of the TYCHO and ROSAT All-Sky 
Survey catalogs, to stars fainter than $V=9.5$\,mag and it is aimed to the identification of sparse populations of young stars in the solar neighborhood.}
  % methods heading (mandatory)
{
We acquired 625 high-resolution spectra for 443 presumabily young stars with four different instruments in the Northern hemisphere. 
The radial and rotational velocity (\vsini) of our targets are measured by means of the cross-correlation technique, 
which is also helpful to discover single-lined (SB1), double-lined spectroscopic binaries (SB2), and multiple systems. We use the 
code ROTFIT for performing an MK spectral classification and for determining the atmospheric parameters (\teff, \logg, 
\feh) and \vsini\ of the single stars and SB1 systems. For these objects, the spectral subtraction of slowly rotating templates
is used to measure the equivalent widths of the H$\alpha$ 
and \ion{Li}{i}\,6708\,\AA\ lines, which enables us to derive their chromospheric activity level and  
lithium abundance. We make use of \textit{Gaia} DR1 parallaxes and proper motions for locating the targets in the HR diagram
and for computing the space velocity components of the youngest objects.
}
  % results heading (mandatory)
{We find a remarkable fraction (at least 35\,\%) of binaries and multiple systems.
On the basis of the lithium abundance, the sample of single stars and SB1 systems appears to be mostly ($\sim$\,60\,\%) 
composed of stars younger than the members of the UMa cluster. 
The remaining sources are in the age range between the UMa and Hyades clusters ($\sim$\,20\,\%) or older ($\sim$\,20\,\%).
In total, we identify 42 very young ({\it PMS-like}) stars, which lie above or very close to the Pleiades
upper envelope of the lithium abundance. 
A significant fraction ($\sim$\,12\,\%) of evolved stars (giants and subgiants) is also present in our sample. Some of them 
($\sim$\,36\,\%) are also lithium rich ($A$(Li)$>$\,1.4).
 }
 % conclusions heading (optional)
{} %leave it empty if necessary

   \titlerunning{A Spectroscopic Survey of Young Field Stars. The RasTyc faint sample}
      \authorrunning{A. Frasca et al.}

\maketitle

%===================================================================
\section{Introduction}
\label{Sec:intro}

It has been shown that open clusters (OCs) cannot account for the total star formation in the Galaxy, but, at most, 
for about 50\%, the remaining occurring in OB associations \citep[see, e.g.,][]{Piskunov2008, Zinnecker2008}.
Although at the end of their parent cloud collapse newly-formed stars are located in the same region of space, many events tend 
to carry them away from each other. 
After being mixed with the stellar population of the galactic plane, young stars are hardly distinguishable
from older ones. Indeed, their magnitudes and colors are similar to the latter ones and there is no gas left from the
parent cloud that can help to identify them. Other properties related to their young age, such as their kinematics, magnetic
activity, infrared (IR) excess, and the presence of lithium in their atmospheres, must be used to spot them among the older population.

The most favorite scenario for the recent history of star formation in the solar neighborhood is that of a large structure 
in which giant molecular clouds generated massive stars about 50 Myr ago. The latter, exploding as supernovae, triggered star formation 
in a ring-like structure, the Gould Belt (see, e.g., \citealt{Comeron1994, Bally2008}, and references therein), and cleaned 
the region around the Sun from the residual gas, generating the local bubble  \citep[see, e.g.,][]{Bonchkarev1984, Gehrels1993}.
Many OB associations and star forming regions (SFRs) within 500--600 pc are known to be part of the Gould Belt \citep{Elias2009,Bobylev2014}.

Most studies on young low-mass stars in the solar vicinity were focused on SFRs or young OCs that represent 
homogeneous samples in terms of age and chemical composition. However, sparse populations of late-type young stars apparently
unrelated to known SFRs and young OCs have been discovered \citep[e.g.,][and references therein]{Guillout1998a}. 
Different scenarios, like ejection from their birth sites by close encounters \citep{Sterzik1995} or by a very fast dispersion of 
small clouds \citep{Feigelson1996}, have been proposed to explain these sources scattered over the sky. Depending on their sky position, 
they could also be related to star formation in the Gould Belt  \citep[e.g.,][and references therein]{Biazzo2012a}.

To pick up isolated young stars scattered over the whole sky, \citet{Guillout1999} cross-correlated the ROSAT All-Sky Survey 
(RASS, \citealt{Voges1999}) and the TYCHO (or Hipparcos) catalogs \citep{HIPPA97}.  This selection produced the so-called  
{\it RasTyc} and {\it RasHip} samples, containing about 14\,000 and 6200 active stars, respectively.
These samples are very suitable for the study of the large-scale distribution of X-ray active 
stars in the solar neighborhood  and led to the discovery of the late-type stellar population of the Gould Belt 
\citep[]{Guillout1998a,Guillout1998b}.	
A similar approach was followed by \citet{Haakonsen2009} who made a statistical cross-association of the ROSAT bright 
source catalog with the 2MASS point source catalog \citep{2MASS}.
However, complementary optical data are absolutely needed for a full and safe exploitation of the huge 
scientific potential of the {\it RasTyc} and {\it RasHip} samples, and to search for young field stars unrelated to the main nearby 
SFRs and young associations. 

To this aim we started a large program for following up spectroscopically a significant fraction of the 
{\it RasTyc} sample observable from the Northern hemisphere. The results of the optically bright ($V_{\rm T} \leq 9.5$\,mag) subsample 
were presented in \citet[][hereafter Paper~I]{Guillout2009}. In that paper, we analyzed high resolution spectra of 426 stars 
with no previous spectroscopic data and found that this sample is mainly composed of stars with ages between 100 and 600 Myr, 
with a minor contribution from an older population (1--2 Gyr).
 A significant fraction  of binaries and lithium-rich giants ($\approx$ 30\,\%), which can be considered 
as contaminants, was also found. 
Based on their high lithium abundance, seven very young ($age \le 30$ Myr) stars, that appear to be unrelated to any know 
SFR, were also uncovered. Some of them are good post-T Tau candidates and five are likely members 
of Pleiades or Castor stellar kinematic groups.   

In the present paper we extend this study to the optically faint ($V > 9.5$\,mag) {\it RasTyc} sources. 
The inclusion of optically fainter sources in the total sample of investigated {\it RasTyc} stars improves the statistics and
allows us to reach a larger distance or intrinsically fainter objects.

A first result of the optically faint sample was the discovery of four lithium-rich stars packed within a few degrees on the sky
in front of an area void of interstellar matter in the Cepheus complex. They form a homogeneous moving group, with a likely age of
10--30 Myr. We  published a preview of the results for these four stars in \citet{Guillout2010}, but we include them in the 
present paper where additional spectra have been also analyzed.

The paper is organized as follows. In Sect.~\ref{Sec:Data} we present the data set, with a brief description of 
the observations and spectra reduction. In Sect.~\ref{Sec:analysis} we describe the analysis aimed at measuring radial and rotational 
velocities (Sects.~\ref{Sec:RV} and \ref{Sec:vsini}), atmospheric parameters (Sect.~\ref{Sec:APs}),  
chromospheric emission fluxes and lithium equivalent widths (Sect.~\ref{Sec:Chrom}). The results are presented in Sect.~\ref{Sec:Results}, 
where the properties and evolutionary status of the targets are discussed on the basis of their position on the Hertzsprung-Russell diagram 
(Sect.~\ref{subsec:HR}), the abundance of lithium and H$\alpha$ emission (Sects.~\ref{subsec:Age} and \ref{subsec:Chromo}), kinematics 
(Sect.~\ref{subsec:kinem}), and spectral energy distribution (Sect.~\ref{subsec:sed}). The main conclusions are summarized in 
Sect.~\ref{Sec:Conclusions}. Individual notes on the youngest stars in our sample can be found in Appendix~\ref{Sec:notes}.

%====================================================================
\section{Observations and reduction}
\label{Sec:Data}

We use the RasTyc catalog \citep[][]{Guillout1999} to obtain a statistically significant sample of the youngest field stars in the solar 
neighborhood. We adopted the criteria described in Paper~I to select the optically fainter sources to be investigated spectroscopically 
for this study. We briefly recall them for the reader's convenience:
\begin{description}
\item[-] declination $\delta\geq 0\degr$ to observe the targets with telescopes located in the Northern hemisphere; 
\item[-] right ascension between 15h and 8h;
\item[-] $0.6 \leq (B-V)_{\rm T} \leq 1.3$, which is the range in which the lithium $\lambda 6707.8$\,\AA\, line can be used as an age indicator; 
\item[-] ROSAT PSPC count rate greater than or equal to 0.03 cts\,s$^{-1}$ to avoid RASS scanning bias \citep[see][]{Guillout1999};
\item[-] $V_{\rm T} \leq 10.5$\,mag, since the Tycho catalog is largely incomplete above this limit.
\end{description}

As we observed many targets from La Palma, the first selection criterion has been slightly relaxed, so that this sample contains also some 
objects with negative declination ($\delta\geq -15\degr$). This allowed us to compare our results with those of the SACY survey 
\citep{Torres2006} of Southern stellar X-ray sources. 

We made a detailed search of previous data and works dealing with these sources in the Simbad and Vizier astronomical databases. 
We did not observe the stars whose physical parameters were already derived in the literature from high resolution spectra. 
However, a few stars with published high-resolution spectra are present in our list either because they were purposely selected 
for doing a comparison with the results of previous surveys or because their data were published after the approval of our observation proposals.    
Moreover, when two stars of similar brightness are located near the X-ray source within the typical ROSAT positional accuracy of 
30\arcsec\  \citep{Voges1999}, we decided to observe both of them.  

We observed 134 optically faint stars from November 2001 to August 2005 with the spectrograph AURELIE at the 1.52-m telescope at the Observatoire 
de Haute Provence (OHP, France; see Paper~I for details) and 178 stars with SARG, the high-resolution
spectrograph at the 3.5-m TNG telescope (Canary Islands, Spain), in 2007\footnote{Proposals TAC67-AOT15/07A and TAC35-AOT16/07B.}. 
We chose the SARG setup with the yellow grism and the slit width of 0$\farcs$8, which provides a resolving power of $R=\lambda/\Delta\lambda\simeq 57\,000$ 
and a spectral coverage in the range 4600--7900\,\AA. 
Exposure times ranged from 4 to 35 minutes, according to the star magnitude, and allowed us to reach a signal-to-noise (S/N) ratio from 
50 to 110, depending also on the seeing and sky conditions. 
Ten stars were observed in August 2009 and June 2011 with SARG and the red grism, which covers the wavelength range 5500--11\,000\,\AA\footnote{Proposals 
TAC71-AOT20/09B and TAC34-AOT23/11A.}. We adopted the same slit as for the previous spectra.

Moreover, 140 spectra of 120 {\it RasTyc} stars were acquired in 2000 and 2001 with the  ELODIE \'echelle spectrograph ($R\simeq$\,42\,000) at the 
1.93-m OHP telescope. Sixty-four of these targets with $V\leq 9.5$\,mag were included in the bright {\it RasTyc} survey (Paper~I). The atmospheric 
parameters reported in that work have been derived with the TGMET online analysis software \citep{Katz1998} running at the 1.93-m OHP telescope. 
However, in the present paper we analyze in a homogeneous way both the optically bright and faint ELODIE targets that
are all included in this study and listed in Tables~\ref{Tab:RV}--\ref{Tab:APs}. Likewise, a few sources (about 25 in total) brighter than $V=9.5$\,mag without previous 
spectroscopic information  or with incomplete data in Paper~I  were also observed with the other spectrographs and included in the present work.

Further observations of 30 sources were conducted in 2008 and 2009  at the 0.91-m telescope of the \textit{Osservatorio Astrofisico di Catania} 
(OAC, Italy) with the FRESCO spectrograph that covers the spectral range 4250--6850\,\AA\  with a resolution $R\simeq\,21\,000$.
The S/N ratio of these spectra is in the range 40--80.

In the present paper we use a total of 625 spectra of 443 different sources taken with the aforementioned instruments, while more than 1660 spectra 
were acquired during about ten years for the full program (bright+faint samples). We present a concise log of the observations in Table~\ref{Tab:Obs_log}.

We were not able to observe the full sample of faint {\it RasTyc} sources selected according to the above criteria and about 40 stars, i.e. $\sim$10\,\% 
of the sample, remained unobserved. Although a few other very young stars could have been found among them, the lack of their data is not expected to significantly 
bias our results.

\begin{table}[htb]
\caption{Observing log for the full program (bright + faint sample). }
\begin{tabular}{ccccr}
\hline
\hline
Year   & Period     & Instrument & Resolution & Spectra\\  
\hline
\noalign{\smallskip}    						  
2000   & Jul--Sep   & ELODIE	 & 42\,000    &  83  \\ 
2001   & Aug	    & ELODIE	 & 42\,000    &  57  \\ 
2001   & Oct--Dec   & AURELIE	 & 38\,000    & 290  \\ 
2002   & Jun--Nov   & AURELIE	 & 38\,000    & 396  \\ 
2003   & Jan	    & AURELIE	 & 38\,000    & 207  \\ 
2004   & Jun--Sep   & AURELIE	 & 38\,000    & 231  \\ 
2005   & Jul--Aug   & AURELIE	 & 38\,000    & 126  \\ 
2007   & Sep--Oct   & FRESCO	 & 21\,000    &  28  \\ 
2007   & Feb--Dec   & SARG	 & 57\,000    & 184  \\ 
2008   & Mar--Jul   & FRESCO	 & 21\,000    &  19  \\ 
2009   & Jun--Oct   & FRESCO	 & 21\,000    &  35  \\ 
2009   & Aug        & SARG	 & 57\,000    &   4  \\ 
2011   & Jun        & SARG	 & 57\,000    &   6  \\ 
\hline
Total  &            &		 &	      & 1666  \\
\hline
\end{tabular}
\label{Tab:Obs_log}
\end{table}

Details on the reduction of spectra acquired at OHP can be found in Paper~I. 

The reduction of the SARG and FRESCO spectra was basically done inside the IRAF\footnote{IRAF is distributed by the 
National Optical Astronomy Observatory, which is operated by the Association of the Universities for Research in 
Astronomy, inc. (AURA) under cooperative agreement with the National Science Foundation.} package, following 
standard steps of overscan and bias subtraction, flat field division, and scattered light subtraction. The extraction of the 
spectra from the pre-reduced images was performed with the {\sc echelle} task of IRAF,	
which allowed us to subtract the sky spectrum, extracted at the two sides of the stellar one.

Due to the gap between the two CCDs of SARG, which causes the lost of more than one \'echelle order, the 
reduction of the ``red'' and ``blue'' chip has been done separately. 
Particular care has been paid to the correction for the fringes at red and near-IR wavelengths and to the merging
of \'echelle orders, which was very important for the analysis of broad lines that are close to the edges of spectral orders,
such as the hydrogen H$\alpha$ line. For these tasks we developed ad-hoc tools in IDL\footnote{IDL (Interactive Data Language) is a 
registered trademark of  Harris Corporation.} environment. In particular, the fringe removal was accomplished by means 
of contemporaneous flat field spectra that were rectified by low-order polynomial fits to obtain the fringing pattern. 
We have optimized the fringe-correction procedure on spectra of hot and rapidly rotating stars.

Spectra of rapidly-rotating AB stars (templates for telluric subtraction) as well as of radial and rotational velocity standard stars 
were also acquired with all spectrographs and have been used for the data analysis. 

The telluric water vapor lines at the H$\alpha$ and Na\,{\sc i}\,D$_2$ wavelengths were subtracted using an interactive 
procedure described by \citet{Frasca2000} and adopting the well-exposed telluric templates acquired during the observing runs. 
The broad and shallow spectral lines of the latter have been removed by high-order polynomial fits.  

%==================================================================
\section{Data analysis}
\label{Sec:analysis}

We made an analysis of the spectra with different purposes: i) to measure the radial velocity (RV) and the projected 
rotational velocity ($v\sin i$); ii) to determine the basic atmospheric parameters ($T_{\rm eff}$, $\log g$, 
and [Fe/H]); iii) to evaluate the photospheric lithium content; and iv) to define the level of chromospheric 
activity. 

We used different techniques and reached different goals and accuracies for single stars, double-lined
spectroscopic binaries (SB2), and triple (SB3) or multiple systems.

\subsection{Radial velocity} 
\label{Sec:RV}

The measurement of RV for AURELIE and ELODIE spectra was already described in Paper~I. It is based on the use 
of a numerical mask or a synthetic spectrum as RV template. For the ELODIE spectra this task is normally done by the
on-line reduction pipeline just after the acquisition of the spectrum.

For the SARG and FRESCO \'echelle spectra, the RV was measured by means of the cross-correlation between the target spectrum and 
a template chosen among a list of spectra of RV standard stars (Table\,\ref{Tab:Standards}) that were observed with the
same instrument and in the same seasons as our targets. All these stars are slow rotators and display a very low
or negligible level of magnetic activity so that they were also used as non-active templates (Sect.~\ref{Sec:Chrom}).
The cross-correlation functions (CCFs) were computed for all \'echelle orders but those with the lowest S/N 
ratio (usually for blue wavelengths).  Very broad lines, such as \ion{Na}{i}\,D$_2$ and  H$\alpha$,
as well as strong telluric features were excluded from the CCF analysis. The latter was performed with an ad-hoc software developed by us in 
the IDL environment.
The CCF peak has been fitted with a Gaussian to evaluate its centroid and full width at half maximum (FWHM). The RV error
for each spectral order, $\sigma_{\rm i}$, was estimated by the fitting procedure accounting for the CCF noise (far from the peak). 
For a few spectra, we compared the results of our code with those from the IRAF task {\sc fxcor}, finding similar results and errors.

The final RV is obtained as the weighted mean (weight $w_i=1/\sigma_{\rm i}$) of the $RV_i$ values measured in the individual orders. 
We adopted as RV uncertainty the standard error of the weighted mean, $\sigma_{RV}$. 
The RV values and errors for single stars or single-lined binaries (SB1) are reported in Table\,\ref{Tab:RV}.
The targets observed more than once were classified as SB1 systems whenever the difference of two RV values is larger than the sum of the
corresponding errors. A note about SB1 systems is added in Table~\ref{Tab:APs}. We note that each target with a single 
RV measurement has been classified as a single star (S), although this does not exclude that it is indeed an SB1 system, or even an 
SB2 system observed close to the conjunction, when the lines of the individual components are fully blended.   

This procedure was also used to measure the RV of the components of SB2 or multiple systems. We considered a secondary CCF peak as 
significant whenever it exceeded the 5$\sigma$ level, where $\sigma$ is the CCF noise calculated far from the peaks.   

\begin{table}[htb]
\caption{Radial/rotational velocity standard stars.}
\begin{tabular}{llccll}
\hline
\hline
Name       & Sp. Type &   RV             & $v\sin i^{\mathrm{d}}$      & Notes\\ 
           &          &   (km\,s$^{-1}$)   & (km\,s$^{-1}$) &      \\  
\hline
\noalign{\smallskip}
\object{HD\,187691} &  F8\,V   &  $-0.0^{\mathrm{a}}~ $  & 2.8  & RV, $v\sin i$ \\  
\object{HD\,102870} &  F8\,V   &  4.3$^{\mathrm{a}}$     & 4.5  & RV \\
\object{HD\,157214} &  G0\,V   &  $-79.2^{\mathrm{b}}~ $ & 1.6  & $v\sin i$ \\  
\object{HD\,32923}  &  G4\,V   &   20.50$^{\mathrm{a}}$  & 1.5  & RV, $v\sin i$ \\  
\object{HD\,117176} &  G5\,V   &    4.6$^{\mathrm{b,*}}$ & 1.2  & $v\sin i$ \\
\object{HD\,10700}  &  G8\,V   &  $-17.1^{\mathrm{b}}~ $ & 0.9  & $v\sin i$ \\  
\object{HD\,145675} &  K0\,V   & $-13.69^{\mathrm{c,*}}$ & 0.8  & $v\sin i$ \\  
\object{HD\,221354} &  K1\,V   &  $-25.20^{\mathrm{a}}$  & 0.6  & RV, $v\sin i$ \\  
\object{HD\,115404} &  K2\,V   &   7.60$^{\mathrm{a}}$   & 3.3  & RV \\  
\object{HD\,182572} &  G8\,IV  & $-100.35^{\mathrm{a}}$  & 1.9  & RV, $v\sin i$ \\    
\object{HD\,12929}  &  K2\,III & $-14.6^{\mathrm{a}}~ $  & 1.6  & RV \\  
\object{HD\,161096} &  K2\,III & $-12.5^{\mathrm{a}}$    & 2.1  & RV  \\
\hline
\end{tabular}
\label{Tab:Standards}
\begin{list}{}{}									
\item[$^{\mathrm{a}}$] \citet{Udry}. $^{\mathrm{b}}$ \citet{Nordstr}. 
$^{\mathrm{c}}$ \citet{Nidever2002}. $^{\mathrm{d}}$ \citet{Glebocki2005}. 
\item $^{\mathrm{*}}$ Exoplanet hosting star. RV variable.
\end{list}
\end{table}

\subsection{Projected rotational velocity}
\label{Sec:vsini}

The $v\sin i$ was measured in two different ways. The first is based on the FWHM of the CCF peak through calibrations
FWHM--$v\sin i$ obtained for several standard stars among those listed in Table\,\ref{Tab:Standards}. For each standard star, 
the calibration relationship was defined by measuring the FWHM of the CCF peak obtained correlating the original spectrum
with the same spectrum after being artificially broadened by the convolution with a rotation profile of increasing $v\sin i$ 
(from 0 to 150 km\,s$^{-1}$). We applied this method only to 21 \'echelle orders of the blue-chip SARG spectra covering the 
wavelength range from about 4900 to 6000 \AA. We rejected the orders with very wide lines, which 
would  make the results worse.  The FWHM measured on the CCF of each \'echelle order of  target spectra and  
RV standard stars was converted to \vsini\ using these calibrations.
We took the average of the values for each order and the standard deviation as $v\sin i$ and error, respectively.
The same procedure was applied to the FRESCO spectra, for which we used nine \'echelle orders in the range 5400--6400\,\AA,
and to the single-order AURELIE spectra, for which synthetic templates were adopted (see Paper~I).
The measure of $v\sin i$ on the ELODIE spectra was also performed with this method. We adopted the relation proposed by \citet{Queloz98}
to convert the $\sigma$ of the Gaussian fitted to the CCF peak into $v\sin i$.
In a few cases, it was necessary to perform a new cross-correlation analysis on the original ELODIE spectra for measuring RV and 
$v\sin i$, because the range of radial velocity encompassed by the CCF made on the fly during the observation was not sufficient. 
This was the case of a few very rapid rotators and SB2 systems.

This method was also used for measuring $v\sin i$ for the components of SB2 and SB3 systems. We performed independent Gaussian fits 
to the CCF peaks of the system components whenever they are well separated, while a double (or multiple) Gaussian fit was applied in cases
of blended CCFs. In the latter case the accuracy of \vsini\  and RV determinations decreases with the increase of line blending and with the
decrease in peak intensity.   
The RV and $v\sin i$ of the individual components of SB2 and SB3 spectroscopic systems are listed in 
Table\,\ref{Tab:RV_SB2} and \ref{Tab:RV_SB3}, respectively. 

The other method for measuring $v\sin i$, which can be applied only to single stars or SB1 systems, is  based on the minimization of the 
residuals $observed - template$, where the $template$ is one of the spectra in Table\,\ref{Tab:Standards} rotationally broadened from 0 
to 150\,km\,s$^{-1}$ with a step of 0.5\,km\,s$^{-1}$.
This is the same algorithm used by ROTFIT for the determination of the atmospheric parameters (see Sect.\,\ref{Sec:APs} for further details) 
by adopting a wider list of reference stars.
For very rapid rotators, we explored a wider range of $v\sin i$ (from 0 to 500\,km\,s$^{-1}$) with a larger sampling (5\,km\,s$^{-1}$).

\begin{figure}[th]
\hspace{-.7cm}
\includegraphics[width=9.0cm]{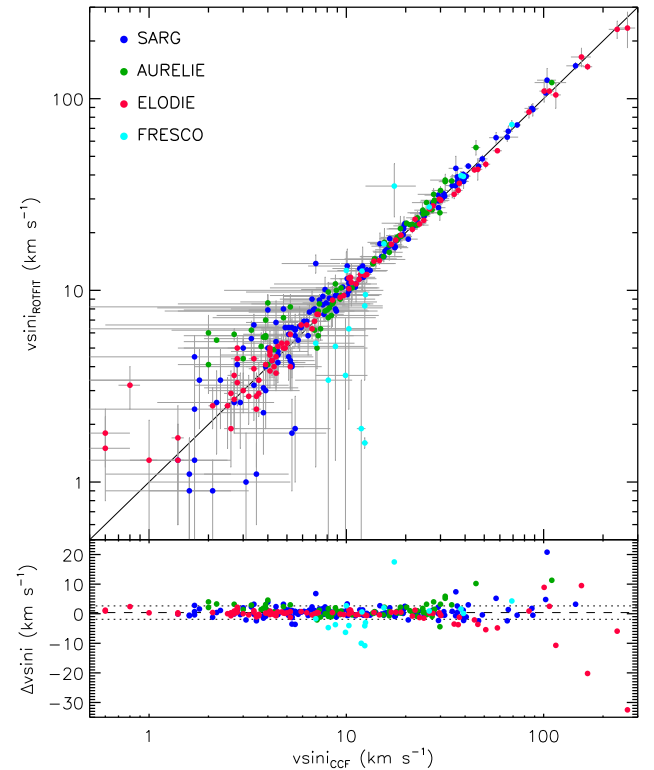}		%fig_vsi_all.eps}
\caption{{\it Top panel)} Comparison between the $v\sin i$ values measured with ROTFIT and those derived from the FWHM of the cross-correlation peak 
 (Table\,\ref{Tab:RV}).  
Dots with different colors have been used for the different instruments, as indicated in the legend.
The continuous line is the one-to-one relationship. {\it Bottom panel)} The residuals from the one-to-one relationship, excluding the values with 
\vsini$\ge 100$\,\kms, have a mean of +0.38~km\,s$^{-1}$ (dashed line), and a standard deviation of 2.26~km\,s$^{-1}$ (dotted lines). }
\label{Fig:vsini}
\end{figure}

The values of $v\sin i$ obtained for single or SB1 stars with both methods, $v\sin i_{\rm CCF}$ and $v\sin i_{\rm ROTFIT}$ are quoted,
along with their errors, in Table\,\ref{Tab:RV}.
They are clearly in very good agreement with each other, as shown in Fig.\,\ref{Fig:vsini}, where the
two data sets follow a close one-to-one relationship. The average difference, calculated for values of $v\sin i \leq 100$\,km\,s$^{-1}$, is only 0.38~km\,s$^{-1}$, 
which is not significant, accounting for the measurement errors. The standard deviation of the residuals is 2.26~km\,s$^{-1}$.
We note that the scatter enhances for the stars rotating faster than about 30\,km\,s$^{-1}$, especially for the ELODIE data. 
This is probably due to the fit of the CCF peak, whose shape is far from a Gaussian for high $v\sin i$, and to 
the \citet{Queloz98} calibration relation adopted for the ELODIE spectra, which has a full validity for $v\sin i$ up to 20\,km\,s$^{-1}$.
A few discrepant values around 10\,\kms\ are visible for the FRESCO data. This is likely the result of the lower resolution of this instrument for which 
the minimum detectable \vsini\  is about 5--7\,\kms.
From the above arguments, we consider as most reliable the values derived with ROTFIT, which are used in the following.

\subsection{Atmospheric parameters and spectral classification}	
\label{Sec:APs}

The MK spectral classification and the determination of the basic atmospheric parameters ($T_{\rm eff}$, $\log g$, and [Fe/H]) were
performed by means of the code ROTFIT \citep{Frasca2006}.
We adopted, as templates, a library of 270 high-resolution spectra of slowly-rotating  stars with well known parameters \citep{PASTEL} spanning 
a wide range of effective temperature, gravity, and iron abundance, which were retrieved from the ELODIE Archive \citep{Moultaka2004}.

The SARG spectra ($R_{\rm SARG}=57\,000$) were degraded to the ELODIE resolution ($R_{\rm ELODIE}=42\,000$)
by means of the convolution with a Gaussian kernel of width $W=\lambda\sqrt{1/R_{\rm ELODIE}^2-1/R_{\rm SARG}^2}$\,\AA.
We applied ROTFIT only to the \'echelle orders with the best S/N ratio, which span from about 4500 to 6800\,\AA, both in the SARG
and FRESCO spectra. To match the lower resolution of FRESCO spectra ($R_{\rm FRESCO}=21\,000$) we degraded the resolution
of the ELODIE templates by convolving them with a Gaussian kernel of width $W=\lambda\sqrt{1/R_{\rm FRESCO}^2-1/R_{\rm ELODIE}^2}$\,\AA.
The same was done for the AURELIE spectra, even though their resolution ($R_{\rm AURELIE}\simeq40\,000$) is close to the
ELODIE spectra.

All the selected \'echelle orders were analyzed independently.  The cores of Balmer lines, which can be contaminated by chromospheric emission, the 
\ion{Li}{i}\,$\lambda 6707.8$ line, and the spectral regions severely affected by telluric line absorption or by CCD defects,  
were excluded from the analysis. 
The final stellar parameters are weighted averages of the results of each $i$-th \'echelle order, where the weight accounts for both the 
$\chi^2_i$ of the fit (more weight is given to the most closely fitted or higher S/N orders) and the amount of information contained in 
each spectral order, which is expressed by the total line absorption,  $f_i=\int(F_{\lambda}/F_{\rm C}-1)d\lambda$, where $F_{\lambda}/F_{\rm C}$
is the continuum-normalized spectrum in the $i$-th \'echelle order.
The uncertainties of atmospheric parameters are the standard errors of the weighted means to which we added in quadrature the average uncertainties 
of the reference stars evaluated as $\pm$50 K, $\pm$0.1 dex, and $\pm$0.1 dex for \teff, \logg, and \feh, respectively.

As already explained in Sect.\,\ref{Sec:RV}, ROTFIT provides us also with a measure of the star projected rotational velocity, because it 
iteratively broadens each template spectrum in a wide range of $v\sin i$ and find the $\chi^2$ minimum. 
However, the values of $v\sin i$ reported in Table\,\ref{Tab:RV} were obtained using as templates the few inactive and slowly rotating standard 
stars of Table\,\ref{Tab:Standards}, whose spectra were acquired with the same instrumentation as for the target ones. This enabled us to 
overcome any possible systematic error due to the different resolution.

Finally, the MK classification of the target star is also provided by the code.  It is defined by the spectral type and the luminosity class of 
the reference star that more frequently matches the target spectrum in the different \'echelle orders.

In Table~\ref{Tab:APs}, we list the stellar parameters, along with their errors ($\sigma_{T_{\rm eff}}$, $\sigma_{\log g}$,
and $\sigma_{[Fe/H]}$), and the MK classification for the single stars and SB1 binaries. The median errors of the atmospheric parameters 
are about 100\,K, 0.16\,dex, and 0.11\,dex, for \teff, \logg, and \feh, respectively.

For double-lined spectroscopic binaries (SB2) and triple systems we do not determine the atmospheric parameters but we only
provide the values of radial and rotational velocities of their components measured thanks to the CCF analysis (cf. Sect.\,\ref{Sec:RV}).
The spectral classification and estimates of the atmospheric parameters of the components of SB2s  via the COMPO2 code \citep[see, e.g.,][]{Frasca2006}
is deferred to a subsequent work devoted to  systems with more observations, for which we will also discuss their orbital parameters. 
However, ten SB2 and one SB3 system for which one component is much brighter than the other have been also analyzed with
ROTFIT.

\subsection{Chromospheric activity and lithium equivalent width}
\label{Sec:Chrom}

As we did for the optically-bright sources (Paper~I), we evaluated the level of chromospheric activity from the emission in the core of the
Balmer H$\alpha$ line for the single stars and SB1 systems.
A rough age classification was performed based on the lithium abundance (see Sect.\,\ref{subsec:Age}). 

\begin{figure}[]
\hspace{-0.5cm}
\includegraphics[width=9.0cm]{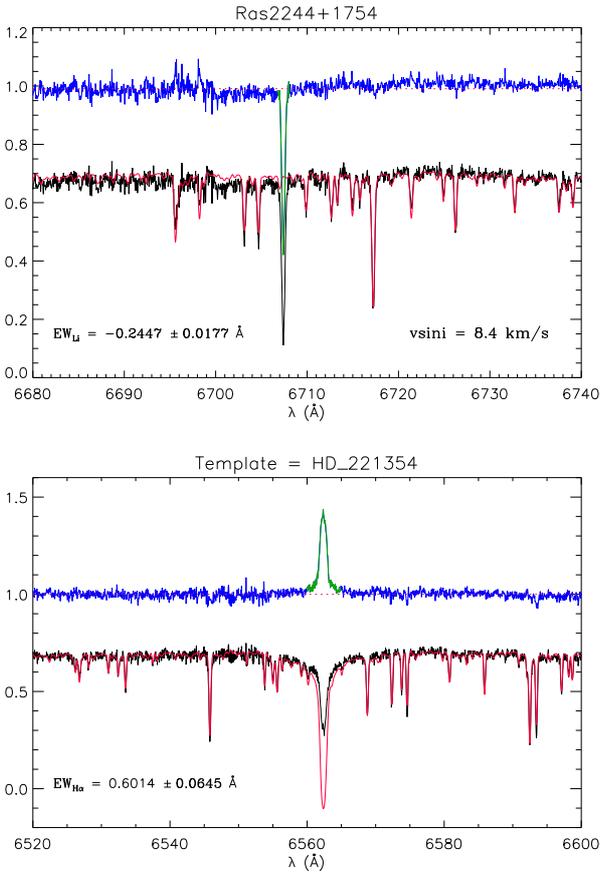}		%Ras2244+1754_red_norm_tell_merg_slin.eps}
\vspace{-1cm}
\caption{Example of the subtraction technique applied to an active and lithium-rich star both in the lithium ({\it upper panel}) and
H$\alpha$ spectral regions ({\it lower panel}). In both panels, the continuum of the residual spectrum (blue line) is set to 1,
while the original spectrum (full black line) and the non-active template 
(thin red line) are shifted downwards by 0.3 in continuum units. The green lines indicate the residual 
absorption (for \ion{Li}{i}) or emission (for H$\alpha$) profile that has been integrated to obtain the equivalent width. }
\label{Fig:HaLisub}
\end{figure}

We measured both the lithium equivalent width and the chromospheric emission level with the ``spectral 
subtraction'' technique \citep[see, e.g.,][]{Herbig85,Frasca1994, Montes95}. This technique is based on the subtraction of 
a  template, which is the spectrum of a slowly rotating star rotationally broadened to the \vsini\ of the target. 
We chose templates with the same spectral type as the target, with a negligible level of chromospheric activity, and without any 
detectable lithium absorption line. 
This method allows us to remove the  \ion{Fe}{i} $\lambda\,6707.4\,$\AA\  absorption line, which is normally blended with the lithium
line, from the target spectrum. Likewise, the subtraction of the photospheric absorption profile of the non-active template allows us to 
measure the net chromospheric emission that fills in the H$\alpha$ line core. 
Therefore, the equivalent width of the lithium line, $W_{\rm Li}$, and the net equivalent width of the 
H$\alpha$ line, $W_{\rm H\alpha}^{em}$, were measured in the residual spectrum obtained by subtracting the non-active template. 
The values of $W_{\rm Li}$ and $W_{\rm H\alpha}^{em}$
have been obtained by integrating the residual emission (absorption for the lithium line) profile (see Fig.\,\ref{Fig:HaLisub}). 
The errors on the equivalent width of H$\alpha$ and lithium lines,  $\sigma_{W_{\rm H\alpha}^{em}}$ and $\sigma_{W_{\rm Li}}$,  were estimated
as the product of the integration range and the mean error per spectral point, which results from the standard deviation of the flux values of the residual 
spectrum measured at the two sides of the line. The equivalent widths of H$\alpha$ and lithium lines are reported in Table~\ref{Tab:APs} together 
with their errors.

We also evaluated the H$\alpha$ surface flux (or radiative losses), $F_{\rm H\alpha}$, and 
the ratio of the H$\alpha$ and bolometric luminosity, $R'_{\rm H\alpha}$\footnote{The prime in $R'_{\rm H\alpha}$ indicates that the photospheric contribution was 
subtracted as usual in the definition of activity indices.}, that are calculated according to the following equations:
\begin{eqnarray}
F_{\rm H\alpha} & = & F_{6563}W_{\rm H\alpha}^{em}, 
\end{eqnarray}
\begin{eqnarray}
R'_{\rm H\alpha}&  = & L_{\rm H\alpha}/L_{\rm bol} = F_{\rm H\alpha}/(\sigma T_{\rm eff}^4),
\end{eqnarray}
{\noindent where $F_{6563}$ is the continuum flux at the stellar surface at the H$\alpha$ wavelength, which has been evaluated from the NextGen synthetic 
low-resolution spectra \citep{Hau99a} at the stellar temperature and surface gravity of the target. We have evaluated the flux error considering the 
$W_{\rm H\alpha}^{em}$ error and the uncertainty in the continuum flux at the line center, $F_{6563}$, which is obtained propagating the \teff\  
and \logg\  errors. }

Finally, as far as the lithium line is concerned, we estimate the threshold for lithium line detection as $\approx$\,10~m\AA,
on the basis of  typical $W_{\rm Li}$ errors and S/N ratios. 

\begin{figure}[th]
\hspace*{-.5cm}
\includegraphics[width=9.5cm]{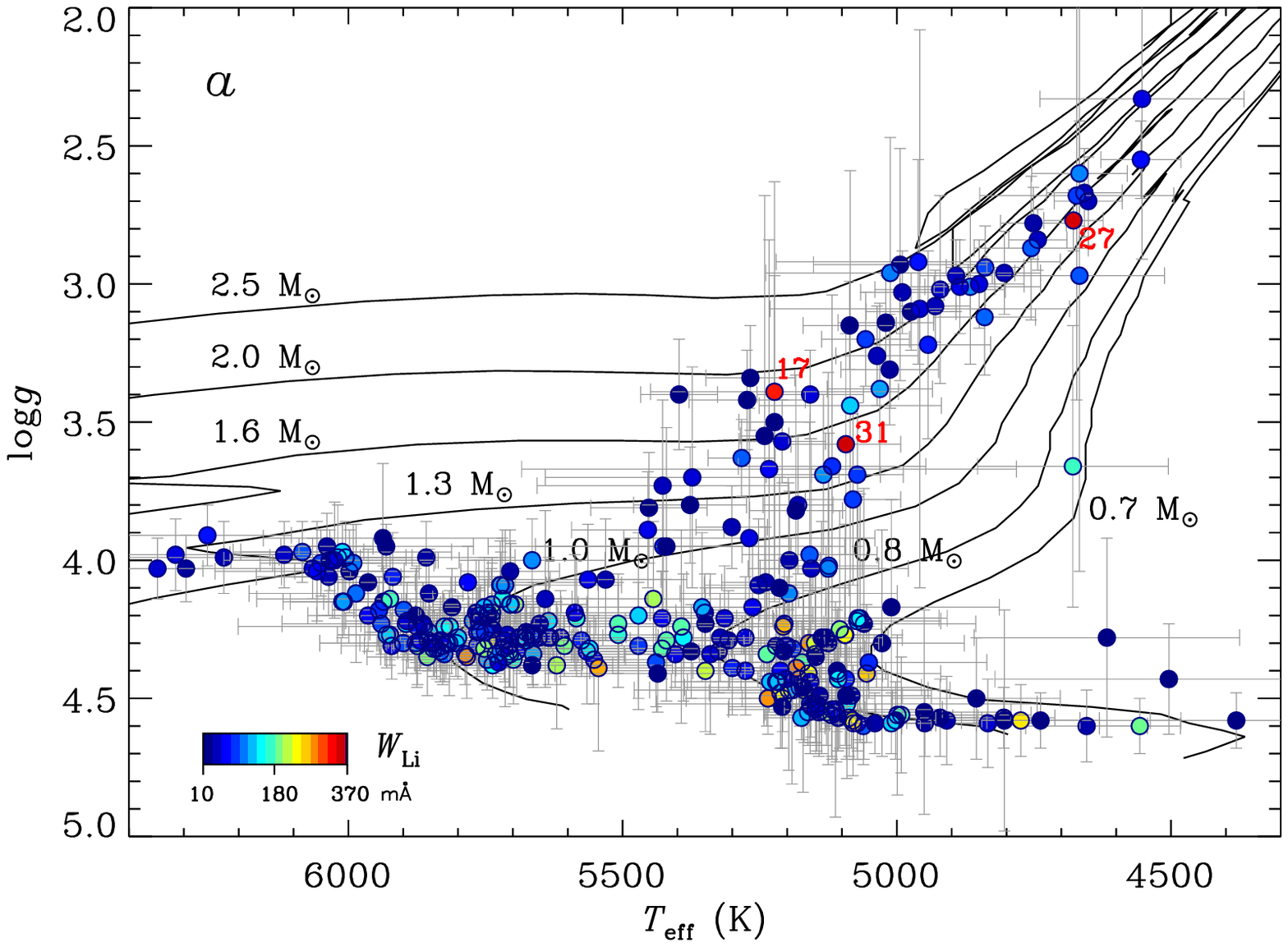}		%logg_teff_gia.eps}\\
\hspace*{-.5cm}
\includegraphics[width=9.5cm]{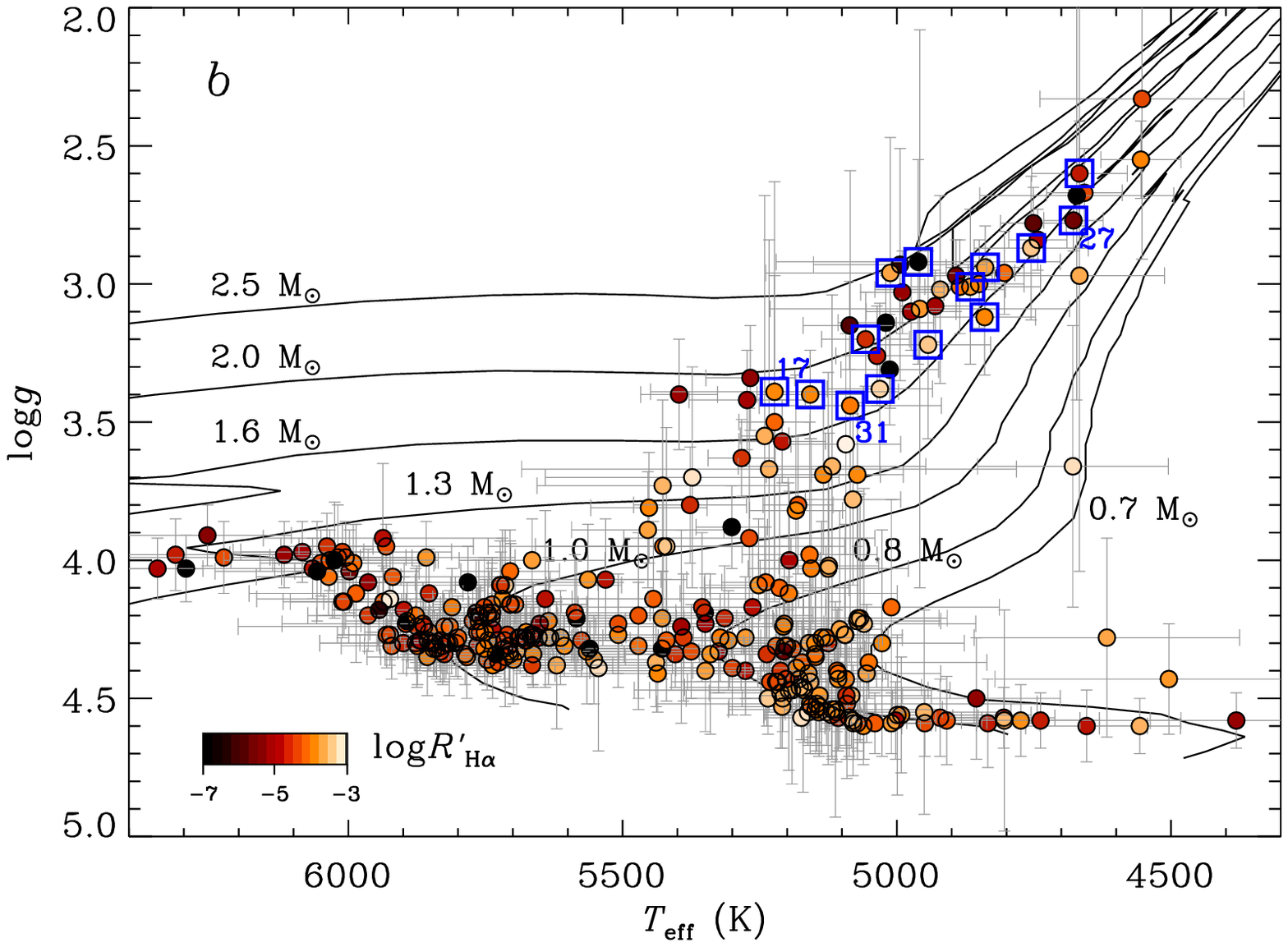}		%logg_teff_gia_halpha.eps}\\
\caption{$T_{\rm eff}$--$\log g$ diagram for all the single stars and SB1 systems. The symbols are color coded by $W_{\rm Li}$ 
({\it upper panel}) and $R'_{\rm H\alpha}$ ({\it lower panel}).
The  black lines are post-MS evolutionary tracks from \citet{Girardi2000} with a metallicity Z=0.019.
In the {\it lower panel}, the lithium-rich giant candidates listed in Table~\ref{Tab:Li-giants} are enclosed in open squares. 
The two very lithium rich giants (\#17 and \#27) and the youngest PMS star (\#31, see Sect.\,\ref{subsec:Age}) are also marked.}
\label{Fig:LoggTeff}
\end{figure}

The $T_{\rm eff}$--$\log g$ diagram of the single and SB1 sources is shown in Fig.~\ref{Fig:LoggTeff}, where the symbols are 
color coded according to the value of $W_{\rm Li}$ (Fig.~\ref{Fig:LoggTeff}a) and $R'_{\rm H\alpha}$ (Fig.~\ref{Fig:LoggTeff}b). 

\section{Results}
\label{Sec:Results}

\subsection{Binaries and multiple systems}
\label{subsec:multiple}

Close binaries are expected to give a large contribution to X-ray selected or chromospherically-active stellar samples
\citep[see, e.g.,][]{Brandner1996}.
In fact, the tidal synchronization of the rotation of the components with the orbital period can greatly intensify the magnetic field, 
and hence the level of activity. Therefore, stellar samples like the {\it RasTyc} one allow us to detect close binaries, which can be 
followed up for the study of their properties. At the same time, they behave as contaminants when looking for young stars.   
Based on the CCF shape and the variation of the peak centroid, we found 12 SB3 (2.7\,\%), 114 SB2 (25.7\,\%), and 38 SB1 (8.6\,\%) systems. 
We note that these numbers, especially that of SB1, must be considered as lower limits, since stars with only one observation or even 
with more spectra could be indeed SB1 systems observed in similar configurations or SB2 close to the conjunctions. 
Indeed, eight SB1 systems were discovered by the comparison of our single RV value with that reported in the SACY catalog (see Sect.~\ref{subsec:SACY}).
Altogether, the close binary systems account for more 
than 37\,\% of the whole sample. A similar contamination (31\,\%) was estimated for the bright sample in Paper\,I.

\subsection{Hertzsprung-Russell diagram}
\label{subsec:HR}

\begin{figure}  
\begin{center}
\hspace{-.5cm}
\includegraphics[width=9.2cm]{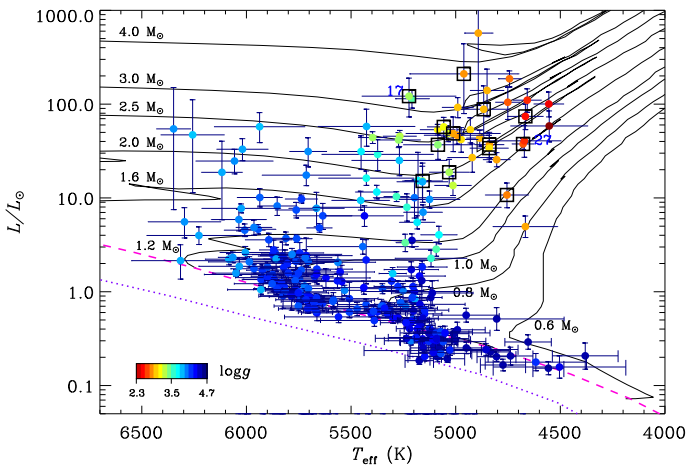}		%hr_rastyc_faint_new.eps}
\caption{Hertzsprung-Russell diagram of the single and SB1 sources with known parallaxes. The symbols are color coded by \logg.
The evolutionary tracks of \citet{Girardi2000} are shown as solid lines with the labels representing their masses. 
The zero-age main sequence (ZAMS) with solar metallicity ($Z=0.019$) and with $Z=0.001$ by the same authors 
are also shown with a dashed and a dotted line, respectively. The lithium-rich giant candidates listed in Table~\ref{Tab:Li-giants}
are enclosed in open squares. The two giants with the highest Li abundance are labeled as in Table~\ref{Tab:LiRich}.}
\label{Fig:HR}
 \end{center}
\end{figure}

In Fig.~\ref{Fig:HR}, we report the position of the single and SB1 targets with known parallaxes in the Hertzsprung-Russell (HR) diagram, which are 263 out 
of a total of 328 such targets. We used the \teff\  values derived with ROTFIT (Table~\ref{Tab:APs}) and the $V$ magnitudes from the TYCHO catalog that are 
listed in Table~\ref{Tab:RV}.
For most stars (243) the parallaxes are retrieved from the TGAS catalog in the first \textit{Gaia} data release \citep{GaiaDR1}, while for 22 objects with 
no entry in that catalog we adopted the values reported by \citet{vanLeeuwen2007}.
The $V$ magnitudes were corrected  for the interstellar extinction as in Paper~I, i.e. assuming a mean extinction of $A_V$\,=\,1.7\,mag/kpc on the galactic plane 
($|b|<5\degr$) and 0.7 mag/kpc out of the plane. 
To convert the $V$ magnitudes to bolometric magnitudes we used the bolometric correction ($BC$) derived by interpolating the relation \teff--$BC$ of 
\citet{PecautMamajek2013}.
 In the same figure we overplot the post-main sequence evolutionary tracks by \citet{Girardi2000}. 

As apparent in Fig.~\ref{Fig:HR}, the values of \logg\ found from the analysis of the spectra are fully consistent with the evolutionary status
of the targets, supporting the reliability of the atmospheric parameters derived by ROTFIT with these spectra. 

\subsection{Lithium abundance and age}	
\label{subsec:Age}

\begin{figure}[]
\includegraphics[width=9.0cm]{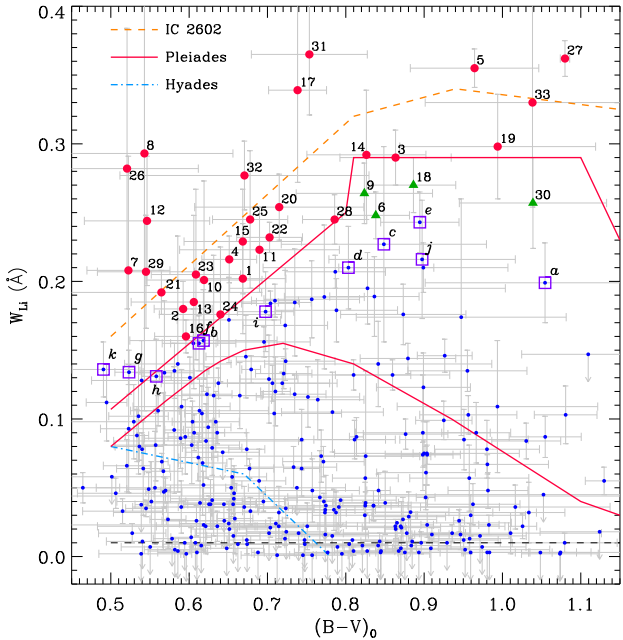}		%Litio_faint_ALL_err.eps}
\caption{Equivalent width of the \ion{Li}{i}\,$\lambda$6707.8 line ($W_{\rm Li}$) plotted as a function of the dereddened  Johnson $(B-V)_0$ color index for 
the single or SB1 stars in our sample (Table\,\ref{Tab:APs}). The lines display the upper boundaries 
for Hyades (blue dash-dotted), Pleiades (red solid for both lower and upper boundary) and IC~2602 (orange dashed) clusters. 
The black dashed line marks the lower limit for positive detection of the  \ion{Li}{i}\,$\lambda$6707.8 line estimated by us (10\,m\AA). 
Twenty-nine stars (red big dots) show lithium equivalent width in excess with respect to the lithium-richest Pleiades stars and ten of them fall above the upper 
envelope of the PMS cluster IC~2602. Four stars (green triangles) have $W_{\rm Li}$ just below the Pleiades upper envelope. 
The open squares enclose other eleven very young objects selected on the basis of their lithium abundance (see text and Fig.~\ref{Fig:NLi}). All 
the lithium-rich stars are labeled according to Table~\ref{Tab:LiRich}.
}
\label{Fig:Li}
\end{figure}

Lithium is burned at relatively low temperatures in stellar interiors ($\sim 2.5\,\times\,10^6$\,K). As a consequence, it is progressively depleted from 
the stellar atmospheres of late-type stars when mixing mechanisms pull it deeply in their convective layers. 
Thus, its abundance can be used as an empirical indicator of age for stars cooler than about 6500\,K. A simple and effective way to get an 
age estimate is a diagram showing the equivalent width of lithium as a function of the color index $(B-V)_0$ together with the upper envelopes of 
young OCs that serve as boundaries to delimit the different age classes (Fig.\,\ref{Fig:Li}). 
Although the stars are not very distant from the Sun, their colors are affected by reddening, which has been calculated from the magnitude extinction,
$A_V$ (see Sect.~\ref{subsec:HR}) according to standard extinction law $E(B-V)=A_V/3.1$.

\begin{figure}[]
\includegraphics[width=9.0cm]{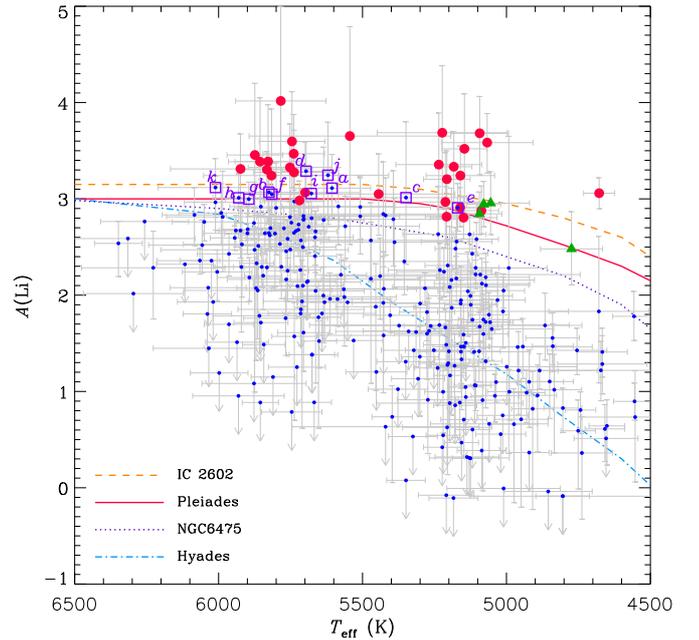}		%NLi_faint_all.eps}
\caption{Lithium abundance as a function of $T_{\rm eff}$ for the stars in Table\,\ref{Tab:APs}. 
The upper envelopes of $A$(Li) for IC~2602, Pleiades,  NGC\,6475 ($age\approx$\,300 Myr), and Hyades clusters adapted from \citet{Sestito2005} are overplotted.
The meaning of the symbols is the same as in Fig.\,\ref{Fig:Li}. 
Other eleven stars with high lithium content are found thanks to this diagram (open squares) and are labeled with letters from {\it a} to {\it k},
as in Table~\ref{Tab:LiRich}.
}
\label{Fig:NLi}
\end{figure}

We adopted, as in Paper~I, the upper envelopes of the following clusters: the Hyades \citep{Soderblom1990},
the Pleiades \citep{Soderblom1993c,Neuhauser1997}, and IC~2602 \citep{Montes2001}, whose ages are of about 650 \citep{White2007AJ133},  125 \citep{White2007AJ133}, and 
30~Myr \citep{Stauffer1997}, respectively. 
As shown by \citet{Soderblom1993b}, the lower envelope of the Pleiades behaves as an upper boundary for the members of the Ursa Major (UMa) cluster 
($age\approx$\,300 Myr).
Therefore, we considered as pre-main sequence (PMS) or very young star candidates (\textit{PMS-like}) those stars located above the Pleiades upper envelope (big dots 
in Fig.\,\ref{Fig:Li}) or very close to it (triangles) and divided the remaining stars into \textit{Pleiades-like}, \textit{UMa-like}, \textit{Hyades-like}, 
and \textit{Old} classes according to their position in the diagram, the \textit{Old} ones being those with a $W_{\rm Li}<10$\,m\AA.
We remark that our selection criterion is not conservative, but it is inclusive; in fact we are not taking into account the errors on $W_{\rm Li}$ and 
$(B-V)_0$. On the other hand, the upper envelopes of OCs are not ``knife edges'' that can sharply separate classes of objects of different ages. 
That is why we used the suffix \textit{like} in our classes. This also means, for example, that the \textit{PMS-like} class can include stars that are indeed closer to the 
Pleiades age.

As a further test of our method to select young stars, we calculated the lithium abundance, $A$(Li), from our values of 
\teff, \logg, and $W_{\rm Li}$ by interpolating the curves of growth of \citet{Soderblom1993c}.
We adopted this calibration in order to treat our data homogeneously to \citet{Sestito2005} for the stars in 22 galactic OCs.
In Fig.\,\ref{Fig:NLi} we display the lithium abundance as a function of $T_{\rm eff}$ along with the upper envelopes of the distributions of some young 
OCs shown by \citet{Sestito2005}. We note that, contrary to the color indices, the effective temperature does not need to be corrected for
interstellar reddening, avoiding to introduce further uncertainties.
Apart from the large errors of $A$(Li), which take into account both the $T_{\rm eff}$ and $W_{\rm Li}$ errors\footnote{We did not consider the errors of 
\logg\  for the evaluation of $A$(Li) errors, because an error $\sigma_{\log g}$\,=\,0.5 dex translates into an uncertainty of only a 0.1 or 0.2 dex, at most, in $A$(Li),
in the \teff\ range of our targets. }, Figure\,\ref{Fig:NLi} clearly
shows that all the 33 \textit{PMS-like} stars already selected from the $W_{\rm Li}$ diagram (Fig.\,\ref{Fig:Li}), with the exception of only two sources, 
are located on or above the Pleiades upper envelope. 
The two stars falling just below the Pleiades upper envelope are RasTyc\,0316+5638 (\#11) and RasTyc\,2039+2644 (\#22). However, considering the errors, they
can be still included in the \textit{PMS-like} class.  
Moreover, the four stars that lie just below the upper envelope of the Pleiades at $0.8 \leq (B-V)_0 \leq 1.1$\,mag in Fig.\,1 (green triangles) are located above or 
exactly superimposed to it in the \teff-$A$(Li) diagram (Fig.\,\ref{Fig:NLi}).
In addition, eleven other stars that were not selected on the basis of $W_{\rm Li}$ show a high lithium abundance, above the Pleiades $A$(Li) upper envelope. 
We have highlighted them with open squares and flagged with letters from {\it a} to {\it k} both in Figs.\,\ref{Fig:Li} and \ref{Fig:NLi} as well as in Table\,\ref{Tab:LiRich}.
The inclusion of these targets leads the number of very young star candidates to 44. 

In the end, the percentage of \textit{PMS-like}, \textit{Pleiades-like}, \textit{UMa-like}, \textit{Hyades-like}, and \textit{Old} stars is about 14\,\%, 13\,\%, 
34\,\%, 19\,\%, and 20\,\%, respectively, while for the bright sample (Paper\,I) they were about 3\%, 7\%, 39\%, 19\%, and 32\%. 
Considering the population of very young stars as defined in Paper\,I, i.e. \textit{PMS-like}, \textit{Pleiades-like} and \textit{UMa-like} stars taken altogether, 
their percentage in the faint sample increases from 49\,\% to 61\,\%, while the fraction of the older population decreases of about 10\,\% accordingly. 
In particular, the fraction of the youngest stars (\textit{PMS-like} and \textit{Pleiades-like}) is almost three times higher in the faint sample than in the bright one.

In an X-ray flux-limited and optically magnitude-limited sample, the fraction of the various populations depends on the X-ray and optical horizons, i.e. the 
maximum distance at which we can detect stars in X-ray and at optical wavelengths, respectively. On the average, when the X-ray horizon is farther than the optical one
and the last is increased by changing the limiting magnitude of the sample (fainter stars detected at larger distance), the fraction of young stars increases as well.
This can be easily understood because young stars are brighter in X-ray and thus also detectable at larger distances compared to older stars. However, this applies only 
in general terms, because of the role of other stellar parameters in the X-ray luminosity, which depends, inter alia, on both age and mass, making the details more complicated. 
X-ray population models are absolutely needed for a meaningful interpretation of the observed properties (sky positions, distance distribution, magnitudes, colors, 
proper motions, etc.) of the {\it RasTyc} sample. This is out of the scope of the present work and will be the subject of a dedicated paper.
We also note that the differences may also arise from the higher incidence of Gould Belt sources in the optically faint sample that reaches beyond the space 
volume sampled in Paper\,I.

\setlength{\tabcolsep}{4pt}

\begin{table*}[t]
\caption{Very young stars and PMS candidates selected from the lithium content.}
\small
\begin{tabular}{llccccccccc}
\hline
\# Name                 & {\it RasTyc}       & $\alpha$ (2000) &  $\delta$ (2000)  & $C_{\rm rate}^{\sharp}$ & Sp. Type & $T_{\rm eff}$ & $\log g$ & [Fe/H] & $A$(Li) & $\log(R'_{H\alpha})$	  \\ 
                        &		   & h m s	     &  $\circ$ ' ''	 & (ct/s)  &	      &   (K)  &  (dex)   & (dex)  &      &  \\ 
\hline
\#1  BD+78 853$^{\dag}$           & 0000+7940 &  00 00 41.14 &  +79 40 39.9  &  0.106  &   G2V    &   5738 &   4.19   &  -0.03 &   3.28    & -4.43 \\ 
\#2 TYC 4496-780-1$^{\dag,\ddag}$ & 0013+7702 &  00 13 40.52 &  +77 02 10.9  &  0.102  &   G0V    &   5924 &   4.14   &  -0.03 &   3.31    & -3.33 \\	
\#3  TYC 4500-1478-1$^{\dag}$     & 0038+7903 &  00 38 06.03 &  +79 03 20.7  &  0.133  &   K1V    &   5160 &   4.30   &  -0.06 &   3.24    & -3.87 \\    
\#4  BD+78 19$^{\dag,\star}$      & 0039+7905 &  00 39 40.13 &  +79 05 30.8  &  0.077  &   G5V    &   5444 &   4.14   &  -0.11 &   3.05    & -4.57 \\ 	 
\#5  TYC 3266-1767-1              & 0046+4808 &  00 46 53.09 &  +48 08 45.2  &  0.278  &   K2V    &   5067 &   4.21   &  -0.09 &   3.58    & -4.02 \\
\#6  BD+49 646                    & 0222+5033 &  02 22 33.82 &  +50 33 37.8  &  0.491  &   K2V    &   5094 &   4.27   &  -0.12 &   2.84    & -3.90 \\
\#7  TYC 3695-2260-1              & 0230+5656 &  02 30 44.81 &  +56 56 13.0  &  0.051  &   G1V    &   5874 &   4.31   &  -0.03 &   3.46    & -4.47 \\
\#8  TYC 2338-35-1                & 0252+3728 &  02 52 24.71 &  +37 28 52.0  &  0.048  &   G1.5V  &   5784 &   4.35   &   0.02 &   4.02    & -4.44 \\
\#9  TYC 4321-507-1               & 0300+7225 &  03 00 14.67 &  +72 25 41.4  &  0.084  &   K2V    &   5079 &   4.59   &  -0.01 &   2.94    & -4.00 \\
\#10 Cl Melotte 20 94             & 0311+4810 &  03 11 16.82 &  +48 10 36.9  &  0.092  &   G1.5V  &   5856 &   4.35   &   0.02 &   3.39    & -4.24 \\
\#11 TYC 3710-406-1               & 0316+5638 &  03 16 28.11 &  +56 38 58.1  &  0.084  &   K1V    &   5208 &   4.50   &  -0.05 &   2.81    & -4.18 \\
\#12 TYC 3715-195-1               & 0323+5843 &  03 23 07.08 &  +58 43 07.4  &  0.148  &   K1V    &   5212 &   4.48   &  -0.07 &   2.97    & -3.92 \\
\#13 Cl Melotte 20 935            & 0331+4859 &  03 31 28.98 &  +48 59 28.6  &  0.081  &   G1.5V  &   5816 &   4.30   &  -0.10 &   3.24    & -4.05 \\
\#14 TYC 2876-1944-1              & 0359+4404 &  03 59 16.70 &  +44 04 17.1  &  0.264  &   K1V    &   5235 &   4.50   &  -0.06 &   3,36    & -3.62 \\
\#15 TYC 3375-720-1               & 0616+4516 &  06 16 46.95 &  +45 16 03.1  &  0.097  &   G2V    &   5739 &   4.21   &   0.11 &   3.47    & -3.83 \\
\#16 TYC 3764-338-1               & 0621+5415 &  06 21 56.92 &  +54 15 49.0  &  0.426  &   G2V    &   5719 &   4.14   &  -0.08 &   2.98    & -4.15 \\
\#17 HD 234808$^{\rm *}$          & 1908+5018 &  19 08 14.03 &  +50 18 49.6  &  0.045  & G8III-IV &   5223 &   3.39   &  -0.37 &   3.69    & -4.20 \\
\#18 KIC 8429280$^{\bullet}$      & 1925+4429 &  19 25 01.98 &  +44 29 50.7  &  0.254  &    K2V   &   5055 &   4.41   &  -0.02 &   2.96    & -3.96 \\
\#19 BD-03 4778 	          & 2004$-$0239 &  20 04 49.35 & $-02$ 39 19.7 &  0.175  &  K1V   &   5183 &   4.39   &  -0.02 &   3.33    & -4.26 \\
\#20 HD 332091                    & 2016+3106 &  20 16 57.83 &  +31 06 55.6  &  0.167  &    K1V   &   5087 &   4.57   &   0.01 &   2.88    & -3.72 \\
\#21 TYC 2694-1627-1              & 2036+3456 &  20 36 16.87 &  +34 56 46.1  &  0.037  &    G1V   &   5832 &   4.33   &  -0.01 &   3.30    & -4.33 \\
\#22 TYC 2178-1225-1              & 2039+2644 &  20 39 40.81 &  +26 44 48.4  &  0.227  &    K1V   &   5149 &   4.30   &  -0.08 &   2.80    & -4.19 \\
\#23 BD+68 1182                   & 2106+6906 &  21 06 21.74 &  +69 06 41.0  &  0.114  &   G1.5V  &   5828 &   4.31   &  -0.01 &   3.39    & -4.31 \\
\#24 TYC 3589-3858-1              & 2120+4636 &  21 20 55.42 &  +46 36 12.4  &  0.142  &   G1.5V  &   5699 &   4.36   &   0.04 &   3.07    & -4.29 \\	    
\#25 BD+76 857a                   & 2223+7741 &  22 23 18.87 &  +77 41 57.7  &  0.114  &   G1.5V  &   5745 &   4.29   &  -0.04 &   3.60    & -4.28 \\
\#26 TYC 1154-1546-1              & 2233+1040 &  22 33 00.37 &  +10 40 34.3  &  0.053$^{\diamond}$ & G1.5V & 5544 & 4.39 &   0.03 &   3.65    & -3.44 \\
\#27 HD 214995$^{\rm **}$         & 2241+1430 &  22 41 57.40 &  +14 30 59.2  &  0.094  &   K0III  &   4678 &   2.77   &  -0.04 &   3.61    & -6.11 \\
\#28 BD+17 4799                   & 2244+1754 &  22 44 41.49 &  +17 54 19.0  &  0.835  &   K1V    &   5161 &   4.41   &  -0.02 &   2.91    & -4.14 \\
\#29 TYC 3992-349-1               & 2246+5749 &  22 46 13.19 &  +57 49 58.0  &  0.034  &   G2V    &   5752 &   4.32   &  -0.05 &   3.33    & -4.31 \\
\#30 TYC 2751-9-1                 & 2307+3150 &  23 07 24.83 &  +31 50 14.1  &  0.462  &   K4V    &   4774 &   4.58   &   0.05 &   2.48    & -4.15 \\
\#31 V395 Cep                     & 2320+7414 &  23 20 52.07 &  +74 14 07.1  &  0.035  &   K0IV   &   5093 &   3.58   &  -0.02 &   3.68    & -3.27 \\
\#32 TYC 584-343-1                & 2321+0721 &  23 21 56.36 &  +07 21 33.0  &  0.067  &   K0V    &   5207 &   4.24   &   0.01 &   3.20    & -4.25 \\
\#33 TYC 4606-740-1$^{\odot}$     & 2351+7739 &  23 51 17.29 &  +77 39 35.3  &  0.169  &   K1V    &   5146 &   4.52   &  -0.07 &   3.51    & -3.84 \\
\noalign{\medskip}
{\it a} TYC 2282-1396-1$^{\oplus}$& 0106+3306 &  01 06 18.71 &  +33 06 01.9  &  0.161  &   G4V    &   5606 &   4.31   &  -0.14 &   3.11    & -4.16 \\
{\it b} BD+42 636	          & 0249+4255 &  02 49 54.69 &  +42 55 27.1  &  0.036  &   G1.5V  &   5825 &   4.24   &  -0.02 &   3.07    & -4.52 \\
{\it c} TYC 3325-98-1	          & 0344+5043 &  03 44 34.50 &  +50 43 47.5  &  0.074  &   K0V    &   5349 &   4.40   &  -0.06 &   3.01    & -3.99 \\
{\it d} TYC 2949-780-1            & 0646+4147 &  06 46 46.74 &  +41 47 12.2  &  0.043  &   G2V    &   5696 &   4.16   &   0.13 &   3.29    & -4.60 \\
{\it e} TYC 2087-1742-1           & 1731+2815 &  17 31 03.33 &  +28 15 06.1  &  0.305  &   K1V    &   5169 &   4.47   &  -0.02 &   2.91    & -3.98 \\
{\it f}  KIC 7985370$^{\otimes}$  & 1956+4345 &  19 56 59.73 &  +43 45 08.2  &  0.036  &   G1.5V  &   5815 &   4.24   &  -0.05 &   3.05    & -4.50 \\
{\it g} BD+35 4198	          & 2038+3546 &  20 38 17.71 &  +35 46 33.3  &  0.058  &   G1.5V  &   5895 &   4.23   &   0.01 &   3.00    & -4.86 \\
{\it h} BD+39 4490	          & 2114+3941 &  21 14 55.24 &  +39 41 11.9  &  0.099  &   G1.5V  &   5930 &   4.27   &  -0.02 &   3.01    & -4.75 \\
{\it i} TYC 3198-1809-1           & 2203+3809 &  22 03 49.83 &  +38 09 42.9  &  0.033  &   G5IV-V &   5678 &   4.28   &   0.04 &   3.06    & -4.95 \\
{\it j} TYC 576-1220-1            & 2308+0000 &  23 08 50.46 &  +00 00 52.8  &  0.104  &   G2V    &   5620 &   4.38   &  -0.00 &   3.25    & -3.91 \\
{\it k} TYC 4283-219-1            & 2324+6215 &  23 24 40.37 &  +62 15 51.1  &  0.037  &   F9V    &   6011 &   3.97   &  -0.21 &   3.12    & -4.84 \\
\hline
\end{tabular}
\begin{list}{}{}									
\item[$^{\sharp}$] X-ray count rate from the second ROSAT all-sky survey source catalog \citep[2RXS,][]{Boller2016}.  
\item[$^{\dag}$] PMS star in Cepheus discovered by \citet{Guillout2010}.
\item[$^{\ddag}$] Close visual pair \citep[$\rho=1\farcs$41, $\Delta V\simeq 2.2$\,mag,][]{Fabricius2002}. Parameters of the brighter component.  
\item[$^{\star}$] SB2 system composed of two nearly identical stars. Parameters derived from spectra taken close to the conjunctions.
\item[$^{\bullet}$] Very young star in the {\it Kepler} field discovered by \citet{Frasca2011}.
\item[$^{\diamond}$] X-ray count rate from the ROSAT All-Sky Bright Source Catalog \citep[1RXS,][]{Voges1999}.  
\item[$^{\odot}$] Close visual pair \citep[$\rho=0\farcs$82, $\Delta V\simeq 0.15$\,mag,][]{Fabricius2002}. 
\item[$^{\oplus}$] Close visual pair \citep[$\rho=0\farcs$85, $\Delta V\simeq 0.29$\,mag,][]{Fabricius2002}. 
\item[$^{\otimes}$] Very young star in the {\it Kepler} field discovered by \citet{Froehlich2012}.
\item[$^{\rm *}$]  Discarded as PMS candidate on the basis of spectral classification and the position in the HR diagram. Likely lithium-rich giant.
\item[$^{\rm **}$] Discarded as PMS candidate on the basis of spectral classification and the very low H$\alpha$ flux. Lithium-rich giant.  
\end{list}
\normalsize
\label{Tab:LiRich}
\end{table*}

\begin{figure}  
\begin{center}
\hspace{-.5cm}
\includegraphics[width=9.2cm]{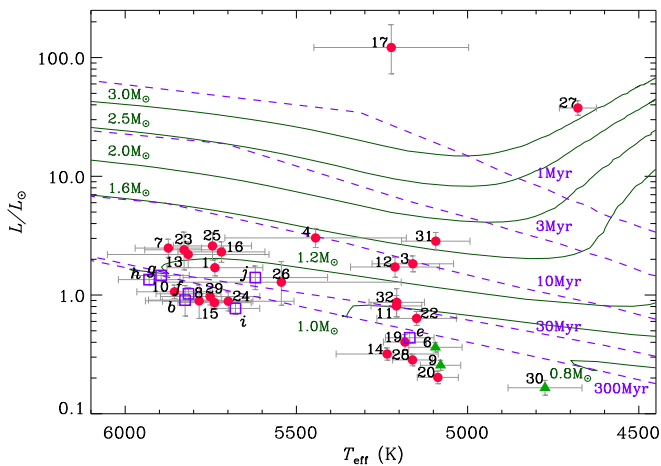}		%HR_Rastyc_Siess_new.eps}
\caption{Hertzsprung-Russell diagram of the {\it PMS-like} sources with known parallaxes. The symbols are as in Fig.~\ref{Fig:Li}.
The pre-main sequence evolutionary tracks of \citet{Siess00} are shown as solid lines with the labels representing their masses. 
The isochrones at ages of 1, 3, 10, 30, and 300 Myr are shown with dashed lines. The stars labeled as \#17 and \#27 lie above
the 1-Myr isochrone and are likely lithium-rich giants. 
}
\label{Fig:HR_PMS}
 \end{center}
\end{figure}

Figure~\ref{Fig:HR_PMS} shows the HR diagram for the {\it PMS-like} objects along with the PMS evolutionary tracks and isochrones by Siess et al. (2000).
We note that stars \#17 and \#27 lie above the isochrone at 1\,Myr and are likely lithium-rich giants, as also shown in Fig.~\ref{Fig:HR} where they lie 
close to the post-main sequence tracks for stellar models with 3.0\,$M_{\sun}$ and 1.5\,$M_{\sun}$, respectively. We have therefore disregarded them as PMS-like candidates
 (see also Table\,\ref{Tab:LiRich} and Appendix\,\ref{Sec:notes}). 

\subsection{Evolved stars}
\label{subsec:Giants}

Besides the two stars mentioned above, a considerable number of giants and subgiants is present among single stars and SB1 systems, 
as apparent from the \teff--\logg\ diagram (Fig.~\ref{Fig:LoggTeff}) and the HR diagram (Fig.~\ref{Fig:HR}). Considering as `evolved stars' 
those ones with \logg$\,\leq 3.5$, we got a sample of 39 out of 328 sources, i.e. 12\,\% of the single+SB1 sources.
Half of them also exhibit a remarkable chromospheric emission ($\log R'_{\rm H\alpha}\geqsim-5$, see Fig.~\ref{Fig:LoggTeff}b), which is normally paired 
to a rapid rotation (\vsini\,$\geqsim$\,9\,\kms). It is worth noticing that all the SB1 and possible SB2 systems (10 in total) among the  39 evolved stars 
also display a high chromospheric activity level. 
The remaining active giants could be either single stars that are rotating quite fast as a result of a particular evolutionary path 
or spectroscopic binaries that are still undetected. 

\setlength{\tabcolsep}{5pt}
\begin{table*}[ht]
\caption{Lithium-rich giant candidates.}
%\scriptsize
\begin{tabular}{llcccrccr}
\hline
 Name                 & {\it RasTyc} & Sp. Type & $T_{\rm eff}$ & $\log g$ & [Fe/H] & $A$(Li) & $\log(R'_{H\alpha})$ & \vsini \\ 
                      &              &          &   (K)         &  (dex)   & (dex)  &	      &        & \scriptsize{(\kms)}  \\ 
\hline
TYC 3676-2444-1                  & 0106+5729 &     G8III  &   5158 &   3.40   &  $-$0.05 &   1.46  &  $-$4.22  &  9.1 \\
TYC 4319-714-1$^{\ast}$          & 0222+7204 &  K0III-IV  &   4866 &   3.01   &  $-$0.09 &   1.69  &  $-$4.06  & 22.3 \\
TYC 4364-1262-1$^{\ast,\dag}$    & 0712+7021 &      K1IV  &   5012 &   2.96   &  $-$0.03 &   1.83  &  $-$3.97  & 10.3 \\
BD+65~601                        & 0755+6509 &     K0IV   &   5085 &   3.44   &  $-$0.07 &   2.12  &  $-$4.35  &  5.0 \\
TYC 3501-626-1$^{\ast,\diamond}$ & 1702+4713 &      K0IV  &   5031 &   3.38   &  $-$0.05 &   1.95  &  $-$3.55  & 15.0 \\
BD+06\,3372a$^{\ast}$            & 1714+0623 &     G9III  &   4839 &   2.94   &  $-$0.10 &   1.47  &  $-$3.96  & 13.0 \\
BD+02\,3384$^{\ast}$             & 1741+0228 &     G9III  &   4840 &   3.12   &  $-$0.12 &   1.56  &  $-$4.23  & 29.3 \\
HD 234808                        & 1908+5018 &  G8III-IV  &   5223 &   3.39   &  $-$0.37 &   3.69  &  $-$4.20  &  8.1 \\
BD-04\,5118$^{\ast}$             & 2025$-$0429 & K0III-IV &   4943 &   3.22   &  $-$0.05 &   1.47  &  $-$3.63  &  9.0 \\
BD+48\,3149$^\bullet$            & 2030+4852 &     K1III  &   4667 &   2.60   &  $-$0.06 &   1.41  &  $-$5.20  & 54.1 \\
HD 205173$^{\ast}$               & 2132+3604 &      K0IV  &   5057 &   3.20   &     0.02 &   1.72  &  $-$5.01  &  4.1 \\
HD 214995                        & 2241+1430 &	   K0III  &   4678 &   2.77   &  $-$0.04 &   3.61  &  $-$6.11  &  5.3 \\
HD 220338$^{\ast,\dag,\bullet}$  & 2323$-$0635 & K1III-IV &   4755 &   2.87   &  $-$0.10 &   1.43  &  $-$3.66  & 26.8 \\
TYC 3638-993-1                   & 2348+4615 &     G8III  &   4961 &   2.92   &  $-$0.03 &   1.46  &    \dots  &  3.4 \\
\hline
\end{tabular}
\begin{list}{}{}								       
\item[$^{\ast}$] Rotationally variable star \citep{aavso,Kiraga2013}.  
\item[$^{\dag}$] SB1.  
\item[$^{\diamond}$] Small-amplitude secondary peak in the CCF. Likely SB2 with a faint component.  
\item[$^{\bullet}$] Likely still undergoing the lithium dilution process. 
\end{list}
\label{Tab:Li-giants}
\end{table*}

Fourteen out of the 39 evolved stars are lithium-rich giant candidates ($A$(Li)\,$>$\,1.4, e.g., \citealt{Charbonnel2000}). 	
Their properties are summarized in Table~\ref{Tab:Li-giants}. Among them, only the bright source HD~214995 is already known as a Li-rich 
giant at the position of the red-giant bump \citep{Charbonnel2000}. The other stars, excepting perhaps BD+48\,3149 and HD~220338, are 
hotter than the boundary corresponding to the deepest penetration of the convective zone during the first dredge up (cf. Fig.~1 of \citealt{Charbonnel2000}), 
which is around 4400--4600\,K for the stars in our mass range. Therefore, these stars have started but not yet completed the standard first dredge-up dilution 
and should not be considered as abnormally lithium rich. They would exhibit a lithium abundance close to its initial value. 
As suggested by \citet{FekelBala1993}, angular momentum may be dredged-up from the
stellar interior along with Li-rich material during this phase, giving rise to a faster rotation and, consequently, to a higher activity level.
This can explain the higher incidence (more than 35\,\%) of lithium-rich giant candidates in our X-ray selected sample compared to the typical fraction of 
1\,\% found in several spectroscopic surveys of giant stars \citep[e.g.,][and references therein]{deLaReza2012}. This is also in line with 
the high percentage (about 50\,\%) of Li-rich candidates among fast rotating K giants (with \vsini$\ge$\,8\,\kms) reported by \citet{Drake2002}. It is 
worth noticing that most of the stars listed in Table~\ref{Tab:Li-giants} are indeed rotating faster than 8\,\kms.

The source TYC~3501-626-1 could be an SB2 system with a faint component, on the basis of the CCF. It has a high chromospheric activity level $\log R'_{\rm H\alpha}=-3.55$ 
and has been reported as a rotationally variable source ($P_{\rm rot}\simeq$\,7.93 days) by \citet{Kiraga2013}. The presence of the 
faint secondary component could have affected the determination of the atmospheric parameters of this star; therefore we cannot exclude that it is a 
binary system containing young Li-rich stars. 
In addition, seven of the stars in Table\,\ref{Tab:Li-giants}, namely TYC~4319-714-1, TYC~4364-1262-1, BD+06\,3372a, BD+02\,3384, BD-04\,5118, HD~205173, and HD~220338, 
are also classified as rotationally variable sources by \citet{Kiraga2013} and  \citet{aavso}, who report photometric periods of 34.5, 35.6, 6.5, 20.4, 39.8, 101.8,
and 15.4 days, respectively. With the exception of TYC\,4364-1262-1 and HD~220338, which have been detected by us as SB1 systems (see Tables\,\ref{Tab:RV} and \ref{Tab:APs}), 
these stars have been observed only once and are therefore preliminarily classified as single stars. We cannot exclude that they are single-lined binaries similar to some 
long-period RS~CVn systems, which also show a fast rotation and lithium enrichment \citep[e.g.,][]{Pallavicini1992,Barrado1998}. \\

\subsection{Comparison with the SACY survey}
\label{subsec:SACY}

As anticipated in Sect.~\ref{Sec:Data}, a similar high-resolution spectroscopic survey of optical counterparts of Southern X-ray sources (SACY) 
was carried out by \citet{Torres2006}. They cross-correlated the RASS catalog with Hipparcos and Tycho-2 catalogs to build their sample of stellar
X-ray sources and considered only stars with $(B-V)\geq 0.6$, as we have done in Paper\,I and in the present work. Unlike us, they observed stars at all right ascensions, 
but excluded all Hipparcos stars having $M_V <2.0$\,mag to reduce the sample contamination by evolved stars as much as possible. Moreover, they excluded all the already 
known giants and active binaries of the RS~CVn and W~UMa classes.
This explains the higher incidence of such objects in our survey with respect to SACY.

As we have observed a few stars with a low or negative declination (mainly with SARG@TNG), we have some sources in common with the SACY survey. 
In Table~\ref{Tab:SACY} we compare the parameters derived by us for the single stars and SB1 systems with those in the SACY survey. 
The SACY catalog \citep{Torres2006} lists the spectral type, RV, $W_{\rm Li}$ for 16 of such sources and \vsini\  for eight of them.
For three other stars  \citet{DaSilva2009} report $W_{\rm Li}$, \vsini, and  \teff\ (instead of spectral type), while \citet{Elliott2014} quote their RV values.   
As shown in Table~\ref{Tab:SACY}, there is a general good agreement between  \teff\ values (within 150\,K), spectral types (within 1--2 spectral subclasses),  
\vsini\ (within 10\,\%), and $W_{\rm Li}$ (within 10\,\%) with only three exceptions. 
For RasTyc\,2155$-$0947 there is a large discrepancy between spectral types. It is an eclipsing binary system according to the ASAS
data analysed by \citet{Kiraga2012}. The presence of a secondary component could have affected the spectra giving rise to this large discrepancy.  
We found a very different spectral type also for RasTyc\,2256+0235. It is classified by \citet{Kiraga2012} as a rotationally variable star with a very short period, 
$P_{\rm rot}=0.8389$\,days.  Different \teff\ determinations are available in the literature, including that of 5316\,K \citep{MunozBermejo2013} based on the same ELODIE 
spectrum taken by us. This determination is midway between \teff$\simeq4900$\,K corresponding to the SACY K1\,IV spectral type 
\citep[according to the relation of][]{PecautMamajek2013} and our value of 5665\,K. The distortion of spectral lines and CCF that is likely due to starspots can be responsible
 for that discrepancy.
RasTyc\,2352$-$1143 shows instead a remarkable difference of \vsini\ values, which is still larger if compared to the value of 8.2\,\kms\ reported by \citet{Jenkins2011}, who also 
quote a value of RV$=-13.4\pm 1.9$\,\kms. This star has been recently discovered to be a close visual binary with a separation of only 0$\farcs$22 \citep{Horch2015}. 

As regards the RV, we note that ten objects have very discrepant values and should be considered as SB1 systems. For eight of them we could collect only one
spectrum; therefore, without literature data, we would have preliminarly classified them as single stars.

\begin{table*}[ht]
\caption{Comparison with SACY parameters for single stars and SB1 systems.}
\begin{tabular}{llcccccccc}
\hline
 {\it RasTyc} & \multicolumn{2}{c}{Sp. Type/\teff} & \multicolumn{2}{c}{\vsini} & \multicolumn{2}{c}{$W_{\rm Li}$}  & \multicolumn{2}{c}{$RV$}  & Notes \\ 
              & \multicolumn{2}{c}{ ~~~~/(K)}    & \multicolumn{2}{c}{(\kms)}   & \multicolumn{2}{c}{(m\AA)}       & \multicolumn{2}{c}{(\kms)} &       \\ 
              &  SACY          & Present        & SACY   & Present &  SACY        & Present      & SACY & Present   &       \\ 
\hline
 0051$-$1306  & G2V     &    G1V     &  \dots &	 4.5   &   50  &   27  &   3.4  &    5.10  &	   \\  
 0242+3837    & 4917    &    5071    &    6.0 &  6.4   &  146  &  132  &  -3.46 &   -3.53  &	   \\  
 1959$-$0432  & 5630    &    5752    &    9.0 &  9.4   &  140  &  126  & -19.41 &  -18.69  &	   \\  
 2004$-$0239  & 5083    &    5183    &    8.0 &  9.0   &  290  &  298  & -16.46 &  -15.92  &	   \\  
 2118$-$0631  & K1IV    &    K1V     &  \dots &	20.9   &   40  &   48  & -23.3  &   22.82  &  SB1  \\  
 2155$-$0947  & K0V     &    G1.5V   &  \dots &	44.4   &    0  &   39  & -20.0  &  -34.35  &  SB1  \\  
 2157$-$0753  & K2IV    &    K1V     &  \dots &	28.2   &    0  &   17  &   6.7  &   14.05  &  SB1  \\  
 2202$-$0406  & K2V     &    K1V     &  \dots &	 8.0   &  200  &  179  &   0.3  &    5.07  &  SB1  \\  
 2204+0236    & K5V	&    K5V     &  \dots & 10.8   &    0  &   18  & -29.1  &  -26.36  &	   \\  
 2236+0010    & K0IV	&    G8IV-V  &  \dots & 31.9   &    0  &   26  &  -3.3  &    7.95  &  SB1  \\  
 2256+0235    & K1IV	&    G1.5V   &  \dots & 33.2   &  140  &  106  &  -9.4  &  -39.66  &  SB1  \\  
 2308+0000    & G8V	&    G2V     &   39.0 & 42.8   &  250  &  216  &   7.4  &    5.30  &	   \\  
 2309$-$0225  & K4Ve    &    K4V     &   13.3 &	12.1   &  200  &  195  & -13.7  &  -12.16  &	   \\  
 2321+0721    & K0V     &    K0V     &   14.6 &	14.3   &  300  &  277  &   6.6  &    6.05  &	   \\  
 2323$-$0635  & G9IIIe  &   K1III-IV &   26.4 &	26.8   &  105  &   87  &  13.4  &    4.06  &  SB1  \\  
 2324$-$0733  & G5V     &   G5IV-V   &    4.0 &	 2.5   &  133  &  136  &   4.0  &    4.00  &	   \\  
 2340$-$0402  & G1V     &    G1V     &   33.6 &	36.2   &  150  &  128  &  17.7  &   15.73  &  SB1  \\  
 2340$-$0228  & G2V     &    G3V     &   31.0 &	29.6   &   0   &   0   &  66.3  &  -20.27  &  SB1  \\  
 2352$-$1143  & G1V     &    G1V     &   27.8 &	19.4   &  115  &   87  &   2.6  &    0.49  &  SB1  \\  
\hline
\end{tabular}
\label{Tab:SACY}
\end{table*}

\subsection{Chromospheric activity}
\label{subsec:Chromo}

\begin{figure*}[htb]  
\begin{center}
\includegraphics[width=8.8cm]{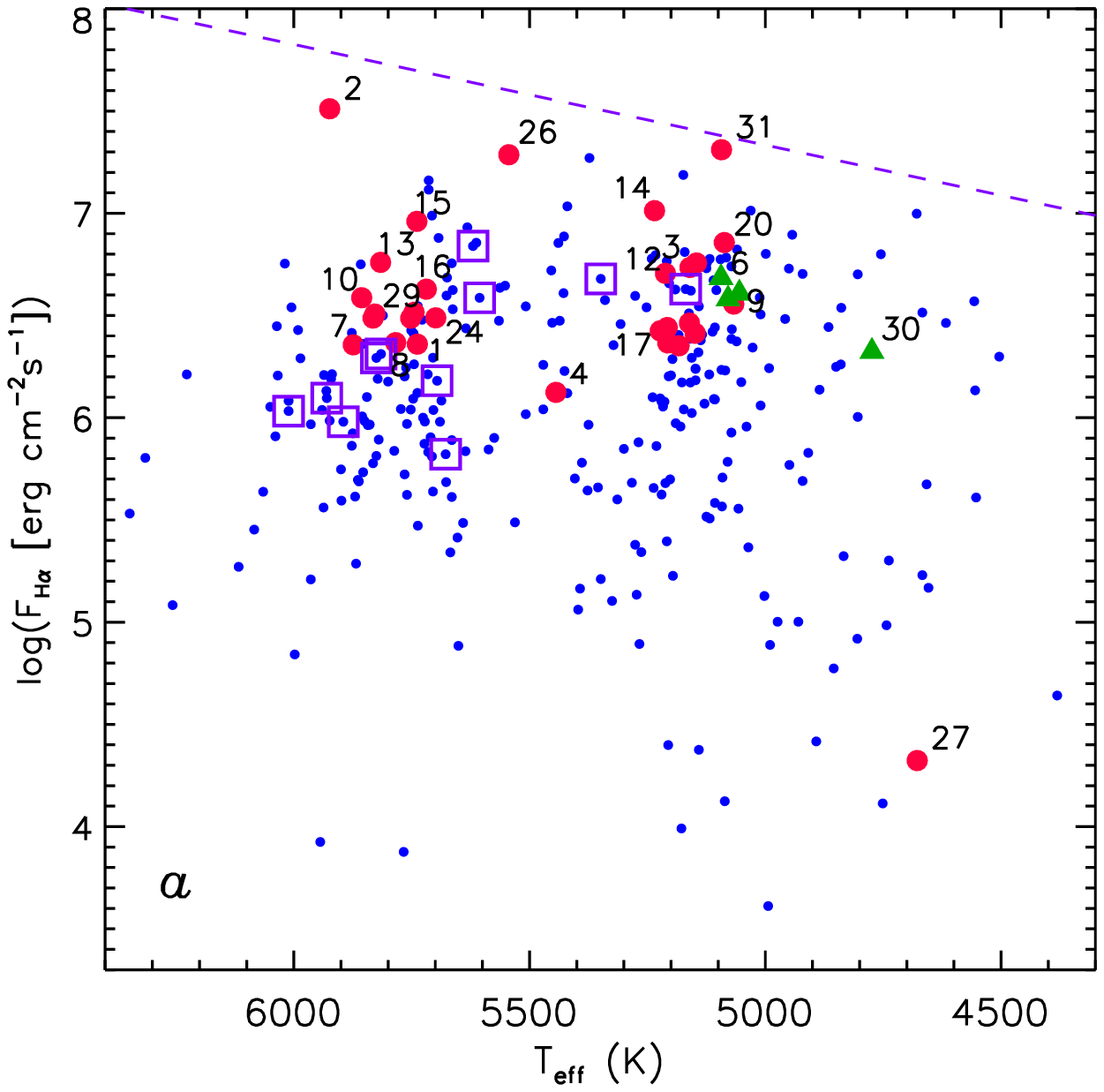}		%teff_fhalpha_rastyc.eps}  
\includegraphics[width=8.8cm]{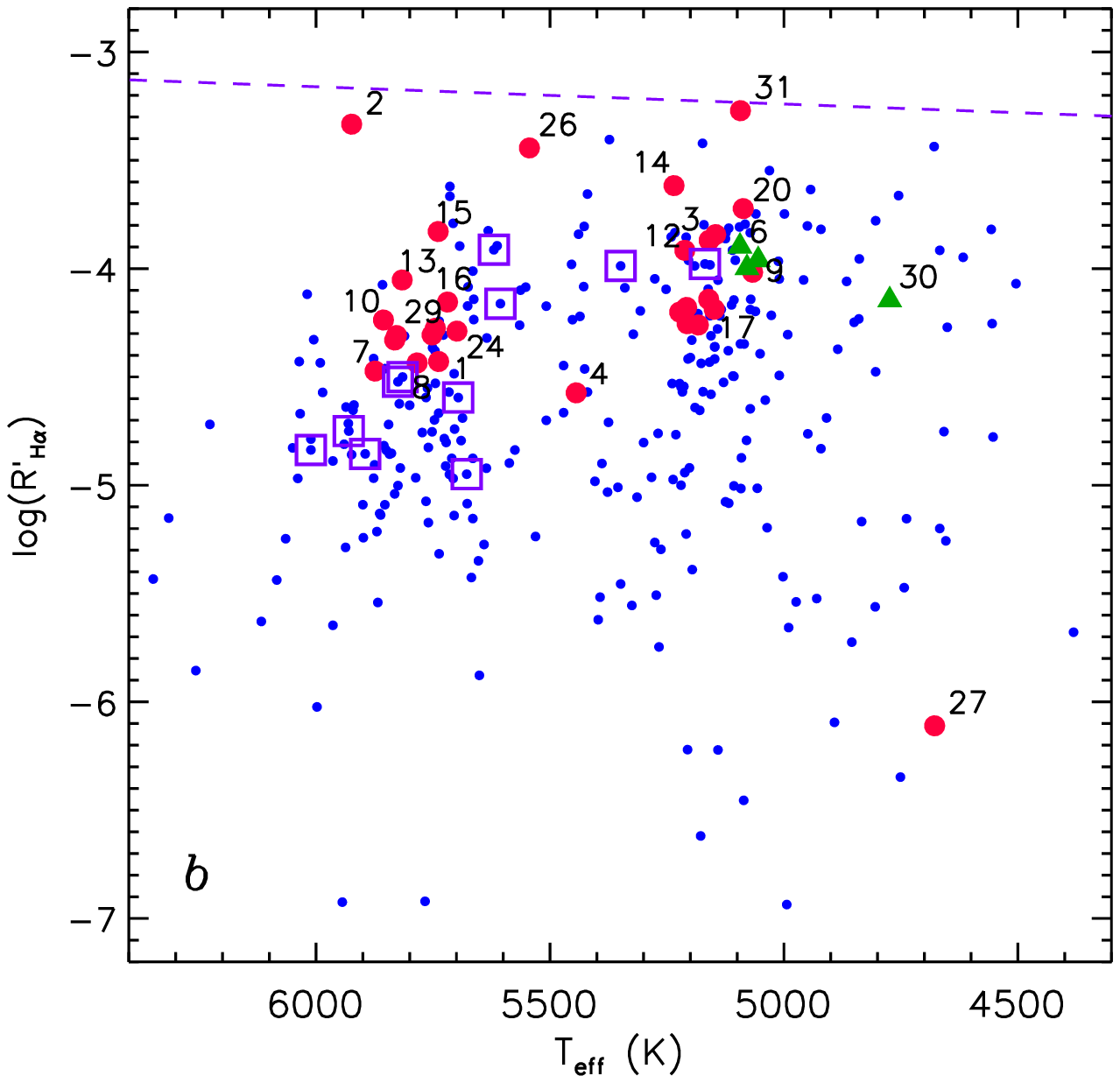}		%teff_lhalpha_lbol_rastyc.eps} 
\vspace{-.3cm}
\caption{H$\alpha$ flux ({\it a}) and $R'_{\rm H\alpha}$ ({\it b}) as a function of $T_{\rm eff}$. The symbols are 
as in Fig.~\ref{Fig:Li}. The lithium-rich stars that are not too crowded are labeled according to Table~\ref{Tab:LiRich}. The dashed straight line is 
the boundary between chromospheric activity (below it) and emission due to accretion as derived by \citet{Frasca2015}.
}
\label{Fig:Flux_Teff}
 \end{center}
\end{figure*}

The H$\alpha$ fluxes and $R'_{\rm H\alpha}$ indices, plotted against \teff, are shown in Fig.~\ref{Fig:Flux_Teff}, where  
the boundary between accreting objects and chromospherically active stars, as defined by \citet{Frasca2015}, is also shown. 

Our targets display a wide range in $R'_{\rm H\alpha}$ with only a few of them close to the saturation of magnetic 
activity, i.e. with $\log(F_{\rm H\alpha})\ge 6.3$ \citep[e.g.,][and references therein]{Martinez2011} or $\log(R'_{\rm H\alpha})< -4.0$
\citep[e.g.,][]{Soderblom1993a}.   
We note that there is no object above the activity/accretion boundary, but only the target with the highest H$\alpha$ flux, namely \#31 (V395\,Cep), 
is nearly superimposed to it.
This star is classified by \citet{Kun2009} as a classical T Tau star (CTTS) belonging to the L1261 cloud. Indeed, the H$\alpha$ emission profile 
displayed by the SARG spectrum is broad and double-peaked, with a central absorption, i.e. it is typical for a CTTS. Moreover, the full width at 
10\% of the line peak ($10\%W_{\rm H\alpha}$) is about 480 \kms, suggesting an accreting object according to the \citet[][]{WhiteBasri2003} 
criterion, $10\%W_{\rm H\alpha}\ge 270$\,\kms. From the H$\alpha$ luminosity, we have also estimated the mass accretion rate as 
$\dot M_{\rm acc}\simeq 3.6\times 10^{-9}$\,\Msun/yr, following the prescriptions of \citet{Biazzo2012b}. This value confirms  \#31 as 
a CTTS that displays a significant accretion \citep[e.g.,][]{Alcala2017}. This object is also a class\,II source, as testified
by the large IR excess (see Fig.\,\ref{Fig:SEDs}).

The other two stars close to the boundary are \#2 (TYC\,4496-780-1) and \#26 (TYC\,1154-1546-1). The former belongs to a small
group of young stars discovered by \citet{Guillout2010} in the surroundings of the Cepheus flare region \citep[e.g.,][and references therein]{Tachihara2005}. 
It is the only object that displays a substantial IR excess among those in \citet{Guillout2010}. Furthermore, its H$\alpha$ profile is similar to that of a 
CTTS, with a double-peaked shape and $10\%W_{\rm H\alpha}\simeq 510$\,\kms. However, this star is a close visual binary \citep[e.g.,][]{Fabricius2002} 
that deserves a deeper analysis, which is deferred to a subsequent work. 
The second one (\#26) is an ultrafast rotator (\vsini$\simeq 234$\,\kms) with a broad and shallow H$\alpha$ profile filled in 
by emission. This object is likely a zero-age main sequence (ZAMS) star, as also suggested by the HR diagram (Fig.~\ref{Fig:HR_PMS}) and by the
lack of IR excess (see Fig.\,\ref{Fig:SEDs}).

One of the candidate lithium-rich giants, namely \#27 (HD~214995) shows a very low H$\alpha$ emission ($\log(R'_{H\alpha})=-6.11$), while the other one 
(\#17\,=\,HD~234808) has $\log(R'_{H\alpha})=-4.20$, which is similar to the values of other very young stars in our sample.
We note that the former star does not display IR excess, while star \#17 shows far-infrared excess (see Sect.\,\ref{subsec:sed}).

\subsection{Spectral energy distributions of the very young sources}
\label{subsec:sed}

We analyzed the spectral energy distribution (SED) of the very young and {\it PMS-like} candidates to check the consistency between the photometry
and the stellar parameters derived from our spectra. 
This analysis also allows us to detect IR excesses and to classify the sources in the \citet{Lada1987} scheme. 

We constructed the SED of the {\it RasTyc} sources using the optical and near-infrared (NIR) photometric data available in the literature.
In particular, we used $BVR$ photometry from the NOMAD catalog (\citealt{zachariasetal2004}), where the magnitudes in the Johnson $B$ and $V$ bands are
obtained by transformations from TYCHO $B_{\rm T}$--$V_{\rm T}$ system 
and the $R$ magnitude is normally taken from the USNO-B1.0 catalog \citep{Monet2003}. 
Whenever available (in a few cases), a more precise simultaneous photometry in the optical bands was adopted instead \citep[e.g.,][]{Guillout2010,Frasca2011,Froehlich2012}.  
For all the sources we retrieved $I$ magnitudes from the TASS catalog \citep{Droege2006} and completed the optical/NIR SED with the 2MASS $JHK_{\rm s}$ photometry \citep{2MASS,skrutskieetal2006}.
The mid-infrared (MIR) photometry was retrieved from the AllWISE Data release \citep{WISE,AllWISE} and, for a few sources, from AKARI \citep{AKARI} and 
IRAS \citep[e.g.,][]{IRAS,Abrahamyan2015}.

We adopted the grid of NextGen low-resolution synthetic spectra, with $\log g$ in the range 3.5--5.0 and solar metallicity by \citet{Hau99a}, to 
fit the optical-NIR portion (from $B$ to $J$ band) of the SEDs,  as done by \citet{frascaetal2009}.
In the fitting procedure we fixed \teff\ and \logg\  of each target to the values found with the code ROTFIT (Table\,\ref{Tab:LiRich}) and let the angular stellar diameter and 
the extinction $A_V$ vary until a minimum $\chi^2$ was reached. For the stars with known distance, this also provides us with a measure of the stellar radius.
We always found low values of $A_V$, ranging from 0.0 to 0.6\,mag, in agreement with the estimates made in Sect.\,\ref{subsec:HR}.

We found a relevant MIR excess only for two stars, RasTyc\,0013+7702 and RasTyc\,2320+7414 (V395~Cep), which behave as class\,II IR sources. This classification 
agrees with their H$\alpha$ profiles, which are typical of CTTS. 
In addition, another star in Cepheus, namely RasTyc\,0000+7940 (\#1) displays a small MIR excess (class\,III), in line with its weak-line T Tauri star (WTTS) nature.

\begin{figure*}[ht]
\includegraphics[width=6.0cm]{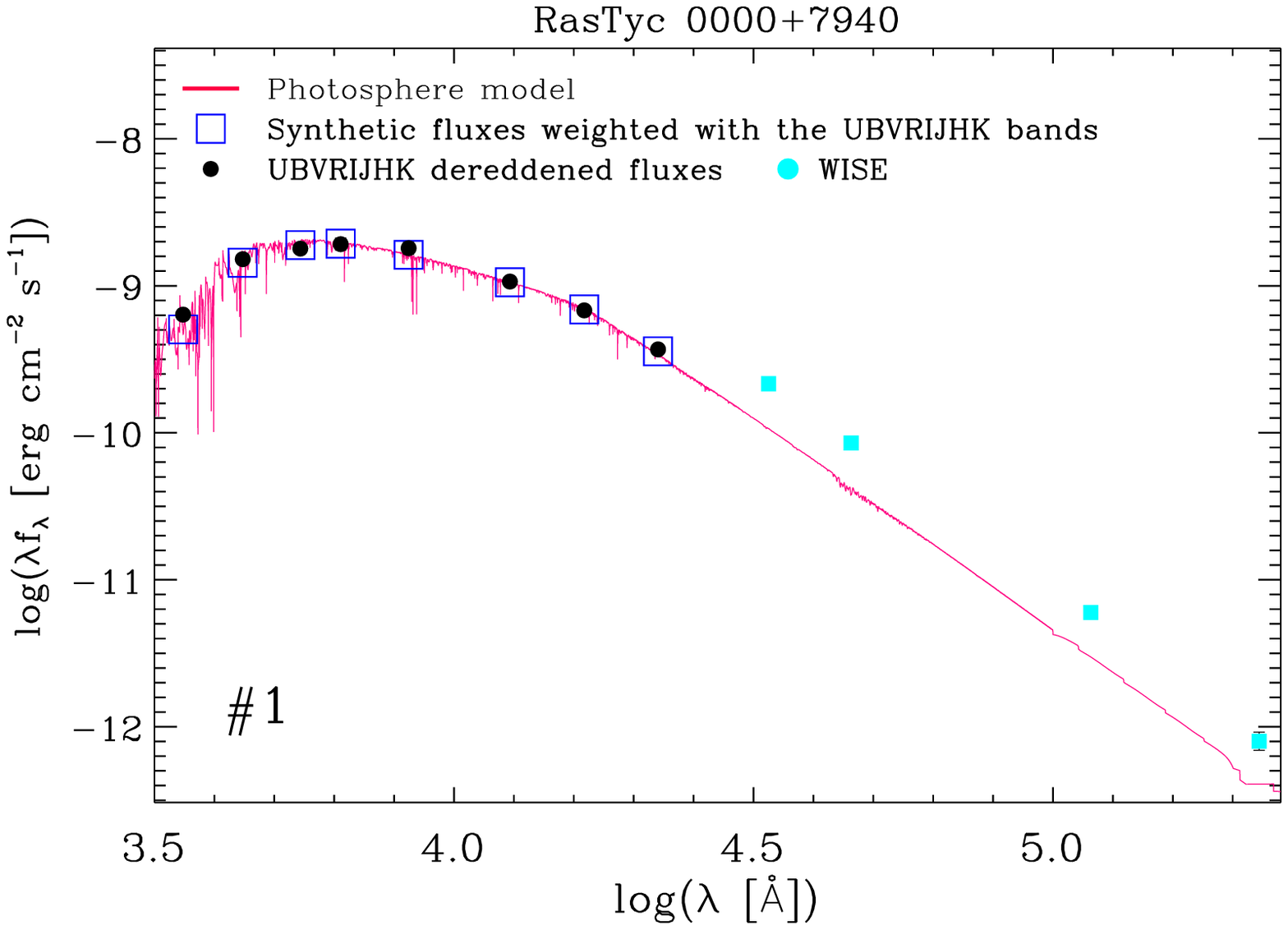}
\includegraphics[width=6.0cm]{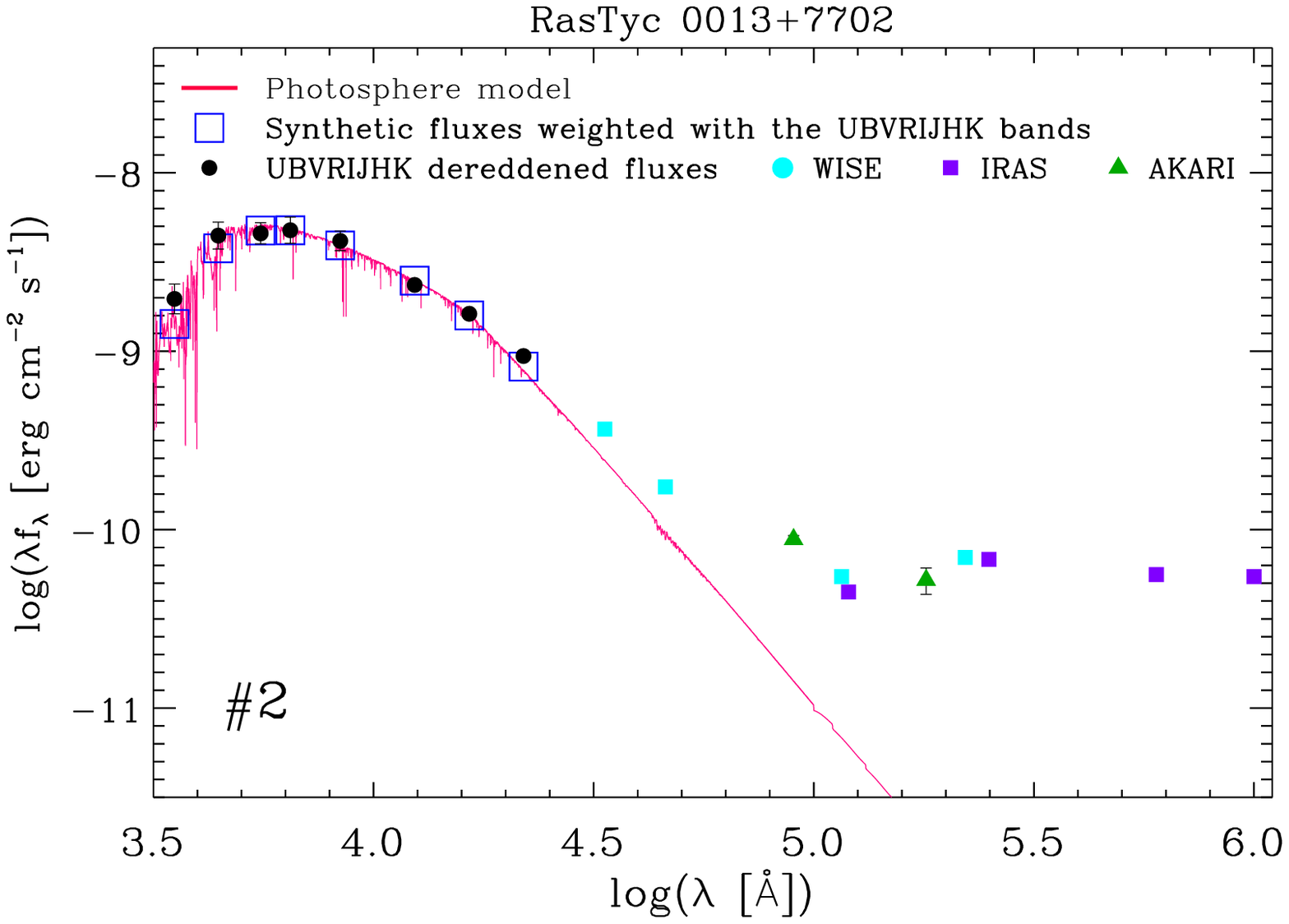}
\includegraphics[width=6.0cm]{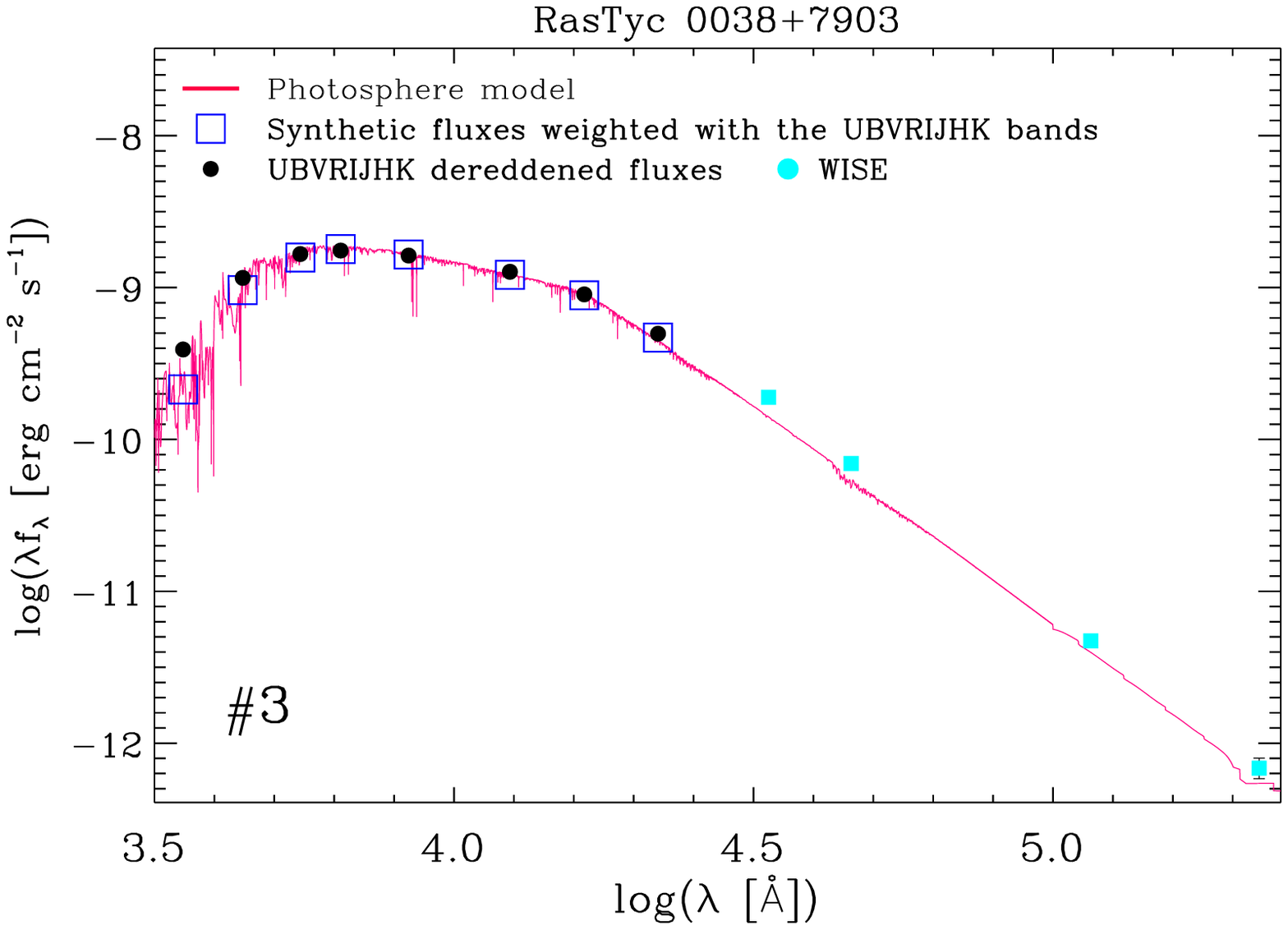}
\includegraphics[width=6.0cm]{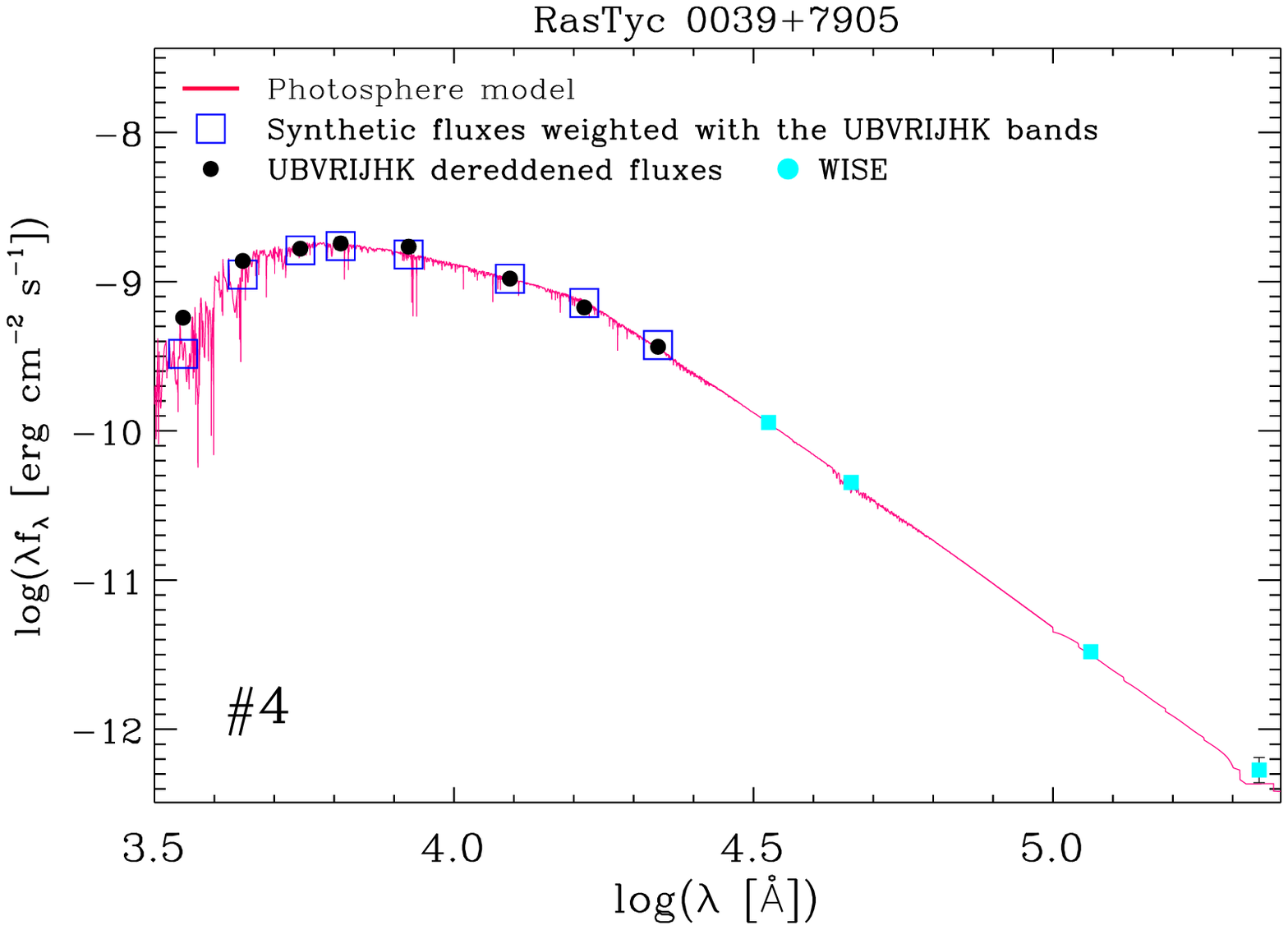}
\includegraphics[width=6.0cm]{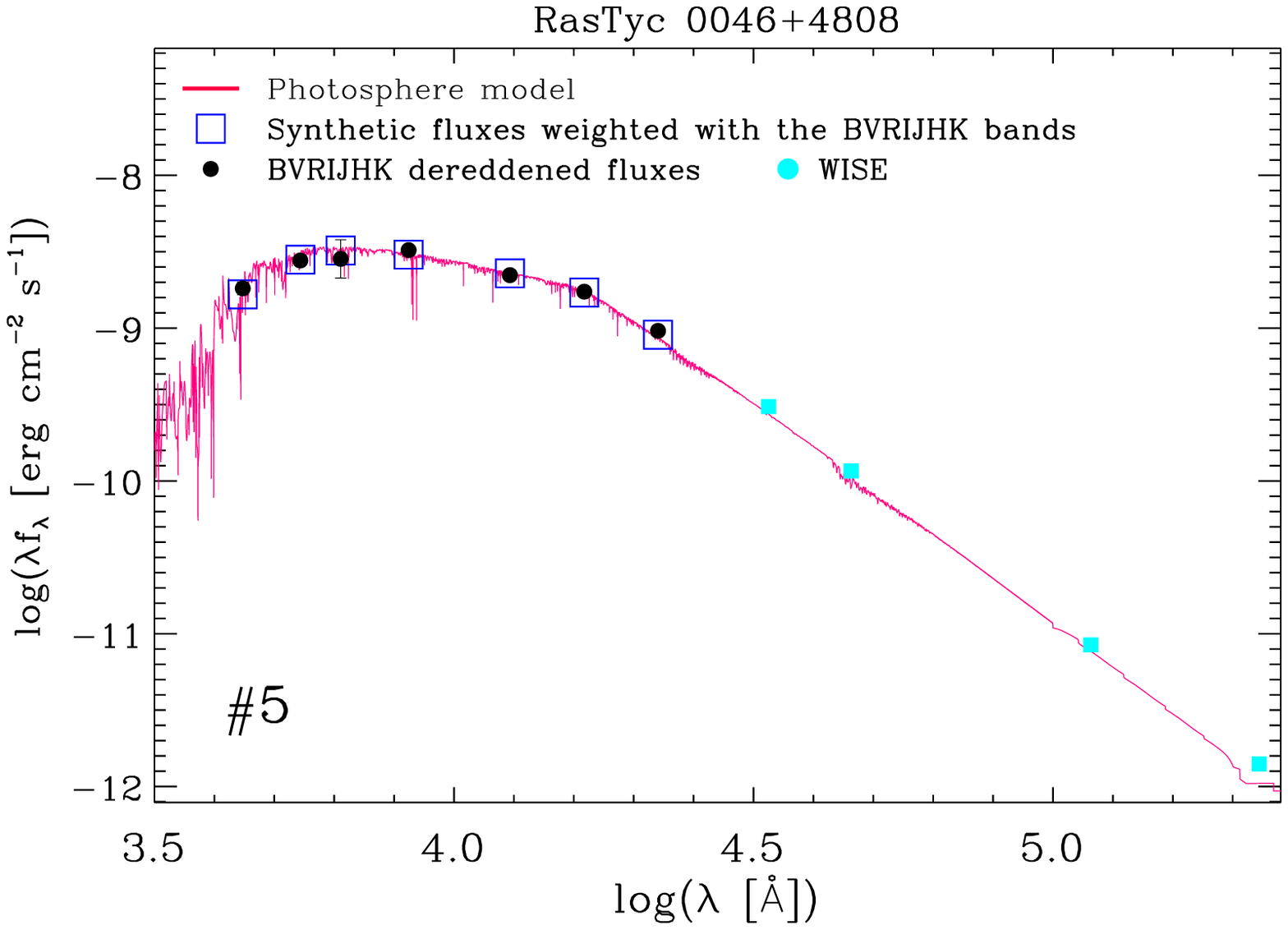}
\includegraphics[width=6.0cm]{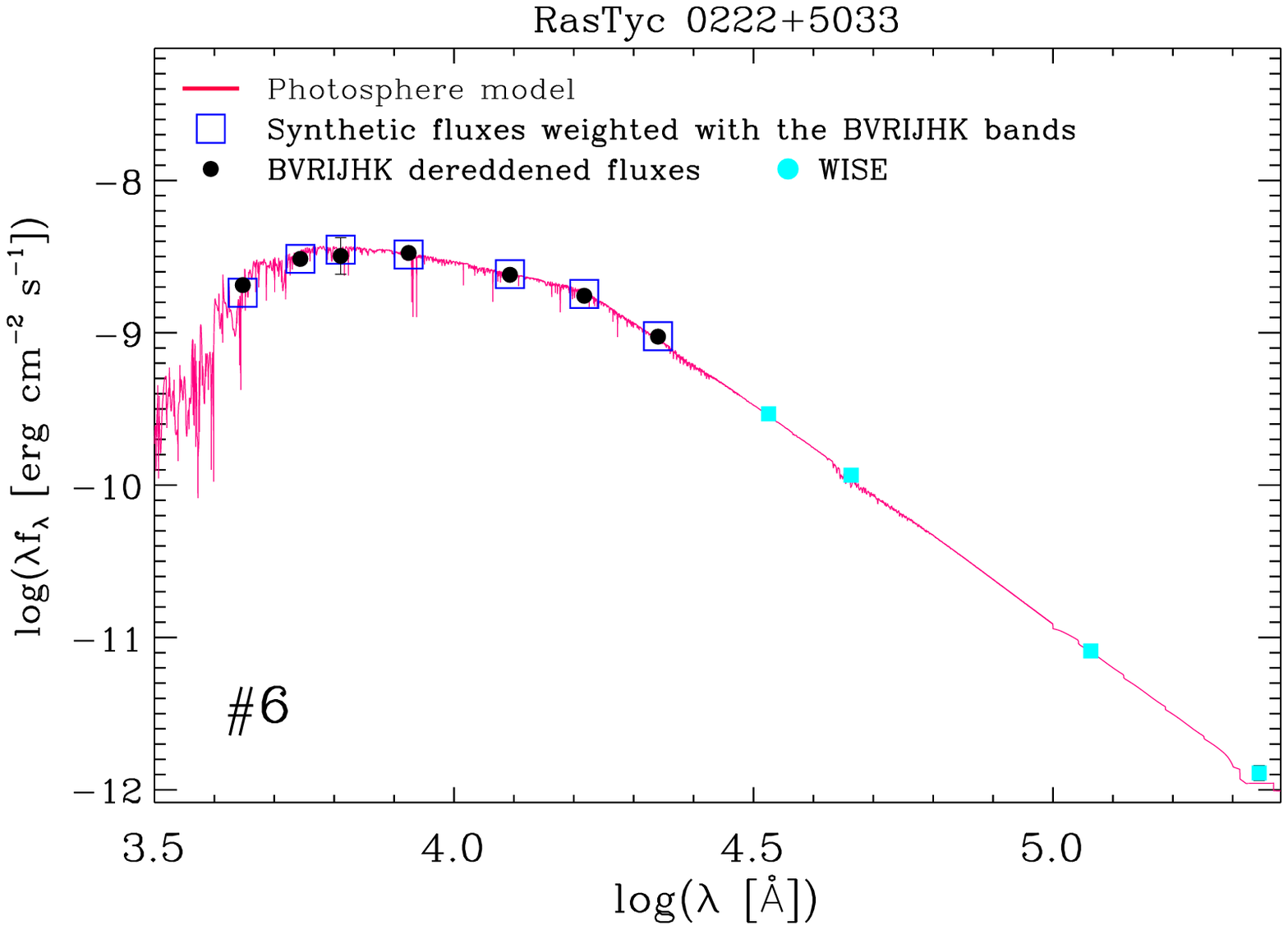}
\includegraphics[width=6.0cm]{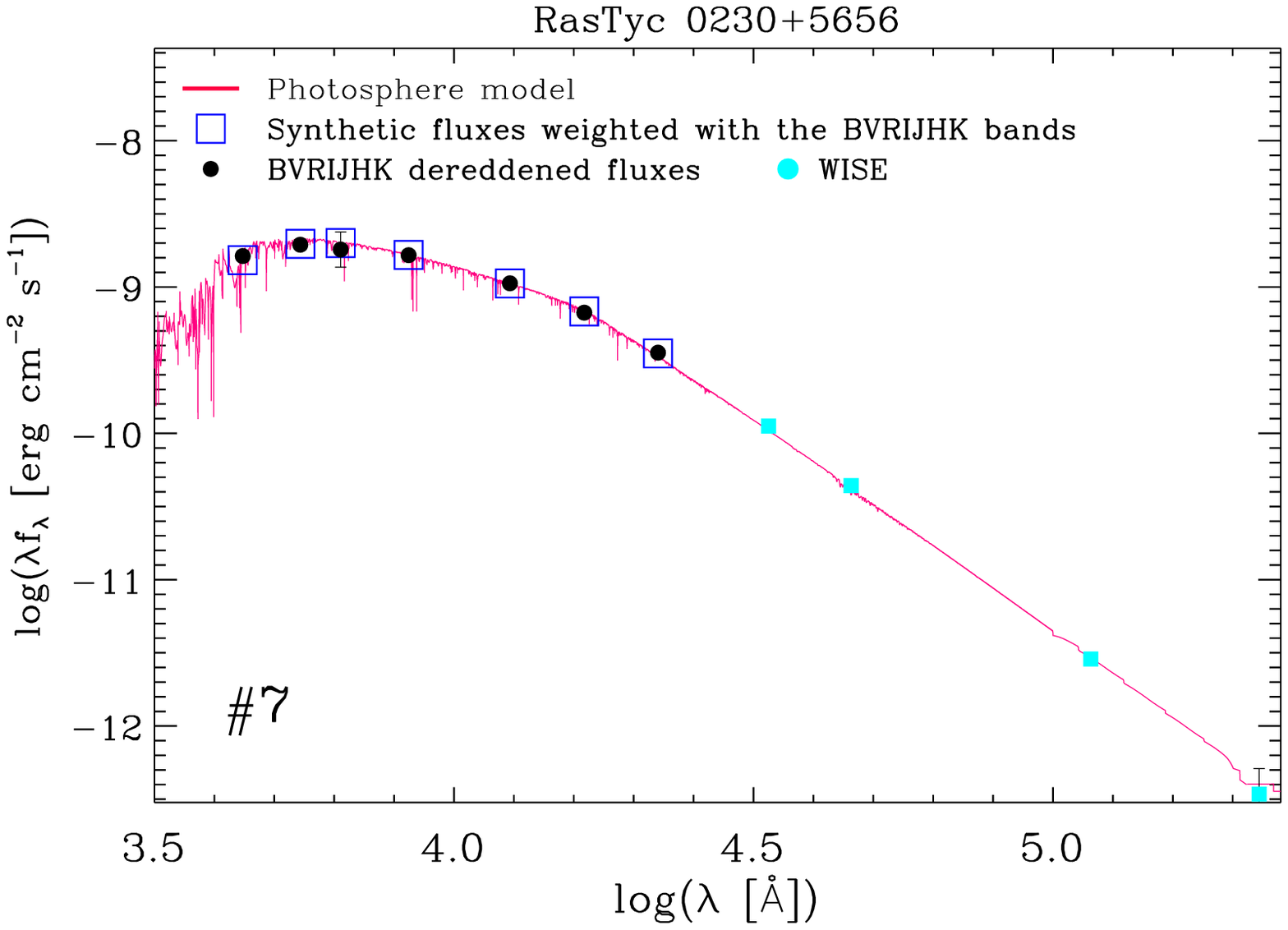}
\includegraphics[width=6.0cm]{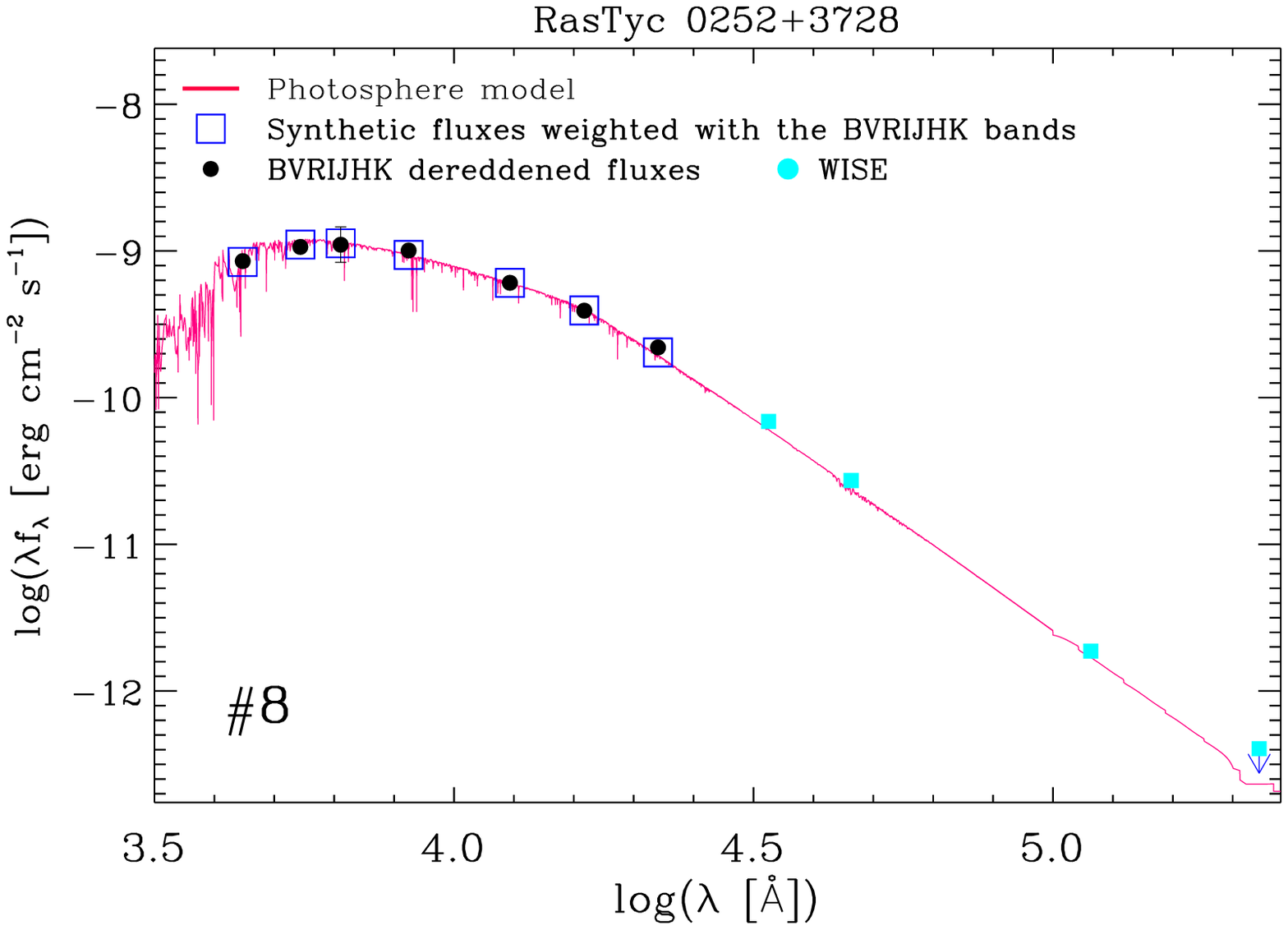}
\includegraphics[width=6.0cm]{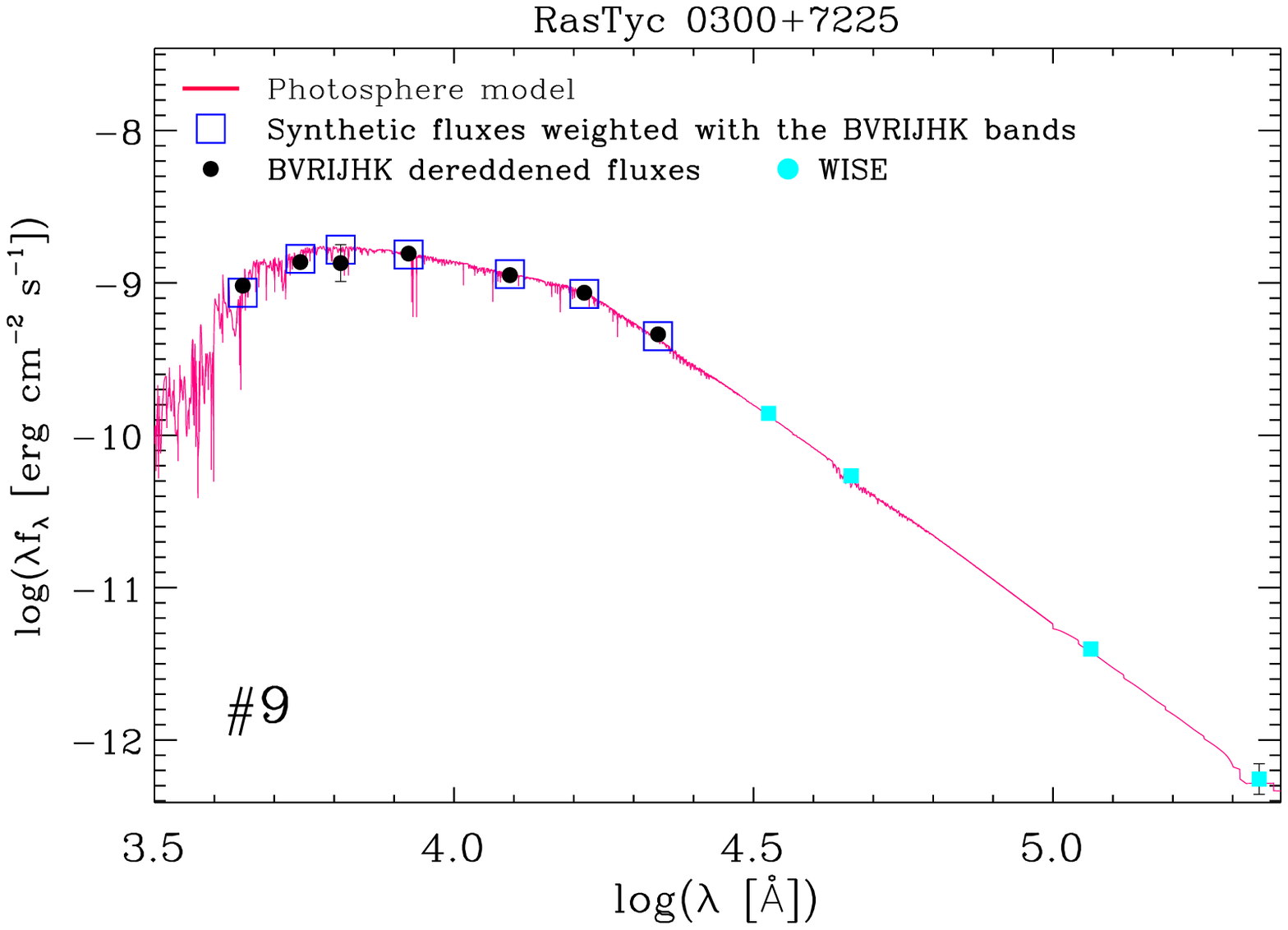}
\includegraphics[width=6.0cm]{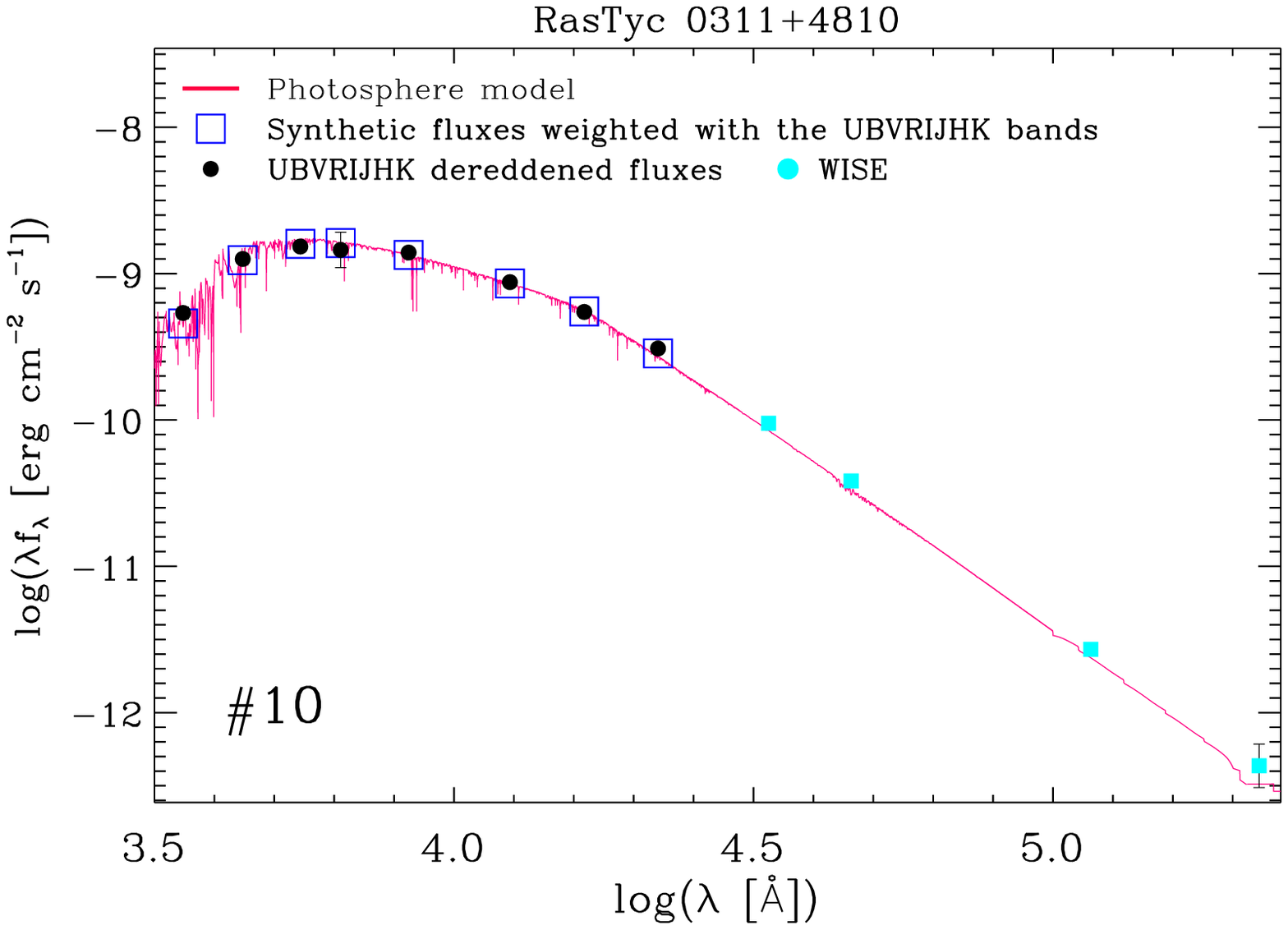}
\includegraphics[width=6.0cm]{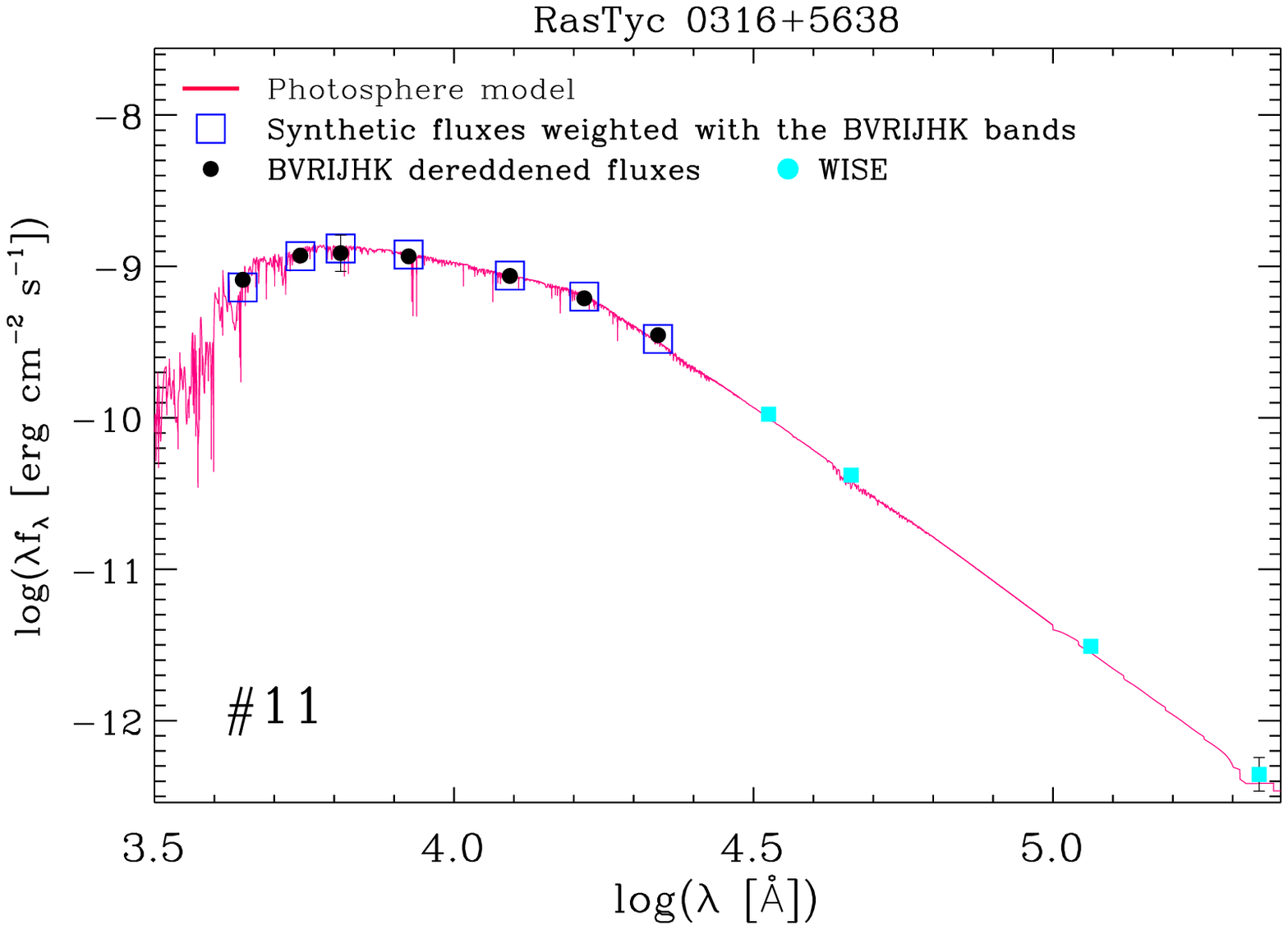}
\includegraphics[width=6.0cm]{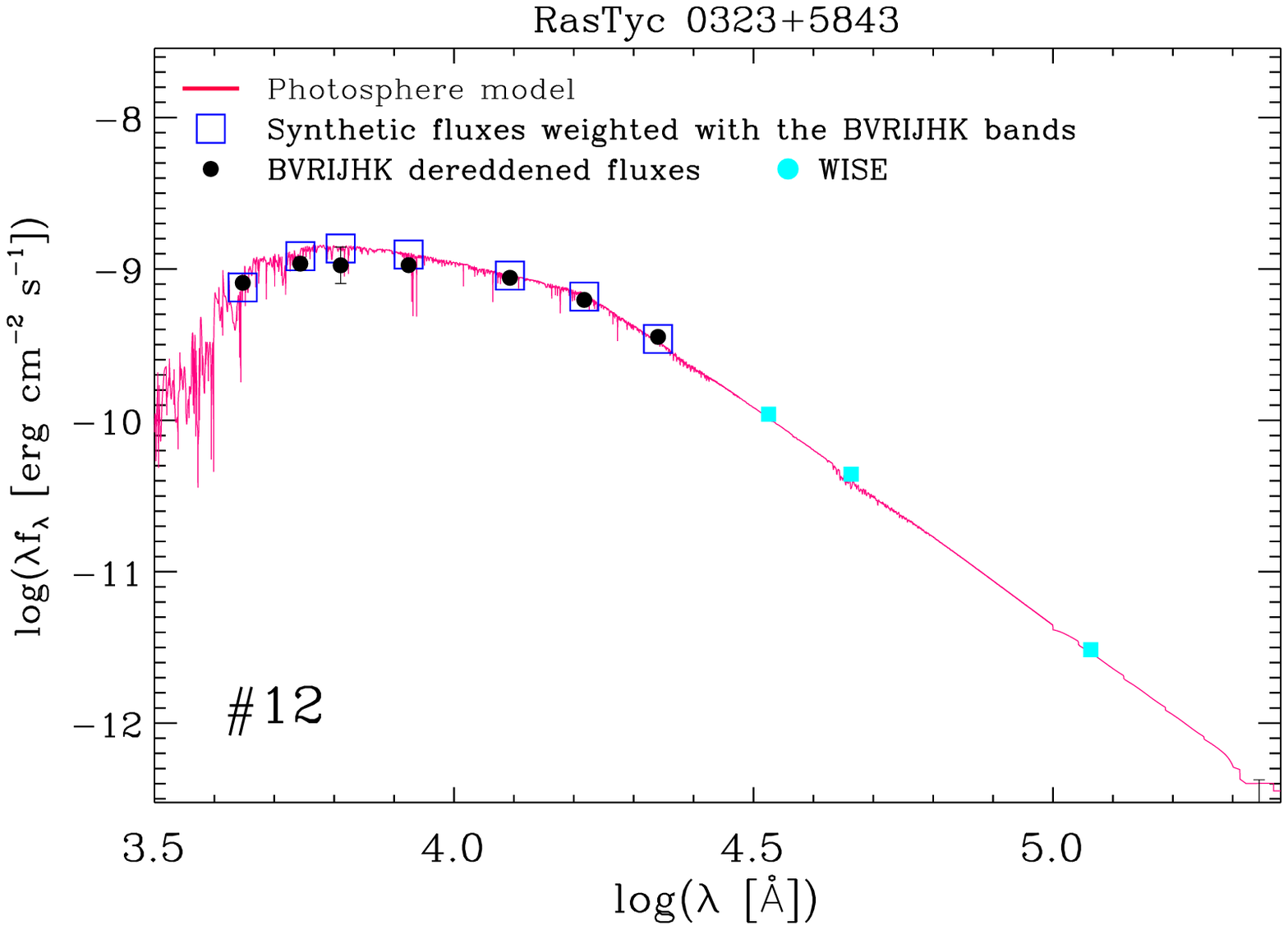}
\includegraphics[width=6.0cm]{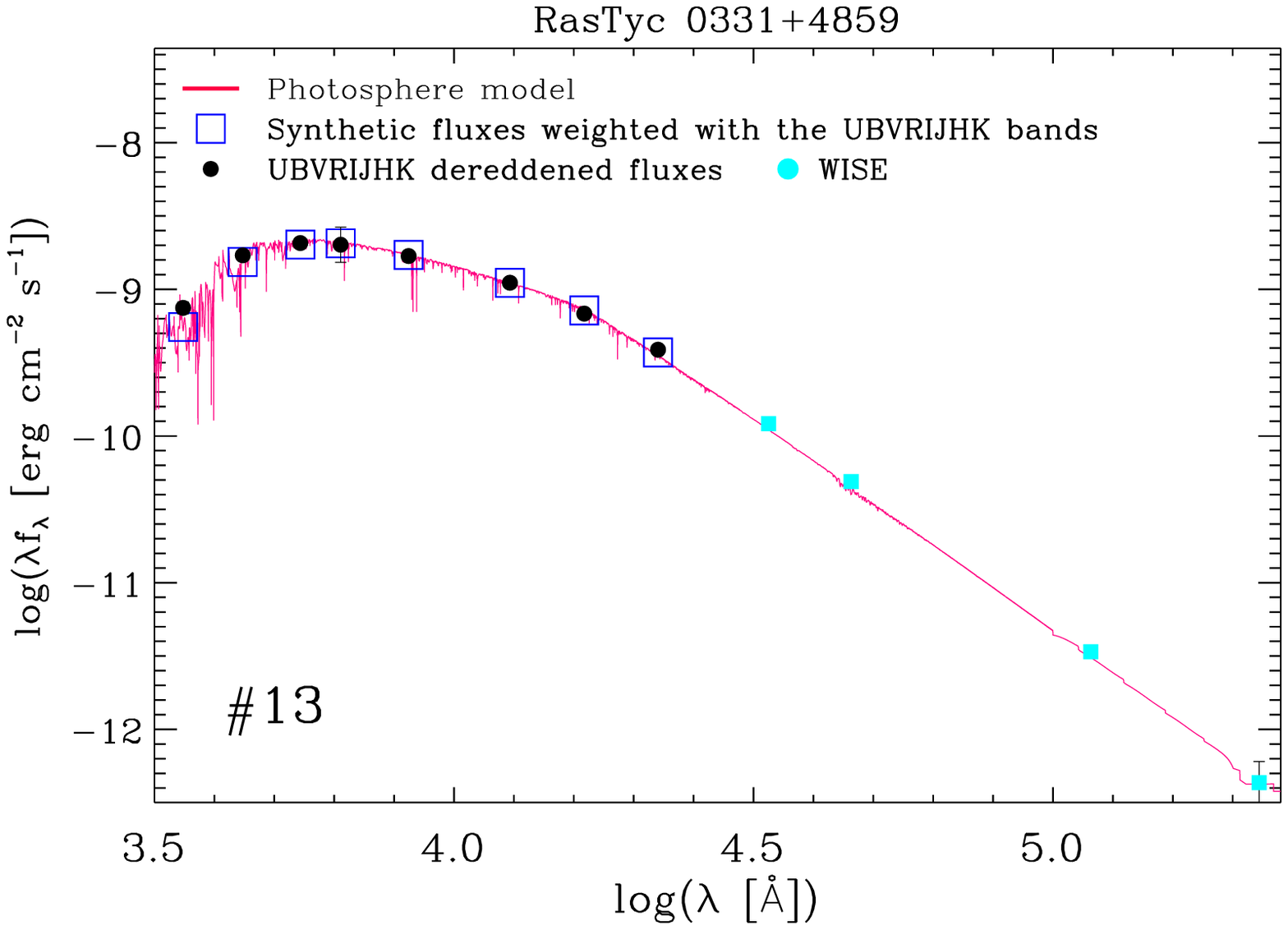}
\includegraphics[width=6.0cm]{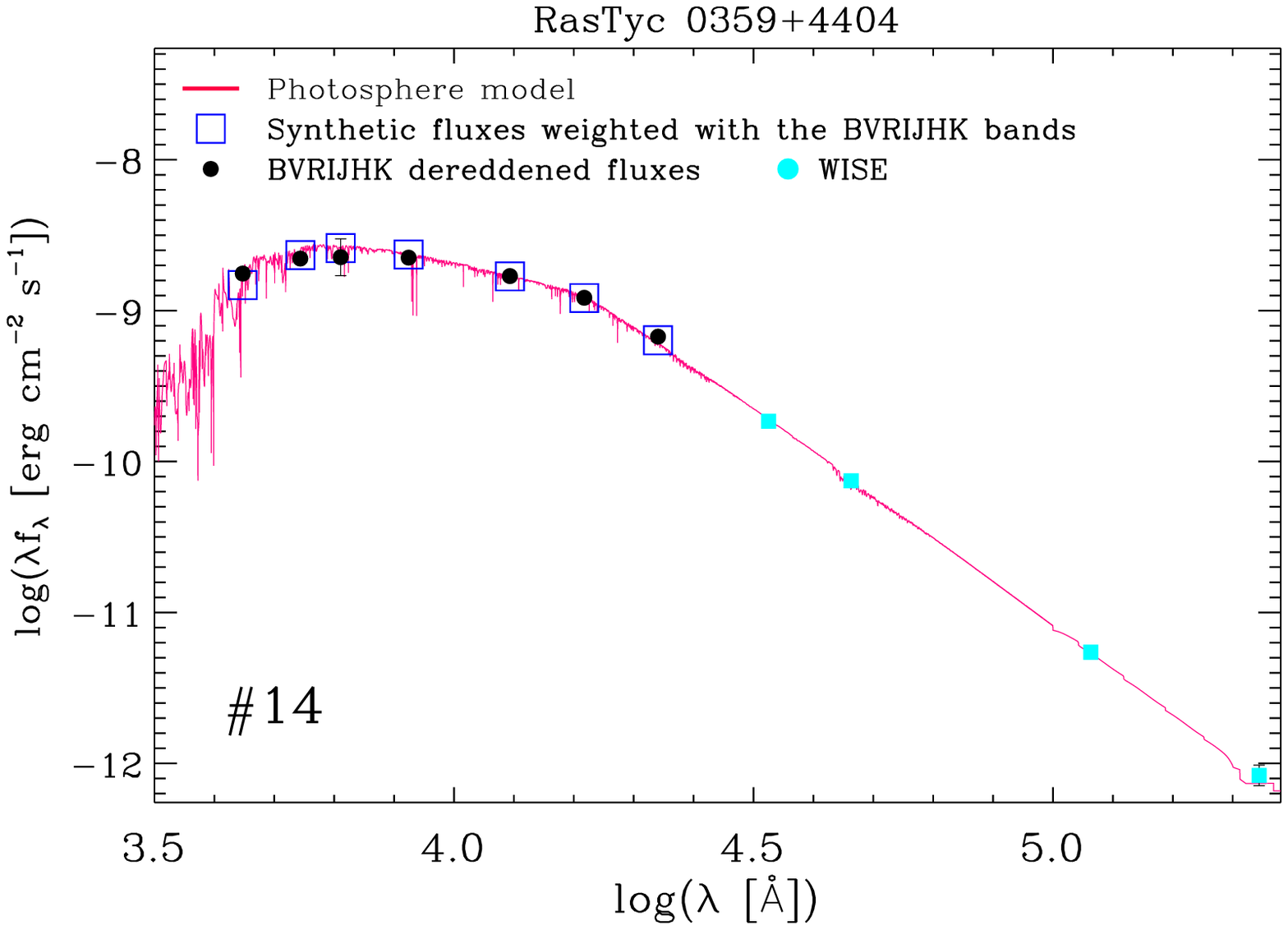}
\includegraphics[width=6.0cm]{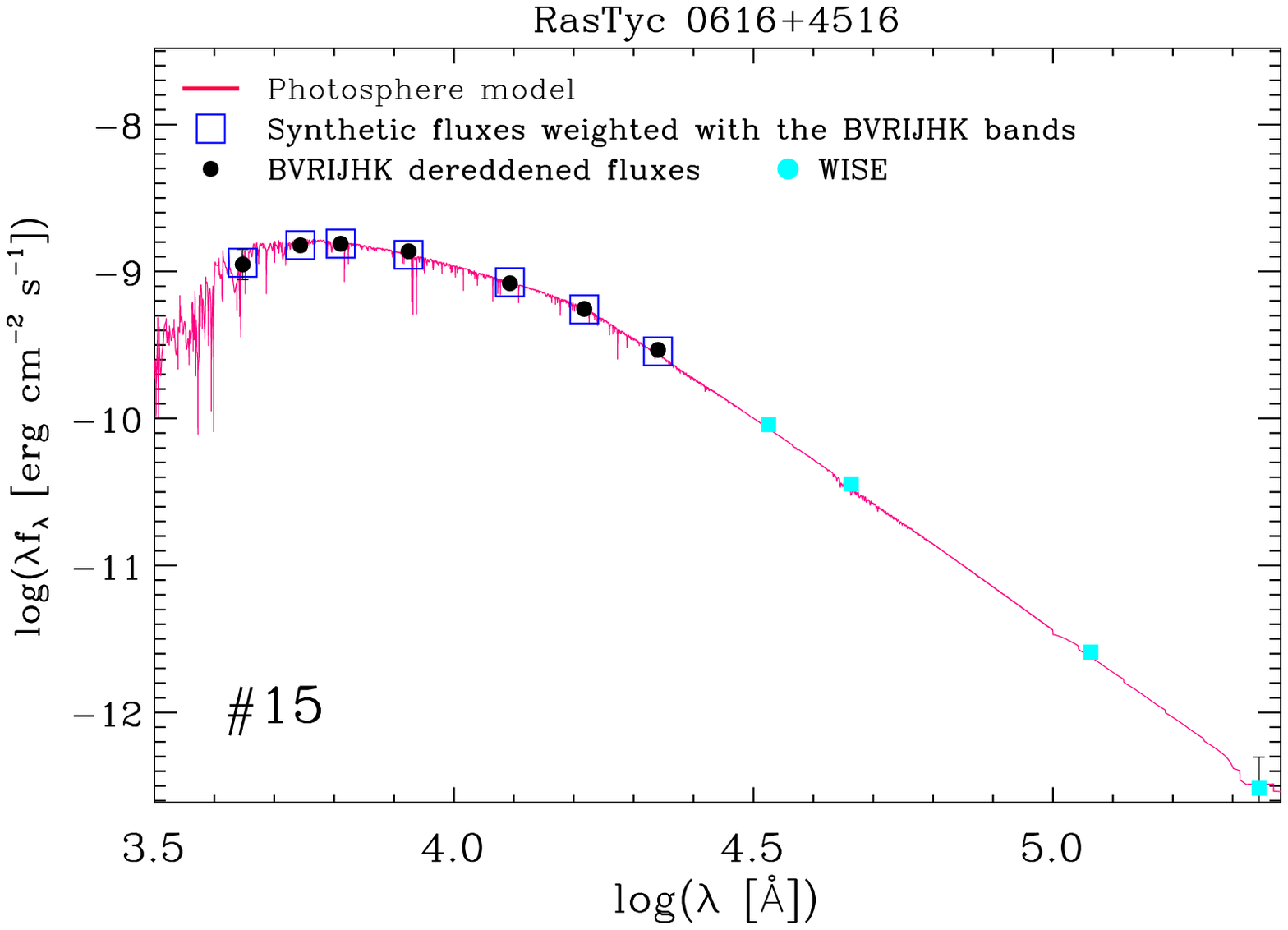}
\caption{Spectral energy distributions (dots) of the {\it PMS-like} sources and the two lithium-rich giants discussed in the paper. 
In each panel, the best fitting low-resolution NextGen spectrum \citep{Hau99a} is displayed by a continuous line. Mid-infrared fluxes are displayed 
with different as indicated in the legends of the plots.}
\label{Fig:SEDs}
\end{figure*}

\addtocounter{figure}{-1}

\begin{figure*}[ht]
\includegraphics[width=6.0cm]{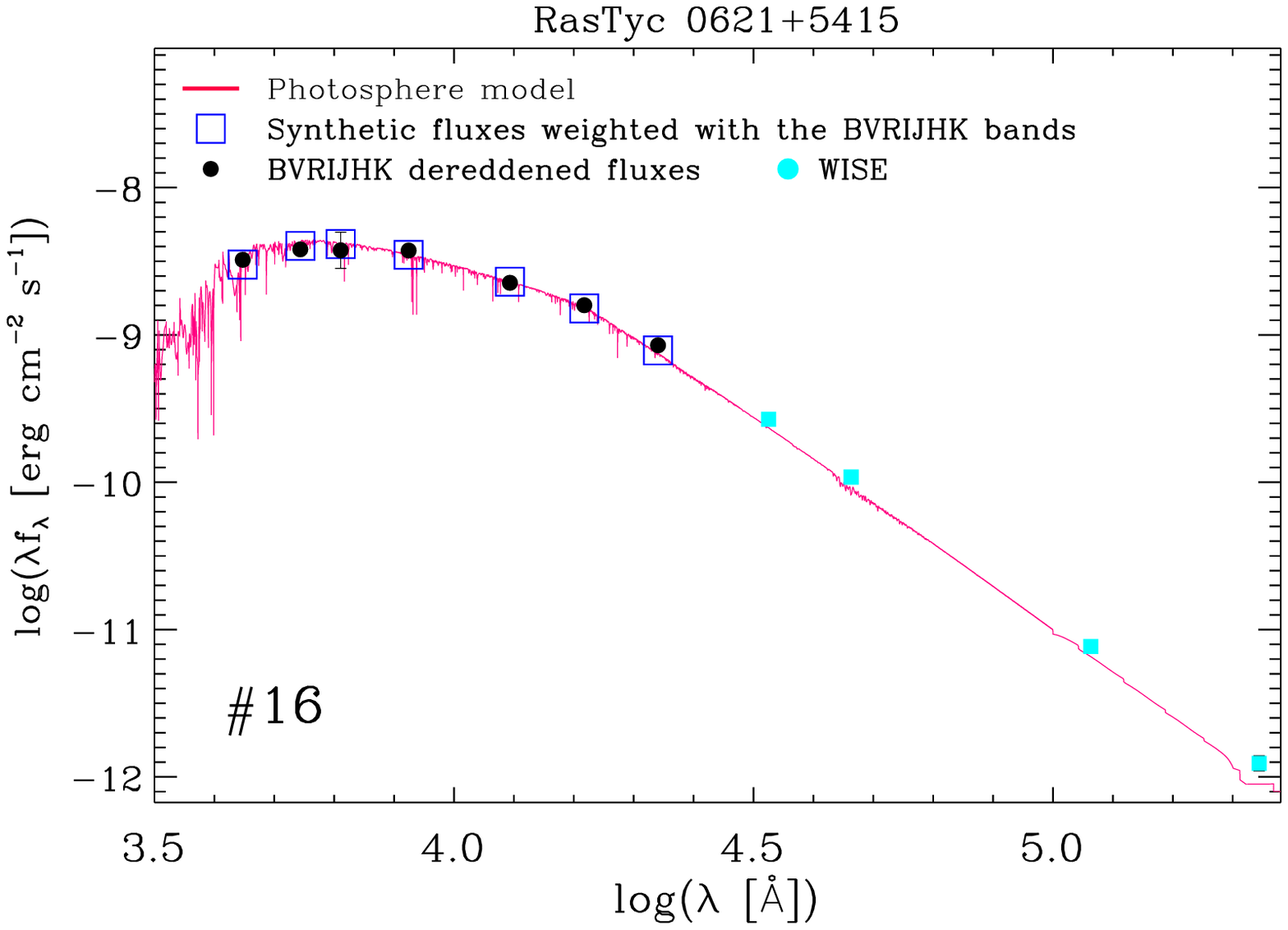}
\includegraphics[width=6.0cm]{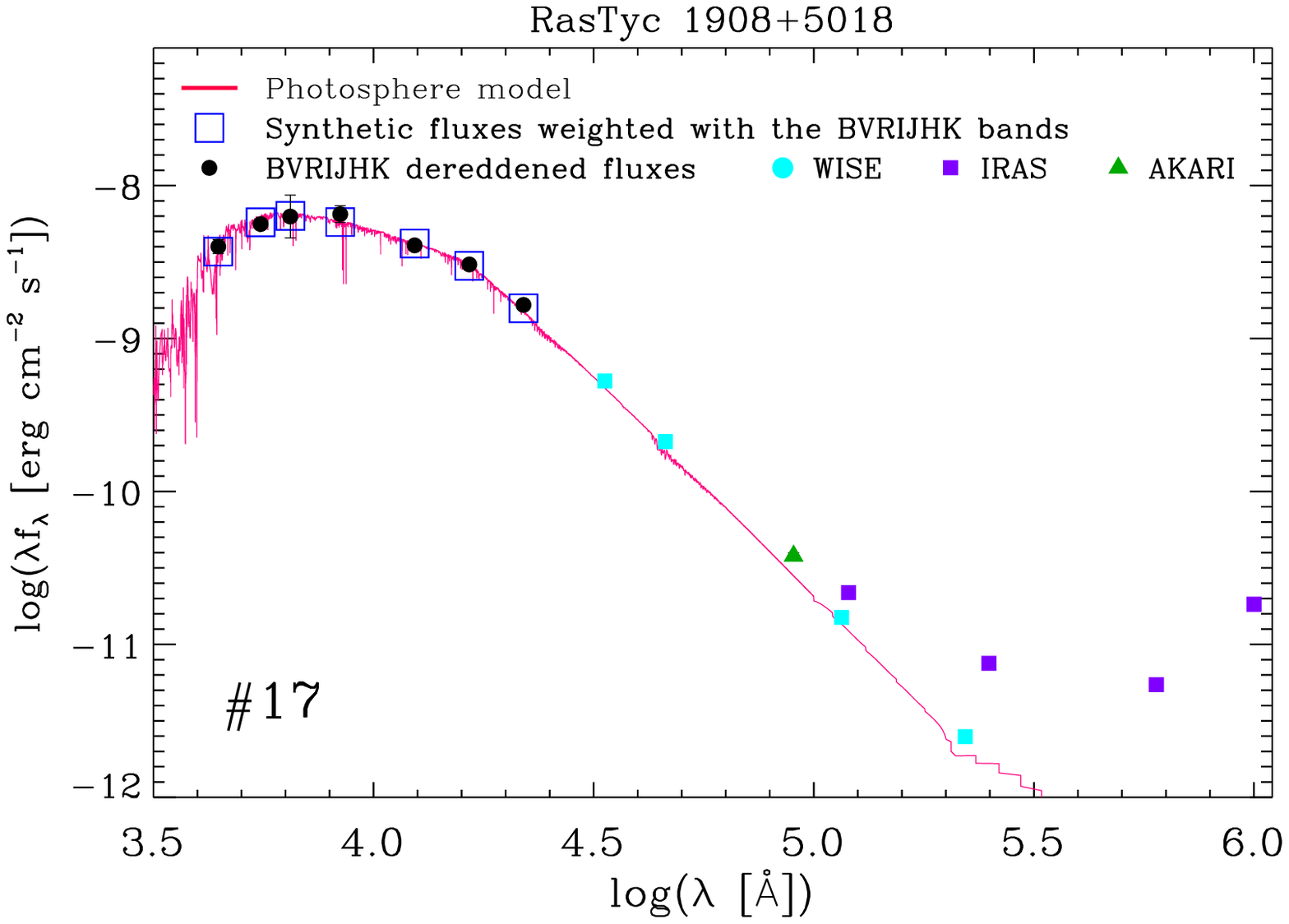}
\includegraphics[width=6.0cm]{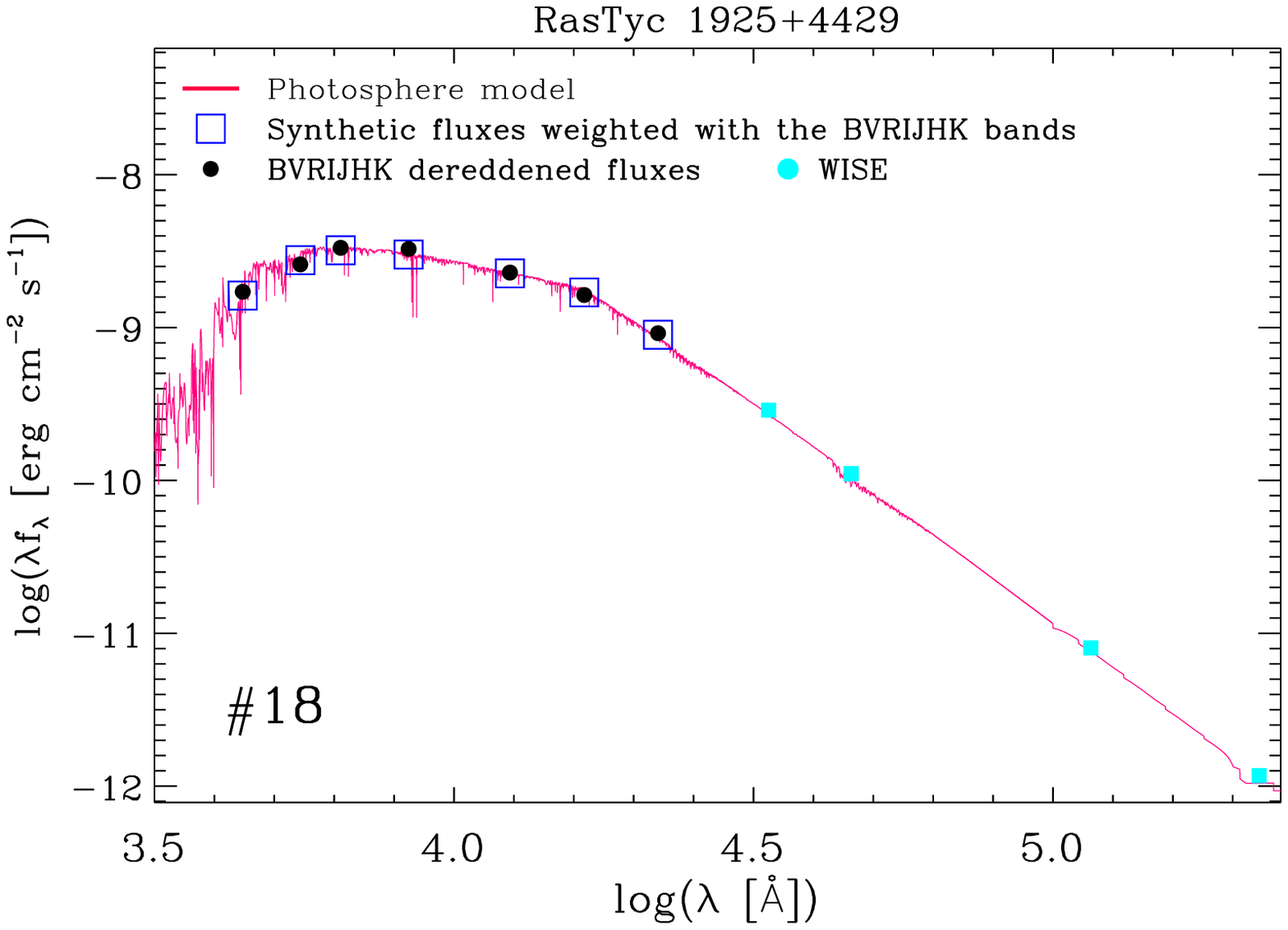}
\includegraphics[width=6.0cm]{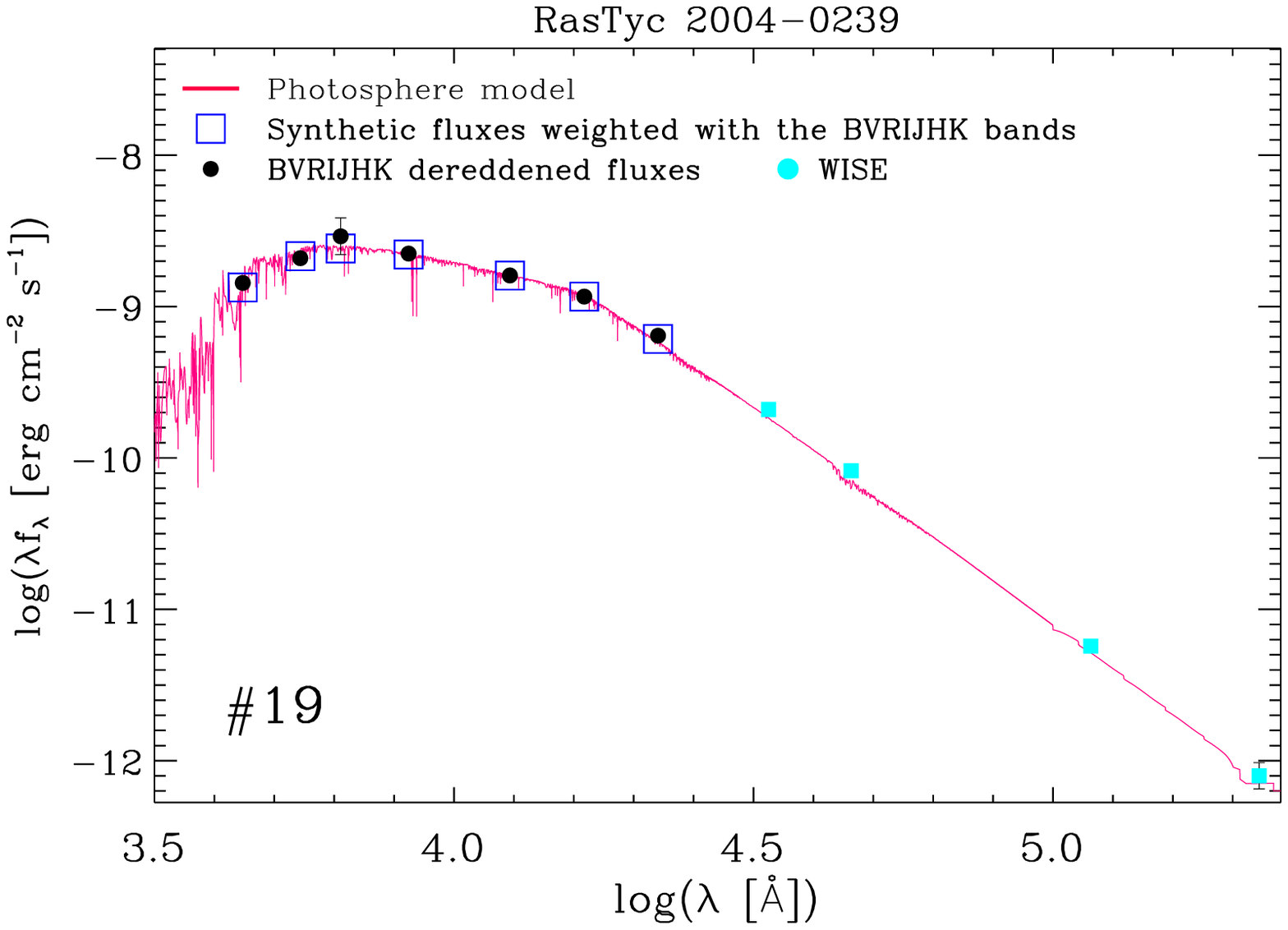}
\includegraphics[width=6.0cm]{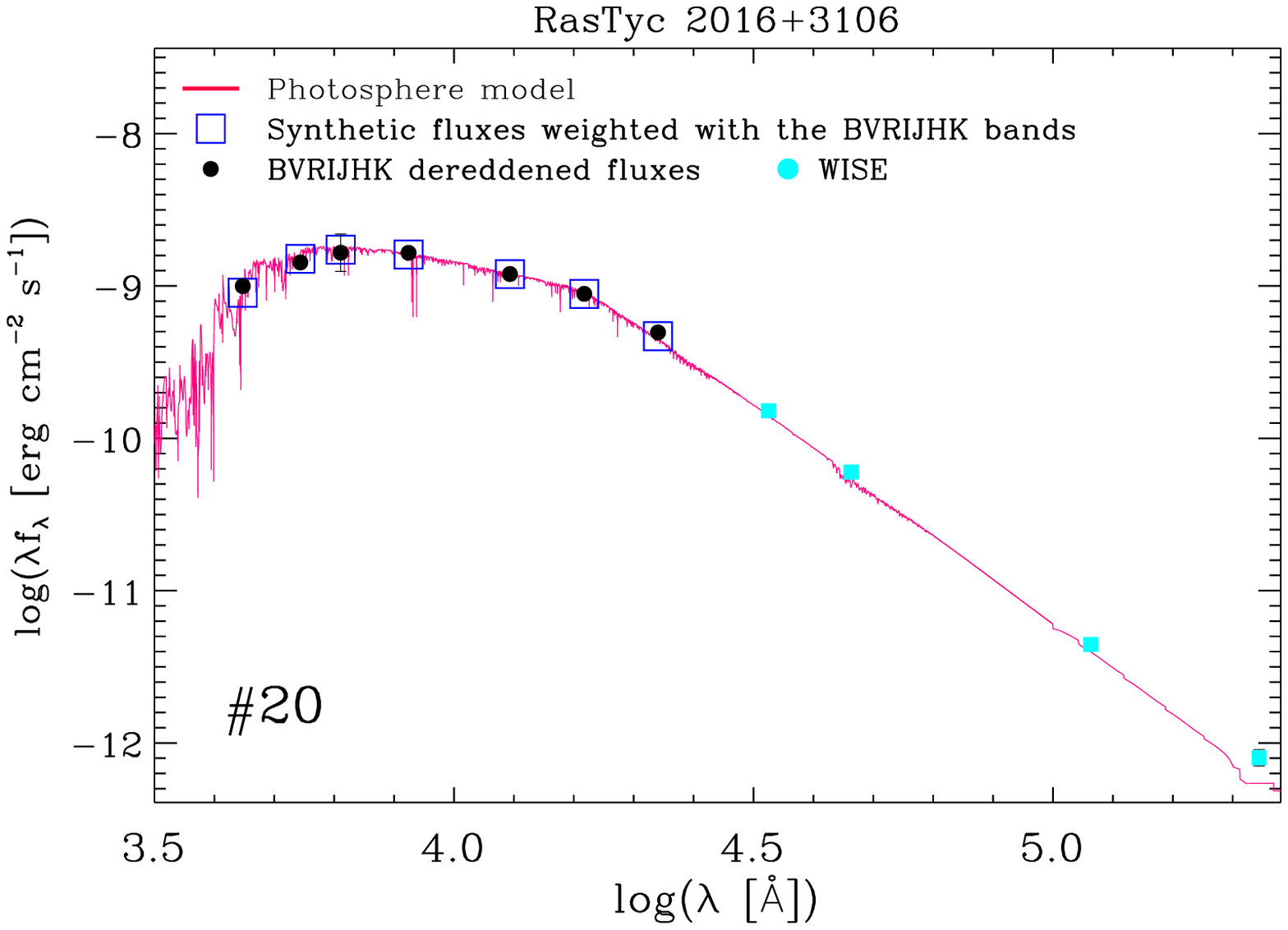}
\includegraphics[width=6.0cm]{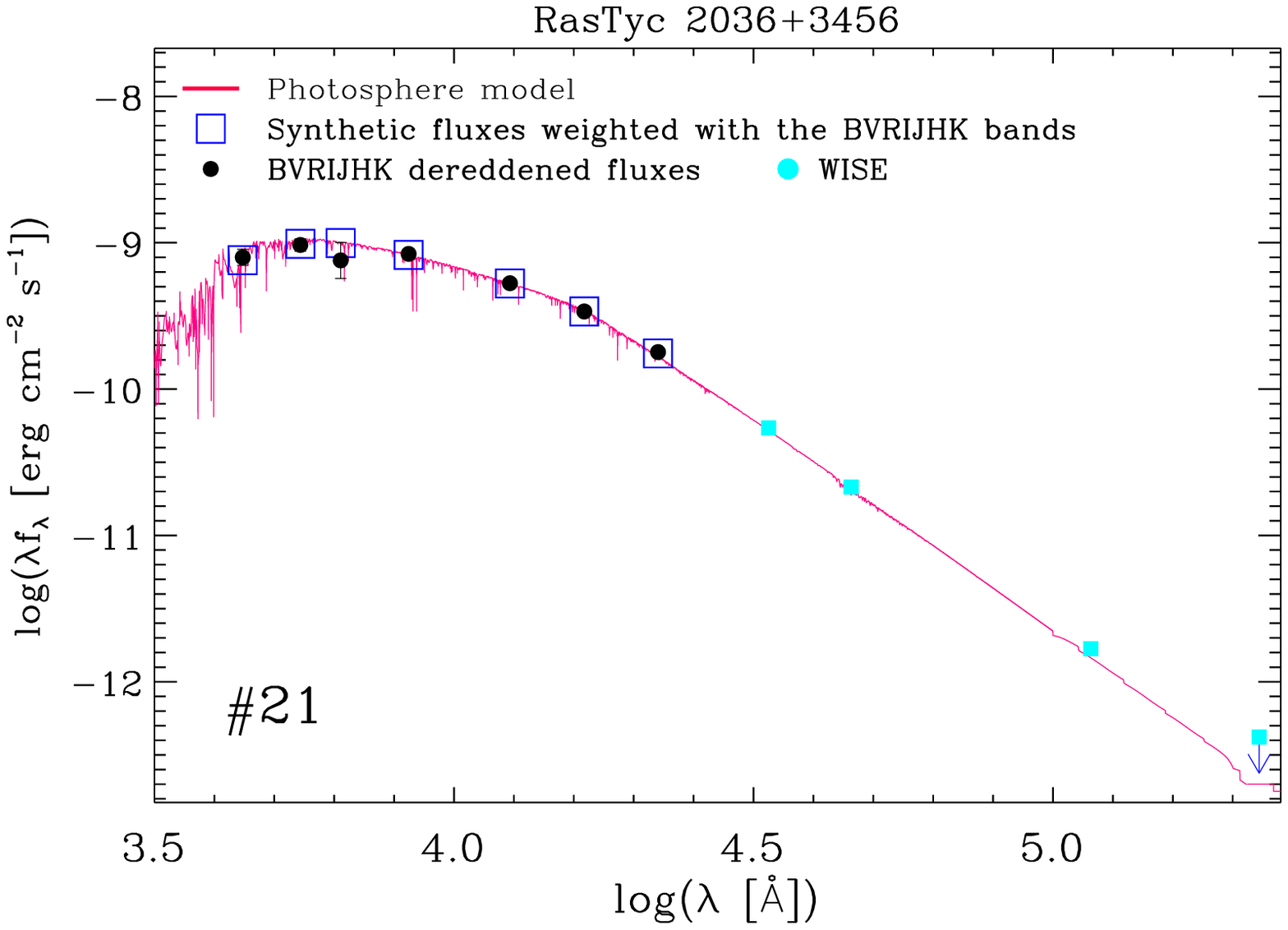}
\includegraphics[width=6.0cm]{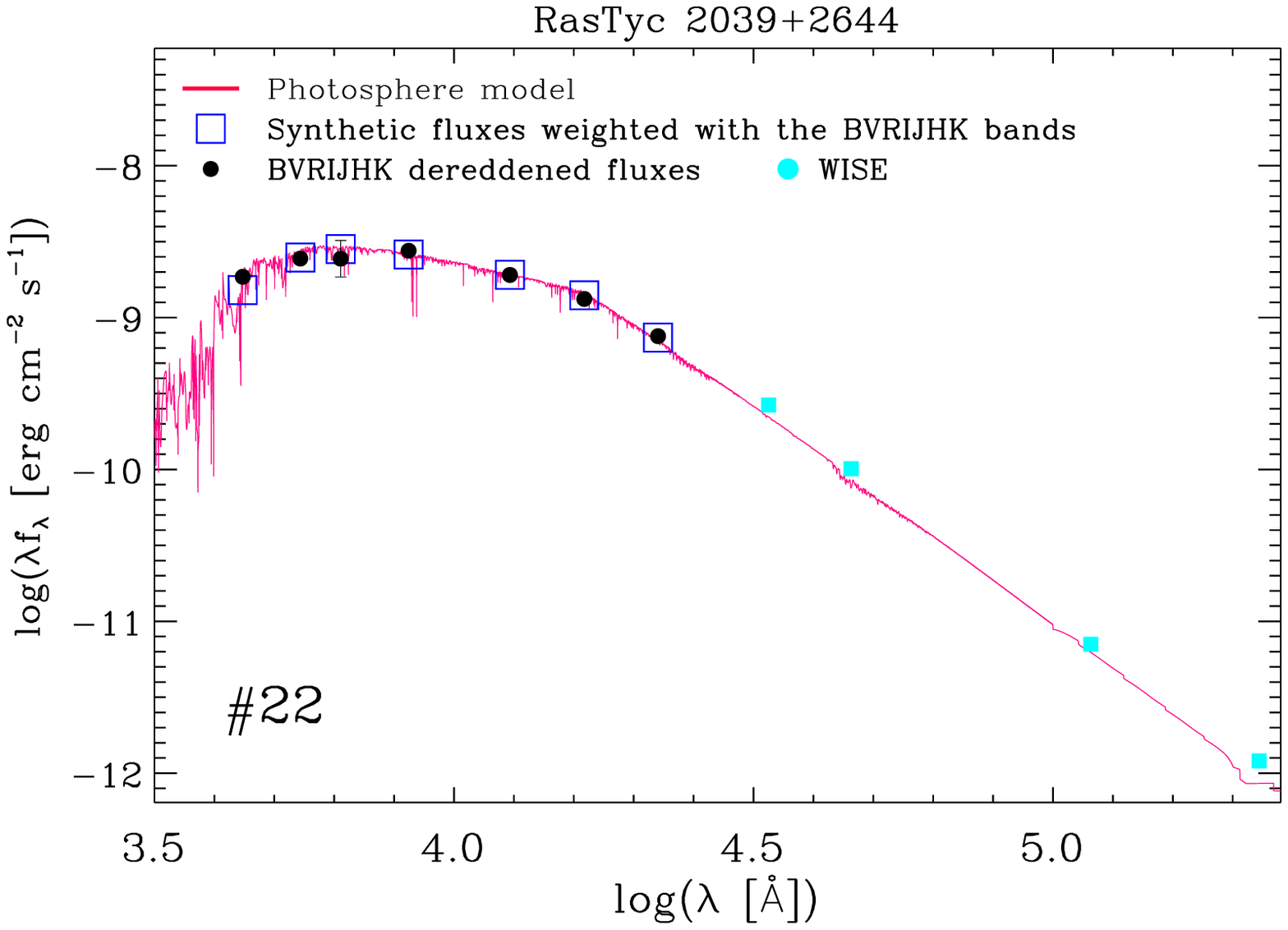}
\includegraphics[width=6.0cm]{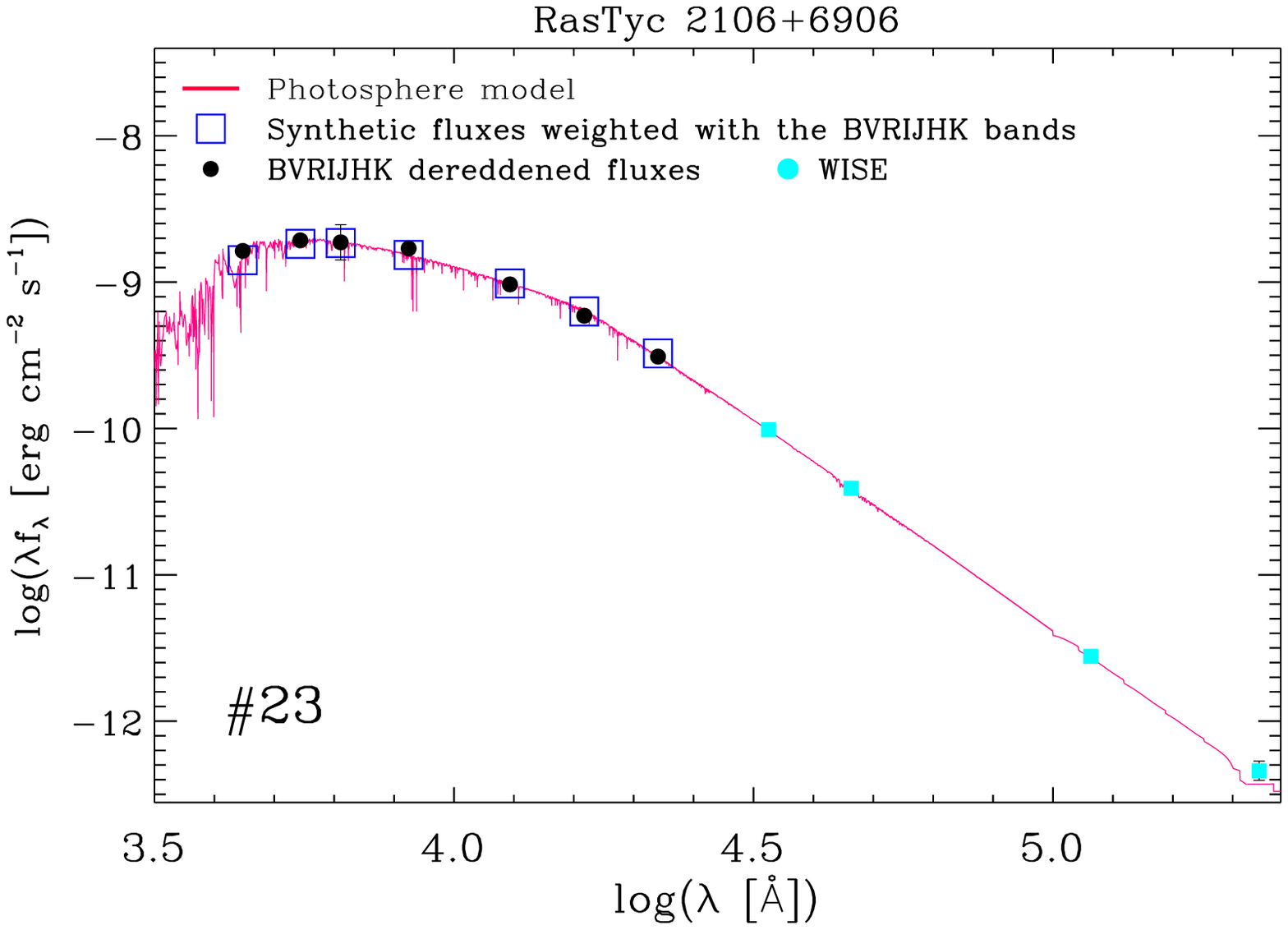}
\includegraphics[width=6.0cm]{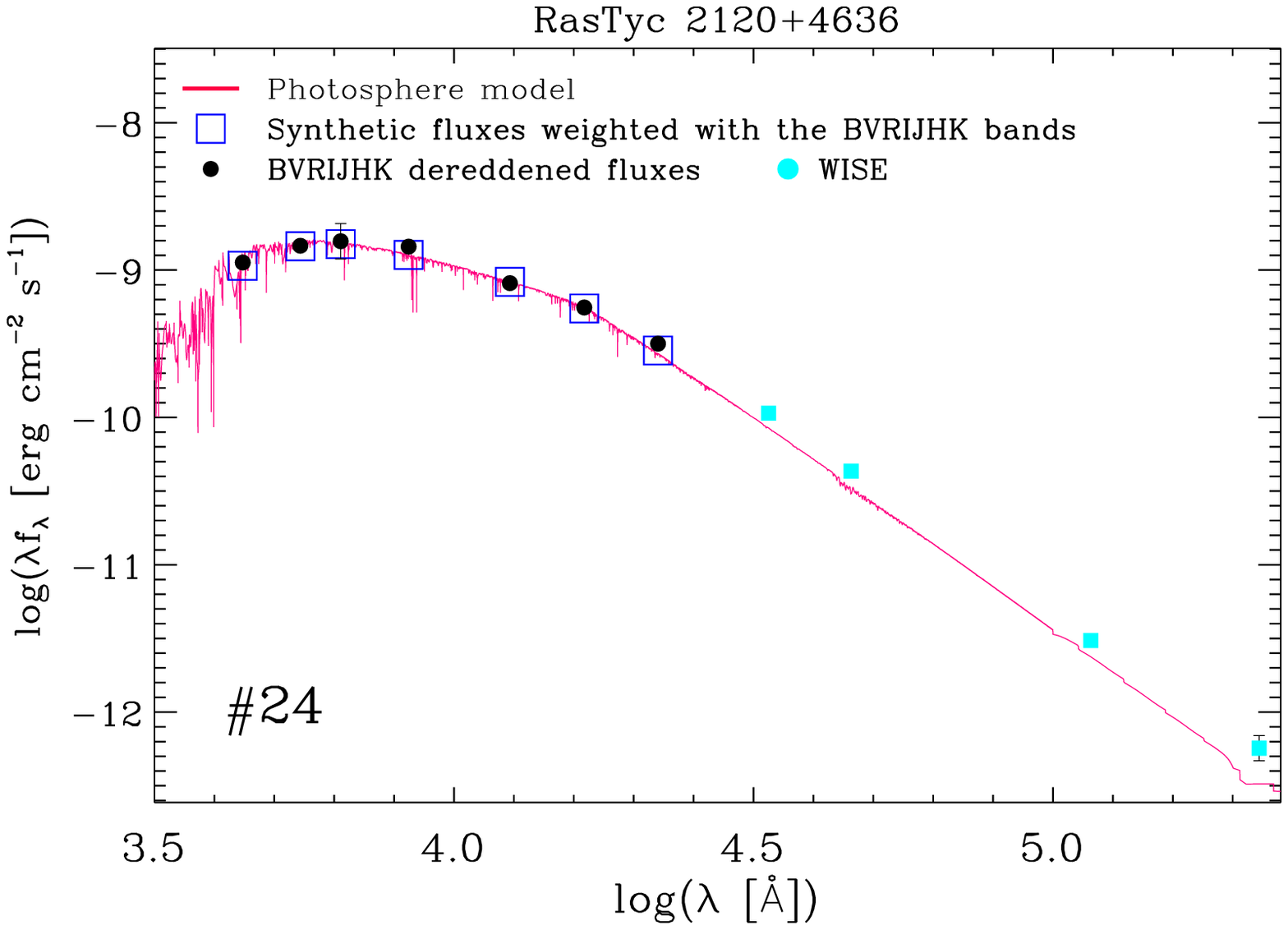}
\includegraphics[width=6.0cm]{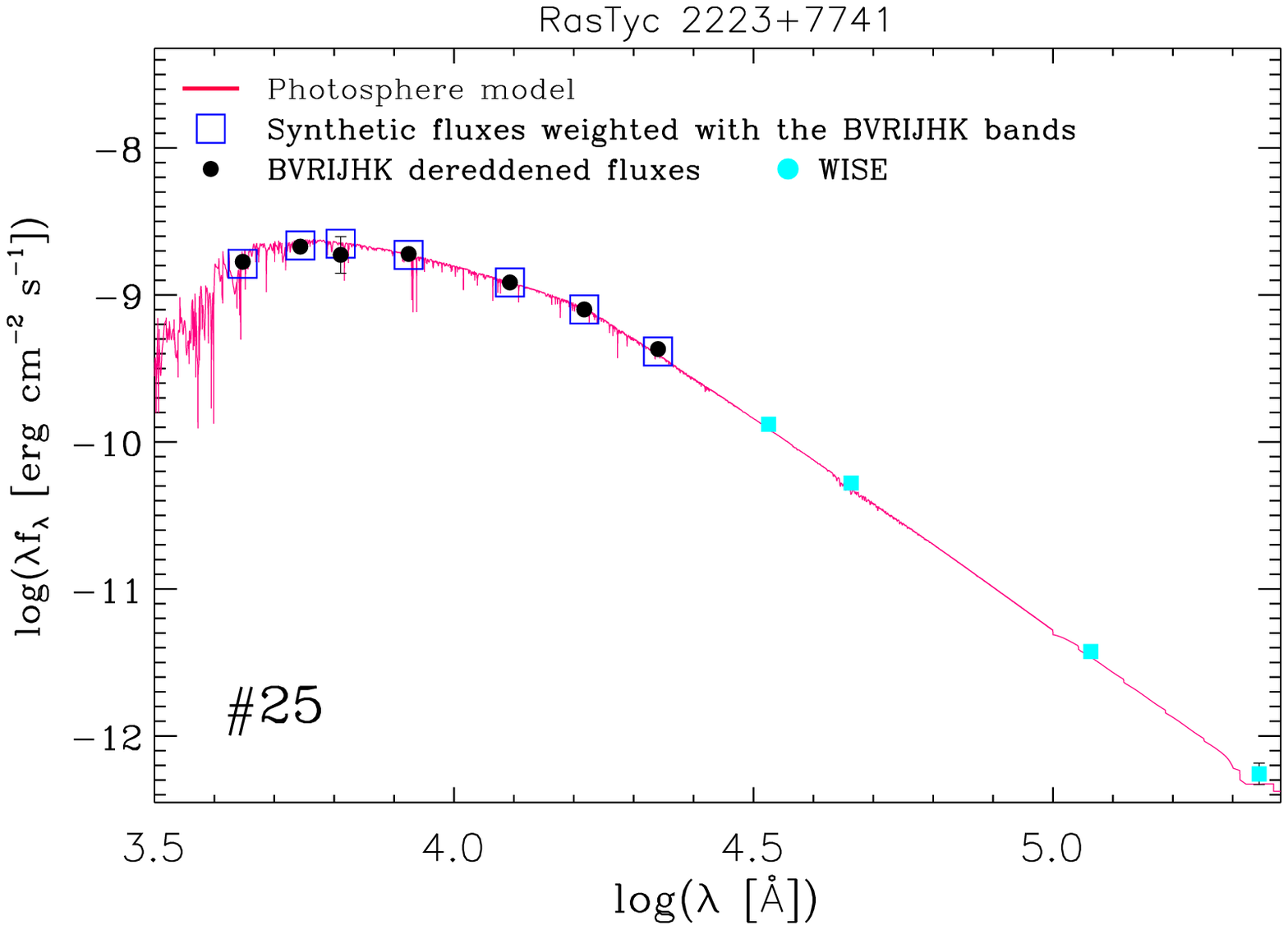}
\includegraphics[width=6.0cm]{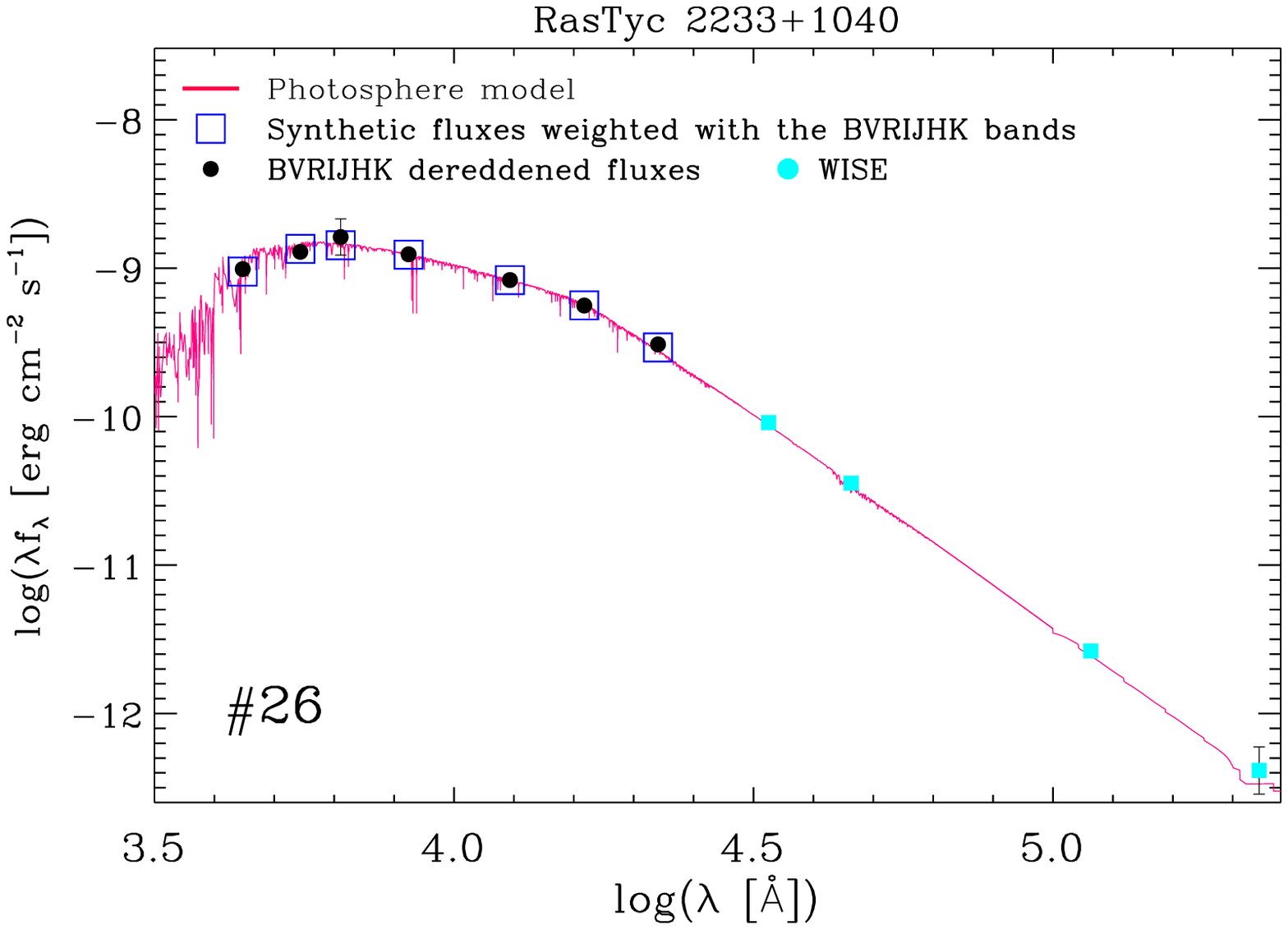}
\includegraphics[width=6.0cm]{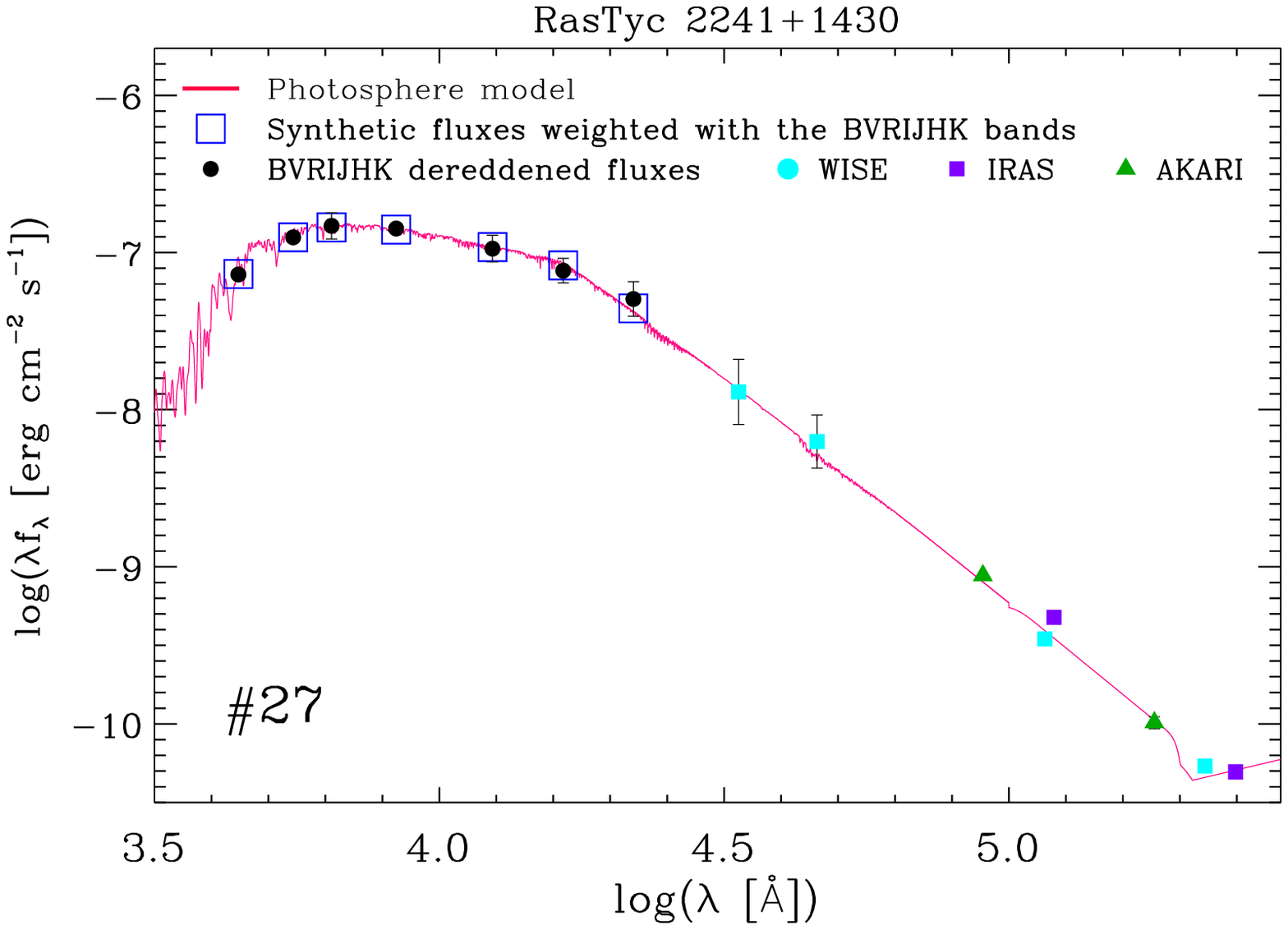}
\includegraphics[width=6.0cm]{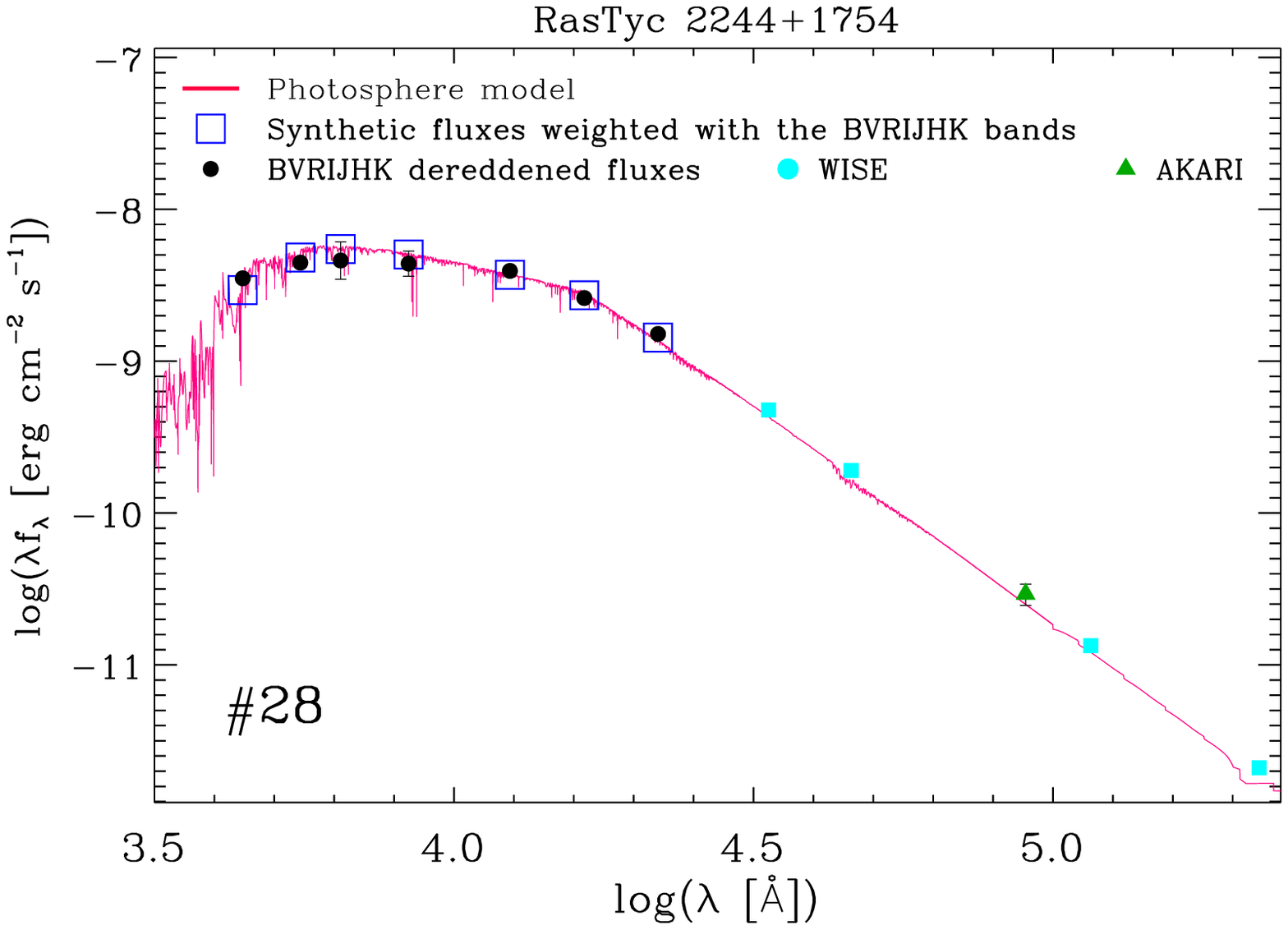}
\includegraphics[width=6.0cm]{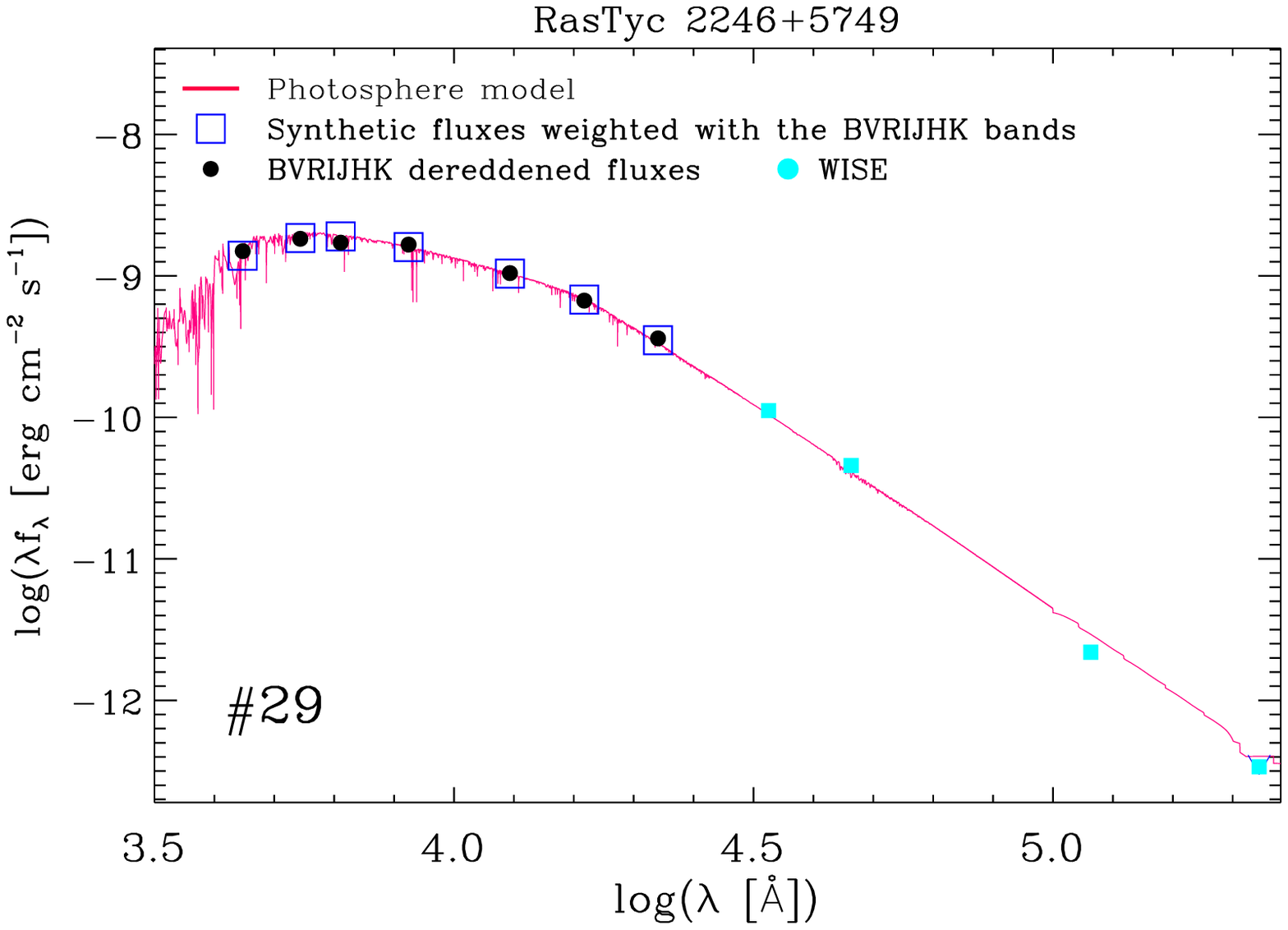}
\includegraphics[width=6.0cm]{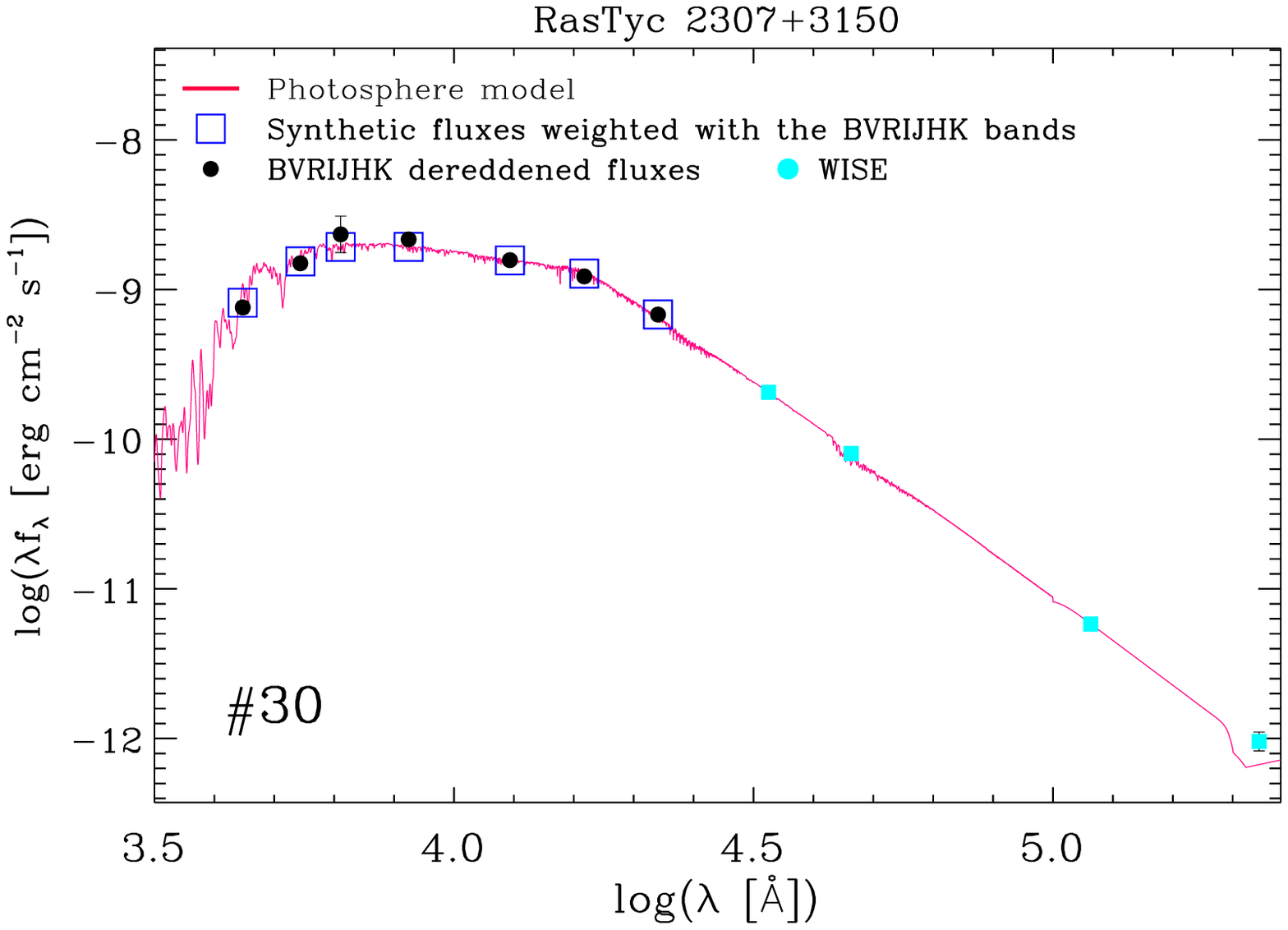}
\caption{continued.}
\end{figure*}

\addtocounter{figure}{-1}

\begin{figure*}[ht]
\includegraphics[width=6.0cm]{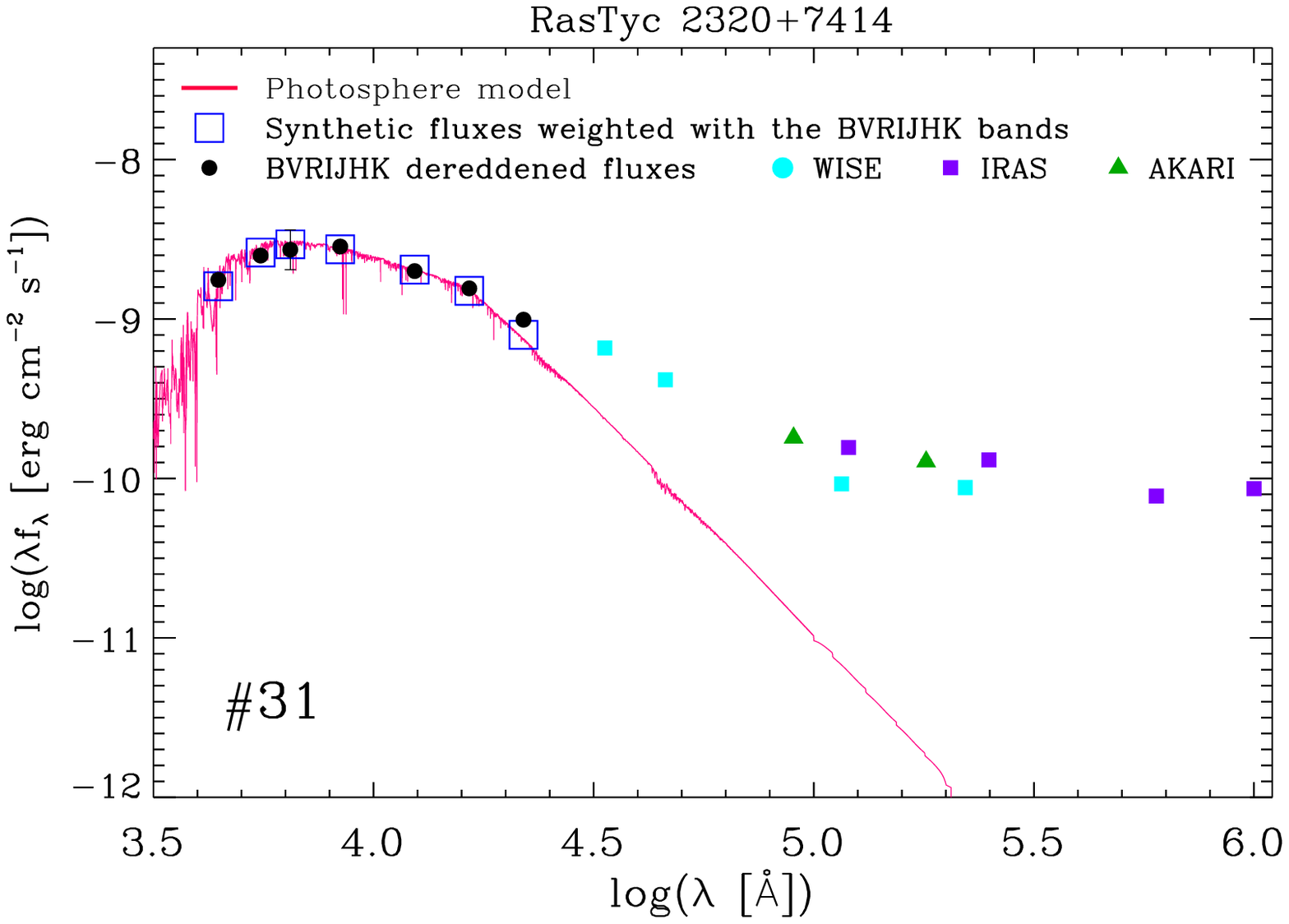}
\includegraphics[width=6.0cm]{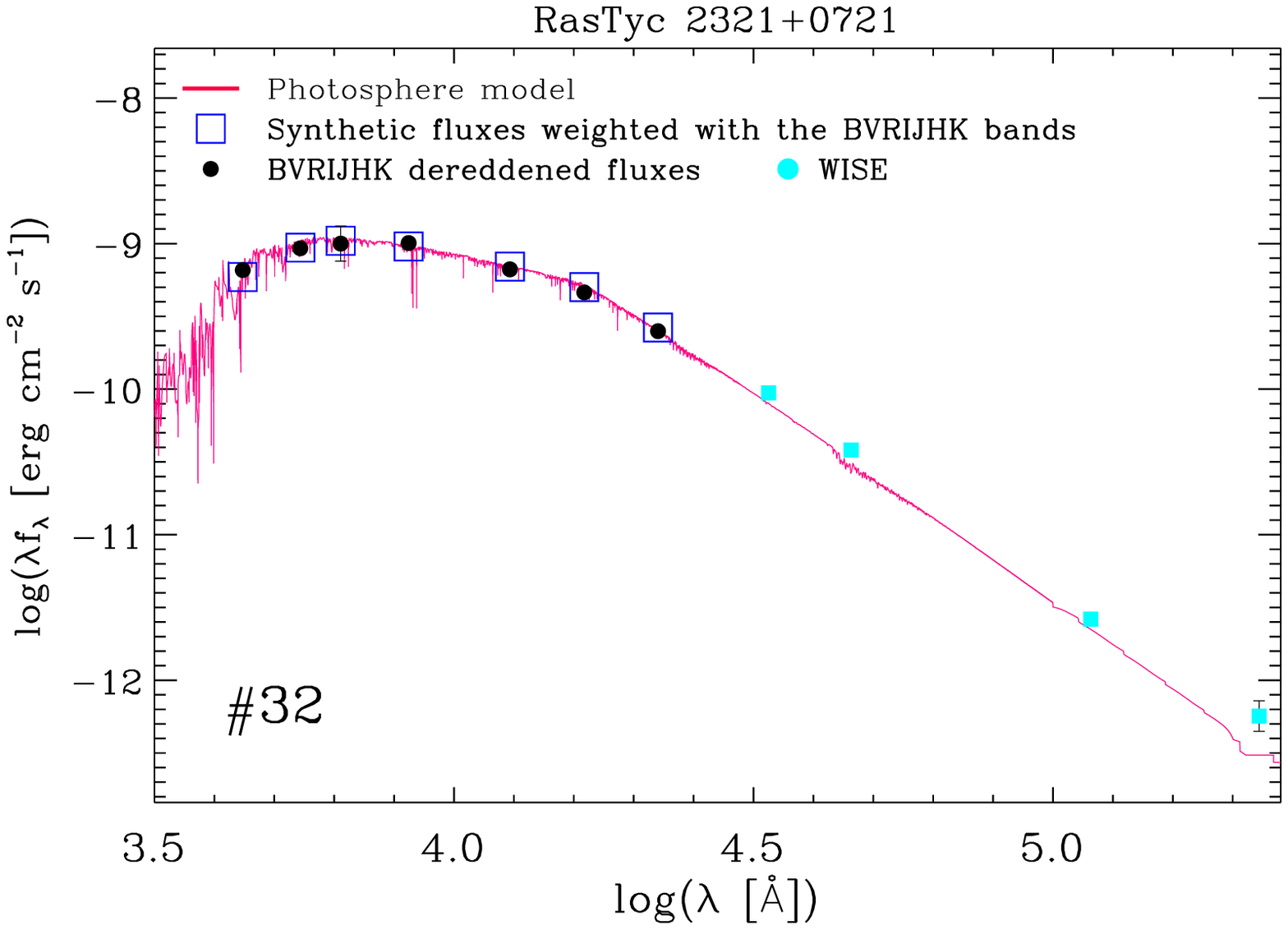}
\includegraphics[width=6.0cm]{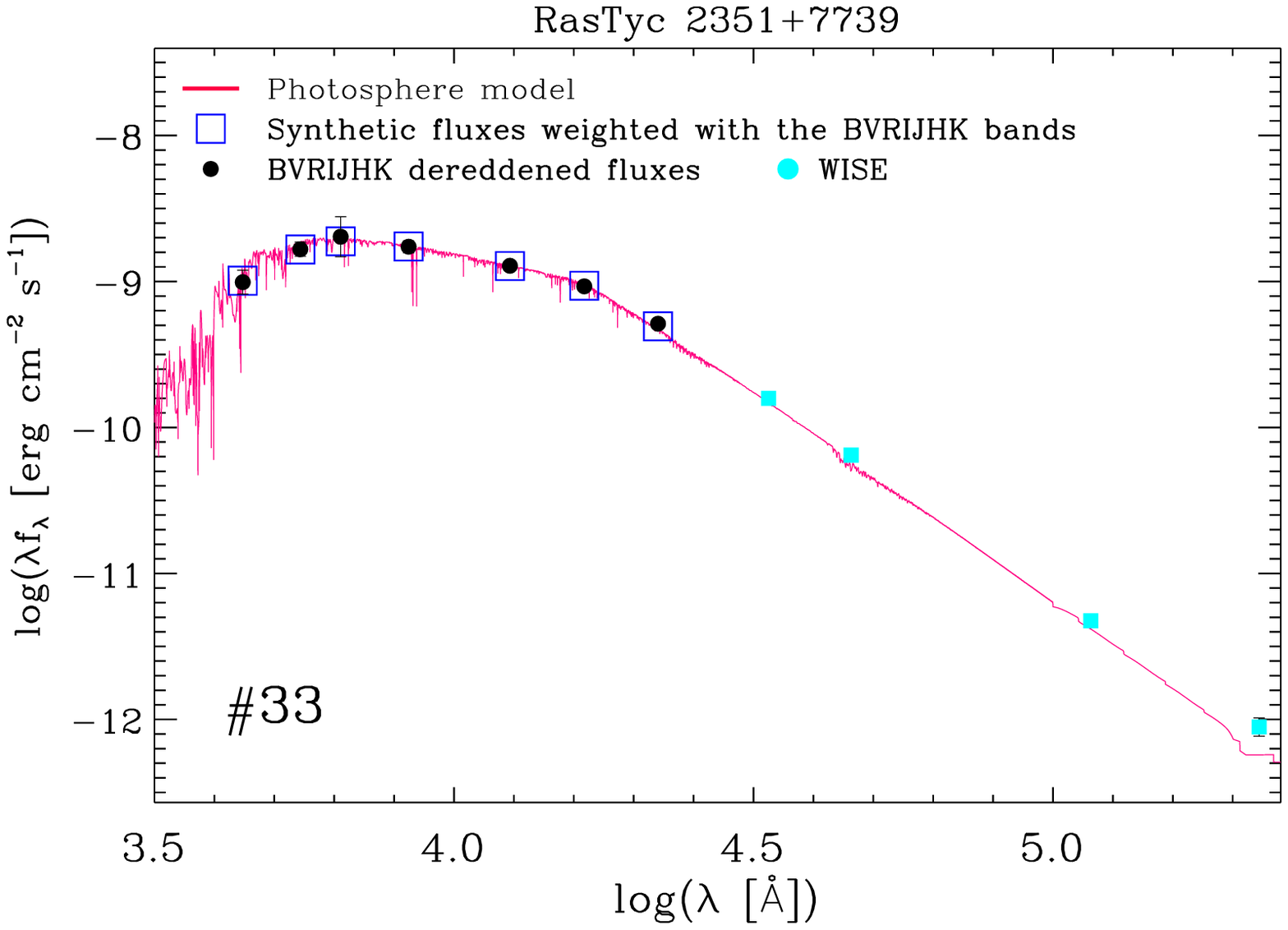}
\includegraphics[width=6.0cm]{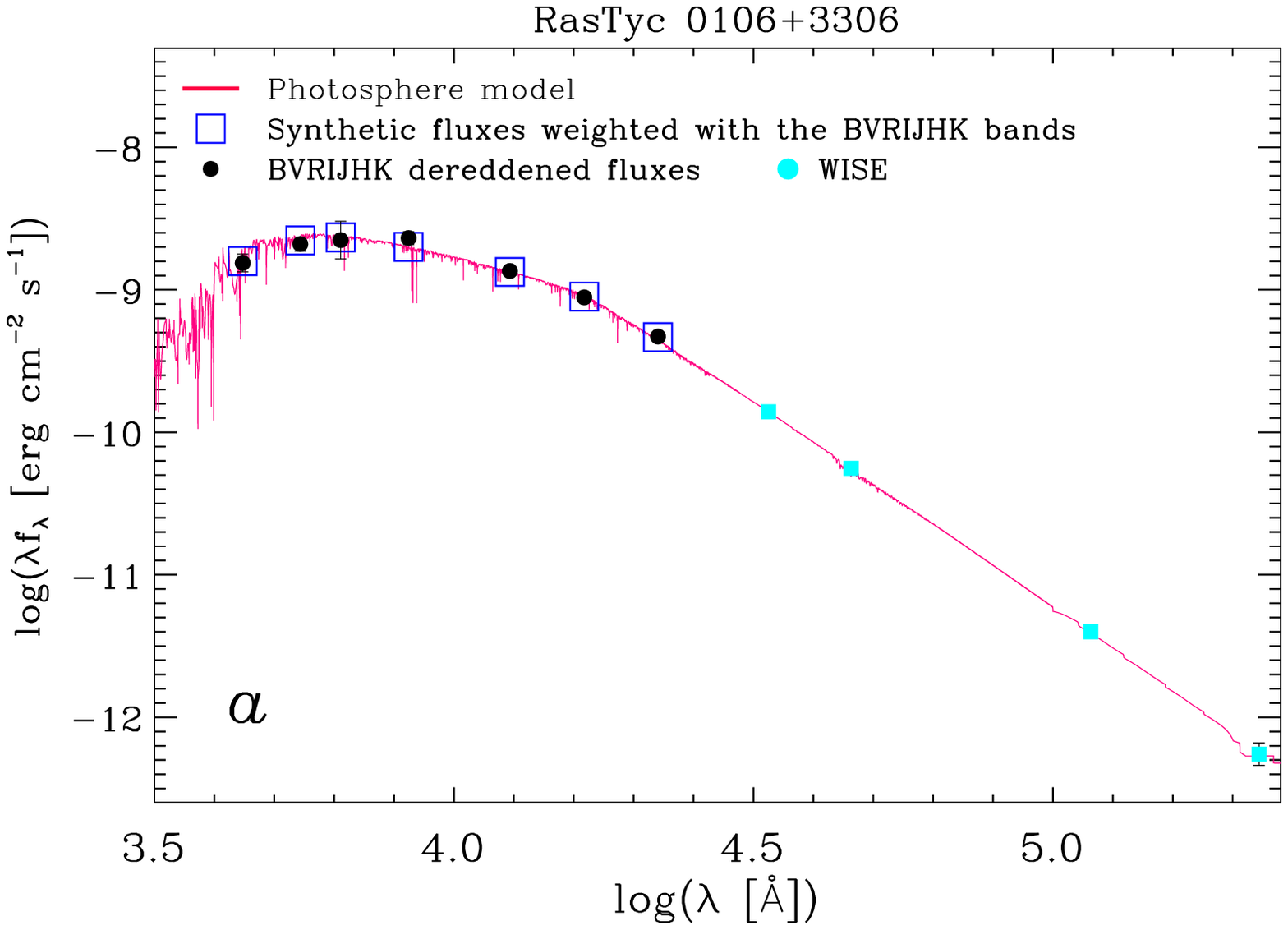}
\includegraphics[width=6.0cm]{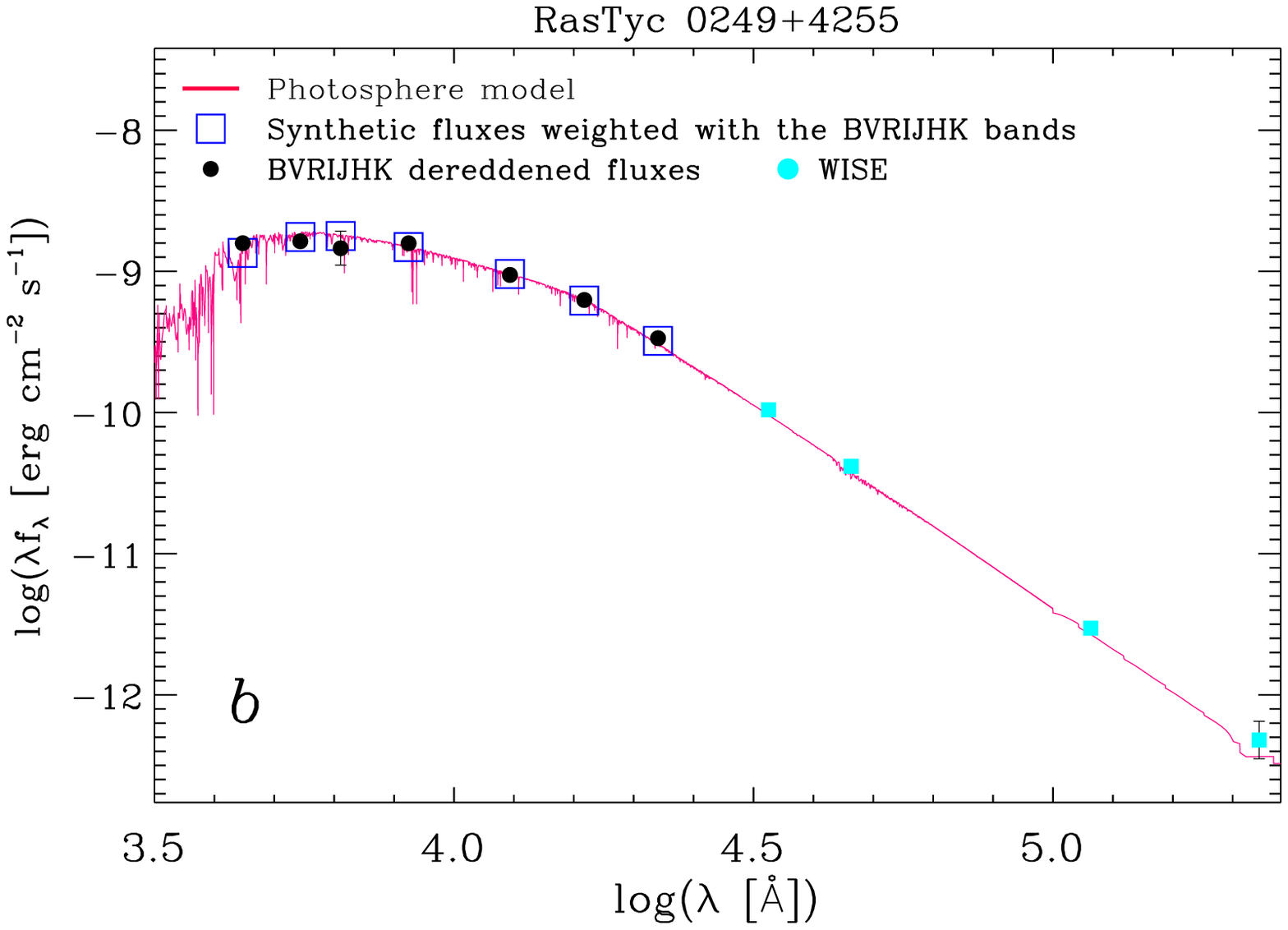}
\includegraphics[width=6.0cm]{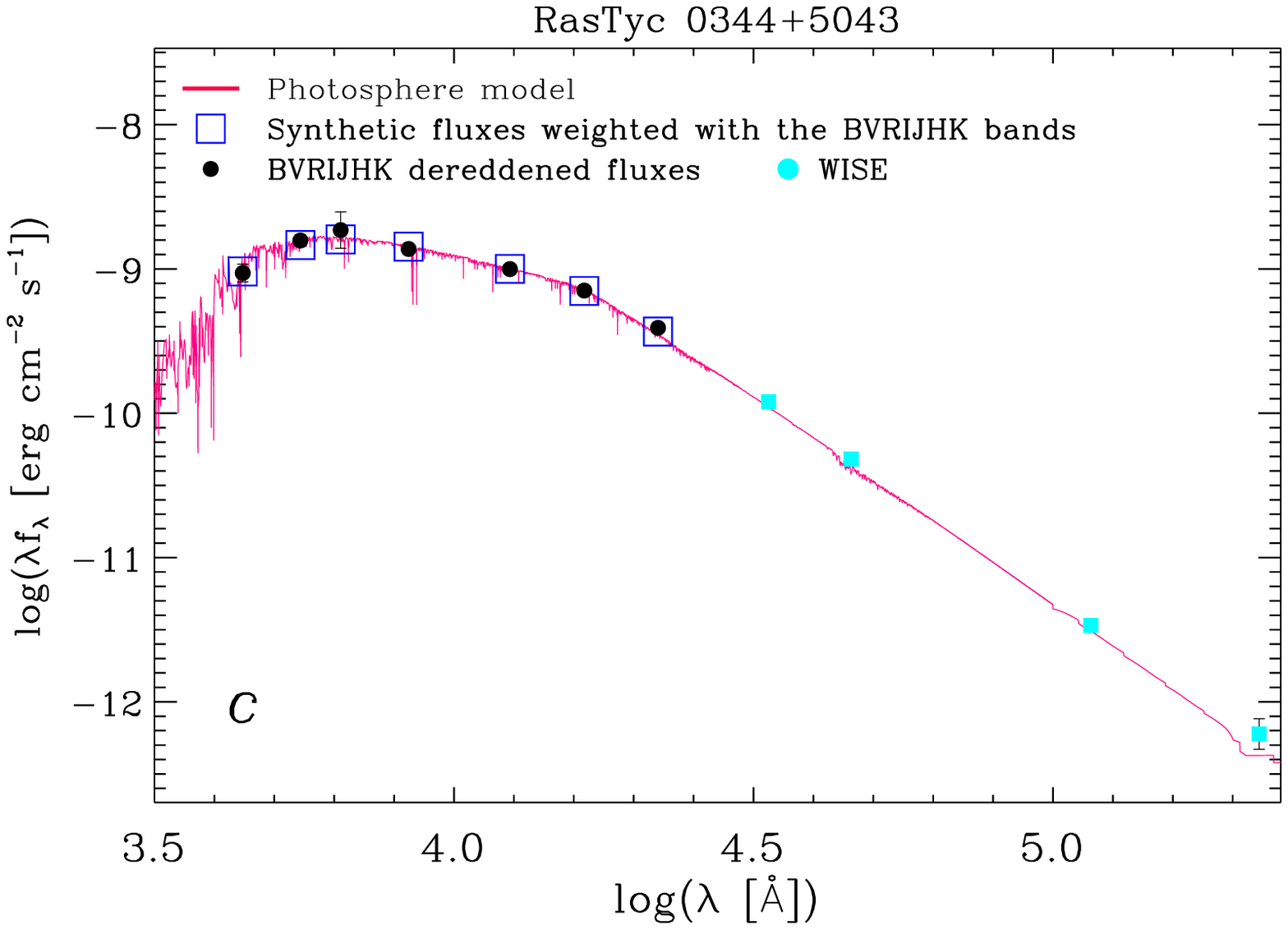}
\includegraphics[width=6.0cm]{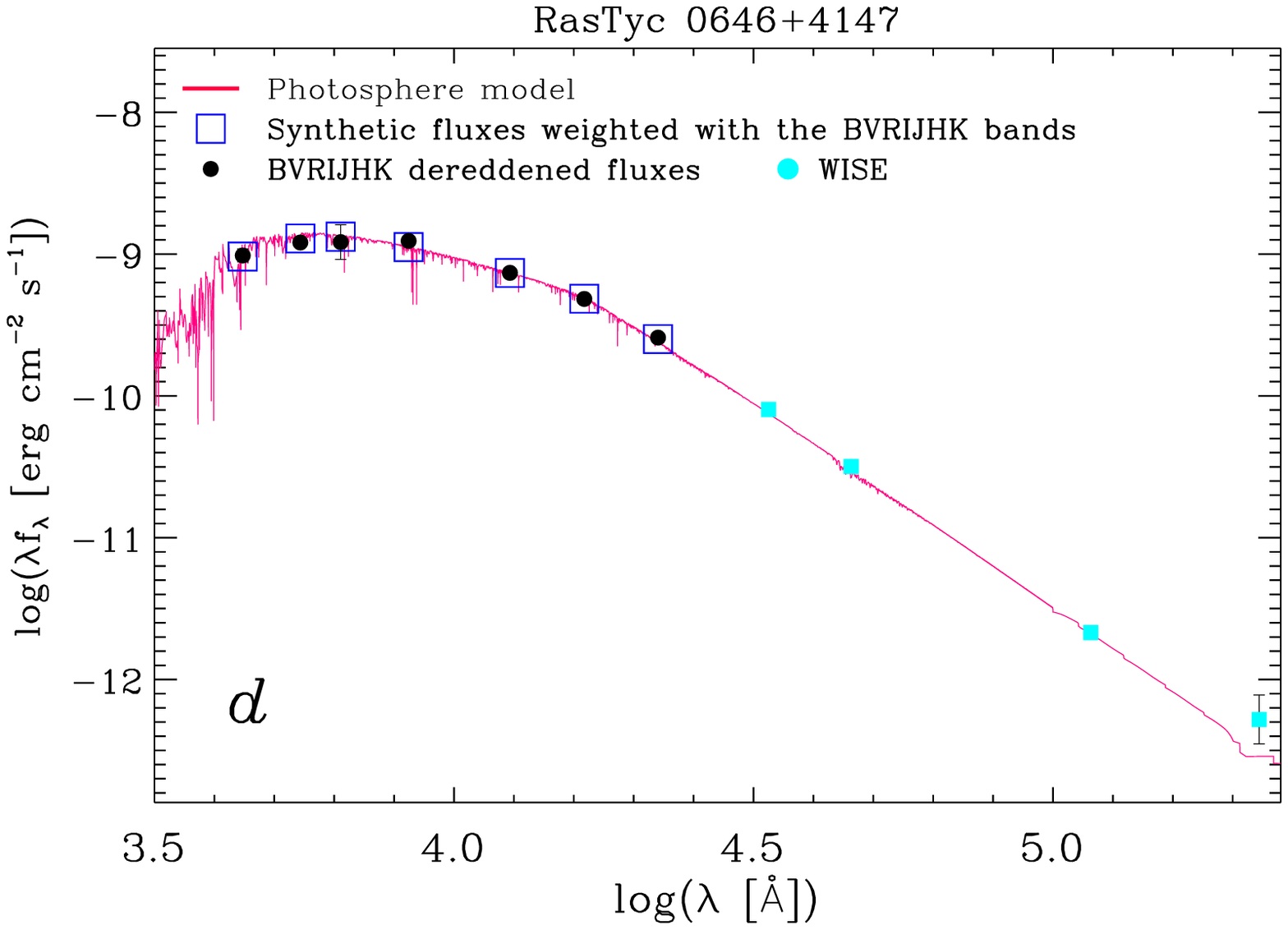}
\includegraphics[width=6.0cm]{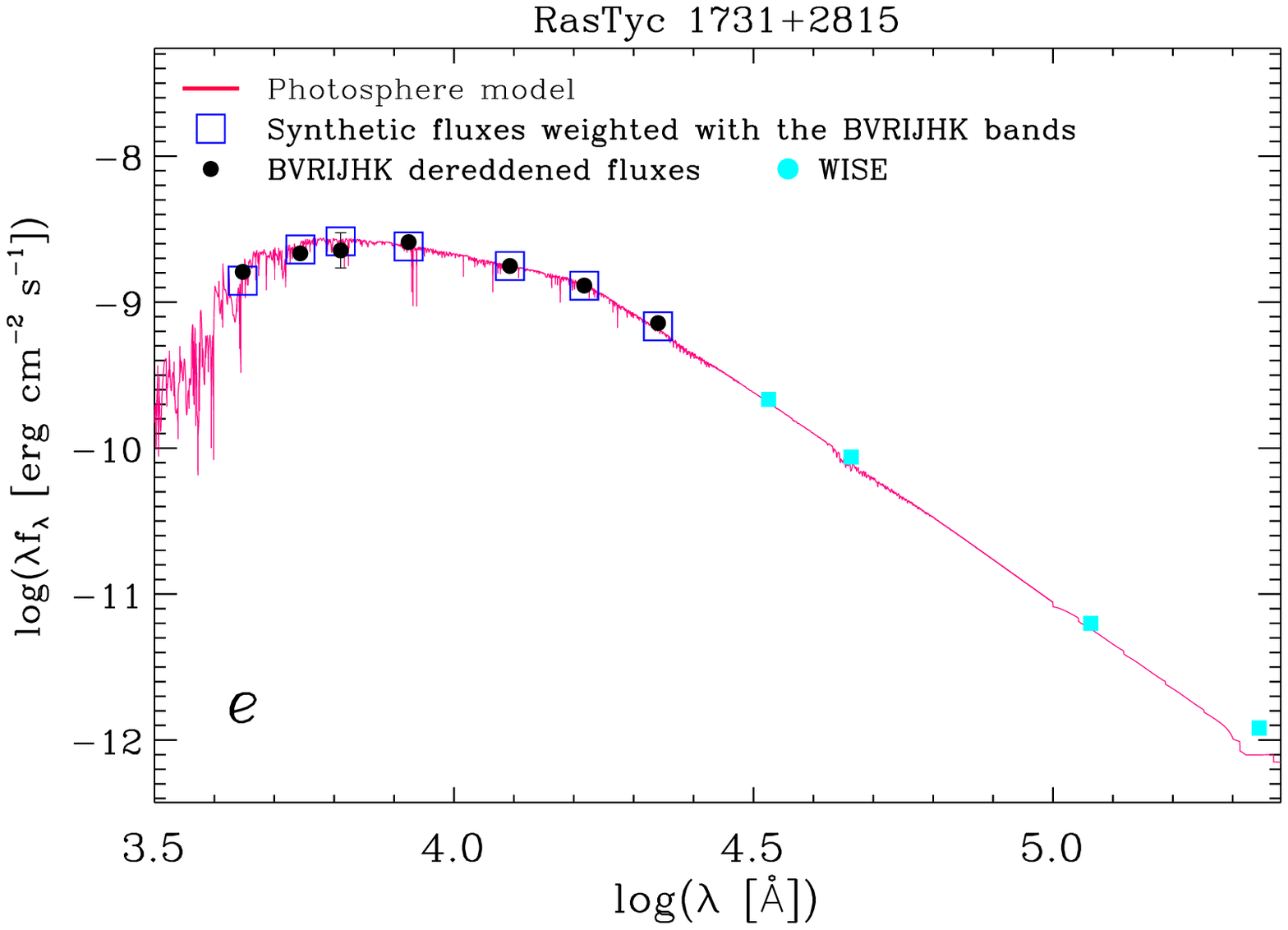}
\includegraphics[width=6.0cm]{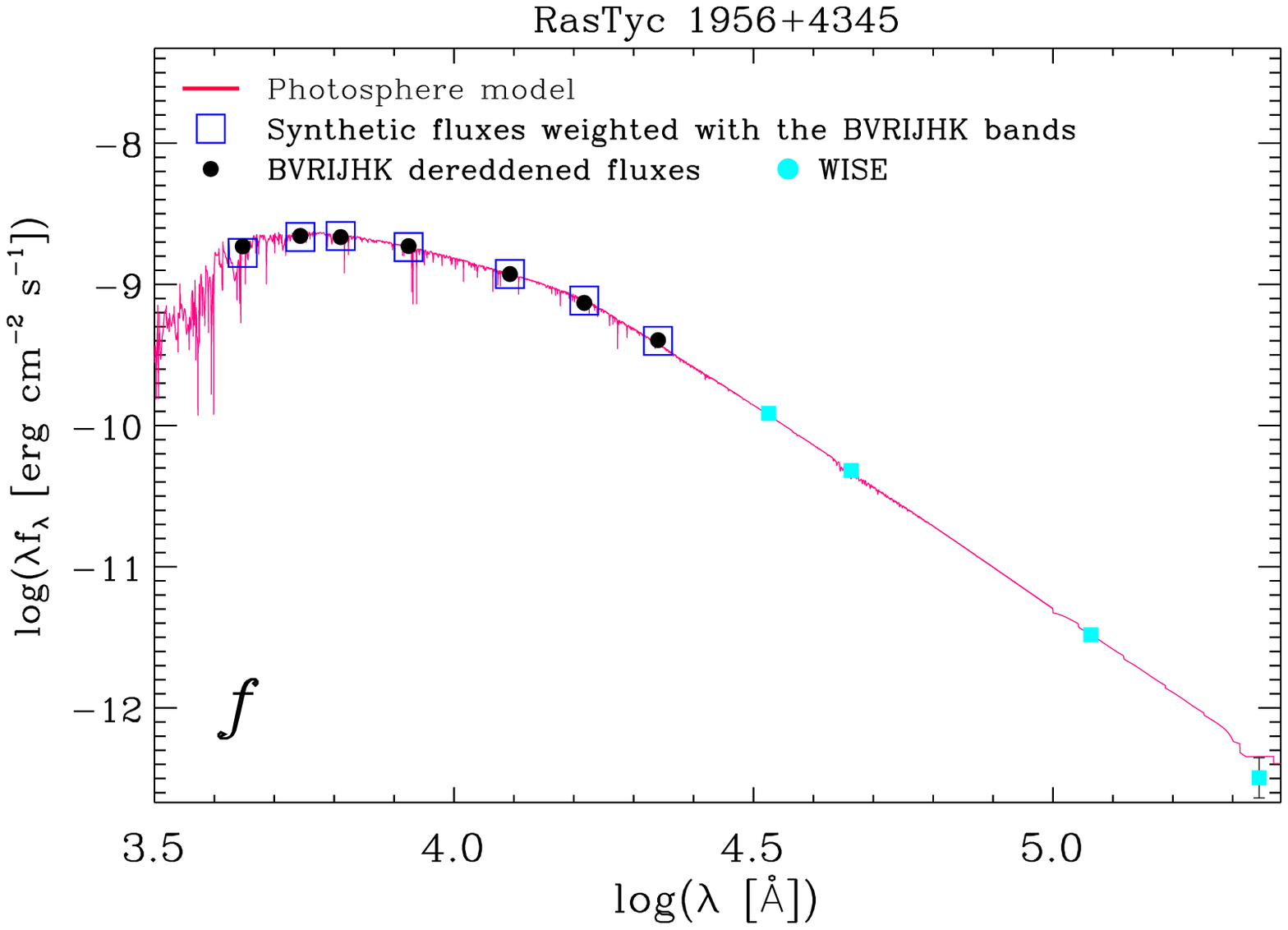}
\includegraphics[width=6.0cm]{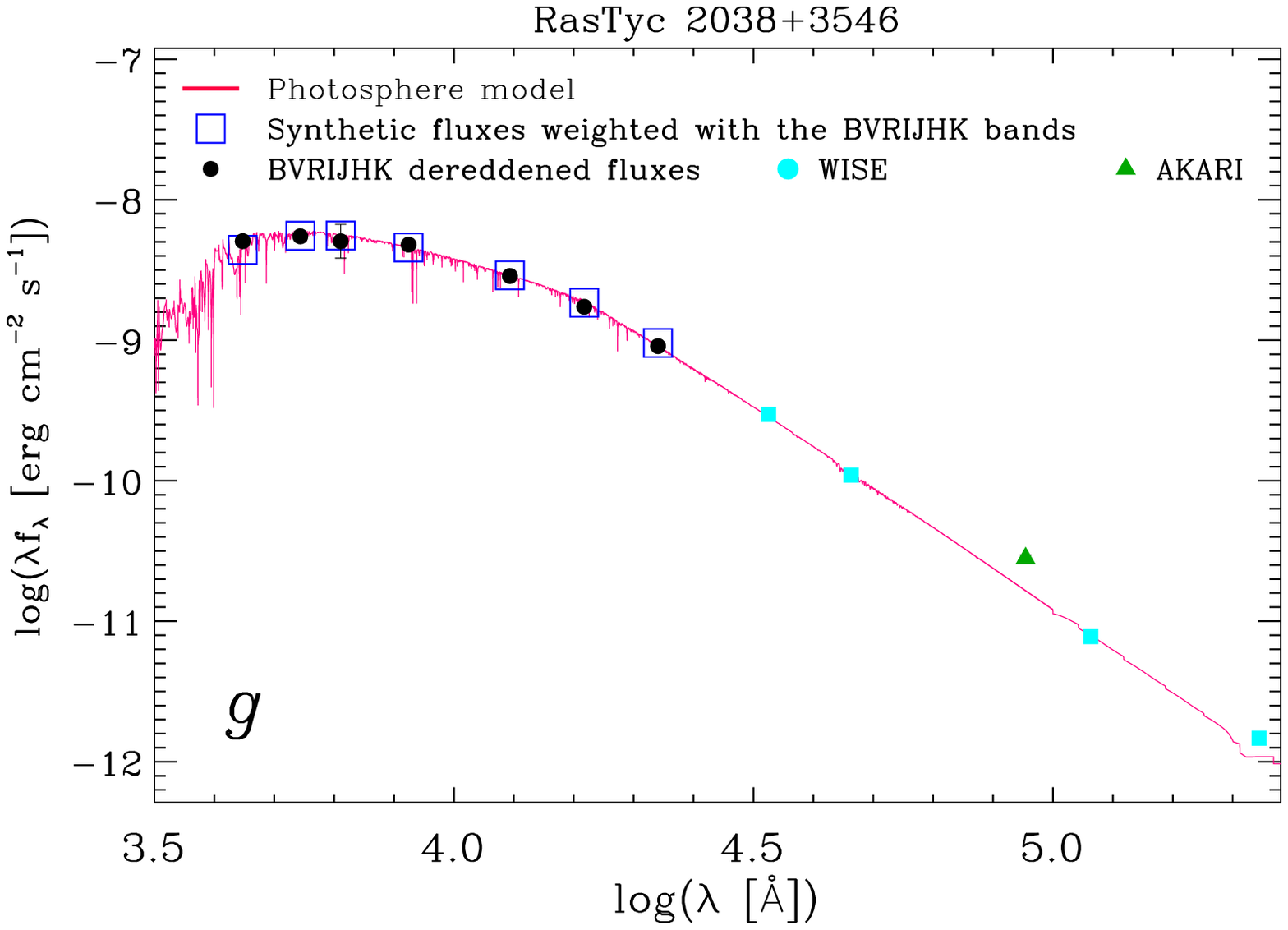}
\includegraphics[width=6.0cm]{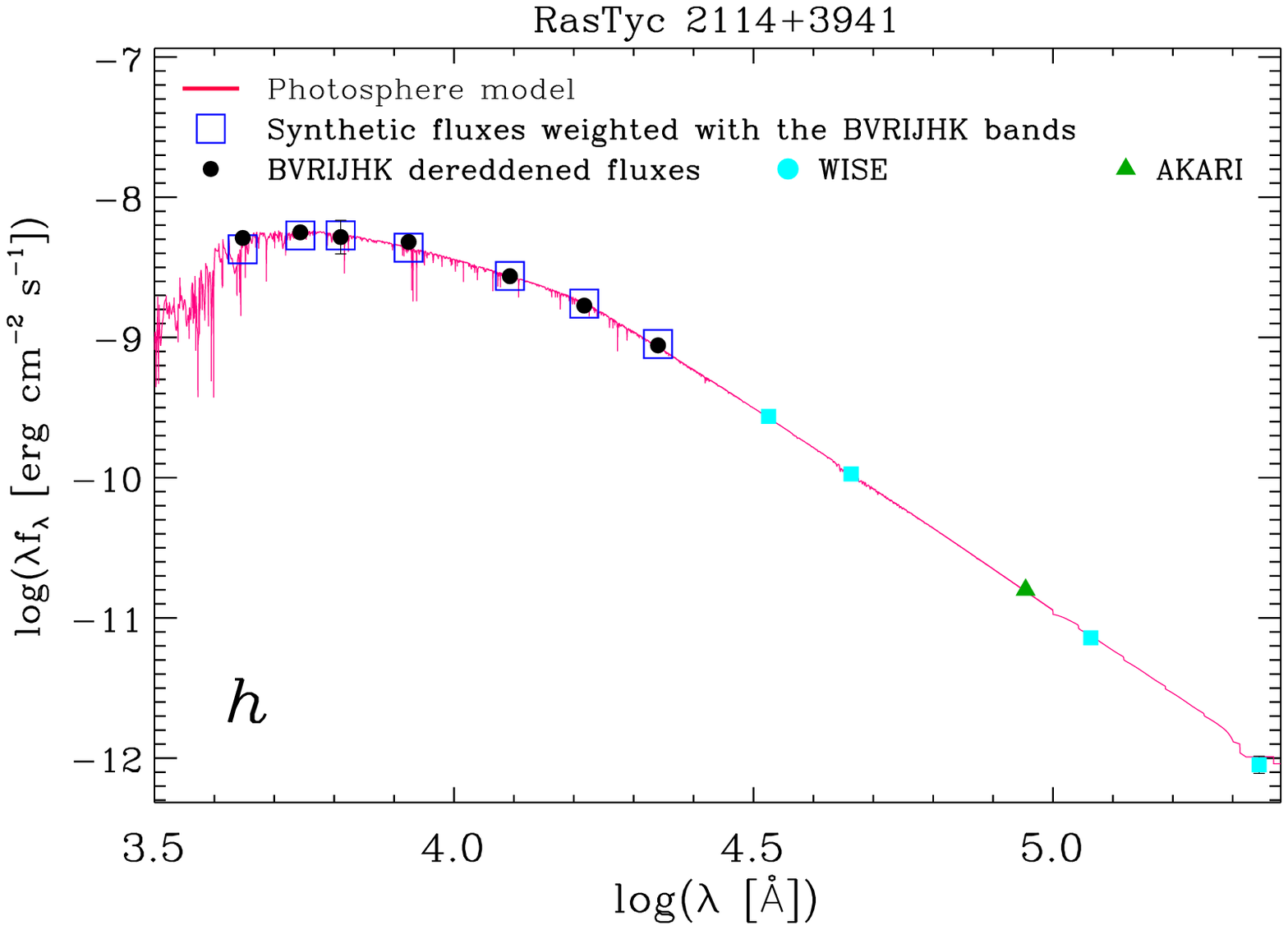}
\includegraphics[width=6.0cm]{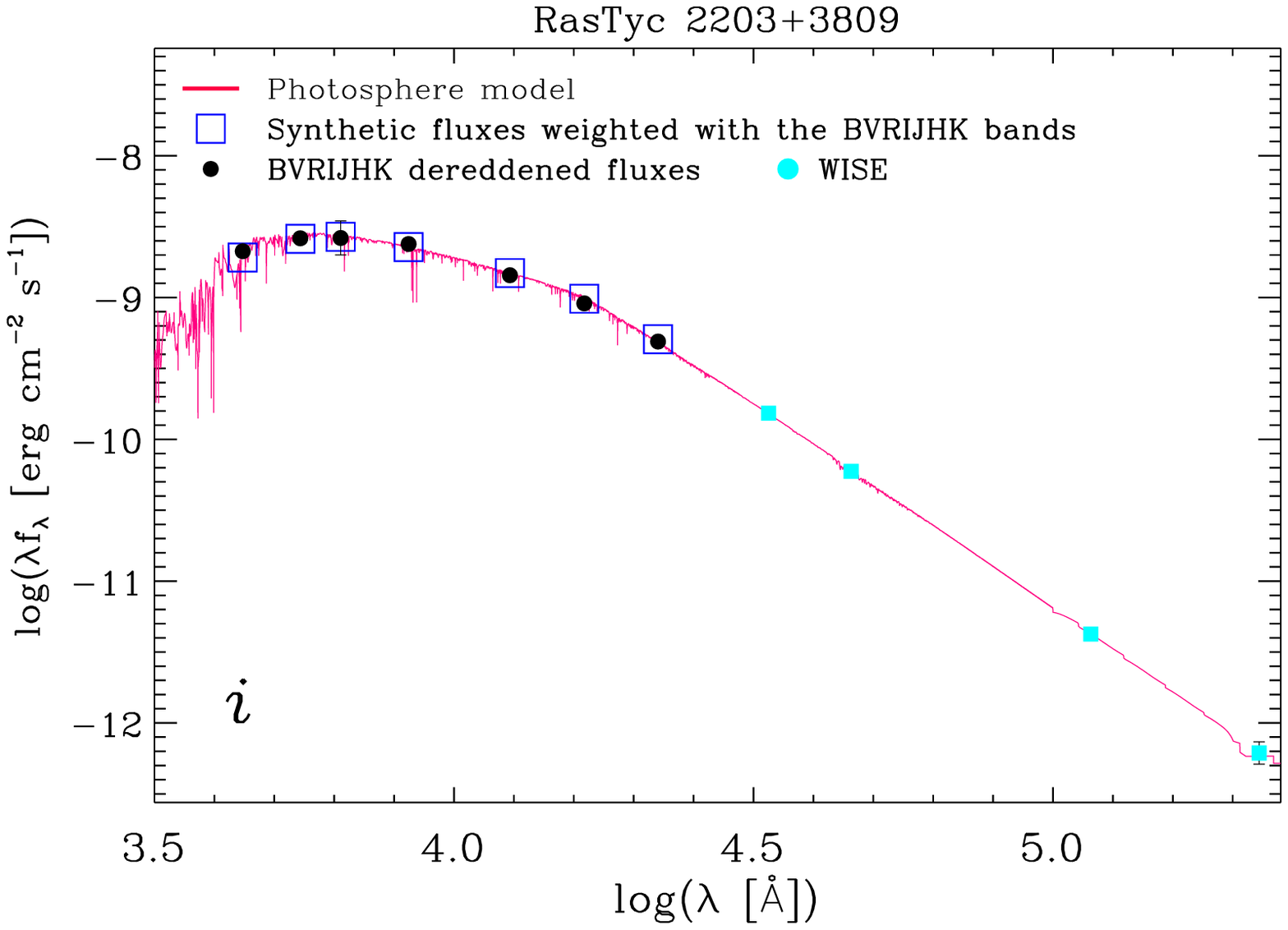}
\includegraphics[width=6.0cm]{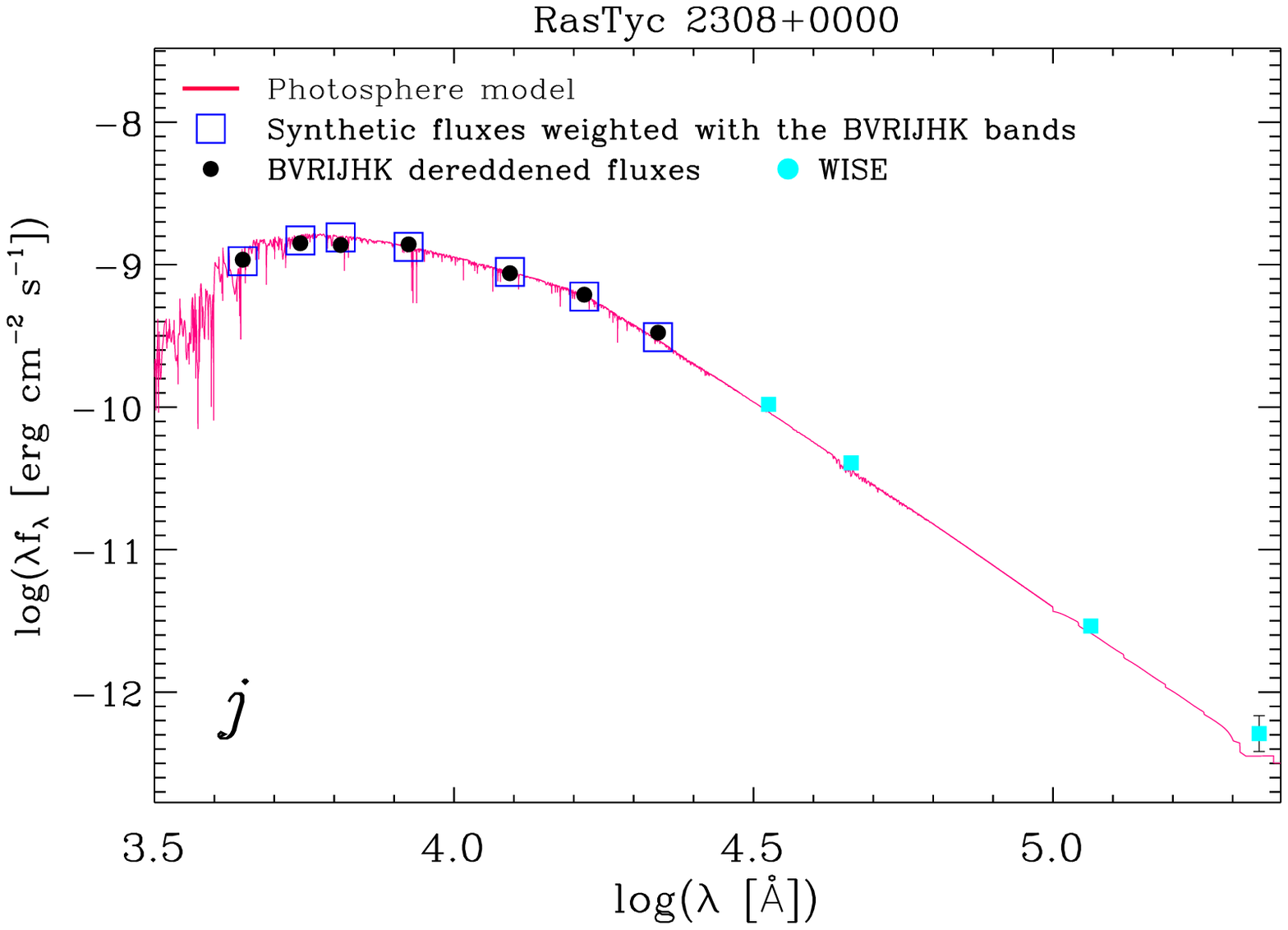}
\includegraphics[width=6.0cm]{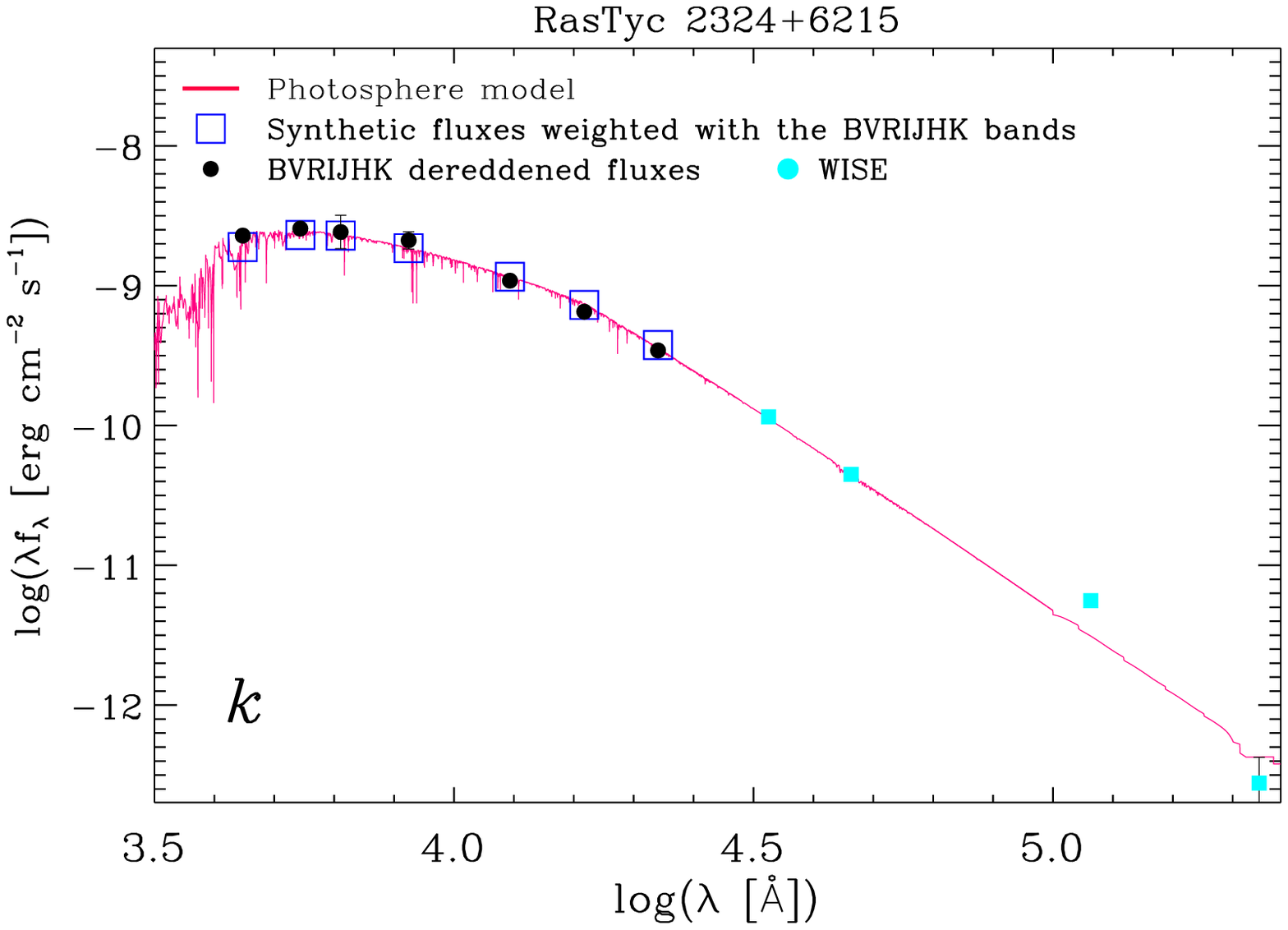}
\caption{continued.}
\end{figure*}

\subsection{Kinematics of the very young sources}
\label{subsec:kinem}

The availability of accurate radial velocities and the \textit{Gaia} DR1 TGAS parallaxes and proper motions \citep{GaiaDR1} allowed us to 
calculate the galactic space-velocity components for most of the sources investigated in the present paper. 
We used the outline of \citet{JohnsonSoderblom1987} to compute the velocity components, $U_{\sun}$, $V_{\sun}$, and $W_{\sun}$ and 
their uncertainties, which we report in a heliocentric, left-handed coordinate system, where $U_{\sun}$ is directed towards the galactic anticenter.

\begin{figure*}[ht]
\includegraphics[width=9.0cm]{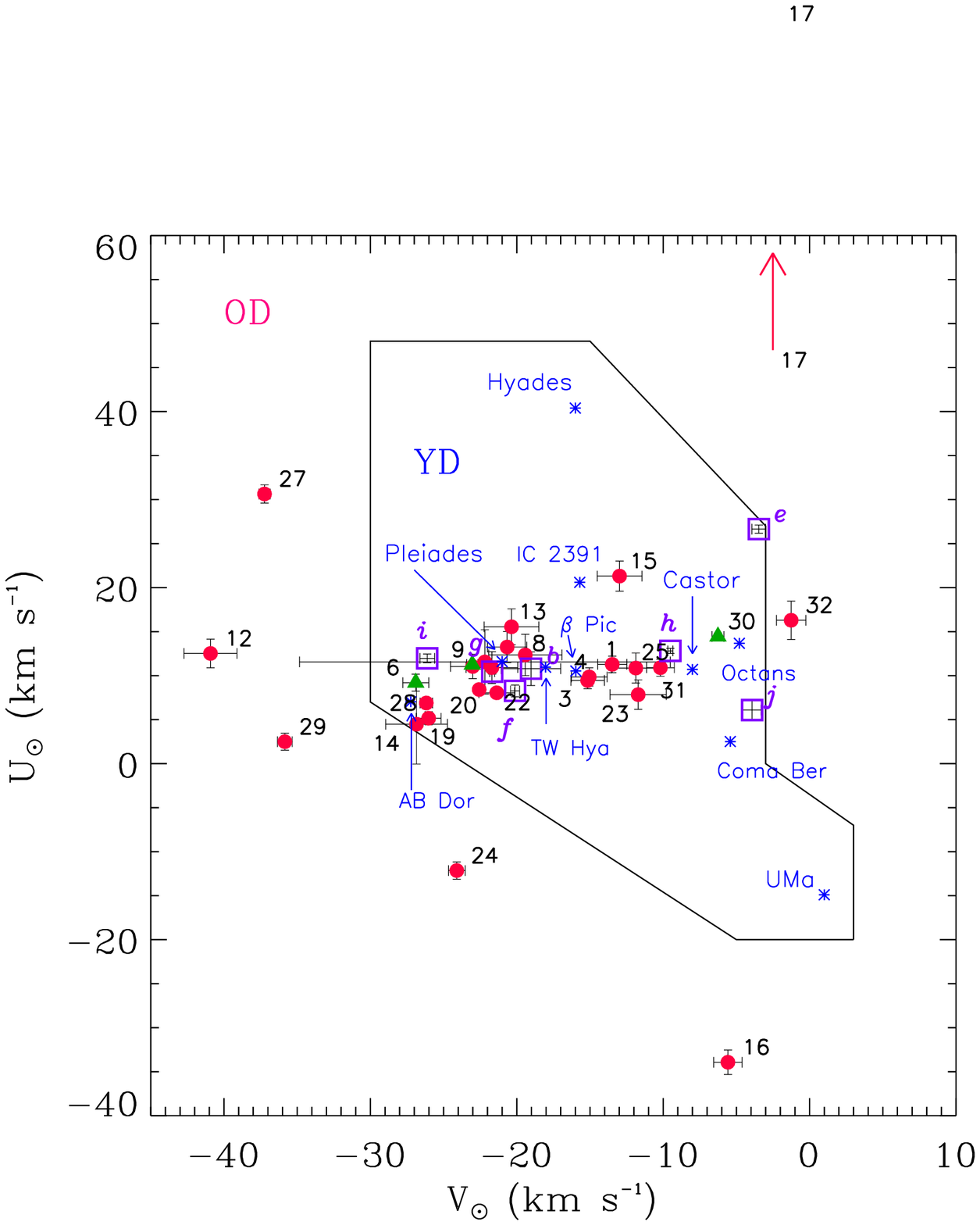}		%uv_rastyc_faint.eps}
\includegraphics[width=9.0cm]{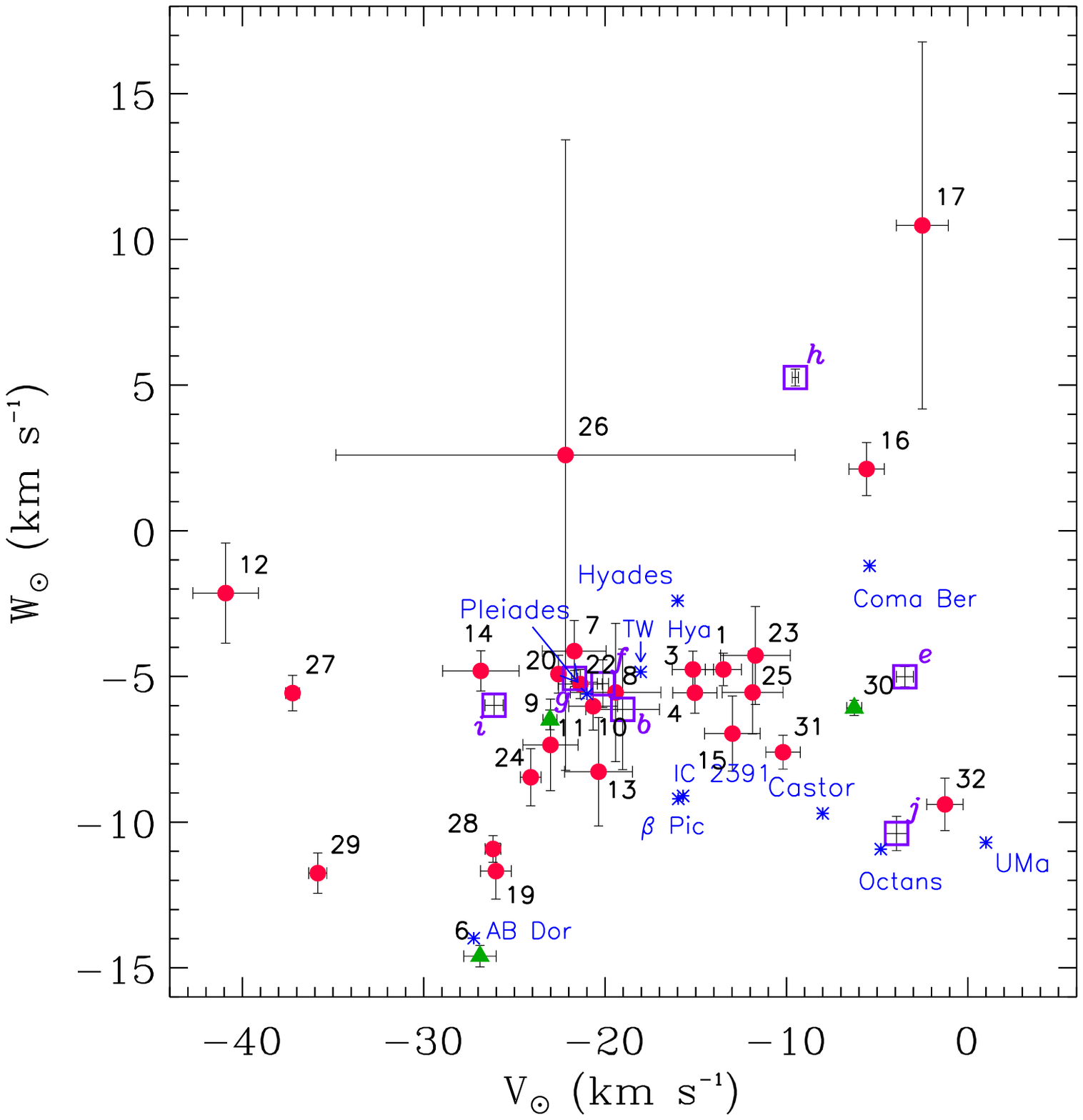}		%vw_rastyc_faint.eps}
\caption{($V_{\sun},U_{\sun}$) (\textit{left panel}) and ($V_{\sun},W_{\sun}$) (\textit{right panel}) diagrams for the {\it PMS-like} sources 
and the two lithium-rich giants discussed in the paper. 
The symbols are as in Fig.~\ref{Fig:Li} and the labels are as in Table~\ref{Tab:LiRich}. The average velocity components (blue asterisks) of 
some young SKGs and the locus of the young-disc (YD) population \citep{Eggen1996} are also marked in the ($V_{\sun},U_{\sun}$) plane. 
The position of the source \#17 ($V_{\sun}=-2.5$\,\kms, $U_{\sun}=83.5$\,\kms) is out of the scale of the plot, as 
indicated by the red arrow. }
\label{Fig:kinematics}
\end{figure*}

Space-velocity components are of great importance to assign membership to a known stellar kinematic group (SKG).
We show in Fig.~\ref{Fig:kinematics} the ($U_{\sun},V_{\sun}$) and ($V_{\sun},W_{\sun}$) diagrams for the {\it PMS-like} stars, along with the average 
position of the five major young SKGs discussed in \citet{Montes2001b}, namely  the IC 2391 supercluster ($\sim$\,50\,Myr), the Local association or Pleiades 
group \citep[$\sim$\,20--150\,Myr][]{Asiain1999,Montes2001b}, the Castor group ($\sim$\,200\,Myr), the Ursa Major (UMa) group ($\sim$\,300\,Myr), and the Hyades supercluster ($\sim$\,600 Myr). 
Additional SKGs and loose associations, such as TW~Hya (3--15\,Myr), $\beta$\,Pic (10--24\,Myr), Octans (20--40\,Myr), AB~Dor (50--150\,Myr), and
Coma Ber ($\sim$\,400\,Myr) have been also considered \citep[e.g.,][and references therein]{Zuckerman2001,Zuckerman2004,Zuckerman2013,Riedel2017}.
The locus of the young-disc population (YD; age $\leq$\,2 Gyr) as defined by \citet[][and references therein]{Eggen1996} is shown as well.

We note that the two lithium-rich giant candidates, \#17 and \#27, lie outside the YD locus, which confirms their nature of evolved stars. 

Most of the {\it PMS-like} sources have $U_{\sun}$, $V_{\sun}$, and $W_{\sun}$ compatible with the Pleiades and/or Castor and/or TW~Hya SKGs, which are 
the youngest SKGs considered in this study. However, \#12, \#16, \#24, and \#29, are located far outside of the YD locus in the ($U_{\sun},V_{\sun}$) plane.
With the exception of \#16, for which we measured the same RV in the two spectra within the errors, the other three sources have been observed only once.
Therefore we cannot exclude that they are SB1 systems observed far from the conjunctions, so that their barycentric RV could be very different from the
value measured by us. This would lead to wrong space velocity components.

Some notes about the possible association to the aforementioned SKGs are given in Appendix\,\ref{Sec:notes}.

\section{Summary}
\label{Sec:Conclusions}

We have presented the results of a high-resolution spectroscopic survey of optical counterparts of X-ray sources. 
Our targets were selected from the {\it RasTyc} sample \citep{Guillout1999}, which is obtained by the
cross-correlation of the TYCHO and RASS catalogs. In particular, we have measured radial (RV) and projected rotational velocities
(\vsini) for 443 stars, most of which are optically faint {\it RasTyc} sources ($V\ge 9.5$\,mag). We found 114 double-lined 
spectroscopic binaries (SB2) and 12 triple systems among these sources.  

For the remaining targets, including 38 single-lined binaries (SB1), we were also able to determine the atmospheric 
parameters (\teff, \logg, and \feh), lithium abundance, and the level of chromospheric activity as measured by the ratio of H$\alpha$ 
and bolometric luminosity.
The trigonometric parallax from the TGAS catalog \citep{GaiaDR1} or from the catalog of \citet{vanLeeuwen2007} is also available
for 263 of the stars with measured parameters.
The position of these stars in the HR diagram is in very good agreement with the gravities derived with our analysis code (ROTFIT).

The equivalent width of the \ion{Li}{i}\,$\lambda$6707.8 and the lithium abundance allowed us to perform an age classification
of our targets that were divided in five classes,  \textit{PMS-like} (14\,\%), \textit{Pleiades-like} (13\,\%), 
\textit{UMa-like} (34\,\%), \textit{Hyades-like} (19\,\%), and \textit{Old} stars (20\,\%). The higher percentage of \textit{PMS-like}
and \textit{Pleiades-like} stars found in comparison with the bright {\it RasTyc} sample is likely the result of the greater distances
reached in the present work for objects optically fainter.
Indeed, at larger distances, the X-ray brighter sources (younger stars) are more easily detected compared to less active 
(older) stars.

We have investigated in more detail the 44 \textit{PMS-like} candidates and found that two of them must be rejected from this
class, because their position in the HR diagram, their spectral classification, and their space velocity components suggest they are 
lithium-rich giants. The remaining \textit{PMS-like} sources display a rather high level of chromospheric activity, $\log(R'_{H\alpha})>-5.0$ 
and, with few exceptions, are located in the domain of the ($V_{\sun},U_{\sun}$) plane occupied by young-disk stars. The two objects with 
the highest values of $R'_{H\alpha}$ (\#2 and \#31) are very close to the dividing line between chromospheric sources and accreting stars
defined by \citet{Frasca2015}. 
They both display a broad and double-peaked H$\alpha$ profile that is typical of CTTS, as well as a remarkable IR excess in their SEDs. 

Another important outcome of this survey is the presence of a significant fraction of giant and subgiant stars ($\sim$\,12\,\%).
Half of them also display a high activity level, comparable to that of some \textit{PMS-like} sources, and a rather rapid rotation, which 
could be the result of a particular evolutionary path or the effect of an undetected binarity (spin-orbit synchronization).  
Some of these evolved stars ($\sim$\,36\,\%) are also rich in lithium ($A$(Li)$>$\,1.4).

\begin{acknowledgements}

This paper is dedicated to the memory of our colleague and friend Rubens Freire Ferrero.\\
We thank the anonymous referee for useful suggestions.
We are grateful to the TNG staff and, particularly, to Aldo F. Fiorenzano and Antonio Magazz\`u for conducting the service observations with SARG.
We also thank the night assistants of the OHP and OAC observatories for their support and help with the observations. 
Support from the Italian {\it Ministero dell'Istruzione, Universit\`a e  Ricerca} (MIUR) is also acknowledged.
D.M. acknowledges financial support from the Universidad Complutense de Madrid (UCM) and the Spanish Ministry of Economy
and Competitiveness (MINECO) from project AYA2016-79425-C3-1-P.
This research made use of SIMBAD and VIZIER databases, operated at the CDS, Strasbourg, France. This publication uses ROSAT data. 
This publication makes use of data products from the Wide-field Infrared Survey Explorer, which is a joint 
project of the University of California, Los Angeles, and the Jet Propulsion Laboratory/California
Institute of Technology, funded by the National Aeronautics and Space Administration.
This work has made use of data from the European Space Agency (ESA)
mission {\it Gaia} ({\tt https://www.cosmos.esa.int/gaia}), processed by
the {\it Gaia} Data Processing and Analysis Consortium (DPAC,
{\tt https://www.cosmos.esa.int/web/gaia/dpac/consortium}). Funding
for the DPAC has been provided by national institutions, in particular
the institutions participating in the {\it Gaia} Multilateral Agreement.
\end{acknowledgements}

\bibliographystyle{aa}

{}

\begin{appendix}
\section{The data}
\label{Sec:Cat}
The data resulting from our analysis are listed in the following tables and are available in electronic format at the CDS. 

\setlength{\tabcolsep}{3pt}

\begin{table*}
\caption{Radial (RV) and rotational ($v\sin i$) velocities for the single stars and SB1 binaries.  }
\begin{center}
\begin{scriptsize} 
\begin{tabular}{llccrrcrrrrrrc}
\hline
\hline
\noalign{\medskip}
\textsl{RasTyc} Name & Name & $\alpha$ (2000) & $\delta$ (2000)            & $V^{\rm a}$~~~ & $B$--$V^{\rm a}$ & HJD	    & RV	    & $\sigma_{RV}$      & $v\sin i^{\rm b}$ & $\sigma_{v\sin i}$  & $v\sin i^{\rm c}$  & $\sigma_{v\sin i}$  & Instr.$^{\rm d}$ \\
                     &      &  h m s          & $\degr ~\arcmin ~\arcsec$  & (mag)          &  (mag)           & (2450000+) & \multicolumn{2}{c}{(km s$^{-1}$)}  & \multicolumn{2}{c}{(km s$^{-1}$)}  & \multicolumn{2}{c}{(km s$^{-1}$)} & \\
\noalign{\medskip}
\hline
\noalign{\medskip}
  RasTyc0000+7940   & \object{BD+78 853}  	 & 00 00 41.14 & +79 40 39.9  & 10.27 &  0.704 &  2224.2896 &  -6.97 &  1.10  &  27.7 & 4.1 & 28.9 & 0.9 & AU \\  
  RasTyc0000+7940   & \object{BD+78 853}  	 & 00 00 41.14 & +79 40 39.8  & 10.27 &  0.704 &  2234.4033 &  -7.46 &  2.93  &  50.3 & 1.6 & \dots & \dots & AU  \\  
  RasTyc0000+7940   & \object{BD+78 853}  	 & 00 00 41.14 & +79 40 39.8  & 10.27 &  0.704 &  5109.4942 &  -7.39 &  0.52  &  33.5 & 0.7 & 32.1 & 1.1 & FR  \\
  RasTyc0000+7940   & \object{BD+78 853}  	 & 00 00 41.14 & +79 40 39.8  & 10.27 &  0.704 &  5110.5890 &  -7.64 &  0.49  &  33.5 & 1.1 & 32.4 & 0.9 & FR  \\
  RasTyc0001+5212   & \object{BD+51 3761}        & 00 01 42.66 & +52 12 51.0 &   9.30  &  0.756 & 4347.6393 &   4.97 &  1.15 &  5.9  &  1.7  &	 6.4 &  0.7  & SA \\	    
  RasTyc0004-0951   & \object{HD 225263}         & 00 04 46.96 & -09 51 53.4 &  10.37  &  0.831 & 2152.5741 & -10.26 &  0.15 &  6.9  &  0.4  &   6.9 &  0.7  & EL \\
\end{tabular}
\end{scriptsize} 
\end{center}
\label{Tab:RV}
\end{table*}

\clearpage

\setlength{\tabcolsep}{2pt}

\addtocounter{table}{-1}
\begin{table*}
\caption{continued.  }
\begin{center}
\begin{scriptsize} 
\begin{tabular}{llccrrcrrrrrrc}
\hline
\hline
\noalign{\medskip}
\textsl{RasTyc} Name & Name & $\alpha$ (2000) & $\delta$ (2000)            & $V^{\rm a}$~~~ & $B$--$V^{\rm a}$ & HJD	    & RV	    & $\sigma_{RV}$      & $v\sin i^{\rm b}$ & $\sigma_{v\sin i}$  & $v\sin i^{\rm c}$  & $\sigma_{v\sin i}$  & Instr.$^{\rm d}$ \\
                     &      &  h m s          & $\degr ~\arcmin ~\arcsec$  & (mag)          &  (mag)           & (2450000+) & \multicolumn{2}{c}{(km s$^{-1}$)}  & \multicolumn{2}{c}{(km s$^{-1}$)}  & \multicolumn{2}{c}{(km s$^{-1}$)} & \\
\noalign{\medskip}
\hline
\noalign{\medskip}
  RasTyc0313+7417   & \object{HD 19438}          & 03 13 27.48 & +74 17 47.0 &   7.31  &  0.559 & 5135.5433 & -51.64 &  1.83 &  69.0 &   1.4 &  73.3 &  3.6  & FR \\
  RasTyc0313+3832   & \object{BD+37 729}         & 03 13 47.40 & +38 32 04.9 &  10.00  &  0.741 & 2993.4424 &  18.59 &  1.33 &   2.0 &  28.9 &   6.0 &  1.4  & AU \\  
  RasTyc0316+4724   & \object{TYC 3315-962-1}    & 03 16 07.07 & +47 24 58.7 &   9.87  &  0.759 & 2601.4526 &  34.37 &  1.53 & $<$ 5 &   \dots &   2.0 &  1.5  & AU \\
  RasTyc0316+4724   & \object{TYC 3315-962-1}    & 03 16 07.07 & +47 24 58.7 &   9.87  &  0.759 & 2977.4292 &  35.20 &  1.57 & $<$ 5 &   \dots &   3.4 &  2.3  & AU \\
  RasTyc0316+4724   & \object{TYC 3315-962-1}    & 03 16 07.07 & +47 24 58.7 &   9.87  &  0.759 & 2978.3332 &  33.82 &  1.57 & $<$ 5 &   \dots &   3.0 &  2.2  & AU \\
  RasTyc0316+4724   & \object{TYC 3315-962-1}    & 03 16 07.07 & +47 24 58.7 &   9.87  &  0.759 & 2979.3315 &  35.24 &  1.53 & $<$ 5 &   \dots &   3.5 &  2.1  & AU \\
\end{tabular}
\end{scriptsize} 
\end{center}
\end{table*}

\clearpage

\addtocounter{table}{-1}
\begin{table*}
\caption{continued.  }
\begin{center}
\begin{scriptsize} 
\begin{tabular}{llccrrcrrrrrrc}
\hline
\hline
\noalign{\medskip}
\textsl{RasTyc} Name & Name & $\alpha$ (2000) & $\delta$ (2000)            & $V^{\rm a}$~~~ & $B$--$V^{\rm a}$ & HJD	    & RV	    & $\sigma_{RV}$      & $v\sin i^{\rm b}$ & $\sigma_{v\sin i}$  & $v\sin i^{\rm c}$  & $\sigma_{v\sin i}$  & Instr.$^{\rm d}$ \\
                     &      &  h m s          & $\degr ~\arcmin ~\arcsec$  & (mag)          &  (mag)           & (2450000+) & \multicolumn{2}{c}{(km s$^{-1}$)}  & \multicolumn{2}{c}{(km s$^{-1}$)}  & \multicolumn{2}{c}{(km s$^{-1}$)} & \\
\noalign{\medskip}
\hline
\noalign{\medskip}
  RasTyc0712+7021   & \object{TYC 4364-1262-1}   & 07 12 50.04 & +70 21 06.8 &   9.87  &  1.052 & 2981.6709 & -23.67 &  1.30 &   9.4 &   5.3 &  10.3 &  1.1  & AU \\ 
  RasTyc0712+7021   & \object{TYC 4364-1262-1}   & 07 12 50.04 & +70 21 06.8 &   9.87  &  1.052 & 2985.5803 & -31.55 &  1.37 &   9.4 &   4.8 &  10.9 &  0.9  & AU \\ 
  RasTyc0713+5106   & \object{TYC 3404-384-1}	 & 07 13 36.53 & +51 06 17.3 &   9.74  &  0.634 & 2235.6692 &  20.22 &  1.38 &  27.2 &   3.0 &  26.3 &  0.9  & AU \\ 
  RasTyc0713+5106   & \object{TYC 3404-384-1}	 & 07 13 36.53 & +51 06 17.3 &   9.74  &  0.634 & 2237.7139 &  20.61 &  1.30 &  20.7 &   2.3 &  22.0 &  2.1  & AU \\ 
  RasTyc0714+5307   & \object{TYC 3780-247-1} & 07 14 36.39 & +53 07 40.4 &  10.03 &  0.575  & 2993.6101 & 11.17  &  1.77 & \dots & \dots  & 32.9 &  2.1  & AU \\
  RasTyc0730+6343   & \object{TYC 4116-276-1}	 & 07 30 55.31 & +63 43 50.1 &   9.93  &  1.009 & 2993.5456 & -37.71 &  1.53 & $<$ 5 &   \dots &   7.7 &  0.7  & AU \\
\end{tabular}
\end{scriptsize} 
\end{center}
\end{table*}

\clearpage

\setlength{\tabcolsep}{3pt}

\addtocounter{table}{-1}
\begin{table*}
\caption{continued.  }
\begin{center}
\begin{scriptsize} 
\begin{tabular}{llccrrcrrrrrrc}
\hline
\hline
\noalign{\medskip}
\textsl{RasTyc} Name & Name & $\alpha$ (2000) & $\delta$ (2000)            & $V^{\rm a}$~~~ & $B$--$V^{\rm a}$ & HJD	    & RV	    & $\sigma_{RV}$      & $v\sin i^{\rm b}$ & $\sigma_{v\sin i}$  & $v\sin i^{\rm c}$  & $\sigma_{v\sin i}$  & Instr.$^{\rm d}$ \\
                     &      &  h m s          & $\degr ~\arcmin ~\arcsec$  & (mag)          &  (mag)           & (2450000+) & \multicolumn{2}{c}{(km s$^{-1}$)}  & \multicolumn{2}{c}{(km s$^{-1}$)}  & \multicolumn{2}{c}{(km s$^{-1}$)} & \\
\noalign{\medskip}
\hline
\noalign{\medskip}
  RasTyc1831+5418   & \object{TYC 3905-319-1}	 & 18 31 37.59 & +54 18 57.9 &   9.96  &  0.842 & 4149.7020 & -28.02 &  0.90 &   5.3 &   2.2 &   4.0 &  1.0  & SA \\   
  RasTyc1838+0224   & \object{BD+02 3633}	 & 18 38 38.19 & +02 24 12.7 &   9.61  &  0.586 & 4210.6953 & -27.13 &  0.52 &  31.5 &   2.1 &  31.0 &  0.8  & SA \\     
RasTyc1842+5751  & \object{HD\,173605}  & 18 42 49.32 & +57 51 54.6 &	7.96  &  0.630 & 2119.3499 &  24.73 &  0.15 &  21.6 &	1.1 &  20.8 &  0.8  & EL \\    
RasTyc1843+4328   & \object{KIC 7730305}       & 18 43 58.66 & +43 28 27.8 &   9.38  &  0.573 & 5046.5341 &   6.78 &  1.33 &  14.1 &   5.8 &  14.5 & 1.5  & SA \\  
RasTyc1843+4328   & \object{KIC 7730305}       & 18 43 58.66 & +43 28 27.8 &   9.38  &  0.573 & 5023.4904 &   5.93 &  0.28 &  17.7 &   1.1 &  16.3 & 3.0  & FR  \\
  RasTyc1847+2930   & \object{BD+29 3349}	 & 18 47 39.96 & +29 30 39.3 &   9.93  &  0.865 & 4249.6876 &   5.20 &  0.55 &   5.5 &   1.8 &   6.2 &  0.7  & SA \\
\end{tabular}
\end{scriptsize} 							 
\end{center}								 
\end{table*}								 
									 
\clearpage

\addtocounter{table}{-1}						 
\begin{table*}								 
\caption{continued.  }						 
\begin{center}								 
\begin{scriptsize} 							 
\begin{tabular}{llccrrcrrrrrrc} 					 
\hline									 
\hline									 
\noalign{\medskip}							 
\textsl{RasTyc} Name & Name & $\alpha$ (2000) & $\delta$ (2000)            & $V^{\rm a}$~~~ & $B$--$V^{\rm a}$ & HJD	    & RV	    & $\sigma_{RV}$	 & $v\sin i^{\rm b}$ & $\sigma_{v\sin i}$  & $v\sin i^{\rm c}$  & $\sigma_{v\sin i}$  & Instr.$^{\rm d}$ \\
                     &      &  h m s          & $\degr ~\arcmin ~\arcsec$  & (mag)	    &  (mag)	       & (2450000+) & \multicolumn{2}{c}{(km s$^{-1}$)}  & \multicolumn{2}{c}{(km s$^{-1}$)}  & \multicolumn{2}{c}{(km s$^{-1}$)} & \\
\noalign{\medskip}
\hline
\noalign{\medskip}
  RasTyc2100+4405   & \object{TYC 3180-1594-1}   & 21 00 25.21 & +44 05 54.7 &  10.52  &  0.877 & 1763.4905 &  35.24 &  0.15 &   4.2 &   0.2 &   4.3 &   0.8 & EL \\
  RasTyc2101+1008   & \object{BD+09 4700}	 & 21 01 44.80 & +10 08 40.9 &   9.73  &  1.021 & 2474.5676 & -29.21 &  1.38 &   9.3 &   4.7 &   9.0 &   0.5 & AU \\
  RasTyc2101+1008   & \object{BD+09 4700}	 & 21 01 44.80 & +10 08 40.9 &   9.73  &  1.021 & 2478.4917 & -29.33 &  1.32 &   9.0 &   6.4 &   7.3 &   0.8 & AU \\
  RasTyc2102+4553   & \object{V2436 Cyg}	 & 21 02 40.42 & +45 53 03.9 &   7.72  &  0.969 & 1732.5386 & -14.19 &  0.15 &   3.6 &   0.2 &   2.9 &   0.7 & EL \\
  RasTyc2103+4104   & \object{TYC 3172-643-1}	 & 21 03 16.76 & +41 04 06.3 &  10.31  &  0.992 & 1799.3572 & -23.22 &  0.15 &   9.3 &   0.5 &   9.2 &   0.6 & EL \\
  RasTyc2105+3949   & \object{TYC 3172-652-1}	 & 21 05 50.98 & +39 49 48.1 &  10.74  &  0.765 & 2118.4740 &   0.78 &  0.17 &  24.7 &   1.3 &  23.2 &   1.3 & EL \\
\end{tabular}
\end{scriptsize} 
\end{center}
\end{table*}

\clearpage

\addtocounter{table}{-1}
\begin{table*}
\caption{continued.  }
\begin{center}
\begin{scriptsize} 
\begin{tabular}{llccrrcrrrrrrc}
\hline
\hline
\noalign{\medskip}
\textsl{RasTyc} Name & Name & $\alpha$ (2000) & $\delta$ (2000)            & $V^{\rm a}$~~~ & $B$--$V^{\rm a}$ & HJD	    & RV	    & $\sigma_{RV}$      & $v\sin i^{\rm b}$ & $\sigma_{v\sin i}$  & $v\sin i^{\rm c}$  & $\sigma_{v\sin i}$  & Instr.$^{\rm d}$ \\
                     &      &  h m s          & $\degr ~\arcmin ~\arcsec$  & (mag)          &  (mag)           & (2450000+) & \multicolumn{2}{c}{(km s$^{-1}$)}  & \multicolumn{2}{c}{(km s$^{-1}$)}  & \multicolumn{2}{c}{(km s$^{-1}$)} & \\
\noalign{\medskip}
\hline
\noalign{\medskip}
  RasTyc2236+0010   & \object{HD 214129}	& 22 36 11.99  & +00 10 07.6 &   9.86  &  0.667 & 2473.6016 & -30.30 &  2.86 &  44.1 &   1.6 &  38.0 &   1.5 & AU \\ 
  RasTyc2236+0010   & \object{HD 214129}        & 22 36 11.99  & +00 10 07.6 &   9.86  &  0.667 & 4348.4851 &   7.95 &  0.87 &  31.2 &   0.9 &  31.9 &   0.7 & SA \\ 
  RasTyc2236+7032   & \object{TYC 4480-965-1}    & 22 36 15.90 & +70 32 04.1 &  10.61  &  0.961 & 4346.5461 &  14.03 &  0.63 &  10.1 &   1.8 &  11.5 &   0.8 & SA \\ 
  RasTyc2238+0217   & \object{HD 214494}         & 22 38 29.22 & +02 17 56.4 &   8.20  &  0.571 & 1734.5729 &  -9.44 &  0.15 &  17.8 &   0.9 &  18.3 &   0.8 & EL \\ 
  RasTyc2239+0406   & \object{V403 Peg}          & 22 39 50.66 & +04 06 57.1 &   8.49  &  0.915 & 2122.4801 &  24.01 &  0.15 &   2.7 &   0.2 &   3.6 &   0.8 & EL \\ 
  RasTyc2240+1432   & \object{HD 214850}         & 22 40 52.53 & +14 32 56.0 &   5.74  &  0.730 & 1762.6330 & -10.21 &  0.15 &   1.4 &   0.1 &   1.7 &   0.8 & EL \\
\end{tabular}

\begin{list}{}{}									
\item[$^{\rm a}$] From the TYCHO catalog \citep{HIPPA97}.  
\item[$^{\rm b}$] $v\sin i$ derived from the FWHM of the CCF.	  
\item[$^{\rm c}$] $v\sin i$ derived with ROTFIT.  
\item[$^{\rm d}$] Spectrograph: SA = SARG, EL = ELODIE, AU = AURELIE, FR = FRESCO.  
\item[$^{\rm e}$] Coordinates from 2MASS catalog \citep{2MASS}. 
\item[$^{\rm f}$] Close visual pair \citep[$\rho=0\farcs$85, $\Delta V\simeq 0.29$\,mag,][]{Fabricius2002}. The $B-V$ color is probably unreliable.
\item[$^{\rm g}$] Close visual binary. Composite spectrum with broad + narrow lines at the same velocity. 
\item[$^{\rm h}$] Close visual pair. Composite spectrum. The primary is a W\,UMa-type binary (V410\,Aur). The radial velocity and vsini are estimates for the B component. 
\item[$^{\rm i}$] Close visual binary. Composite spectrum. 
\item[$^{\rm *}$] Double-peaked CCF. Possible SB2 with blended lines. See Table\,\ref{Tab:RV_SB2} for the estimates of RV and $v\sin i$ of the two components.  
\item[$^{\dag}$]  Asymmetric CCF. Possible SB2 with blended lines. See Table\,\ref{Tab:RV_SB2} for the estimates of RV and $v\sin i$ of the two components.  
\item[$^{\ddag}$] Hierarchical triple system with a very faint inner binary. The velocities of all the components are listed in Table\,\ref{Tab:RV_SB3}.
\item[$^{\diamond}$] Small-amplitude secondary peak in the CCF. Likely SB2 with a faint component. See Table\,\ref{Tab:RV_SB2}. 
\item[$^{\rm **}$] Newly discovered close visual binary. Both stars observed simultaneously with SARG. Separation $\rho=1\farcs$9, $\Delta V=1.15$\,mag.	
\item[$^{\rm ***}$] Tertiary component of a W\,UMa-type inner binary \citep{Rucinski2008}.
\end{list}
\end{scriptsize} 
\end{center}
\end{table*}

\setlength{\tabcolsep}{2pt}

\begin{table*}
\caption{Radial (RV) and rotational ($v\sin i$) velocities for the components of SB2 binaries.  }
\begin{center}
\begin{scriptsize} 
\begin{tabular}{llccrcrrrrrrrrc}
\hline
\hline
\noalign{\medskip}
\textsl{RasTyc} Name & Name & $\alpha$ (2000) & $\delta$ (2000)           & $V^{\rm a}$   & HJD        & $RV_1$          & $\sigma$  &  $RV_2$  	& $\sigma$        & $v\sin i_1$  & $\sigma$  & $v\sin i_2$  & $\sigma$ &  Instr.$^{\rm b}$ \\
                     &      &  h m s          & $\degr ~\arcmin ~\arcsec$ & (mag) & (2450000+) & \multicolumn{2}{c}{(km s$^{-1}$)}  &  \multicolumn{2}{c}{(km s$^{-1}$)} & \multicolumn{2}{c}{(km s$^{-1}$)}    & \multicolumn{2}{c}{(km s$^{-1}$)}  & \\
\noalign{\medskip}
\hline
\noalign{\medskip} 
    RasTyc0013+7702  & \object{TYC 4496-780-1}  & 00 13 40.52  & +77 02 10.9  &  9.76 & 2226.2918 &   -1.30 &  1.57  & \dots~~	&\dots~~ &   31.7  &  2.3  & \dots~~  &\dots~~  & AU \\ 
    RasTyc0013+7702  & \object{TYC 4496-780-1}  & 00 13 40.52  & +77 02 10.9  &  9.76 & 2235.3276 &   -5.30 &  1.96  &    45.27 &  1.34  &   41.8  &  1.4  & $<$ 5  & \dots~~ & AU  \\  
    RasTyc0013+7702  & \object{TYC 4496-780-1}  & 00 13 40.52  & +77 02 10.9  &  9.76 & 5108.5549 &   -5.78 &  0.70  &    44.73 &  2.74  &   28.2  &  9.9  &   15.6   &   9.0 & FR \\
    RasTyc0013+7702  & \object{TYC 4496-780-1}  & 00 13 40.52  & +77 02 10.9  &  9.76 & 5118.4417 &   -6.78 &  0.62  &    47.27 &  3.87  &   30.0  &  6.7  &   26.2   &   7.8 & FR \\
    RasTyc0020+5711  & \object{TYC 3661-1206-1} & 00 20 47.59  & +57 11 45.1  & 10.06 & 2987.2483 &  -11.78 &  1.27  &   -69.15 &  1.52  &   22.0  &  1.4  & $<$ 5  & \dots~~ & AU  \\  
    RasTyc0032+5558  & \object{HD 236456}       & 00 32 09.35  & +55 58 35.2  &  9.53 & 2666.2915 &    2.75 &  4.13  &   -23.93 &  2.95  &  $<$ 5  & \dots~~ &  55.8  &  1.2  & AU  \\  
    RasTyc0032+5558  & \object{HD 236456}       & 00 32 09.35  & +55 58 35.2  &  9.53 & 2601.2847 &    0.85 &  1.66  &   -19.42 &  1.39  &  $<$ 5  & \dots~~ &  43.3  &  0.5  & AU  \\  
 RasTyc0033+5315$^{\diamond}$ &\object{TYC 3654-1907-1} &00 33 55.90 &+53 15 41.5 & 9.94 &4347.6497 & -9.45 &  0.23  &    36.87 &  9.45  &    3.4  &  2.9  &  14.0  &  8.5  & SA  \\  
    RasTyc0038+3325  & \object{TYC 2279-359-1}  & 00 38 44.96  & +33 25 34.5  & 10.27 & 2987.3413 &    6.32 &  1.50  & \dots~~	&\dots~~ &   28.7  &  2.0  & \dots~~  &\dots~~  & AU  \\  
    RasTyc0039+7905   & \object{BD+78 19}   	& 00 39 40.13  & +79 05 30.8  &  9.60 & 2581.3381 &   -8.66 &  1.38  & \dots~~  &\dots~~ &   20.2  &  5.1  & \dots~~ &\dots~~ & AU \\ 
    RasTyc0039+7905   & \object{BD+78 19}   	& 00 39 40.13  & +79 05 30.8  &  9.60 & 2584.3640 &   -9.68 &  1.37  & \dots~~  &\dots~~ &   12.0  &  3.5  & \dots~~ &\dots~~ &  AU \\
    RasTyc0039+7905   & \object{BD+78 19}   	& 00 39 40.13  & +79 05 30.8  &  9.60 & 5077.6191 &   -9.59 &  0.41  & \dots~~  &\dots~~ &  \dots~~ &\dots~~ & \dots~~ &\dots~~ & FR \\
    RasTyc0039+7905   & \object{BD+78 19}   	& 00 39 40.13  & +79 05 30.8  &  9.60 & 5078.6177 &  -23.00 &  0.71  &	4.47    &  0.77  &   6.8  &  5.0  &  11.9  &  5.1 & FR \\
    RasTyc0039+7905   & \object{BD+78 19}   	& 00 39 40.13  & +79 05 30.8  &  9.60 & 5108.5172 &  	7.96 &  0.71 & -30.30   &  0.82  &  15.1 &  5.9  &  11.8  &  2.6 & FR \\
    RasTyc0039+7905   & \object{BD+78 19}   	& 00 39 40.13  & +79 05 30.8  &  9.60 & 5110.5424 &  -20.37&  0.72 &	4.26    &  0.70  &    7.3 &  3.4  &  10.4  &  4.1 & FR \\
    RasTyc0055+5046  & \object{TYC 3274-955-1}  & 00 55 50.28  & +50 46 12.6  &  9.99 & 2988.3284 &   53.25 &  2.01  &    23.46 &  2.81  &  $<$ 5  & \dots~~ &  61.7  &  1.3  & AU \\ 
    RasTyc0104+5533  & \object{TYC 3672-1578-1} & 01 04 33.02  & +55 33 29.0  & 10.46 & 4348.5993 &  -14.77 &  0.52  &   -47.40 &  1.29  &    3.7  &  2.8  &  27.3  &  6.0  & SA  \\ 
    RasTyc0110+5100  & \object{TYC 3276-1291-1} & 01 10 00.27  & +51 00 27.8  & 10.19 & 4346.6661 &  -23.02 &  0.34  &    17.74 &  0.91  &    3.7  &  2.1  &   7.4  &  3.7  & SA  \\ 
    RasTyc0129+3303  & \object{HN Psc}          & 01 29 47.94  & +33 03 36.1  & 10.65 & 4348.6915 &   43.73 & 17.18  &  -112.86 & 20.63  &   81.4  & 27.6  & 112.9  & 28.1  & SA  \\ 
    RasTyc0139+7018  & \object{TYC 4314-1757-1} & 01 39 25.89  & +70 18 50.1  & 10.25 & 4347.6776 &   21.65 &  0.31  &   -21.85 &  2.03  &    5.7  &  3.4  &  16.3  &  6.6  & SA  \\ 
    RasTyc0146+3317  & \object{TYC 2298-964-1}  & 01 46 33.48  & +33 17 11.7  & 10.00 & 2990.2585 &  -17.75 &  6.05  &     9.10 &  2.99  &  $<$ 5  & \dots~~ &  40.1  &  1.7  & AU  \\ 
    RasTyc0315+5741  & \object{TYC 3710-247-1}  & 03 15 32.73  & +57 41 23.8  & 10.21 & 4348.7382 &    6.75 &  0.67  &   -82.50 &  1.78  &   34.1  &  6.7  &  24.2  &  4.0  & SA  \\ 
    RasTyc0326+4243  & \object{BD+42 765}       & 03 26 38.82  & +42 43 25.3  &  9.82 & 2665.3008 &   65.90 & 11.30  &  -226.90 & 18.00  &  \dots  & \dots & \dots &  \dots  & AU \\ 
    RasTyc0409+5558  & \object{BD+55 849}	& 04 09 58.60  & +55 58 52.6  & 10.44 & 4459.4190 &  -40.67 &  5.52  &    95.25 &  8.03  &   84.6  &  6.5  &  51.1  & 12.2  & SA  \\ 
   RasTyc0412+4616A  & \object{TYC 3328-2470-1} & 04 12 14.74  & +46 16 15.5  & 10.52 & 4459.4776 &  -97.70 &  2.25  &   104.25 &  3.81  &   52.6  &  9.1  &  46.7  & 10.0  & SA  \\ 
   RasTyc0413+3832   & \object{BD+38 859}       & 04 13 28.59  & +38 32 26.6  &  9.67 & 2669.3423 &   -4.05 &  1.43  & \dots~~	&\dots~~   &   54.4  &  2.4  & \dots~~  &\dots~~  & AU \\ 
   RasTyc0413+3832   & \object{BD+38 859}       & 04 13 28.59  & +38 32 26.6  &  9.67 & 2673.2935 &   69.14 &  1.11  & \dots~~	&\dots~~   &   36.4  &  6.1  & \dots~~  &\dots~~  & AU \\ 
   RasTyc0415+3129   & \object{HD 281777}	& 04 15 28.92  & +31 29 55.8  &  9.68 & 2240.5223 & -118.60  & 15.00 & 79.50 & 18.00 &  \dots &   \dots &   \dots &  \dots  & AU \\ 
   RasTyc0423+5556   & \object{TYC 3723-1082-1} & 04 23 15.14  & +55 56 53.0  &  9.77 & 2222.6230 &   12.92 &  1.52  &     0.98 &  2.02  &  $<$ 5  & \dots~~ &  43.5  &  0.5  & AU \\
   RasTyc0423+5556   & \object{TYC 3723-1082-1} & 04 23 15.14  & +55 56 53.0  &  9.77 & 2230.4290 &    0.45 &  1.11  & \dots~~	&\dots~~   &   35.5  &  5.1  & \dots~~  &\dots~~  & AU \\  
   RasTyc0423+5556   & \object{TYC 3723-1082-1} & 04 23 15.14  & +55 56 53.0  &  9.77 & 2234.5657 &   19.18 &  2.60  &    -0.17 &  3.11  &  $<$ 5  & \dots~~ &  63.7  &  1.5  & AU \\ 
   RasTyc0511+3707   & \object{HD 280583}       & 05 11 51.55  & +37 07 36.4  &  9.51 & 2218.5610 & -35.82  &  2.92  &  122.08  &  8.29  &  \dots  & \dots &  \dots & \dots & AU  \\
   RasTyc0511+3707   & \object{HD 280583}       & 05 11 51.55  & +37 07 36.4  &  9.51 & 2224.6148 &   24.61 &  6.73  &  -87.94  &  2.18  &  \dots  & \dots &  \dots & \dots & AU  \\
   RasTyc0511+3707   & \object{HD 280583}       & 05 11 51.55  & +37 07 36.4  &  9.51 & 2225.6592 & -37.52  &  5.67  &  120.61  &  5.28  &  \dots  & \dots &  \dots & \dots & AU  \\
   RasTyc0524+5439   & \object{V607 Aur}        & 05 24 24.58  & +54 39 22.2  & 10.33 & 4459.6818 &  -94.15 &  1.58  &    98.77 &  2.38  &   30.2  &  7.4  &  27.9  &  6.4  & SA  \\  
   RasTyc0604+5142   & \object{TYC 3386-868-1}  & 06 04 51.40  & +51 42 00.8  &  9.35 & 4459.7215 &    2.34 &  1.50  &    53.24 &  0.61  &   28.7  &  4.5  &   4.9  &  2.5  & SA  \\  
   RasTyc0608+6804   & \object{TYC 4345-1307-1} & 06 08 04.79  & +68 04 59.7  & 10.08 & 4459.7316 &  112.43 &  4.63  &   -93.74 &  3.09  &   32.6  &  5.5  &  37.5  &  6.7  & SA  \\  
   RasTyc0609+3229   & \object{TYC 2424-396-1}	& 06 09 51.09  & +32 29 48.7  & 10.05 & 2990.5056 &   14.78 &  1.49  &    -7.32 &  1.45  &  $<$ 5  & \dots~~ & $<$ 5  & \dots~~ & AU  \\  
   RasTyc0653+3851   & \object{TYC 2942-2009-1} & 06 53 00.11  & +38 51 52.9  &  9.86 & 2985.5321 &  -60.34 &  7.81  &   187.42 & 13.59  &  \dots~~  &\dots~~  & \dots~~  &\dots~~  & AU  \\ 
   RasTyc0702+4448   & \object{TYC 2955-479-1}  & 07 02 45.77  & +44 48 05.6  &  9.94 & 2993.6623 &  -26.06 &  1.49  &  -135.95 &  1.34  &    8.5  &  3.0  &   1.4  & 22.7  & AU  \\ 
   RasTyc0707+4017   & \object{BD+40 1796}      & 07 07 52.87  & +40 17 07.7  & 10.01 & 4459.7978 &   15.21 &  0.53  &   -35.07 &  3.62  &   17.2  &  5.0  &   7.5  &  4.9  & SA  \\ 
   RasTyc1502+1546   & \object{HD 133162}	& 15 02 38.99  & +15 46 29.3  &  9.84 & 2465.4009 &  -81.98 &  2.08  &   -49.88 &  4.55  &   62.6  &  1.4  & $<$ 5  & \dots~~ & AU \\
   RasTyc1502+1546   & \object{HD 133162}	& 15 02 38.99  & +15 46 29.3  &  9.84 & 2467.3860 &    0.05 &  5.24  &   -31.73 &  3.05  &  $<$ 5  & \dots~~ &  71.7  &  1.3  & AU \\
   RasTyc1525+6142   & \object{TYC 4181-507-1}  & 15 25 52.78  & +61 42 21.5  & 10.89 & 5726.4030 & -120.05 &  1.86  &    98.46 & 11.66  &   69.2  & 21.0  &  80.3  & 20.0  & SA  \\ 
   RasTyc1527+6515   & \object{TYC 4187-96-1}   & 15 27 56.19  & +65 15 33.3  & 10.17 & 5726.3798 &   34.94 &  0.78  &   -55.55 &  4.05  &   20.9  &  2.8  &  22.4  &  6.7  & SA  \\ 
   RasTyc1533+0809   & \object{BD+08 3048}      & 15 33 45.30  & +08 08 59.6  & 10.16 & 5727.4255 &   29.67 &  1.81  &   -82.51 &  1.99  &   37.8  &  6.0  &  28.5  &  4.8  & SA  \\ 
   RasTyc1538+4911   & \object{BD+49 2412}      & 15 38 12.25  & +49 11 49.7  &  9.70 & 2438.3887 & -61.85 &  5.59  &	  69.62 &  6.49  &  \dots  & \dots &  \dots & \dots  & AU  \\
   RasTyc1538+4911   & \object{BD+49 2412}      & 15 38 12.25  & +49 11 49.7  &  9.70 & 2440.4763 & -15.53 &  1.52  &	  56.73 &  3.00  &   27.0  &   1.3   & 31.7 &  1.2  & AU   \\
   RasTyc1547+1509   & \object{G 137-52}	& 15 47 11.90  & +15 09 14.9  &  9.74 & 2439.3928 &   57.43 &  1.29  &   -10.19 &  1.48  &  $<$ 5  & \dots~~ & $<$ 5  & \dots~~ & AU  \\ 
   RasTyc1547+1509   & \object{G 137-52}	& 15 47 11.90  & +15 09 14.9  &  9.74 & 2441.3860 &   -7.34 &  1.43  &    54.13 &  1.36  &  $<$ 5  & \dots~~ & $<$ 5  & \dots~~ & AU  \\ 
  RasTyc1550+1440B   & \object{HIP 77628}       & 15 50 54.36  & +14 40 33.9  & 10.06 & 4144.7518 &  -39.01 &  0.35  &     5.31 &  0.90  &   10.0  &  2.3  &   9.4  &  2.9  & SA  \\ 	
   RasTyc1605+1028   & \object{BD+10 2953}      & 16 05 02.23  & +10 28 54.6  &  9.21 & 4148.6430 &  -50.91 &  0.80  &   -15.19 &  0.92  &   10.5  &  3.0  &   5.0  &  3.6  & SA  \\ 
   RasTyc1606+0919   & \object{TYC 945-949-1}   & 16 06 29.27  & +09 19 02.5  & 10.06 & 4144.7694 &   -0.65 &  0.24  &   110.35 &  4.36  &    6.4  &  3.9  &  15.2  & 11.8  & SA  \\ 
   RasTyc1607+0238   & \object{TYC 370-538-1}	& 16 07 04.36  & +02 38 23.8  &  9.53 & 2438.4416 &  -16.41 &  1.32  &   -53.21 &  1.52  &   13.5  &  1.6  &   7.1  &  3.4  & AU  \\ 
   RasTyc1607+0238   & \object{TYC 370-538-1}	& 16 07 04.36  & +02 38 23.8  &  9.53 & 2466.4475 &  -12.68 &  1.45  &   -58.67 &  1.31  &   11.2  &  2.4  &   1.4  & 23.8  & AU  \\ 
   RasTyc1620+2436   & \object{V1079 Her}	& 16 20 13.72  & +24 36 11.1  &  9.48 & 2439.4846 &   17.91 &  1.57  &   -45.20 &  1.38  &   18.2  &  1.8  &  19.6  &  1.6  & AU  \\ 
   RasTyc1620+2436   & \object{V1079 Her}	& 16 20 13.72  & +24 36 11.1  &  9.48 & 2441.5015 &   41.48 &  1.52  &   -69.37 &  1.31  &   16.4  &  1.9  &  16.4  &  1.9  & AU  \\ 
   RasTyc1620+2436   & \object{V1079 Her}	& 16 20 13.72  & +24 36 11.1  &  9.48 & 2443.3980 &   40.14 &  1.40  &   -69.29 &  1.30  &   18.8  &  1.8  &  18.8  &  1.8  & AU  \\ 
   RasTyc1624+4555   & \object{BD+46 2173} 	& 16 24 10.45  & +45 55 26.0  &  9.94 & 4148.6916 & \dots~~ & \dots~~ & \dots~~ & \dots~~ & 250$^{\rm c}$ & \dots~~ & \dots~~  & \dots~~ & SA  \\ 
   RasTyc1629+3212   & \object{BD+32 2733}   	& 16 29 12.03  & +32 12 29.9  &  9.14 & 4148.7118 &  -80.96 &  1.24  &    28.04 &  1.91  &   21.2  &  3.4  &   6.6  &  4.8  & SA  \\  
   RasTyc1639+2243   & \object{BD+23 2969}   	& 16 39 29.33  & +22 43 59.6  &  9.99 & 4993.3633 &  -38.86 &  1.36  &     7.20 &  2.72  &   31.8  &  8.5  &  33.0  &  8.5  & FR  \\  
   RasTyc1700+2001   & \object{BD+20 3376}   	& 17 00 33.85  & +20 01 33.9  &  9.99 & 4993.4370 &  -23.59 &  0.65  &    17.44 &  3.18  &   24.3  &   4.8  & 18.5  & 12.8  & FR  \\
   RasTyc1702+4713$^{\diamond}$ &\object{TYC 3501-626-1} &17 02 48.85 &+47 13 06.5& 10.08 & 4215.4693 &  -7.85 &  0.21  &   -38.01 &  0.47  &   13.2  &  3.8  &   1.9  &  1.4  & SA \\
   RasTyc1703+2453   & \object{TYC 2064-1273-1} & 17 03 13.52  & +24 53 21.1  &  9.83 & 4149.7208 &   30.22 &  0.89  &   -86.97 &  0.83  &   11.5  &  4.2  &   6.1  &  6.3  & SA  \\  
   RasTyc1718+2128   & \object{TYC 1548-2040-1} & 17 18 00.31  & +21 28 09.4  &  9.96 & 4210.6407 &   20.21 &  1.50  &   -69.82 &  2.58  &   34.3  &  3.5  &  36.7  &  5.2  & SA  \\  
   RasTyc1736+5523   & \object{HD 238727} & 17 36 42.35  &  +55 23 02.5  &  10.11 & 5111.3185  & -45.34  & 0.61 &  36.66 &   2.89  &  27.0   &  3.1  &  28.6 & 13.3 & FR  \\
   RasTyc1744+7452   & \object{BD+74 736}	& 17 44 49.84  & +74 52 26.9  &  9.59 & 2438.4775 &   88.42 &  0.97  &  -106.88 &  0.77  &   38.1  &  0.8  &  12.5  &  1.1  & AU  \\ 
   RasTyc1744+7452   & \object{BD+74 736}	& 17 44 49.84  & +74 52 26.9  &  9.59 & 2443.4412 &  -38.02 &  1.57  &    42.19 &  1.76  &   31.6  &  0.6  &  31.6  &  0.6  & AU  \\ 
   RasTyc1746+0358   & \object{BD+04 3503}	& 17 46 25.47  & +03 58 49.4  &  9.60 & 2442.4309 &  -81.44 &  1.84  & \dots~~	&\dots~~   &   30.8  &  2.5  & \dots~~  &\dots~~  & AU  \\  
   RasTyc1746+0358   & \object{BD+04 3503}	& 17 46 25.47  & +03 58 49.4  &  9.60 & 2443.5190 &  -41.36 &  1.48  &  -118.10 &  1.36  &   19.2  &  1.6  &   8.8  &  3.0  & AU  \\
   RasTyc1749+2328   & \object{V1298 Her}       & 17 49 51.67  & +23 28 07.4  &  9.84 & 5111.2817 &   73.64 &  1.04  &  -119.81 &  1.88  &   34.1  &  2.1  &  30.2  &  3.6  & FR  \\
   RasTyc1808-0858$^{\dag}$  & \object{BD-08 4562} & 18 08 01.34 & -08 58 57.5&  9.54 & 4150.7652 &    0.64 &  0.93  &    -6.44 &  1.28  &    4.4  &  2.9  &  21.7  &  7.1  & SA  \\
   RasTyc1811+3823   & \object{BD+38 3104}      & 18 11 47.87  & +38 23 07.5  &  9.33 & 4249.6612 &   -3.11 &  0.47  &   -49.85 &  0.52  &    3.9  &  2.2  &   3.8  &  1.8  & SA  \\
   RasTyc1824+2818   & \object{BD+28 2992}	& 18 24 09.98  & +28 18 21.2  &  9.81 & 2480.4585 &   -9.43 &  1.32  &   -78.43 &  1.98  &  $<$ 5  & \dots~~ &  17.1  &  1.5  & AU  \\
   RasTyc1824+2818   & \object{BD+28 2992}	& 18 24 09.98  & +28 18 21.2  &  9.81 & 2481.3809 &   -9.56 &  1.33  &    35.02 &  1.48  &    1.1  & 24.2  & $<$ 5  & \dots~~ & AU  \\
   RasTyc1825+1817   & \object{AW Her}  	& 18 25 38.71  & +18 17 40.1  &  9.68 & 3583.4434 &  -12.75 &  1.63  &   -78.11 &  1.36  &   21.9  &  1.4  &   8.6  &  2.9  & AU  \\  
   RasTyc1825+1817   & \object{AW Her}  	& 18 25 38.71  & +18 17 40.1  &  9.68 & 3588.4092 &   10.31 &  1.43  &  -101.97 &  1.72  &   11.0  &  2.2  &  25.7  &  1.9  & AU  \\  
   RasTyc1842+0742   & \object{G 141-30}	& 18 42 58.62  & +07 42 52.6  &  9.85 & 2442.4753 &   -2.05 &  1.53  & \dots~~	&\dots~~   &    2.6  & 17.5  & \dots~~  &\dots~~  & AU  \\ 
   RasTyc1842+0742   & \object{G 141-30}	& 18 42 58.62  & +07 42 52.6  &  9.85 & 2444.4751 &   25.79 &  1.56  &   -11.01 &  1.46  &   24.8  &  0.3  &   9.4  &  4.3  & AU  \\ 
\hline
\noalign{\medskip}
\end{tabular}
\end{scriptsize} 
\end{center}
\label{Tab:RV_SB2}
\end{table*}

\clearpage

\addtocounter{table}{-1}
\begin{table*}
\caption{continued.  }
\begin{center}
\begin{scriptsize} 
\begin{tabular}{llccrcrrrrrrrrc}
\hline
\hline
\noalign{\medskip}
\textsl{RasTyc} Name & Name & $\alpha$ (2000) & $\delta$ (2000)           & $V^{\rm a}$   & HJD        & $RV_1$          & $\sigma$  &  $RV_2$  	& $\sigma$        & $v\sin i_1$  & $\sigma$  & $v\sin i_2$  & $\sigma$ &  Instr.$^{\rm b}$ \\
                     &      &  h m s          & $\degr ~\arcmin ~\arcsec$ & (mag) & (2450000+) & \multicolumn{2}{c}{(km s$^{-1}$)}  &  \multicolumn{2}{c}{(km s$^{-1}$)} & \multicolumn{2}{c}{(km s$^{-1}$)}    & \multicolumn{2}{c}{(km s$^{-1}$)}  & \\
\noalign{\medskip}
\hline
\noalign{\medskip} 
   RasTyc1845+2841   & \object{TYC 2120-388-1} & 18 45 10.88 & +28 41 10.4 &   9.89  &  3578.4612 &  24.35 &  3.48 & -54.93 & 4.62 &  \dots & \dots & \dots &  \dots  & AU \\  
   RasTyc1845+2841   & \object{TYC 2120-388-1} & 18 45 10.88 & +28 41 10.4 &   9.89  &  3581.4358 &-75.30 &  3.94 &   20.55 & 6.19 &  \dots & \dots & \dots &  \dots  & AU \\  
   RasTyc1852+6223   & \object{BD+62 1656}      & 18 52 00.58  & +62 23 59.9  &  9.31 & 4301.6923 &  -50.96 &  0.25  &    39.14 &  1.36  &    4.5  &  1.9  &   6.6  &  5.5  & SA  \\ 
   RasTyc1856+0217   & \object{TYC 453-327-1}   & 18 56 09.59  & +02 17 43.9  &  9.95 & 4210.7100 &  -12.82 &  0.58  &   -63.21 &  1.51  &    9.3  &  4.1  &  10.2  &  8.2  & SA  \\ 
   RasTyc1900+0435   & \object{BD+04 3943}      & 19 00 46.23  & +04 35 57.9  &  9.35 & 4210.7196 &   -1.71 &  0.71  &  -115.94 &  1.35  &    4.8  &  3.3  &  12.2  &  9.1  & SA  \\ 
   RasTyc1902+1416   & \object{BD+14 3751}      & 19 02 13.43  & +14 16 53.6  & 10.12 & 4249.6741 &  -57.69 &  0.97  &     5.12 &  1.76  &   14.9  &  4.1  &  12.6  &  4.0  & SA  \\ 
   RasTyc1905+2319   & \object{BD+23 3557}      & 19 05 47.78  & +23 19 22.0  &  9.59 & 4249.6982 &  -66.64 & 16.16  &   199.68 & 13.16  &  145.8  &  9.5  & 117.8  & 41.9  & SA  \\ 
   RasTyc1914+6229   & \object{BD+62 1699}      & 19 14 19.78  & +62 29 57.2  &  9.95 & 4270.5264 &  -19.18 &  0.48  &   -71.25 &  1.76  &    4.7  &  3.8  &  29.6  &  8.1  & SA  \\ 
   RasTyc1931+1143$^{\rm d}$ & \object{HD 183957}& 19 31 36.98 & +11 43 26.2  &  8.24 & 2120.3379 &  -73.93 &  0.15  &    19.73 &  0.15  &    3.8  &  0.2  &   3.9  &  0.5  & EL  \\ 
   RasTyc1947+0105   & \object{HD 187003}	& 19 47 33.32  & +01 05 20.0  &  6.78 & 1797.3101 &  -21.11 &  0.15  &    38.75 &  0.15  &    5.1  &  0.3  &   4.9  &  0.6  & EL  \\ 
   RasTyc1958+4301$^{\rm *}$ & \object{KIC 7477572} & 19 58 37.62 & +43 01 02.5&  9.54 & 4346.3606 &  -36.59 &  2.20 &   -68.66 &  1.41  &   24.5  &  5.5  &  13.8  &  6.3  & SA  \\ 
   RasTyc2018+5636   & \object{V2477 Cyg}       & 20 18 58.95  & +56 36 19.0  &  9.94 & 4348.3539 &   56.32 & 13.91  &  -222.12 &  7.94  &  167.0  & 26.3  & 105.1  & 33.5  & SA  \\ 
   RasTyc2025+3631   & \object{TYC 2697-941-1}  & 20 25 09.60  & +36 31 15.4  & 10.36 & 2120.4759 &    6.60 &  0.15  &   -42.75 &  0.16  &    7.3  &  0.4  &   1.2  &  0.5  & EL  \\ 
   RasTyc2029+1227   & \object{TYC 1095-349-1}  & 20 29 32.84  & +12 27 31.1  &  9.74 & 3581.4871 &  -93.25 &  1.29  &    34.93 &  2.47  &   17.0  &  1.5  &  14.8  &  1.7  & AU  \\ 
   RasTyc2029+1227   & \object{TYC 1095-349-1}  & 20 29 32.84  & +12 27 31.1  &  9.74 & 3582.5264 &   -3.01 &  1.34  &   -64.48 &  1.64  &   18.4  &  1.5  &  10.4  &  2.2  & AU  \\ 
   RasTyc2031+3332   & \object{V2425 Cyg}	& 20 31 07.72  & +33 32 33.6  &  8.34 & 2147.4005 &  -31.25 &  0.15  &   -16.88 &  0.15  &    7.2  &  0.4  &   5.6  &  0.6  & EL  \\ 
   RasTyc2044+2916   & \object{HD 335070}	& 20 44 58.09  & +29 16 21.4  & 10.80 & 2147.4449 &   96.48 &  0.69  &  -107.43 &  1.03  &   84.4  &  4.6  &  82.9  &  4.6  & EL  \\ 
   RasTyc2044+2916   & \object{HD 335070}	& 20 44 58.09  & +29 16 21.4  & 10.80 & 1798.4077 &  -11.95 &  0.16  & \dots~~  & \dots~~ &  80.3$^{\mathrm{c}}$    &  6.7  & \dots~~  &  \dots~~ & EL \\  
   RasTyc2051+4547   & \object{BD+45 3306}  	& 20 51 31.46  & +45 47 14.3  &  9.86 & 4347.3761 &  -79.84 &  0.60  & \dots~~  & \dots~~ &   7.8  &  1.9  & \dots~~ & \dots~~ & SA \\         
   RasTyc2051+4547   & \object{BD+45 3306}  	& 20 51 31.46  & +45 47 14.3  &  9.86 & 5077.5466 &  -28.37 &  0.28  &  -95.06  &   2.60 & 16.6 &   2.2 &   23.8      &	8.0    &  FR  \\
   RasTyc2054-0808   & \object{BD-08 5514}      & 20 54 27.81  & -08 08 33.2  &  9.61 & 4301.6708 &   67.75 &  3.56  &  -103.92 &  2.47  &   50.9  &  5.4  &  30.7  &  6.8  & SA  \\  
   RasTyc2058+3510   & \object{CG Cyg}  	& 20 58 13.45  & +35 10 29.6  & 10.07 & 1796.4274 &   -50.0 &  4.0   &    89.0  &  5.0   &   119.0 &  6.3  &  84.3  &  4.4  & EL  \\  
   RasTyc2058+6317   & \object{BD+62 1880}      & 20 58 16.40  & +63 17 38.8  &  9.75 & 4346.3885 &   37.19 &  0.77  &    -8.25 &  1.29  &   11.4  &  4.7  &  19.0  &  5.1  & SA  \\  
   RasTyc2102+2748   & \object{ER Vul}  	& 21 02 25.85  & +27 48 26.4  &  7.36 & 1733.4096 &   93.1  &  7.3   &  -152.4  &  6.9   &  105.8  &  5.7  & 102.3  &  6.1  & EL  \\   
   RasTyc2103+3413   & \object{BD+33 4140}      & 21 03 48.71  & +34 13 13.4  & 10.08 & 1761.3788 &    6.32 &  0.17  &   -28.80 &  0.17  &   32.4  &  1.7  &  21.9  &  1.2  & EL  \\  
   RasTyc2103+3413   & \object{BD+33 4140}      & 21 03 48.71  & +34 13 13.4  & 10.08 & 4347.3874 &    5.54 &  1.15  &   -32.11 &  0.58  &   28.5  &  6.3  &  12.0  &  5.9  & SA  \\  
   RasTyc2103+3413   & \object{BD+33 4140}      & 21 03 48.71  & +34 13 13.4  & 10.08 & 5078.4865 &   -5.55 &  0.84  & \dots~~  & \dots~~ &    44.9  &   2.1 & \dots~~   & \dots~~  &  FR \\
   RasTyc2109+6253   & \object{BD+62 1902}      & 21 09 07.65  & +62 53 19.5  &  9.88 & 4347.3966 & -144.31 & 11.77  &   140.70 & 17.11  &  143.4  & 10.6  & 111.1  & 20.9  & SA  \\ 
   RasTyc2110+3323   & \object{TYC 2706-22-1}	& 21 10 56.16  & +33 23 13.1  & 10.20 & 1736.5037 &   54.31 &  0.15  &   -64.53 &  0.16  &   13.9  &  0.7  &   4.7  &  0.6  & EL  \\ 
   RasTyc2117+4330   & \object{TYC 3181-914-1}  & 21 17 35.44  & +43 30 52.3  & 10.50 & 1762.5273 &  -21.98 &  0.82  &    35.67 &  0.94  &    9.2  &  1.3  &   0.0  &  1.6  & EL  \\ 
   RasTyc2129+3355   & \object{TYC 2712-2338-1} & 21 29 02.69  & +33 55 55.6  & 10.85 & 2150.4887 &   53.81 & 49.89  &  -237.26 & 47.03  &  244.2  & 12.9  & 157.2  &  8.3  & EL  \\  
   RasTyc2131+2320   & \object{LO Peg}  	& 21 31 01.63  & +23 20 08.6  &  9.24 & 1733.4379 &   -2.08 &  0.22  &   -57.79 &  0.23  &   48.2  &  2.5  &  39.3  &  2.1  & EL  \\ 
   RasTyc2134+5632   & \object{HD 239702}	& 21 34 34.34  & +56 32 48.0  &  9.72 & 2443.5735 &  -42.16 &  1.88  &   -64.11 &  2.10  &   45.5  &  0.7  & $<$ 5  & \dots~~ & AU  \\ 
   RasTyc2134+5632   & \object{HD 239702}	& 21 34 34.34  & +56 32 48.0  &  9.72 & 2462.5950 &    1.07 &  1.49  &   -22.53 &  1.80  &  $<$ 5  & \dots~~ &  60.8  &  1.7  & AU  \\ 
   RasTyc2140+2748   & \object{BD+27 4129}	& 21 40 05.41  & +27 48 29.5  & 10.60 & 2119.4738 &   19.36 &  0.15  &   -41.06 &  0.18  &   16.3  &  0.9  &   3.2  &  0.5  & EL  \\  
   RasTyc2151+0956   & \object{BD+09 4914}      & 21 51 24.88  & +09 56 22.2  & 10.03 & 4347.4248 &   23.10 &  0.34  &   -19.91 &  2.58  &    8.8  &  2.1  &   7.2  &  5.3  & SA  \\ 
   RasTyc2156+0515   & \object{TYC 553-33-1}	& 21 56 27.19  & +05 15 56.7  &  9.74 & 2479.4976 &  -87.68 &  1.29  &    64.39 &  1.31  &   13.5  &  1.6  &   2.0  & 14.1  & AU  \\ 
   RasTyc2156+0515   & \object{TYC 553-33-1}	& 21 56 27.19  & +05 15 56.7  &  9.74 & 2482.5432 &  -82.59 &  1.29  &    61.33 &  1.31  &   14.5  &  1.7  &  14.6  &  1.7  & AU  \\ 
   RasTyc2159+1602$^{\rm d}$ & \object{OT Peg}  & 21 59 40.16  & +16 02 18.5  &  9.77 & 1735.5212 &  -15.44 &  0.15  &   -48.01 &  0.17  &    9.0  &  0.5  &   9.4  &  0.5  & EL  \\  
   RasTyc2159+1602$^{\rm d}$ & \object{OT Peg}  & 21 59 40.16  & +16 02 18.5  &  9.77 & 2478.5796 &  -28.32 &  1.38 &  \dots~~ &  \dots~~ & 10.4 &   2.7  &\dots~~ &  \dots~~ & AU \\  
   RasTyc2159+1602$^{\rm d}$ & \object{OT Peg}  & 21 59 40.16  & +16 02 18.5  &  9.77 & 2483.5872 &    8.15 &  1.45 &  \dots~~ &  \dots~~ &  6.0 &   3.7  &  \dots~~ & \dots~~ & AU \\ 
   RasTyc2159+0302   & \object{BD+02 4456}	& 21 59 59.90  & +03 02 25.0  &  9.77 & 2479.5481 &   42.44 &  1.29  &   -68.37 &  1.56  &   13.5  &  1.6  &  13.5  &  1.6  & AU  \\ 
   RasTyc2159+0302   & \object{BD+02 4456}	& 21 59 59.90  & +03 02 25.0  &  9.77 & 2482.5803 &   49.28 &  1.32  &   -75.89 &  1.34  &   14.5  &  1.7  &  14.6  &  1.7  & AU  \\ 
   RasTyc2202+4831$^{\dag}$ & \object{BD+47 3668} & 22 02 57.33 & +31 08 46.8 &  9.89 & 4347.4381 &  -81.47 & 21.50  &    42.89 & 16.94  &   78.1  & 26.1  &  79.6  & 23.1  & SA  \\ 
  RasTyc2206+1005A   & \object{HD 209845} & 22 06 10.58 & +10 05 36.5 &  7.11 & 2150.4603 &   32.72 &  0.17  &   -25.09 &  0.32  &   28.0  &  1.5  &  27.0  &  1.5  & EL  \\  
  RasTyc2206+1005A   & \object{HD209845}	 & 22 06 10.58 & +10 05 36.5 &  7.11  &  2475.5757 &  14.44 &  1.11 & -45.00  & 1.77   & 20.9 &   4.9 &  \dots & \dots & AU \\  
  RasTyc2206+1005A   & \object{HD209845}	 & 22 06 10.58 & +10 05 36.5 &  7.11  &  2476.5798 &    2.00 &  1.38 &  \dots   &  \dots & 24.8 &   1.9 &  \dots &  \dots & AU \\ 
   RasTyc2212+3754   & \object{TYC 3199-3329-1} & 22 12 14.56  & +37 54 56.4  & 10.63 & 4347.5005 &  -74.09 &  1.15  &    53.33 &  2.19  &    9.5  &  4.8  &  24.2  &  6.9  & SA  \\ 
   RasTyc2213+2015$^{\rm *}$ &\object{TYC 1689-910-1} &22 13 18.14 &+20 15 35.5& 10.33 & 4346.4906 &    9.66 &  1.72  &   -29.87 &  1.86  &   25.5  &  6.7  &  22.7  &  4.5  & SA  \\ 
   RasTyc2213+8445   & \object{BD+84 507}	& 22 13 19.94  & +84 45 36.9  &  9.65 & 2482.4724 &   26.20 &  1.38  &   -11.49 &  1.38  &   30.1  &  0.8  &  19.8  &\dots~~  & AU  \\  
   RasTyc2213+8445   & \object{BD+84 507}	& 22 13 19.94  & +84 45 36.9  &  9.65 & 3216.5720 &   32.55 &  1.50  &   -12.12 &  1.52  &   25.8  &  1.9  &  23.7  &  1.6  & AU  \\ 
   RasTyc2214+3356$^{\rm d}$ & \object{BD+33 4462}& 22 14 28.27 & +33 56 29.6 &  8.97 & 1733.5008 &  -65.18 &  0.16  &    20.47 &  0.23  &    9.2  &  0.5  &  30.1  &  1.6  & EL  \\ 
   RasTyc2222+2814B  & \object{BD+27 4302p}	& 22 22 29.06  & +28 14 38.6  &  9.95 & 1733.5518 &   24.38 &  0.16  &   -32.71 &  0.19  &   22.8  &  1.2  &  18.5  &  1.0  & EL  \\  
   RasTyc2222+2814B  & \object{BD+27 4302p}     & 22 22 29.06  & +28 14 38.6  &  9.95 & 4347.5607 &   70.51 &  0.93  &   -80.05 &  2.47  &   21.8  &  3.7  &  13.5  &  4.3  & SA  \\ 
   RasTyc2222+3021   & \object{KX Peg}  	& 22 22 32.57  & +30 21 27.0  &  7.55 & 1734.5085 &   26.52 &  0.15  &   -22.53 &  0.19  &    9.0  &  0.5  &  30.2  &  1.6  & EL  \\ 
   RasTyc2224+1653   & \object{HD 212525}	& 22 24 37.33  & +16 53 48.8  &  8.30 & 2119.6232 &  -42.52 &  0.15  &    -8.52 &  0.15  &    5.1  &  0.3  &   4.8  &  0.6  & EL  \\ 
   RasTyc2224+0637   & \object{BD+05 5019}	& 22 24 58.42  & +06 37 01.4  &  9.57 & 1761.5927 & -100.39 &  0.36  &   111.99 &  0.56  &   52.1  &  2.8  &  18.5  &  1.3  & EL  \\ 
   RasTyc2224+0637   & \object{BD+05 5019}	& 22 24 58.42  & +06 37 01.4  &  9.57 & 5110.4077 &  -68.60  &  0.94  &     74.11 &  3.38  &   49.5  &  8.9  &  35.6  &  6.7  & FR  \\
   RasTyc2226-0050   & \object{BD-01 4295}      & 22 26 53.26  & -00 50 39.2  & 10.36 & 4347.5390 &    3.97 &  0.28  &    33.41 &  0.79  &    7.2  &  3.0  &   5.3  &  3.9  & SA  \\ 
   RasTyc2233+5001   & \object{BD+49 3885}	& 22 33 02.45  & +50 01 10.9  &  9.84 & 3262.4890 &   -1.35 &  1.51  & \dots~~	&\dots~~   &   16.7  &  3.0  & \dots~~  &\dots~~  & AU  \\ 
   RasTyc2233+5001   & \object{BD+49 3885}	& 22 33 02.45  & +50 01 10.9  &  9.84 & 3264.4143 &   -4.03 &  1.33  & \dots~~	&\dots~~   &   12.7  &  3.8  & \dots~~  &\dots~~  & AU  \\ 
   RasTyc2233+1639   & \object{BD+15 4671}	& 22 33 28.44  & +16 39 01.7  &  9.43 & 1761.5433 &   56.29 &  0.41  &   -82.78 &  0.43  &   62.7  &  3.4  &  55.2  &  3.0  & EL  \\  
   RasTyc2233+1639   & \object{BD+15 4671}	& 22 33 28.44  & +16 39 01.7  &  9.43 & 5109.4380 &  -9.74  &  0.73  & \dots~~	&\dots~~   &   48.9  &  3.9 & \dots~~	&\dots~~ & FR \\ 
   RasTyc2236+3318   & \object{TYC 2739-689-1}  & 22 36 16.71  & +33 18 56.4  & 10.57 & 4347.5727 & -103.29 & 19.62  &   103.36 & 21.52  &  126.0  & 20.1  & 132.5  & 26.5  & SA  \\  
   RasTyc2236+4526   & \object{HD 214261}	& 22 36 16.80  & +45 26 46.4  &  9.55 & 3585.5730 &    4.62 &  1.41  & \dots~~	&\dots~~   &   15.4  &  3.7  & \dots~~  &\dots~~  & AU  \\  
   RasTyc2236+4526   & \object{HD 214261}	& 22 36 16.80  & +45 26 46.4  &  9.55 & 3588.5452 &   63.65 &  1.33  &   -53.41 &  1.30  &   13.4  &  1.6  &  10.6  &  2.1  & AU  \\  
RasTyc2242+1900$^{\rm *}$ & \object{TYC 1705-265-1} & 22 42 04.91 & +19 00 49.8& 11.00 &2121.5637 &   19.51 & 19.64  &  -152.40 & 34.071 &  152.7  & 14.0  & 151.8  & 16.0  & EL  \\  
   RasTyc2247+4743   & \object{TYC 3625-1314-1} & 22 47 38.01  & +47 43 02.0  & 10.08 & 4347.5852 &   34.01 &  0.24  &   -50.51 &  0.69  &    1.4  &  1.0  &   2.4  &  1.8  & SA  \\  
   RasTyc2250+1431   & \object{BD+13 5000}	& 22 50 24.03  & +14 31 43.2  & 10.75 & 1798.5803 &   37.25 &  0.15  &   -56.24 &  0.17  &   12.6  &  0.7  &   4.8  &  0.6  & EL  \\  
   RasTyc2304+0949   & \object{BD+09 5155}	& 23 04 58.04  & +09 49 15.7  &  9.60 & 2148.6526 & -119.95 &  0.46  &   108.19 &  0.45  &  47.2   &  2.6  &  42.2  &  2.3  & EL  \\  
   RasTyc2313+0240   & \object{SZ Psc}  	& 23 13 23.78  & +02 40 31.4  &  7.43 & 1734.5952 &  107.32 &  2.6   &   -49.36 & 11.64  &  16.3   &  2.0  &  93.1  &  7.5  & EL  \\  
   RasTyc2317+0551   & \object{BD+05 5154}	& 23 17 13.46  & +05 51 08.1  & 10.20 & 2123.5783 &   11.22 &  0.21  &   -86.90 &  0.23  &   47.6  &  2.5  &  33.0  &  1.8  & EL  \\  
\hline
\noalign{\medskip}
\end{tabular}
\end{scriptsize} 
\end{center}
\end{table*}

\clearpage

\addtocounter{table}{-1}
\begin{table*}
\caption{continued.  }
\begin{center}
\begin{scriptsize} 
\begin{tabular}{llccrcrrrrrrrrc}
\hline
\hline
\noalign{\medskip}
\textsl{RasTyc} Name & Name & $\alpha$ (2000) & $\delta$ (2000)           & $V^{\rm a}$   & HJD        & $RV_1$          & $\sigma$  &  $RV_2$  	& $\sigma$        & $v\sin i_1$  & $\sigma$  & $v\sin i_2$  & $\sigma$ &  Instr.$^{\rm b}$ \\
                     &      &  h m s          & $\degr ~\arcmin ~\arcsec$ & (mag) & (2450000+) & \multicolumn{2}{c}{(km s$^{-1}$)}  &  \multicolumn{2}{c}{(km s$^{-1}$)} & \multicolumn{2}{c}{(km s$^{-1}$)}    & \multicolumn{2}{c}{(km s$^{-1}$)}  & \\
\noalign{\medskip}
\hline
\noalign{\medskip} 
   RasTyc2322+6113   & \object{TYC 4279-1821-1} & 23 22 40.03  & +61 13 33.3  &  9.90 & 2230.3281 &   33.13 &  1.49  &   -53.17 &  1.53  &   11.8  &  2.4  &   1.8  & 17.5  & AU \\ 
   RasTyc2322+6113   & \object{TYC 4279-1821-1} & 23 22 40.03  & +61 13 33.3  &  9.90 & 2234.3389 &  -19.19 &  1.59  &     4.75 &  1.80  &   10.6  &  2.2  &   8.2  &  2.7  & AU \\ 
   RasTyc2330+0126   & \object{TYC 585-897-1}	& 23 30 07.59  & +01 26 04.4  & 10.53 & 1799.5573 &   13.30 &  0.24  &   -60.23 &  0.42  &   48.1  &  2.5  &  29.3  &  1.7  & EL \\ 
   RasTyc2339-0310   & \object{BD-03 5681}	& 23 39 17.55  & -03 10 38.7  & 10.66 & 1797.6419 &  -41.19 &  0.15  &    32.02 &  0.17  &    9.0  &  0.5  &   4.2  &  0.6  & EL \\ 
   RasTyc2348+5744   & \object{V651 Cas}        & 23 48 33.48  & +57 44 56.7  & 10.21 & 2984.3020 &    5.61 &  2.49  &   -59.20 &  5.87  &   40.0  &  3.0  &  46.2  &  4.2  & AU \\  
   RasTyc2348+5744   & \object{V651 Cas}        & 23 48 33.48  & +57 44 56.7  & 10.21 & 2986.3049 &   -8.43 &  4.55  &   -58.90 &  6.43  &   45.5  &  4.0  &  41.2  &  3.5  & AU \\  
RasTyc2351-0636$^{\bullet}$ & \object{HD 223688}& 23 51 46.78  & -06 36 47.0  &  8.73 & 2121.6117 &   15.85 &  0.30  &    26.18 &  0.30  &    2.6  &  0.7  &   2.6  &  0.7  & EL \\ 
   RasTyc2351-0636   & \object{HD 223688}	& 23 51 46.78  & -06 36 47.0  &  8.73 & 2209.4329 &   -1.67 &  1.38  & \dots~~	&\dots~~   &   22.0$^{\mathrm{c}}$  &  4.0  & \dots~~  &\dots~~  & AU  \\
   RasTyc2351-0636   & \object{HD 223688}	& 23 51 46.78  & -06 36 47.0  &  8.73 & 2212.3741 &   15.98 &  1.45  & \dots~~	&\dots~~   &    5.7$^{\mathrm{c}}$  &  0.8  & \dots~~  &\dots~~  & AU  \\
   RasTyc2351-0636   & \object{HD 223688}	& 23 51 46.78  & -06 36 47.0  &  8.73 & 2214.3555 &   15.31 &  1.57  & \dots~~	&\dots~~   &    8.1$^{\mathrm{c}}$  &  1.8  & \dots~~  &\dots~~  & AU  \\
   RasTyc2354+3645   & \object{TYC 2780-2053-1} & 23 54 39.11  & +36 45 16.1  & 10.18 & 4348.5692 &   17.13 &  0.34  &  -215.12 & 14.58  &  164.7  & 15.6  & 100.5  & 24.3  & SA  \\ 
\hline
\noalign{\medskip}
\end{tabular}
\begin{list}{}{}
\item[$^{\mathrm{a}}$] From the TYCHO catalog \citep{HIPPA97}.  
\item[$^{\mathrm{b}}$] Spectrograph: SA = SARG, EL = ELODIE, AU = AURELIE, FR = FRESCO. 
\item[$^{\mathrm{c}}$] Blended components; combined $v\sin i$ . 
\item[$^{\mathrm{d}}$] The full set of RV data and orbital solution are reported in \citet{Frasca2006}. Unpublished AURELIE data have been used here.
\item[$^{\mathrm{*}}$] Double-peaked CCF. Possible SB2 with blended lines.   
\item[$^{\dag}$] Asymmetric CCF. Possible SB2 with blended lines.  
\item[$^{\diamond}$] Small-amplitude secondary peak in the CCF. Likely SB2 with a faint component. 
\item[$^{\bullet}$] A small-amplitude third peak in the CCF at $RV\simeq -1.0$\,km\,s$^{-1}$. SB3? 
\end{list}
\end{scriptsize} 
\end{center}
\end{table*}

\setlength{\tabcolsep}{2pt}

\begin{table*}
\caption{Radial (RV) and rotational ($v\sin i$) velocities for the components of SB3 binaries.  }
\begin{center}
\begin{scriptsize} 
\begin{tabular}{llcccrrrrrrrr}
\hline
\hline
\noalign{\medskip}
\textsl{RasTyc} Name & Name & $\alpha$ (2000) & $\delta$ (2000)           & HJD        & $RV_1$        & $RV_2$        & $RV_3$        & $v\sin i_1$   & $v\sin i_2$   & $v\sin i_3$   \\
                     &      &  h m s          & $\degr ~\arcmin ~\arcsec$ & {\tiny (2450000+)} & (km s$^{-1}$) & (km s$^{-1}$) & (km s$^{-1}$) & (km s$^{-1}$) & (km s$^{-1}$) & (km s$^{-1}$) \\
\noalign{\medskip}
\hline
\noalign{\medskip}
                     &    \multicolumn{6}{c}{\bf SARG}   & \multicolumn{2}{c}{} \\
\noalign{\medskip}
     RasTyc1831+2515      & \object{TYC 2110-348-1} &  18 31 01.05 & +25 15 31.5 & 4210.7296 & -44.89\,$\pm$\,0.83  &  71.33\,$\pm$\,2.10 &  15.42\,$\pm$\,0.22 &  11.2\,$\pm$\,5.6 &  17.6\,$\pm$\,5.8  &  2.0\,$\pm$\,2.3 \\    
RasTyc2052+4258$^{\rm a}$ & \object{BD+42 3895}     &  20 52 57.27 & +42 58 25.7 & 4348.3670 &  10.18\,$\pm$\,4.42  & -48.46\,$\pm$\,5.58 & -19.20\,$\pm$\,0.36 &   9.0\,$\pm$\,4.3 &  14.0\,$\pm$\,12.0 &  3.4\,$\pm$\,2.2 \\
    RasTyc2259-0431       & \object{TYC 5238-1223-1}&  22 59 02.13 & -04 31 35.3 & 4347.6046 &   8.68\,$\pm$\,0.29  & -32.54\,$\pm$\,0.76 & -11.60\,$\pm$\,0.83 &   2.7\,$\pm$\,2.0 &  6.6\,$\pm$\,2.6   &  4.2\,$\pm$\,1.6 \\
\noalign{\medskip}
                       &    \multicolumn{6}{c}{\bf ELODIE}   & \multicolumn{2}{c}{} \\
\noalign{\medskip}
 RasTyc2052+4258$^{\rm a}$ & \object{BD+42 3895} &  20 52 57.27 & +42 58 25.7 & 1797.3516 & -20.38\,$\pm$\,0.15  & -52.74\,$\pm$\,0.17 &   9.98\,$\pm$\,0.16 &   0.0\,$\pm$\,0.2 &   9.6\,$\pm$\,0.5  &  6.5\,$\pm$\,0.4 \\  
 RasTyc2206+1005B      & \object{BD+09 4984B}	 &  22 06 11.82 & +10 05 28.7 & 1795.5746 & -40.00\,$\pm$\,0.54  &  65.75\,$\pm$\,0.48 &	 \dots	     &  52.2\,$\pm$\,2.9 &  42.2\,$\pm$\,2.3  &    \dots. 	 \\  
 RasTyc2206+1005B      & \object{BD+09 4984B}	 &  22 06 11.82 & +10 05 28.7 & 2148.6228 & -26.40\,$\pm$\,0.40  &  18.67\,$\pm$\,0.28 &  57.95\,$\pm$\,0.64 &  28.9\,$\pm$\,1.6 &  30.0\,$\pm$\,1.6  & 28.9\,$\pm$\,1.8 \\  
 RasTyc2206+1005B      & \object{BD+09 4984B}	 &  22 06 11.82 & +10 05 28.7 & 2149.4504 & -29.63\,$\pm$\,0.64  &  34.71\,$\pm$\,0.77 &  74.12\,$\pm$\,0.81 &  38.6\,$\pm$\,2.2 &  43.7\,$\pm$\,2.6  & 29.3\,$\pm$\,2.0 \\  
 RasTyc2206+1005B      & \object{BD+09 4984B}	 &  22 06 11.82 & +10 05 28.7 & 2150.4430 & -25.35\,$\pm$\,0.35  &  16.05\,$\pm$\,0.25 &  60.28\,$\pm$\,0.34 &  29.4\,$\pm$\,1.6 &  30.7\,$\pm$\,1.6  & 28.9\,$\pm$\,1.6 \\  
 RasTyc2233+0506       & \object{TYC 573-566-1}  &  22 33 39.30 & +05 06 20.0 & 1799.4782 &-143.86\,$\pm$\,0.88  &  37.90\,$\pm$\,0.17 &  87.47\,$\pm$\,1.06 &  62.9\,$\pm$\,3.6 &  16.5\,$\pm$\,0.9  & 44.5\,$\pm$\,2.8 \\  
 RasTyc2309-0049$^{\rm b}$ & \object{BD-01 4397} &  23 09 47.09 & -00 49 34.4 & 2149.5545 & -86.39\,$\pm$\,0.24  & -12.92\,$\pm$\,0.15 &  62.51\,$\pm$\,0.23 &  14.0\,$\pm$\,0.8 &   3.1\,$\pm$\,0.2  & 13.2\,$\pm$\,0.7 \\  
\noalign{\medskip}
                       &    \multicolumn{6}{c}{\bf AURELIE}   & \multicolumn{4}{c}{} \\
\noalign{\medskip}
RasTyc0157+3310        & \object{BD+32 354}  	 & 01 57 52.61 & +33 10 19.6 & 2216.3262 & -53.38\,$\pm$\,1.45  & -97.21\,$\pm$\,3.91  &   1.33\,$\pm$\,4.09  &  7.1\,$\pm$\,1.6 & 17.8\,$\pm$\,2.5         &   14.4\,$\pm$\,2.6  \\
RasTyc0157+3310        & \object{BD+32 354}  	 & 01 57 52.61 & +33 10 19.6 & 2217.3308 & -52.78\,$\pm$\,1.83  &-100.53\,$\pm$\,3.82 & -3.76\,$\pm$\,4.50  &  7.3\,$\pm$\,1.4 &  18.8\,$\pm$\,2.3  &   15.6\,$\pm$\,2.3  \\
RasTyc0749+5346        & \object{TYC 3783-646-1} & 07 49 21.54 & +53 46 07.1 & 2986.5720 &   -62.75\,$\pm$\,1.80  &   69.08\,$\pm$\,1.50  & 16.37\,$\pm$\,0.76 &  29.1\,$\pm$\,1.8  & 31.0\,$\pm$\,1.2 & 7.4\,$\pm$\,1.5  \\
RasTyc1932+5433        & \object{HD 234928} 	 & 19 32 34.66 & +54 33 06.4 & 3580.5229 &  56.33\,$\pm$\,1.53  &  -38.31\,$\pm$\,1.61 & 17.81\,$\pm$\,1.31 & 20.1\,$\pm$\,1.3  & \dots &  10.4\,$\pm$\,2.2 	\\
RasTyc1932+5433        & \object{HD 234928} 	 & 19 32 34.66 & +54 33 06.4 & 3583.4944 &  88.24\,$\pm$\,1.14  &  -74.74\,$\pm$\,1.83  &  17.93\,$\pm$\,1.33 & 20.3\,$\pm$\,1.6  & 12.5\,$\pm$\,2.0 & 9.5\,$\pm$\,2.0  \\
RasTyc1932+5433        & \object{HD 234928} 	 & 19 32 34.66 & +54 33 06.4 & 5002.5686 &   30.21\,$\pm$\,0.81  &     \dots	   &      \dots   & 30.4\,$\pm$\,1.7 &      \dots	   &      \dots   \\
RasTyc2034+8253$^{\rm c}$ & \object{BD+82 622}   & 20 34 27.27 & +82 53 35.2 & 2475.5159 &   20.57\,$\pm$\,1.27 & -43.24\,$\pm$\,1.41 & -10.96\,$\pm$\,1.26  &  $<$ 5 &  $<$ 5 & $<$ 5  \\
RasTyc2034+8253$^{\rm c}$ & \object{BD+82 622}   & 20 34 27.27 & +82 53 35.2 & 2482.3916 & -50.12\,$\pm$\,1.42  &  29.29\,$\pm$\,1.28  &  -9.99\,$\pm$\,1.39 & $<$ 5 &  $<$ 5 &  $<$ 5   \\
\noalign{\medskip}
                       &    \multicolumn{6}{c}{\bf FRESCO}   & \multicolumn{2}{c}{} \\
\noalign{\medskip}

RasTyc0221+7811        &  \object{BD+77 80}      & 02 21 52.44 & +78 11 16.6 & 5108.6061 &  -50.73\,$\pm$\,1.14  &  50.55\,$\pm$\,1.81  &  0.59\,$\pm$\,0.48 & 19.4\,$\pm$\,4.0 & 23.7\,$\pm$\,10.0 & 13.7\,$\pm$\,1.5 \\
RasTyc1631+5755        &  \object{HD 238571}     & 16 31 05.95 & +57 55 50.9 & 5110.2851 &  -71.92\,$\pm$\,1.87  & 35.68\,$\pm$\,2.62  &  -19.20\,$\pm$\,0.33 & 19.3\,$\pm$\,5.6 &  30.0\,$\pm$\,8.7 & 12.1\,$\pm$\,3.8 \\

\hline		       
\noalign{\medskip      }
\end{tabular}	       
\begin{list}{}{}									
\item[$^{\rm a}$] The tertiary component is much brighter than the inner binary and was also analyzed as a single star. 
\item[$^{\rm b}$] Close visual pair \citep[sep$\simeq1\farcs$5, $\Delta V\simeq 0.6$\,mag,][]{Hog2000}. 
\item[$^{\rm c}$] The full set of RV data and orbital solution are reported in \citet{Klutsch2008}.
\end{list}
\end{scriptsize} 
\end{center}
\label{Tab:RV_SB3}
\end{table*}

\setlength{\tabcolsep}{3pt}

\begin{table*}[p]
\caption{Spectral types, atmospheric parameters, and equivalent widths of \ion{Li}{i} and H$\alpha$ lines for the single stars (S) and SB1 binaries.}
\begin{center} 
\begin{scriptsize} 
\begin{tabular}{lccccrccrcrrrrcc}
\hline
\hline
\noalign{\medskip}
\textsl{RasTyc} Name & $\alpha$ (2000) & $\delta$ (2000)	      & Sp. Type &  $T_{\rm eff}$ & $\sigma_{T_{\rm eff}}$ & $\log g$ & $\sigma_{\log g}$ & $[Fe/H]$ & $\sigma_{[Fe/H]}$ & $W_{\rm Li}$ & $\sigma_{W_{\rm Li}}$ & $W_{\rm H\alpha}^{em}$ $^{\bullet}$ & $\sigma_{W_{\rm H\alpha}^{em}}$ &  Bin$^{\rm a}$ &  Instr.$^{\rm b}$  \\
                     &  h m s	       & $\degr ~\arcmin ~\arcsec$    & 	 &   \multicolumn{2}{c}{(K)}		   &	      \multicolumn{2}{c}{(dex)}	   &	    \multicolumn{2}{c}{(dex)}		 &	   \multicolumn{2}{c}{(m\AA)}			      &       \multicolumn{2}{c}{(m\AA)}   &  & \\
\noalign{\medskip}
\hline
\noalign{\medskip}
    RasTyc0000+7940  & 00 00 41.14 & +79 40 39.9   & G2V  & 5738  & 140  &  4.19 &  0.20 &  -0.03 &  0.09 & 202 & 40  & 310 &  50 & S & AU+FR  \\ 
    RasTyc0001+5212  & 00 01 42.66 & +52 12 51.0   &  K1V      &   5300   &   103  &  4.39  &  0.17 & -0.03 & 0.11 &  64 &  20 &   130 &   37  & S & SA \\ 
    RasTyc0004-0951  & 00 04 46.96 & -09 51 53.4   &   K1V     &   5141   &    63  &  4.53  &  0.14 & -0.01 & 0.11 &  41 &  46 &     5 &   41  & S & EL \\ 
    RasTyc0008+5347  & 00 08 04.62 & +53 47 47.0	&    F8V  &   5944   &    85  &  4.18  &  0.12 & -0.12 & 0.11 &  69 &	6 &	1 &    6  & S & EL \\	
    RasTyc0013+7702$^{\rm c}$ & 00 13 40.52 & +77 02 10.9 &  G0V & 5924 &  243 & 4.14 & 0.22 & -0.03 & 0.09 & 175 & 10 & 3900 & 90 & SB2? &  AU  \\   
    RasTyc0013+3946  & 00 13 58.01 & +39 46 02.3   &    G3V	&   5707   &   140  &  4.31  &  0.15 & -0.07 & 0.12 &  40 &   7 &    89 &    9 &  S &  AU   \\
    RasTyc0016+4104  & 00 16 44.88 & +41 04 08.6   &    K3V	&   5010   &   197  &  4.17  &  0.39 &  0.03 & 0.14 &	2 &   5 &   272 &   38 & S &  AU   \\
    RasTyc0023+7503  & 00 23 41.24 & +75 03 16.9   &    K2V	&   5091   &   347  &  4.52  &  0.27 & -0.06 & 0.15 & 147 & 176 &   112 &  289  & S & SA \\    
    RasTyc0032+7806  & 00 32 42.87 & +78 06 47.4   &    G3V     &   5715   &   145  &  4.29  &  0.15 & -0.02 & 0.11 &  21 &   7 &	93 &   14 & S & AU   \\
    RasTyc0033+5315$^{\rm c}$ & 00 33 55.90 & +53 15 41.5 & G6V &   5716   &	70  &  4.32  &  0.11 & -0.07 & 0.11 &  55 &  23 &   223 &   55  & SB2? &  SA \\ 
    RasTyc0033+6126  & 00 33 57.58 & +61 26 33.3   &    K0V     &   5209   &    97  &  4.53  &  0.15 &  0.05 & 0.13 &   1 &  24 &  1160 &  117 & S &  AU   \\  
    RasTyc0038+7903  & 00 38 06.03 & +79 03 20.7   &    K1V     &   5160   &   119  &  4.30  &  0.19 & -0.06 & 0.08 & 290 &  20 &  1125  & 157  & S &  AU+FR   \\
    RasTyc0039+7905$^{\rm *}$  & 00 39 40.13 & +79 05 30.8  &    G5V     &   5444   &   265  &  4.14  &  0.28 & -0.11 & 0.11 & 216 &  17 &   221 &   19 & SB2 &  AU  \\
    RasTyc0040+4343  & 00 40 20.90 & +43 43 25.4   &    K2V	&   5095   &	74  &  4.27  &  0.18 & -0.04 & 0.10 &  38 &  15 &  1303 &  108 & S &  SA \\    
    RasTyc0041+3425  & 00 41 17.30 & +34 25 17.2   &    K1V     &   5174   &    68  &  4.57  &  0.08 & -0.01 & 0.07 & 123 &  25 &  3153 &  160 & S & AU \\  
    RasTyc0045+7943  & 00 45 22.95 & +79 43 49.6   &    G3V     &   5705   &   140  &  4.16  &  0.23 & -0.32 & 0.09 & 135 &  20 &  \dots  &  \dots & S & AU \\
    RasTyc0046+4808  & 00 46 53.09 & +48 08 45.2   &    K2V     &   5067   &   159  &  4.21  &  0.45 & -0.09 & 0.07 & 355 &  14 &   809 &   42 &  S  &  AU   \\
    RasTyc0050+4651  & 00 50 38.39 & +46 51 57.5   &    G2V     &   5900   &   152  &  4.30  &  0.11 &  0.05 & 0.07 &  80 &  10 &    68 &   12 &  S  &  AU   \\
    RasTyc0051+5425  & 00 51 03.32 & +54 25 18.6   &    K0IV	&   5191   &   288  &  4.47  &  0.34 &  0.01 & 0.16 & 117 & 102 &   856 &  210 &  S  &  SA \\  
    RasTyc0051-1306  & 00 51 17.10 & -13 06 52.5   &    G1V     &   5870   &    67  &  4.29  &  0.11 & -0.02 & 0.11 &  27 &  21 &    51 &   40 &  S  &  EL \\	
    RasTyc0100+5411  & 01 00 11.71 & +54 11 01.7   &    G1V     &   5832   &    80  &  4.29  &  0.11 &  0.03 & 0.06 &  47 &  11 &    76 &   13 &  S  &  AU   \\
    RasTyc0103+5727  & 01 03 04.48 & +57 27 17.7   &    K3V	&   4909   &	73  &  4.58  &  0.16 &  0.09 & 0.11 &  17 &  15 &   175 &   40 &  S  &  SA \\  
    RasTyc0106+3306  & 01 06 18.71 & +33 06 01.9   &    G4V     &   5606   &   134  &  4.31  &  0.11 & -0.14 & 0.12 & 199 &  29 &   569 &   64 &  S  &  AU   \\
    RasTyc0106+5729  & 01 06 27.36 & +57 29 44.1   &    G8III	&   5158   &	97  &  3.40  &  0.16 & -0.05 & 0.11 &  36 &  16 &   510 &   97 &  S  &  SA \\  
    RasTyc0116+3938  & 01 16 50.67 & +39 38 18.9   &    K1V     &   5148   &    80  &  4.02  &  0.45 & -0.10 & 0.06 &   7 &   7 &   319 &   29 &  S  &  AU   \\
    RasTyc0137+3900  & 01 37 27.16 & +39 00 08.6   &    G8IV    &   5241   &    84  &  3.55  &  0.87 & -0.26 & 0.26 &\dots & \dots & 1177 & 69 &  S  &  AU   \\
    RasTyc0140+4212  & 01 40 28.78 & +42 12 01.6   &    K2V     &   5060   &   204  &  4.23  &  0.49 & -0.09 & 0.10 &   8 &  14 &  1505 &   83 & SB1  &  AU   \\
    RasTyc0140+4952  & 01 40 51.62 & +49 52 31.2   &    G5IV	&   5508   &   137  &  4.27  &  0.19 & -0.05 & 0.12 & 168 &  40 &   553 &  118 & S  & SA \\  
    RasTyc0144+6508  & 01 44 22.37 & +65 08 47.2   &    G5III	&   5454   &   139  &  3.89  &  0.21 &  0.08 & 0.11 &  43 &  34 &   866 &  172 & S  & SA \\  
    RasTyc0156+4354  & 01 56 38.45 & +43 54 46.8   &    K2V     &   5129   &   139  &  4.28  &  0.33 & -0.15 & 0.11 &   3 &   3 &   249 &   19 &  SB1  & AU  \\
    RasTyc0158+3601  & 01 58 26.62 & +36 01 19.6   &    G5V     &   5760   &   145  &  4.26  &  0.15 & -0.14 & 0.18 &  90 &   8 &   124 &   17 & S  & AU   \\
    RasTyc0218+4346  & 02 18 13.49 & +43 46 30.2   &    K2V     &   5124   &    91  &  4.02  &  0.48 & -0.10 & 0.08 &  89 &  11 &  1225 &   61 & SB1  &  AU   \\
    RasTyc0221+3404  & 02 21 33.30 & +34 04 45.6   &    K0IV    &   5184   &    85  &  3.82  &  0.49 &  0.06 & 0.18 &   1 &   8 &   517 &   61 & SB1  & AU \\
    RasTyc0222+7204  & 02 22 12.91 & +72 04 58.0   & K0III-IV	&   4866   &   128  &  3.01  &  0.28 & -0.09 & 0.12 & 105 &  29 &   763 &  122  & S  & SA \\      
    RasTyc0222+5033  & 02 22 33.82 & +50 33 37.8   &    K2V     &   5094   &    78  &  4.27  &  0.40 & -0.12 & 0.09 & 248 &  17 &  1060 &   42 & S  & AU   \\
    RasTyc0227+4554  & 02 27 40.47 & +45 54 37.3   &    G2V	&   5663   &	74  &  4.34  &  0.11 & -0.02 & 0.11 & 116 &  22 &   481 &  137  &  S  &  SA \\  	
    RasTyc0229+7206  & 02 29 44.65 & +72 06 03.8   &    K0IV    &   5156   &    93  &  4.03  &  0.47 & -0.06 & 0.10 &  27 &  16 &   409 &   20 &   SB1  &  AU \\
    RasTyc0230+5656  & 02 30 44.81 & +56 56 13.0   &    G1V	&   5874   &	97  &  4.31  &  0.14 & -0.03 & 0.14 & 208 &  83 &   281 &   76  &  S  & SA \\  	
    RasTyc0230+5533  & 02 30 48.51 & +55 33 07.2   &    K1V     &   5107   &    81  &  4.41  &  0.16 & -0.10 & 0.07 & 156 &  11 &   266 &   18 &   S  &  AU   \\
    RasTyc0235+3139  & 02 35 03.79 & +31 39 22.3   &    G5V     &   5584   &   196  &  4.21  &  0.21 & -0.20 & 0.20 & 172 &  31 &  \dots  &  \dots &  S  & AU   \\
    RasTyc0240+6143  & 02 40 12.74 & +61 43 59.0   &    K1V	&   5276   &   114  &  4.28  &  0.22 &  0.04 & 0.11 &  36 &  19 &   522--742 &   93  &  SB1 &  SA \\ 	
    RasTyc0242+4527  & 02 42 20.50 & +45 27 43.8   &   K0V      &  5389    &  257   &  4.28  &  0.23 &  0.02 & 0.14 &  145 &  30 &   104 &   26 &  S  &  FR \\
    RasTyc0242+3837  & 02 42 20.89 & +38 37 22.2   &    K2V     &   5071   &   148  &  4.21  &  0.44 & -0.07 & 0.08 & 132 &  11 &   607 &   22 &  S  & AU   \\
    RasTyc0249+4255  & 02 49 54.69 & +42 55 27.1   &   G1.5V	&   5825   &   107  &  4.24  &  0.12 & -0.02 & 0.12 & 157 &  37 &   251 &   65  & S  &  SA \\ 
    RasTyc0252+3616  & 02 52 17.56 & +36 16 48.5   &   K1IV	&   4679   &   174  &  3.66  &  0.51 & -0.08 & 0.14 & 174 &  86 &  3179 &  237  &  S  &  SA \\
    RasTyc0252+3728  & 02 52 24.71 & +37 28 52.0   &   G1.5V	&   5784   &   157  &  4.35  &  0.15 &  0.02 & 0.14 & 293 & 187 &   305 &  108  &  S  &  SA \\   
    RasTyc0256+7253  & 02 56 11.33 & +72 53 10.4   &   G1V      &   5804   &   175  &  4.30  &  0.12 & -0.15 & 0.18 &  92 &	 46&    \dots  &  \dots & S  &  FR \\
    RasTyc0256+6033  & 02 56 58.03 & +60 33 52.7   &   K2V	&   5072   &	57  &  4.59  &  0.13 & -0.02 & 0.10 &  32 &  19 &   189 &   70  &  S  &  SA \\     
    RasTyc0300+7225  & 03 00 14.67 & +72 25 41.4   &   K2V	&   5079   &	59  &  4.59  &  0.13 & -0.01 & 0.10 & 264 &  22 &   845 &  105  &  S  &  SA \\     
    RasTyc0302+4421  & 03 02 46.57 & +44 21 03.1   &   K1V      &   5214   &   127  &  4.10  &  0.40 & -0.08 & 0.06 &   0 &   8 &   238 &   13 &  S  &  AU   \\
    RasTyc0311+4810  & 03 11 16.82 & +48 10 36.9   &   G1.5V	&   5856   &   110  &  4.35  &  0.14 &  0.02 & 0.15 & 201 &  72 &   483 &  136  &  S  &  SA \\  
    RasTyc0313+7417  &	03 13 27.48 & +74 17 47.0  &   G0IV     &   6025   &  189   &  4.00  &  0.21 &   0.02 & 0.12 &   38 &  92 & \dots  &  \dots &  S  &  FR \\
    RasTyc0313+3832  & 03 13 47.40 & +38 32 04.9   &    K0V     &   5560   &   258  &  4.32  &  0.10 &  0.11 & 0.10 & 142 &   9 &  \dots  &  \dots &  S  &  AU  \\
    RasTyc0316+4724  & 03 16 07.07 & +47 24 58.7   &    G4V     &   5665   &   147  &  4.38  &  0.07 &  0.07 & 0.23 &   0 &   8 &   110 &   22 &  S  &  AU   \\
    RasTyc0316+5638  & 03 16 28.11 & +56 38 58.1   &    K1V	&   5208   &	74  &  4.50  &  0.12 & -0.05 & 0.11 & 223 &  31 &   550 &   83  &  S  &  SA \\     
    RasTyc0316+6049  & 03 16 59.73 & +60 49 10.0   &    K1III    &   4553   &   186  &  2.33  &  0.58 & -0.08 & 0.13 &  39 &  23 &   155 &   14 &  SB1  &  AU  \\
    RasTyc0319+4328  & 03 19 59.33 & +43 28 08.1   &    F8V     &   5931   &   159  &  3.95  &  0.08 & -0.26 & 0.15 &   2 &   8 &   162 &   22 &  SB1  &  AU  \\
    RasTyc0323+5843  & 03 23 07.08 & +58 43 07.4   &    K1V	&   5212   &   130  &  4.48  &  0.20 & -0.07 & 0.14 & 244 &  47 &  1009 &  135  & S  & SA \\     
    RasTyc0325+3647  & 03 25 02.39 & +36 47 56.8   &    K0V     &   5375   &   212  &  4.33  &  0.17 & -0.15 & 0.12 &\dots  & \dots &   161 &   18 & S  & AU  \\
    RasTyc0328+3114  & 03 28 57.21 & +31 14 19.2   &    G4V     &   5762   &   101  &  4.20  &  0.20 & -0.04 & 0.26 &\dots  & \dots &   234 &   24 &  S  & AU  \\
    RasTyc0331+4859  & 03 31 28.98 & +48 59 28.6   &   G1.5V	&   5816   &   236  &  4.30  &  0.19 & -0.10 & 0.18 & 185 &  63 &   738 &  120  &  S  &  SA \\     
    RasTyc0334+3846  & 03 34 38.28 & +38 46 28.2   &    G5V     &   5355   &   226  &  4.17  &  0.27 & -0.08 & 0.10 & 118 &  14 &	81 &   14 &  S  &  AU  \\
    RasTyc0336+4816  & 03 36 40.23 & +48 16 13.4   &    G3V     &   5641   &   187  &  4.14  &  0.32 & -0.23 & 0.17 &   24  &  47  &  44  &  59   &  S  &  FR \\
    RasTyc0339+6639  & 03 39 14.22 & +66 39 40.3   &    K2V     &   5093   &    77  &  4.43  &  0.18 & -0.10 & 0.07 &  64 &   7 &   376 &   22 &  S  &  AU   \\
    RasTyc0344+5043  & 03 44 34.50 & +50 43 47.5   &    K0V	&   5349   &   211  &  4.40  &  0.22 & -0.06 & 0.16 & 227 &  71 &   849 &  143  & S  &  SA \\  	
    RasTyc0348+6840  & 03 48 00.50 & +68 40 58.2   &    K1V	&   5200   &	78  &  4.31  &  0.25 & -0.09 & 0.11 &  29 &  18 &   323 &   95  &  S  &  SA \\ 		 
    RasTyc0357+5051  & 03 57 19.91 & +50 51 19.2   &    K2V     &   5104   &    76  &  4.25  &  0.43 & -0.13 & 0.06 & 207 &  12 &   914 &   31 &  S  &  AU  \\
    RasTyc0359+4404  & 03 59 16.70 & +44 04 17.1   &    K1V     &   5235   &   149  &  4.50 &  0.13 & -0.06 & 0.13 &  292 & 47 &  2004 & 125 & S?  &  AU \\
   RasTyc0412+7318A  & 04 12 01.13 & +73 18 34.2  &   G3V	&   5531   &   101  &  4.07  &  0.22 & -0.01 & 0.13 &  17 &  22 &    48 &   32  &  S  &  SA \\     
   RasTyc0412+7318B  & 04 12 01.87 & +73 18 38.4  &    K1V	&   4999   &   244  &  4.56  &  0.29 &  0.00 & 0.16 &  34 &  92 &  1513 &  345  &  S  &  SA \\        
   RasTyc0412+4616B  & 04 12 14.87 & +46 16 12.8  &    K2V	&   4992   &   133  &  4.56  &  0.21 & -0.01 & 0.12 & 180 & 173 &   420 & 1581  &  S  &  SA \\     
    RasTyc0414+4218  & 04 14 39.01 & +42 18 54.5   &    K1V     &   5190   &   132  &  4.32  &  0.23 & -0.10 & 0.13 &  74 &   7 &   190 &   13 &  S  &  AU   \\
RasTyc0439+3407$^{\rm **}$& 04 39 30.98& +34 07 45.0 &    G7V     &   5428   &   326  &  4.32  &  0.11 & -0.07 & 0.17 & 187 &  28 &   717 &   74 &  S  &  AU   \\ 
    RasTyc0449+4504  & 04 49 42.49 & +45 04 54.4   &    F8V	&   6084   &   112  &  3.97  &  0.11 & -0.30 & 0.12 &  84 &  27 &    31 &   27  &  S  &  SA \\  
    RasTyc0449+4902  & 04 49 43.13 & +49 02 55.1   &    G9IV    &   5269   &   229  &  3.92  &  0.47 & -0.06 & 0.10 &  39 &   8 &   144 &   13  &  S  &  AU   \\
    RasTyc0454+3410  & 04 54 56.37 & +34 10 07.8   &    K2V     &   5109   &    89  &  4.40  &  0.22 & -0.15 & 0.08 &   0 &   7 &   267 &   30 &  S  &  AU   \\
\hline
\noalign{\medskip}
\end{tabular}
\end{scriptsize} 
\end{center}
\label{Tab:APs}
\end{table*}

\addtocounter{table}{-1}
\begin{table*}[p]
\caption{continued.}
\begin{center} 
\begin{scriptsize} 
\begin{tabular}{lccccrccrcrrrrcc}
\hline
\hline
\noalign{\medskip}
\textsl{RasTyc} Name & $\alpha$ (2000) & $\delta$ (2000)	       & Sp. Type &  $T_{\rm eff}$ & $\sigma_{T_{\rm eff}}$ & $\log g$ & $\sigma_{\log g}$ & $[Fe/H]$ & $\sigma_{[Fe/H]}$ & $W_{\rm Li}$ & $\sigma_{W_{\rm Li}}$ & $W_{\rm H\alpha}^{em}$ $^{\bullet}$ & $\sigma_{W_{\rm H\alpha}^{em}}$ & Bin$^{\rm a}$  & Instr.$^{\rm b}$   \\
                     &  h m s	       & $\degr ~\arcmin ~\arcsec$     &	  &   \multicolumn{2}{c}{(K)}		    &	    \multicolumn{2}{c}{(dex)}	   &	    \multicolumn{2}{c}{(dex)}	  &	    \multicolumn{2}{c}{(m\AA)}  		       &       \multicolumn{2}{c}{(m\AA)}   &  &	\\
\noalign{\medskip}
\hline
\noalign{\medskip}
RasTyc0501+3430$^{\ddag}$ & 05 01 10.83 & +34 30 26.5&    \dots   &   \dots  & \dots  &  \dots & \dots & \dots &\dots & \dots & \dots & \dots &\dots &  S?  & AU   \\
    RasTyc0507+4720  & 05 07 12.41 & +47 20 37.4   &    G2IV    &   5893   &   112  &  4.22  &  0.14 &  0.12 & 0.09 &  33 &   9 & \dots &\dots &  S  & AU   \\
    RasTyc0512+4119  & 05 12 22.96 & +41 19 40.3   &    K1V     &   5107   &    70  &  4.43  &  0.26 & -0.11 & 0.06 & 144 &  11 &   599 &   30 &  S  &  AU   \\
    RasTyc0519+6303  & 05 19 04.44 & +63 03 34.5   &    G3V     &   5767   &   146  &  4.19  &  0.19 & -0.13 & 0.18 &  17 &   7 &	 1 &	6 &  S  &  AU   \\
    RasTyc0535+3946  & 05 35 05.64 & +39 46 31.9   &    K0IV	&   5134   &   104  &  3.69  &  0.29 &  0.03 & 0.13 & 109 &  48 &   545 &  168  &  S  &  SA \\   
    RasTyc0537+5231  & 05 37 03.91 & +52 31 26.1   &    K7V     &   4504   &   317  &  4.43  &  0.20 & -0.04 & 0.07 &\dots&\dots&   807 &   39 &  S  &  AU   \\
    RasTyc0544+4024  & 05 44 26.50 & +40 24 26.1      &    K1V     &   5206   &   155  &  4.33  &  0.22 & -0.09 & 0.06 &   5 &  10 &	 5 &   10  &  S  &  AU   \\
    RasTyc0609+5801  & 06 09 00.42 & +58 01 09.0      &    G0V	   &   5746   &   105  &  4.19  &  0.12 & -0.45 & 0.13 &   2 &  13 &   348 &   86  &  S  &  SA \\   
    RasTyc0612+4733  & 06 12 30.81 & +47 33 09.1      &    G1V     &   5726   &   153  &  4.37  &  0.04 & -0.05 & 0.10 &  38 &   6 &   136 &   12  &  S  &  AU   \\
    RasTyc0616+4516  & 06 16 46.95 & +45 16 03.1      &    G2V	   &   5739   &   107  &  4.21  &  0.18 &  0.11 & 0.11 & 229 &  38 &  1235 &  220  &  S  &  SA \\  	
    RasTyc0620+7353  & 06 20 07.99 & +73 53 30.5      &    G5V     &   5587   &   184  &  4.19  &  0.21 & -0.12 & 0.12 &  38 &   7 &   105 &   16  & SB1 &  AU   \\
    RasTyc0621+5415  & 06 21 56.92 & +54 15 49.0      &    G2V     &   5719   &   127  &  4.14  &  0.22 & -0.08 & 0.13 & 160 &  12 &   583 &   22  &  S  &  AU   \\
    RasTyc0624+5940  & 06 24 43.95 & +59 40 10.8      &    G8III   &   4994   &   158  &  2.93  &  0.42 &  0.05 & 0.07 &\dots&\dots&	 1 &	4  &  S  &  AU   \\
    RasTyc0628+3115A & 06 28 23.61 & +31 15 51.2      &    K1V     &   5217   &   165  &  4.44  &  0.09 & -0.14 & 0.10 & 126 &  11 &   224 &   20  &  S  &  AU   \\
    RasTyc0628+3115B & 06 28 22.99 & +31 15 59.2      &    K1V     &   5231   &   180  &  4.44  &  0.09 & -0.15 & 0.10 & 126 &  13 &   142 &   15  &  S  &  AU   \\
    RasTyc0634+3404  & 06 34 34.67 & +34 04 33.0      &    G9IV    &   5420   &   325  &  3.95  &  0.43 &  0.02 & 0.13 &   3 &   5 &  1829 &  128  &  S  &  AU   \\
    RasTyc0638+3153  & 06 38 12.12 & +31 53 11.8      &    F9IV-V  &   5864   &   257  &  4.23  &  0.15 & -0.05 & 0.17 &  26 &  11 &	62 &   11  &  S  &  AU   \\
    RasTyc0646+3124  & 06 46 25.20 & +31 24 46.0      &    G8IV    &   5223   &    65  &  3.50  &  0.53 & -0.15 & 0.22 &   3 &   8 &   247 &  137  & SB1 &  AU \\
    RasTyc0646+4147  & 06 46 46.74 & +41 47 12.2      &    G2V     &   5696   &	   84  &  4.16  &  0.13 &  0.13 & 0.11 & 210 &  30 &   211 &   65  &  S  & SA \\      
    RasTyc0652+5720  & 06 52 29.28 & +57 20 46.8      &    K0IV	   &   5125   &   146  &  4.03  &  0.71 & -0.11 & 0.11 & 103 &  14 &  1151 &   66  & SB1 &  AU   \\
    RasTyc0712+7021  & 07 12 50.04 & +70 21 06.8      &    K1IV    &   5012   &   227  &  2.96  &  0.49 & -0.03 & 0.08 &  99 &  14 &   931 &   40  & SB1 &  AU   \\
    RasTyc0713+5106  & 07 13 36.53 & +51 06 17.3      &    G4V     &   5877   &   181  &  4.20  &  0.17 & -0.11 & 0.20 &   3 &  11 &   322 &   23  &  S  & AU	\\
    RasTyc0714+5307  & 07 14 36.39 & +53 07 40.4      &    F7IV    &   6057   &   186  &  4.04  &  0.18 & -0.14 & 0.17 &  50 &  11 & \dots &\dots  &  S  & AU \\
    RasTyc0730+6343  & 07 30 55.31 & +63 43 50.1      &    K1III   &   4672   &   177  &  2.68  &  0.74 & -0.07 & 0.13 &  74 &   9 & \dots &\dots  &  S  & AU	\\
    RasTyc0734+3518  & 07 34 34.76 & +35 18 59.7      &    G5V     &   5850   &    98  &  4.30  &  0.10 &  0.01 & 0.09 &  46 &  11 &   121 &   17  &  S  &  AU   \\
    RasTyc0755+4040  & 07 55 04.52 & +40 40 25.0      &    F9IV-V  &   6034   &   193  &  4.00  &  0.11 & -0.14 & 0.18 &   5 &   6 &   181 &   21  &  S  &  AU   \\
    RasTyc0755+6509  & 07 55 54.19 & +65 09 11.1      &    K0IV    &   5085   &   166  &  3.44  &  0.85 & -0.07 & 0.15 & 132 &  11 &   379 &   21  &  S  &  AU   \\
    RasTyc1505+4626  & 15 05 32.49 & +46 26 38.6      &    K3V     &   4950   &   158  &  4.55  &  0.37 &  0.04 & 0.12 &   0 &  32 &  1340 &  141  &  S  & SA \\
    RasTyc1507+5515  & 15 07 27.92 & +55 15 56.5      &     G2V    &   5710   &   134  &  4.27  &  0.12 &  0.01 & 0.13 &  28 &  26 &   110 &   49  &  S  & FR \\ 
    RasTyc1507+8629  & 15 07 53.09 & +86 29 29.4      &    G2V	   &   5787   &	   82  &  4.34  &  0.11 &  0.01 & 0.11 &  79 &  20 &    90 &   29  &  S  &  SA \\ 
    RasTyc1507+0415  & 15 07 59.63 & +04 15 21.1      &    K2V     &   5072   &   152  &  4.22  &  0.47 & -0.10 & 0.09 & 146 &  12 &  1231 &   81  &  S  &  AU \\
    RasTyc1519+3940  & 15 19 38.40 & +39 40 05.5      &    G0IV    &   5853   &	   94  &  4.12  &  0.16 &  0.12 & 0.11 &  13 &  43 &    68 &   37  &  S  &  SA \\    
    RasTyc1521-0920  & 15 21 52.71 & -09 20 18.3      &    K1V     &   5276   &	   96  &  4.40  &  0.16 &  0.01 & 0.11 &  57 &  18 &    45 &   19  &  S  &  SA \\ 
    RasTyc1522+4137  & 15 22 11.06 & +41 37 08.9      &   G5IV-V   &   5636   &    71  &  4.28  &  0.12 &  0.03 & 0.11 &  37 &  22 &    99 &   44  &  S  &  SA \\    
    RasTyc1522+2745  & 15 22 35.40 & +27 45 31.0      &    G9IV    &   5196   &    56  &  4.00  &  0.25 &  0.14 & 0.10 &  17 &  17 &    34 &   22  &  S  &  SA \\
    RasTyc1527+1801  & 15 27 38.20 & +18 01 35.3      &	   G2V	   &   5765   &	   71  &  4.31  &  0.11 &  0.04 & 0.10 &  72 &  17 &    70 &   24  &  S  &  SA \\ 
   RasTyc1529+6712A  & 15 29 24.15 & +67 12 15.9      &	   F8V	   &   5998   &   100  &  4.04  &  0.15 &  0.03 & 0.12 &   3 &  19 &     8 &   16  &  S  &  SA \\    
    RasTyc1529+4836  & 15 29 42.46 & +48 36 13.1      &    G3V     &   5782   &   237  &  4.08  &  0.13 & -0.25 & 0.17 &  52 &  10 &	 0 &    5  &  S  &  AU \\
   RasTyc1540+4027A  & 15 40 58.90 & +40 27 00.3      &    G8IV    &   5563   &   180  &  4.07  &  0.20 & -0.04 & 0.14 &  42 &  21 &   659 &   96  &  S  &  SA \\
   RasTyc1540+4027B  & 15 40 58.90 & +40 27 00.3      &    K1V     &   5263   &    93  &  4.17  &  0.22 & -0.03 & 0.11 &  44 &  58 &    42 &   57  &  S  &  SA \\
    RasTyc1541+7702  & 15 41 59.71 & +77 02 41.8      &    G5IV    &   5722   &	   83  &  4.09  &  0.12 & -0.06 & 0.11 & 118 &  20 &   131 &   40  &  S  &  SA \\ 
   RasTyc1547+5302B  & 15 47 09.38 & +53 01 26.5      &    K1V	   &   5178   &	   75  &  4.43  &  0.15 &  0.07 & 0.10 &   9 &  14 &	 2 &   41  &  S  &  SA \\ 
   RasTyc1549+4608A  & 15 49 24.84 & +46 08 22.2      &    K1V     &   5158   &    75  &  4.44  &  0.22 & -0.11 & 0.11 &  32 &  15 &  1183 &  152  &  S  &  SA \\
   RasTyc1549+4608B  & 15 49 24.32 & +46 08 10.3      &	   K7V     &   4381   &   157  &  4.58  &  0.10 & -0.14 & 0.11 &   0 &  46 &    21 &   62  &  S  &  SA \\
    RasTyc1550+5250  & 15 50 36.72 & +52 50 22.0      &    G1.5V   &   5738   &	   94  &  4.16  &  0.15 & -0.13 & 0.12 & 102 &  42 &   179 &  124  &  S  &  SA \\ 
   RasTyc1550+1440A  & 15 50 55.23 & +14 40 42.8      &    F8V	   &   6008   &	   98  &  4.15  &  0.12 & -0.20 & 0.12 &   2 &   6 &   -31 &   31  &  S  &  SA \\ 
    RasTyc1558+0833  & 15 58 54.66 & +08 33 55.3      &	   K1V	   &   5325   &	   85  &  4.33  &  0.15 &  0.03 & 0.11 &   3 &  12 &	23 &   14  &  S  &  SA \\ 
    RasTyc1603+8142  & 16 03 26.61 & +81 42 20.4      &	   G2V	   &   5845   &	   65  &  4.32  &  0.10 &  0.09 & 0.10 &  38 &  18 &   159 &   38  &  S  &  SA \\ 
    RasTyc1604+0358  & 16 04 16.06 & +03 58 09.4      &	   G1.5V   &   5728   &	   99  &  4.29  &  0.11 &  0.06 & 0.12 &   6 &  14 &   409 &   68  &  S  &  SA \\ 
    RasTyc1617-0406  & 16 17 21.68 & -04 06 50.0      &    F9IV-V  &   6005   &   121  &  3.99  &  0.13 & -0.25 & 0.12 &  30 &  10 &   397 &   81  &  S  &  SA \\    
    RasTyc1620+0707 & 16 20 03.26 & +07 07 29.6       &    G9V     &   5340   &   253  &  4.34  &  0.29 & -0.40 & 0.36 &  32 & 155 &   673 &  173  & SB1 &  FR \\ 
    RasTyc1620+4841  & 16 20 38.67 & +48 41 13.7      &    G2V	   &   5705   &   124  &  4.34  &  0.12 &  0.06 & 0.11 & 129 &  32 &    60 &   24  &  S  &  SA \\  
    RasTyc1623+3535  & 16 23 09.20 & +35 35 18.3      &    F8V	   &   5875   &   326  &  4.24  &  0.22 & -0.12 & 0.16 &  62 &  57 &   104 &  115  &  S  &  SA \\  
    RasTyc1626+3350  & 16 26 41.35 & +33 50 42.1      &    G1V	   &   6019   &   124  &  3.99  &  0.19 & -0.80 & 0.19 &  15 &  32 &   644 &  298  &  S  &  SA \\  
    RasTyc1626+0823  & 16 26 48.87 & +08 23 26.0      &	  G1.5V    &   5687   &   106  &  4.28  &  0.13 &  0.08 & 0.10 & 120 &  25 &   169 &   72  &  S  &  SA \\  
    RasTyc1628+7400  & 16 28 21.48 & +74 00 55.9      &    G2V	   &   5737   &   103  &  4.38  &  0.12 &  0.01 & 0.12 & 155 &  25 &   474 &   85  &  S  &  SA \\  
    RasTyc1629+1728  & 16 29 29.20 & +17 28 16.2      &    G1V	   &   5820   &	   75  &  4.30  &  0.11 & -0.01 & 0.11 &  98 &  22 &   100 &   39  &  S  &  SA \\  
    RasTyc1631+0849  & 16 31 13.87 & +08 49 14.4      &    K1V     &   5118   &	   60  &  4.54  &  0.18 &  0.03 & 0.10 &   6 &  15 &	69 &   26  &  S  &  SA \\  
   RasTyc1631+1916A  & 16 31 34.45 & +19 16 38.8      &   G9.5IV   &   5350   &   100  &  4.19  &  0.14 &  0.01 & 0.11 &  16 &  31 &   -13 &   21  &  S  &  SA \\  
   RasTyc1631+1916B  & 16 31 34.09 & +19 16 38.5      &    K0IV	   &   5307   &	   92  &  4.29  &  0.18 & -0.01 & 0.12 &  12 &  19 &   527 &   67  &  S  &  SA \\  
    RasTyc1637+0717  & 16 37 35.43 & +07 17 03.3      &    G2V     &   5747   &	   87  &  4.36  &  0.11 & -0.01 & 0.11 & 117 &  24 &   166 &   38  &  S  &  SA \\  
    RasTyc1641+1520  & 16 41 26.95 & +15 20 39.7      &    K0IV    &   5013   &	   89  &  3.31  &  0.20 & -0.04 & 0.11 &  11 &  17 &	 0 &   29  &  S  &  SA \\  
    RasTyc1641+0118  & 16 41 29.18 & +01 18 48.6      &    G8V     &   5125   &	   63  &  4.30  &  0.24 &  0.14 & 0.10 &   3 &  12 &	70 &   28  &  S  &  SA \\  
    RasTyc1641+1140  & 16 41 53.08 & +11 40 21.1      &   K0III-IV &   4921   &   119  &  3.02  &  0.34 & -0.14 & 0.11 &  30 &  15 &  1318 &   96  &  S  &  SA \\  
   RasTyc1645+3000A  & 16 45 43.47 & +30 00 17.2      &   F5IV-V   &   6315   &   233  &  3.98  &  0.13 & -0.23 & 0.11 &  37 &  45 &    61 &  162  &  S  &  SA \\  
   RasTyc1645+3000B  & 16 45 44.10 & +30 00 05.6      &    F8V     &   5862   &	   66  &  4.24  &  0.12 &  0.01 & 0.11 &  68 &  38 &    61 &   35  &  S  &  SA \\  
    RasTyc1649+5816  & 16 49 40.31 & +58 16 08.4      &    G2V     &   5800   &   146  &  4.28  &  0.15 &  0.03 & 0.10 & 134 &  52 &   195 &   62  &  S  &  FR \\ 
    RasTyc1652+0121  & 16 52 29.11 & +01 21 20.1      &   G0.5IV   &   5924   &    75  &  4.27  &  0.11 &  0.07 & 0.11 &  81 &  26 &   116 &   41  &  S  &  SA \\  
    RasTyc1658+0547  & 16 58 03.46 & +05 47 05.2      &    G3V     &   5613   &   112  &  4.28  &  0.17 & -0.32 & 0.12 &  36 &  15 &  1053 &  140  &  S  &  SA \\  
    RasTyc1658+3333  & 16 58 20.65 & +33 33 53.1      &    G2V     &   5690   &	   97  &  4.28  &  0.12 &  0.03 & 0.11 &  46 &  20 &   133 &   40  &  S  &  SA \\  
    RasTyc1702+4713$^{\rm c}$  & 17 02 48.85 & +47 13 06.5 & K0IV  &   5031   &	   96  &  3.38  &  0.31 & -0.05 & 0.11 & 114 &  29 &  2406 &  160  & SB2? & SA \\  
    RasTyc1703+2052  & 17 03 59.59 & +20 52 49.1      &    G1.5V   &   5773   &    98  &  4.22  &  0.12 & -0.04 & 0.11 & 104 &  20 &   146 &   27  &  S  &  SA \\  
    RasTyc1705-0147  & 17 05 08.46 & -01 47 09.6      &    K3V     &   4617   &   242  &  4.28  &  0.36 & -0.12 & 0.15 & \dots & \dots &  1030 & 18 & S  & AU \\
    RasTyc1706+0647  & 17 06 56.79 & +06 47 49.0      &    K2V     &   5027   &   150  &  4.30  &  0.41 & -0.04 & 0.10 & \dots & \dots &   513 & 25 & SB1 & AU \\
    RasTyc1714+0623  & 17 14 11.83 & +06 23 33.9      &    G9III   &   4839   &    71  &  2.94  &  0.15 & -0.10 & 0.11 &  78 &  21 &   973 &  117  &  S  &  SA \\  	
    RasTyc1718-0117  & 17 18 05.03 & -01 17 04.8      &    K1V     &   5092   &   155  &  4.49  &  0.23 &  0.12 & 0.20 &  16 &  53 &	  81 &  48   &  S  &  FR \\
    RasTyc1723+1931  & 17 23 25.73 & +19 31 22.1      &    K3V     &   4949   &    94  &  4.59  &  0.12 &  0.00 & 0.11 &  12 &  10 &   147 &   60  &  S  &  SA \\  
\hline
\noalign{\medskip}
\end{tabular}
\end{scriptsize} 
\end{center}
\end{table*}

\clearpage

\addtocounter{table}{-1}
\begin{table*}[p]
\caption{continued.}
\begin{center} 
\begin{scriptsize} 
\begin{tabular}{lccccrccrcrrrrcc}
\hline
\hline
\noalign{\medskip}
\textsl{RasTyc} Name & $\alpha$ (2000) & $\delta$ (2000)	       & Sp. Type &  $T_{\rm eff}$ & $\sigma_{T_{\rm eff}}$ & $\log g$ & $\sigma_{\log g}$ & $[Fe/H]$ & $\sigma_{[Fe/H]}$ & $W_{\rm Li}$ & $\sigma_{W_{\rm Li}}$ & $W_{\rm H\alpha}^{em}$ $^{\bullet}$ & $\sigma_{W_{\rm H\alpha}^{em}}$ &  Bin$^{\rm a}$ & Instr.$^{\rm b}$   \\
                     &  h m s	       & $\degr ~\arcmin ~\arcsec$     &	  &   \multicolumn{2}{c}{(K)}		    &	     \multicolumn{2}{c}{(dex)}	   &	    \multicolumn{2}{c}{(dex)}		  &	    \multicolumn{2}{c}{(m\AA)}  		       &       \multicolumn{2}{c}{(m\AA)}   &  &  \\
\noalign{\medskip}
\hline
\noalign{\medskip}
    RasTyc1728-0128  & 17 28 09.38 & -01 28 52.0      &    K1V     &   5083   &    93  &  4.49  &  0.19 &  0.00 & 0.11 &  13 &  11 &  1343 &  151  &  S  &  SA \\ 
    RasTyc1731+2815  & 17 31 03.33 & +28 15 06.1      &    K1V     &   5169   &    71  &  4.47  &  0.19 & -0.02 & 0.11 & 243 &  22 &   875 &   79  &  S  &  SA \\ 
    RasTyc1740+0554  & 17 40 57.39 & +05 54 46.5      &    K0V     &   5393   &   223  &  4.24  &  0.35 &  0.09 & 0.17 & 185 &  39 &	 25 &	 24 &  S  &  FR \\
    RasTyc1741+0228  & 17 41 46.17 & +02 28 56.2      &    G9III   &   4840   &    93  &  3.12  &  0.21 & -0.12 & 0.11 &  90 &  22 &   514 &   50  &  S  &  SA \\ 
    RasTyc1743+6606  & 17 43 01.88 & +66 06 43.1      &    F8V     &   5936   &   109  &  4.15  &  0.13 & -0.22 & 0.12 &   7 &  15 &   193 &   40  &  S  &  SA \\ 
    RasTyc1758+0155  & 17 58 55.52 & +01 55 06.3      &    K0V     &   5142   &    61  &  4.55  &  0.16 & -0.01 & 0.10 &  15 &  24 &   439 &   68  &  S  &  SA \\ 
    RasTyc1802+3356  & 18 02 38.80 & +33 56 34.5      &    G3V     &   5705   &    90  &  4.04  &  0.15 & -0.29 & 0.13 &  14 &   7 &   272 &   21 &  S  &  AU   \\
    RasTyc1808-0858$^{\dag}$  & 18 08 01.34 & -08 58 57.5 & K1V    &   5173   &   123  &  4.37  &  0.15 & -0.06 & 0.12 & 189 &  29 &   225 &   38  & SB2?  &  SA \\  
    RasTyc1816+2848  & 18 16 31.34 & +28 48 11.2      &    G3V     &   5704   &   123  &  4.16  &  0.22 & -0.19 & 0.17 & 130 &  15 &   151 &   17 &  SB1?  & AU  \\
    RasTyc1831+5418  & 18 31 37.59 & +54 18 57.9      &    G2.5V   &   5676   &    79  &  4.26  &  0.13 &  0.05 & 0.12 &   9 &  21 &   556 &  121  &  S  & SA \\ 
    RasTyc1838+0224  & 18 38 38.19 & +02 24 12.7      &   F9IV-V   &   6011   &   108  &  4.15  &  0.13 &  0.00 & 0.12 & 112 &  28 &   138 &   38  &  S  &  SA \\
    RasTyc1842+5751  & 18 42 49.32 & +57 51 54.6      &    G0.5IV  &   5843   &    74  &  4.30  &  0.12 &  0.07 & 0.10 &   8 &  13 &   116 &   26  &  S  &  EL \\
    RasTyc1843+4328  & 18 43 58.66 & +43 28 27.8      &    G1.5V   &   5964   &   110  &  4.20  &  0.13 &  0.02 & 0.12 &  49 &  12 &   109 &   31  &  S  &  SA \\
    RasTyc1847+2930  & 18 47 39.96 & +29 30 39.3   &    K1V	&   5220   &	59  &  4.31  &  0.14 &  0.02 & 0.10 &	4 &   9 &   236 &   31  &  S  &  SA \\ 
    RasTyc1857+6406  & 18 57 13.08 & +64 06 21.3   &    K3V	&   4834   &	95  &  4.59  &  0.16 &  0.04 & 0.10 &  32 &  22 &    59 &   30  &  S  &  SA \\ 
    RasTyc1857+6223  & 18 57 19.45 & +62 23 39.5   &    G8III	&   5020   &	87  &  3.14  &  0.17 &  0.03 & 0.10 &	7 &   8 &    -3 &   13  &  S  &  SA \\ 
    RasTyc1857+0120  & 18 57 19.46 & +01 20 33.3   &    G5V     &   5420   &   290  &  4.29  &  0.22 & -0.03 & 0.07 & 186 &  12 &   222 &   13 &  S  &  AU   \\
    RasTyc1908+5018  & 19 08 14.03 & +50 18 49.6   &  G8III-IV  &   5223   &   226  &  3.39  &  0.76 & -0.37 & 0.23 & 339 &  67 &   529 &  113  &  S  &  FR  \\
    RasTyc1918+3408  & 19 18 12.13 & +34 08 10.1   &    K1III   &   4658   &	69  &  2.67  &  0.16 & -0.10 & 0.11 &  22 &   9 &   160 &   25  &  S  &  EL \\ 
    RasTyc1919+0628  & 19 19 26.29 & +06 28 42.2   &    F8V	&   5986   &  195   &  4.12  &  0.25 & -0.08 & 0.19 &  93 &  47 &  226 & 78 &  S  &  FR \\
    RasTyc1925+4429  & 19 25 01.98 & +44 29 50.7   &    K2V	&   5055   &   135  &  4.41  &  0.25 & -0.02 & 0.10 & 270 &  20 &   922 &  107  &  S  &  SA \\ 
    RasTyc1926+7840  & 19 26 35.09 & +78 40 04.9   &    K1V	&   5205   &	72  &  4.23  &  0.20 &  0.03 & 0.11 &  21 &  19 &   319 &   61  &  S  &  SA \\ 
    RasTyc1928+4856  & 19 28 10.64 & +48 56 37.3   &	K0V  	&   5197   &   254  &  4.47  &  0.29 & -0.03 & 0.14 &	9 &  45 &   550 &  118  &  S  &  SA \\ 
    RasTyc1928+1232  & 19 28 15.43 & +12 32 09.6   &	K3V  	&   4805   &	93  &  4.57  &  0.11 &  0.07 & 0.10 &	3 &  14 &    24 &   14  &  S  &  SA \\ 
    RasTyc1928+2731  & 19 28 22.32 & +27 31 18.5   &    G2V     &   5745   &   127  &  4.20  &  0.20 &  0.14 & 0.14 &\dots  & \dots &   246 &   17 &  S  &  AU   \\
    RasTyc1930+4932  & 19 30 15.81 & +49 32 08.2   &	  K1V	&   5180   &	64  &  3.80  &  0.25 & -0.01 & 0.11 &  18 &  17 &   185--330 &   71  &  SB1  & SA \\ 
    RasTyc1940+2535  & 19 40 44.79 & +25 35 46.9   &	K3V	&   5051   &   215  &  4.37  &  0.13 & -0.06 & 0.07 &  67 &   7 &   340 &   30 &  S  &  AU   \\
    RasTyc1949-0804  & 19 49 35.22 & -08 04 50.3   &	G8III	&   5233   &   451  &  3.67  &  0.83 & -0.06 & 0.28 &  48 &  13 &  1229 &   38 &  SB1  &  AU   \\
    RasTyc1956+4345  & 19 56 59.73 & +43 45 08.2   &   G1.5V    &   5815   &    95  &  4.24  &  0.12 & -0.05 & 0.10 & 155 &  20 &   264 &   30  & S?  & SA \\  
   RasTyc1958+5355A  & 19 58 04.00 & +53 55 29.6  &  F5.5IV-V	&   6296   &   174  &  4.03  &  0.12 & -0.10 & 0.14 &  11 &  16 &   -13 &   35  &  S  &  SA \\ 
   RasTyc1958+5355B  & 19 58 03.39 & +53 55 24.5  &	  F8V	&   5899   &	91  &  4.18  &  0.12 & -0.10 & 0.12 &  86 &  22 &    48 &   28  &  S  &  SA \\ 
    RasTyc1958+4301$^{\rm *}$  & 19 58 37.62 & +43 01 02.5 & F9IV-V & 5919 &   203  &  4.06  &  0.16 & -0.21 & 0.17 &  64 &  34 &   197 &   92  &  SB2?  &  SA \\  	
    RasTyc1959-0432  & 19 59 24.10 & -04 32 05.7   &    G5V     &   5752   &   116  &  4.17  &  0.18 &  0.06 & 0.10 & 126 &   8 &   147 &   17 &  S  &  AU   \\
    RasTyc2000+5921  & 20 00 31.23 & +59 21 44.5   &    G0V	&   5653   &	77  &  4.28  &  0.12 & -0.49 & 0.12 &  22 &  20 &    37 &   38  &  S  &  SA \\ 
    RasTyc2000+3256  & 20 00 45.52 & +32 56 59.4   &    K0V     &   5212   &    65  &  4.40  &  0.13 &  0.02 & 0.10 &  48 &  12 &    95 &   22  &   S  & EL \\ 
    RasTyc2004-0239  & 20 04 49.35 & -02 39 19.7   &	K1V	&   5183   &	63  &  4.39  &  0.17 & -0.02 & 0.11 & 298 &  38 &   457 &   76  &  S  &  SA \\ 
    RasTyc2013+1625  & 20 13 54.27 & +16 25 25.4   &    K1III   &   4555   &    73  &  2.55  &  0.14 & -0.11 & 0.11 &  55 &  10 &   486--517 &   50  &  S  &  EL \\ 
    RasTyc2015+3316  & 20 15 20.54 & +33 16 11.8   &    K3V     &   4921   &    69  &  4.57  &  0.14 &  0.10 & 0.10 &	6 &  19 &   126 &   31  &   S  &  EL \\ 
    RasTyc2015+3343  & 20 15 23.79 & +33 43 45.7   &	 G8III  &   5086   &    85  &  3.15  &  0.16 &  0.01 & 0.10 &	4 &   5 &     3 &    8  &  S  &  EL \\  
    RasTyc2016+3106  & 20 16 57.83 & +31 06 55.6   &	   K1V  &   5087   &    60  &  4.57  &  0.14 &  0.03 & 0.11 & 254 &  24 &  1168--1585 &  163  &  S  &  EL \\ 
    RasTyc2019+3203  & 20 19 47.52 & +32 03 08.8   &	   F8V  &   6036   &   132  &  4.06  &  0.14 &  0.10 & 0.12 &  20 &  29 &   315 &   60  &  S  &  EL \\   
    RasTyc2021+3150  & 20 21 30.45 & +31 50 32.3   &	 G1.5V  &   5714   &   219  &  4.35  &  0.18 & -0.02 & 0.15 &  23 &  29 &  1983 &  225  &  S  &  EL \\   
    RasTyc2021+3218  & 20 21 33.04 & +32 18 50.7   &	 G8III  &   4885   &    64  &  3.01  &  0.17 & -0.07 & 0.10 &  43 &   9 &   370 &   56  &   S  &   EL \\ 
    RasTyc2021+0616  & 20 21 45.50 & +06 16 13.4   &	 G5IV	&   5426   &   233  &  3.95  &  0.23 & -0.03 & 0.13 &  66 &  19 &   285 &   33  &  S  &  SA \\    
    RasTyc2025-0429  & 20 25 04.26 & -04 29 15.2   & K0III-IV	&   4943   &   114  &  3.22  &  0.34 & -0.05 & 0.12 &  61 &  19 &  2008 &  212  &  S  &  SA \\    
    RasTyc2028+2510  & 20 28 03.50 & +25 10 42.7   &	  K1V	&   5314   &	88  &  4.21  &  0.19 &  0.02 & 0.12 &  34 &  15 &    73 &   27  &  S  &  SA \\ 
    RasTyc2028+1131  & 20 28 23.91 & +11 31 11.3   &         G2V       &    5436   &  187  &  4.41  &  0.14 & -0.16 &  0.25 & \dots & \dots &  496 & 88 & SB?  &  AU \\  
    RasTyc2028-0943  & 20 28 42.28 & -09 43 16.9   &	G5III	&   5373   &   268  &  3.70  &  0.40 &  0.07 & 0.14 &  34 &  47 &  3293 &  313  &  S  &  SA \\ 
    RasTyc2030+4852  & 20 30 59.56 & +48 52 08.1   &	 K1III  &   4667   &    87  &  2.60  &  0.19 & -0.06 & 0.12 &  103 &  21 &    57--229 &   98  &   SB1  &  EL \\ 
    RasTyc2033+3128  & 20 33 24.39 & +31 28 12.3   &	   G2V  &   5439   &   196  &  4.37  &  0.16 & -0.06 & 0.14 &  81 &  28 &  1188 &  135  &  S  &  EL \\ 
    RasTyc2034-0713  & 20 34 47.43 & -07 13 56.7   &	K0III	&   4751   &	60  &  2.78  &  0.13 & -0.06 & 0.11 &  16 &  11 &     4 &   24  &  S  &  SA \\ 
    RasTyc2036+3456  & 20 36 16.87 & +34 56 46.1   &	G1V     &   5832   &    78  &  4.33  &  0.11 & -0.01 & 0.11 & 192 &  16 &   390 &   37  &  S  &  EL \\ 
    RasTyc2037+5106  & 20 37 55.03 & +51 06 23.1   &    G5V     &   5471   &   318  &  4.20  &  0.30 & -0.07 & 0.11 & 133 &  11 &   179 &   15 &  S  &  AU   \\
    RasTyc2038+3546  & 20 38 17.71 & +35 46 33.3   &	G1.5V   &   5895   &    82  &  4.23  &  0.11 &  0.01 & 0.11 & 134 &  16 &   117 &   26  &  S  &  EL \\ 
    RasTyc2039+2644  & 20 39 40.81 & +26 44 48.4   &    K1V     &   5149   &   116  &  4.30  &  0.27 & -0.08 & 0.08 & 232 &  11 &   541 &   22 &  S  &  AU   \\
    RasTyc2046+2815  & 20 46 18.86 & +28 15 44.2   &    G1.5V   &   5765   &   205  &  4.26  &  0.16 &  0.14 & 0.13 &  41 &  51 &   211 &  185  &  SB2  &  EL \\   
    RasTyc2048-0644  & 20 48 59.58 & -06 44 53.4   &    G8IV    &   5118   &   171  &  3.66  &  0.62 & -0.24 & 0.26 &  69 &  11 &  1295 &   44 &  SB1?  &  AU   \\
    RasTyc2052+2705  & 20 52 07.73 & +27 05 49.7   &    G5III   &   5267   &    81  &  3.34  &  0.19 &  0.08 & 0.11 &	0 &   4 &    15 &   13  &  S  &  EL \\   
    RasTyc2052+4258$^{\ddag}$ & 20 52 57.27 & +42 58 25.7 & K1V &   5162   &	78  &  4.55  &  0.12 & -0.09 & 0.11 &  10 &  15 &   670 &  104  &   SB &  SA \\  
    RasTyc2052+4407  & 20 52 58.28 & +44 07 20.1   &	F5IV-V  &   6227   &   249  &  3.99  &  0.13 & -0.01 & 0.13 &  22 &  11 &   164 &   26  &   S  &  EL \\  
    RasTyc2053+2629  & 20 53 04.20 & +26 29 32.1   &   G0.5IV   &   5964   &	86  &  4.08  &  0.12 &  0.12 & 0.11 &  10 &  27 &    19 &   24  &  S  &  EL \\   
    RasTyc2053+3641  & 20 53 35.39 & +36 41 49.7   &	  G1V   &   5877   &	68  &  4.30  &  0.11 & -0.09 & 0.11 &  98 &  15 &    90 &   25  &  S  &  EL \\   
    RasTyc2053+4423  & 20 53 53.63 & +44 23 11.1   &	G5III   &   5427   &   229  &  3.73  &  0.31 &  0.14 & 0.15 &	0 &  11 &  1241--1308 &  158  &  S?  &  EL \\
    RasTyc2055+5348  & 20 55 42.25 & +53 48 21.4   &	 G8III  &   4990   &    76  &  3.03  &  0.15 &  0.00 & 0.11 &  20 &   7 &    19 &   10  &  S  &  EL \\ 
    RasTyc2057+2624  & 20 57 39.50 & +26 24 17.4   &	  G1V   &   5868   &    59  &  4.29  &  0.10 &  0.02 & 0.10 &  48 &   9 &    24 &    8  &  S  &  EL \\
    RasTyc2058+4403  & 20 58 55.63 & +44 03 37.5   &   K0.5III  &   4743   &	53  &  2.84  &  0.11 & -0.02 & 0.10 &  26 &   8 &    30 &   13  &  S  &  EL \\ 
    RasTyc2059+4447  & 20 59 11.79 & +44 47 25.6   &  K0III-IV  &   5209   &	79  &  3.57  &  0.15 &  0.01 & 0.10 &  33 &  10 &    50 &   11  &  S  &  EL \\ 
    RasTyc2100+4405  & 21 00 25.21 & +44 05 54.7   &	  G5IV  &   5452   &   106  &  3.81  &  0.20 & -0.14 & 0.11 &  10 &  12 &   481 &   50  &  S  &  EL \\ 
    RasTyc2101+1008  & 21 01 44.80 & +10 08 40.9   &     K2V    &   5137   &   111  &  4.28  &  0.42 & -0.11 & 0.11 &   3 &   6 &   505 &   29  &  S  &  AU \\
    RasTyc2102+4553  & 21 02 40.42 & +45 53 03.9   &	   K3V  &   5002   &    73  &  4.58  &  0.14 &  0.09 & 0.10 &	0 &  12 &    32 &   13  &  S  &  EL \\ 
    RasTyc2103+4104  & 21 03 16.76 & +41 04 06.3   &	   K3V  &   5010   &    73  &  4.59  &  0.11 &  0.01 & 0.11 & 135 &  15 &   757 &   72  &  S  &  EL \\ 
    RasTyc2105+3949  & 21 05 50.98 & +39 49 48.1   &	 G1.5V  &   5675   &    89  &  4.29  &  0.13 & -0.05 & 0.11 &	9 &  33 &   681 &  100  &  S  &  EL \\ 
    RasTyc2106+0217  & 21 06 04.29 & +02 17 02.4   &	 G5IV	&   5301   &	65  &  3.88  &  0.15 &  0.12 & 0.10 &  20 &  14 &   -13 &   16  &  S  &  SA \\ 
    RasTyc2106+6906  & 21 06 21.74 & +69 06 41.0   &   G1.5V	&   5828   &   115  &  4.31  &  0.14 & -0.01 & 0.12 & 205 &  62 &   409 &   96  &  S  &  SA \\ 
    RasTyc2107+0632  & 21 07 07.13 & +06 32 32.2   &	 G5IV	&   5635   &   140  &  4.22  &  0.23 & -0.09 & 0.16 & 140 &  93 &   397 &  204  &  S  &  SA \\ 
    RasTyc2107+3423  & 21 07 30.01 & +34 23 33.7   &	   K1V  &   5128   &    57  &  4.55  &  0.12 & -0.02 & 0.10 & 210 &  18 &  1200 &  101  &  S  &  EL \\ 
\hline		
\noalign{\medskip}
\end{tabular}
\end{scriptsize} 
\end{center}
\end{table*}

\clearpage

\addtocounter{table}{-1}
\begin{table*}[p]
\caption{continued.}
\begin{center} 
\begin{scriptsize} 
\begin{tabular}{lccccrccrcrrrrcc}
\hline
\hline
\noalign{\medskip}
\textsl{RasTyc} Name & $\alpha$ (2000) & $\delta$ (2000)	      & Sp. Type &  $T_{\rm eff}$ & $\sigma_{T_{\rm eff}}$ & $\log g$ & $\sigma_{\log g}$ & $[Fe/H]$ & $\sigma_{[Fe/H]}$ & $W_{\rm Li}$ & $\sigma_{W_{\rm Li}}$ & $W_{\rm H\alpha}^{em}$ $^{\bullet}$ & $\sigma_{W_{\rm H\alpha}^{em}}$ &  Bin$^{\rm a}$  &  Instr.$^{\rm b}$   \\
                     &  h m s	       & $\degr ~\arcmin ~\arcsec$    & 	 &   \multicolumn{2}{c}{(K)}		   &	     \multicolumn{2}{c}{(dex)}	   &	    \multicolumn{2}{c}{(dex)}		 &	   \multicolumn{2}{c}{(m\AA)}			      &       \multicolumn{2}{c}{(m\AA)}   &   &  \\
\noalign{\medskip}
\hline
\noalign{\medskip}
    RasTyc2109+4029  & 21 09 48.50 & +40 29 23.4   &	 G1.5V  &   5707   &   115  &  4.33  &  0.13 & -0.04 & 0.12 &  85 &  22 &  1181--1340 &  182  &  SB1  &  EL \\  
    RasTyc2114-0058  & 21 14 32.47 & -00 58 52.8   &   G1.5V	&   5714   &   212  &  4.09  &  0.20 & -0.11 & 0.14 &  94 &  29 &  1794 &  201  &  S  &  SA \\ 
    RasTyc2114+3941  & 21 14 55.24 & +39 41 11.9   &	 G1.5V  &   5930   &    90  &  4.27  &  0.12 & -0.02 & 0.12 & 131 &  27 &   149 &   34  &  S  &  EL \\ 
    RasTyc2115+4437  & 21 15 23.98 & +44 37 42.8   &	  K0IV  &   5072   &    84  &  3.69  &  0.26 & -0.01 & 0.11 &  73 &  19 &   547 &   41  &  S  &  EL \\ 
    RasTyc2117+3153  & 21 17 21.28 & +31 53 58.0   &	  G5IV  &   5283   &    61  &  3.63  &  0.14 &  0.07 & 0.10 &  87 &  14 &    91 &   18  &  S  &  EL \\ 
    RasTyc2118-0631  & 21 18 33.53 & -06 31 43.7   &	  K1V	&   5149   &	83  &  4.34  &  0.23 &  0.10 & 0.11 &  48 &  23 &   553 &   69  &  SB1$^{\rm d}$  &  SA \\  
    RasTyc2118+2613  & 21 18 58.13 & +26 13 49.9   &	   K0V  &   5404   &    93  &  4.34  &  0.12 &  0.03 & 0.10 &  40 &   6 &    86 &   11  &  S  &  EL \\ 
    RasTyc2120+4636  & 21 20 55.42 & +46 36 12.4   &	 G1.5V  &   5699   &   192  &  4.36  &  0.16 &  0.04 & 0.14 & 176 &  49 &   426 &  117  &  S  &  EL \\ 
    RasTyc2121+4020  & 21 21 01.44 & +40 20 44.1   &	  F8IV  &   6050   &   111  &  4.01  &  0.12 & -0.25 & 0.13 &  78 &  12 &   115--126 &   24  & SB1  &   EL \\ 
    RasTyc2128+6423  & 21 28 00.96 & +64 23 15.1   &	  K1V   &   5220   &	64  &  4.44  &  0.12 &  0.01 & 0.10 &  11 &   6 &    83 &   14  &  S  &  EL \\ 
    RasTyc2131+5925  & 21 31 01.20 & +59 25 05.0   &	G2.5V   &   5760   &	74  &  4.20  &  0.11 &  0.09 & 0.10 &  50 &  21 &    56 &   21  &  S  &  EL \\ 
    RasTyc2132+3604  & 21 32 39.29 & +36 04 46.0   &	 K0IV   &   5057   &	80  &  3.20  &  0.17 &  0.02 & 0.11 &  75 &   9 &    83 &   11  &  S  &  EL \\ 
    RasTyc2137+1946  & 21 37 03.18 & +19 46 58.3   &	G8III   &   4930   &	64  &  3.08  &  0.13 &  0.02 & 0.10 &  23 &   9 &    26 &   13  &  S  &  EL \\ 
    RasTyc2141+2645  & 21 41 05.97 & +26 45 03.2   &   G5IV-V   &   5668   &	60  &  4.27  &  0.10 &  0.00 & 0.10 &	3 &   3 &    31 &   16  &  S  &  EL \\ 
    RasTyc2141+2658  & 21 41 16.74 & +26 58 58.1   &	 K0III  &   4850   &    82  &  3.00  &  0.16 & -0.03 & 0.11 &  37 &  17 & 387--495 &   92  &  SB1  &  EL \\  
    RasTyc2142+3814  & 21 42 59.83 & +38 14 54.8   &	 G8III  &   4892   &    66  &  2.97  &  0.13 &  0.01 & 0.10 &  24 &  16 &     7 &   21  &   S  &  EL \\ 
    RasTyc2143+2720  & 21 43 31.67 & +27 20 37.6   &	   F8V  &   6065   &    69  &  4.03  &  0.11 & -0.30 & 0.11 &  35 &   6 &    48 &   15  &  S  &  EL \\  
    RasTyc2146+2446  & 21 46 50.43 & +24 46 04.2   &	   G2V  &   6039   &    84  &  3.95  &  0.12 & -0.26 & 0.11 &  11 &  11 & 63--91 &   21  &  S  &   EL \\
    RasTyc2147+4950  & 21 47 48.72 & +49 50 08.0   &	  K2V	&   5061   &	70  &  4.60  &  0.14 & -0.02 & 0.11 &  85 &  22 &   533 &   67  & S  &   SA \\ 	
    RasTyc2147+4949  & 21 47 54.42 & +49 49 29.2   &	  K2V	&   5080   &	85  &  4.58  &  0.20 & -0.04 & 0.11 &  53 &  14 &   135--297 &   65  & S  &   SA \\  
    RasTyc2149-0350  & 21 49 13.15 & -03 50 32.6   &	  G2V	&   5723   &	90  &  4.09  &  0.16 & -0.02 & 0.14 & 109 &  27 &   102 &   49  &  S  &  SA \\ 
    RasTyc2149+3125  & 21 49 16.02 & +31 25 02.1   &	  K1V   &   5141   &    59  &  4.49  &  0.13 &  0.02 & 0.10 &  12 &  20 &   738 &   95  &  S  &  EL \\ 
    RasTyc2149+4028  & 21 49 26.94 & +40 28 00.9   &	   F8V  &   5858   &   162  &  3.99  &  0.13 & -0.44 & 0.14 &  15 &  18 &   705 &  129  &  S  &  EL \\ 
    RasTyc2152+3850  & 21 52 48.07 & +38 50 08.0   &  K0III-IV  &   5036   &	75  &  3.26  &  0.18 & -0.04 & 0.10 &  22 &   5 &    54 &    8  & S  &   EL \\ 
    RasTyc2153+2055  & 21 53 05.36 & +20 55 50.8   &	  K1V   &   5107   &    53  &  4.57  &  0.12 & -0.03 & 0.10 &  17 &   6 &    83 &   18  &  S  &  EL \\
    RasTyc2155-0947  & 21 55 07.21 & -09 47 57.2   &   G1.5V	&   5663   &   169  &  4.27  &  0.19 & -0.07 & 0.13 &  39 &  32 &   597 &  121  &  SB1$^{\rm d}$  &  SA \\     
    RasTyc2157-0753  & 21 57 51.37 & -07 53 47.4   &	  K1V	&   5110   &	72  &  4.55  &  0.18 &  0.02 & 0.11 &  17 &  33 &  1016 &  189  &  SB1$^{\rm d}$  &   SA \\ 	
    RasTyc2159+0041  & 21 59 05.33 & +00 41 10.2   &	G2.5V	&   5693   &   186  &  4.33  &  0.16 &  0.04 & 0.14 &  45 &  52 &  1052 &  216  &  S  &  SA \\ 
    RasTyc2202+1520  & 22 02 13.96 & +15 20 14.0   &	   K0V  &   5148   &    57  &  4.54  &  0.12 & -0.07 & 0.10 &  37 &  11 & 266--363 &   75  &  S  &  EL \\ 
    RasTyc2202+4831$^{\dag}$ & 22 02 26.05 & +48 31 14.9 &  K3V &   4804   &   322  &  4.58  &  0.40 & -0.02 & 0.14 &	3 &  64 &  1455 &  261  & S?  &   SA \\ 	
    RasTyc2202-0406  & 22 02 30.14 & -04 06 11.6   &	  K1V	&   5119   &	54  &  4.56  &  0.15 & -0.04 & 0.10 & 179 &  23 &   349 &   69  &  SB1$^{\rm d}$  &  SA \\     
    RasTyc2202+3108  & 22 02 57.33 & +31 08 46.8   &	  G1V	&   5921   &	92  &  4.31  &  0.12 &  0.02 & 0.11 &  40 &  29 & 126--186 &   42  &  SB1  &  SA \\ 
    RasTyc2203+3809  & 22 03 49.83 & +38 09 42.9   &   G5IV-V	&   5678   &	71  &  4.28  &  0.12 &  0.04 & 0.11 & 178 &  29 &    93 &   35  &  S  &  SA \\ 
    RasTyc2204+0236  & 22 04 17.52 & +02 36 21.0   &	  K5V	&   4654   &   111  &  4.60  &  0.13 &  0.01 & 0.10 &  18 &  18 &    50 &   22  &  S  &  SA \\ 
    RasTyc2206-0102  & 22 06 25.65 & -01 02 43.9   &	  G0V	&   5632   &	96  &  4.28  &  0.13 & -0.33 & 0.16 &  32 &  32 &  1236 &  220  &  S  &  SA \\ 
    RasTyc2206+5153  & 22 06 34.43 & +51 53 11.8   &	  K1V	&   5204   &	79  &  4.31  &  0.17 & -0.00 & 0.11 & 140 &  31 &   908 &  164  &  S  &  SA \\ 
    RasTyc2208+2208  & 22 08 50.40 & +22 08 19.6   &	G5III  &   5397   &    70  &  3.40  &  0.20 & -0.07 & 0.11 &   4 &   3 &    20 &    9  &  S  &  EL \\
    RasTyc2210+1936  & 22 10 18.96 & +19 36 59.6   &	  G1V  &   5737   &    78  &  4.25  &  0.13 & -0.04 & 0.12 &  12 &  10 &    40 &   15  &  S  &  EL \\
    RasTyc2212+1329  & 22 12 13.38 & +13 29 19.8   &	G2.5V  &   5665   &    75  &  4.28  &  0.12 &  0.10 & 0.10 &  78 &  12 &    58 &   20  &  S  &  EL \\
    RasTyc2213+2015$^{\rm *}$ & 22 13 18.14 & +20 15 35.5 & G9.5IV & 5197  &   198  &  4.12  &  0.40 & -0.00 & 0.14 & 104 &  68 &   390--705 &   88  &  SB2?  &  SA \\ 	
    RasTyc2218+6951  & 22 18 27.61 & +69 51 40.0   &   G9.5IV	&   5322   &   113  &  4.28  &  0.16 &  0.01 & 0.11 &  22 &  24 &   411 &   62  &  SB  &  SA \\     
    RasTyc2221-0008  & 22 21 16.29 & -00 08 35.6   &	G1.5V	&   5811   &	76  &  4.17  &  0.14 &  0.09 & 0.11 &	4 &  21 &   408 &   92  &  S  &  SA \\     
    RasTyc2222+2922  & 22 22 28.77 & +29 22 13.0   &	   K1V  &   5171   &   216  &  4.46  &  0.38 & -0.13 & 0.20 &	0 &  26 &  1014--1327 &  164  &  S?  &  EL \\ 
   RasTyc2222+2814A  & 22 22 30.10 & +28 14 24.9   &	 F6IV	&   6257   &   153  &  3.91  &  0.11 & -0.17 & 0.12 &  58 &  72 &    12 &   62  &  S  &  SA \\ 
    RasTyc2223+7741  & 22 23 18.87 & +77 41 57.7   &   G1.5V	&   5745   &   114  &  4.29  &  0.13 & -0.04 & 0.13 & 245 &  50 &   440 &  110  &  S  &   SA \\
    RasTyc2224+2016  & 22 24 33.35 & +20 16 34.5   &	G5IV-V  &   5377   &   103  &  3.80  &  0.21 & -0.01 & 0.11 &	8 &  11 &    77 &   13  &  S  &  EL \\ 
    RasTyc2225+4607  & 22 25 12.03 & +46 07 01.7   &    G1.5V   &  5751    &  103 &  4.26 &  0.19 &  0.09 & 0.09 &  76 &  78  &  358  &  159  &  S  &  FR \\
    RasTyc2226+1814  & 22 26 34.99 & +18 14 00.4   &    F8V     &   5991   &   147  &  4.01  &  0.14 & -0.18 & 0.14 &  98 &  27 &   310 &   56  &  S  &  EL \\ 
    RasTyc2227+1509  & 22 27 12.27 & +15 09 15.3   &	   G1V  &   5825   &    68  &  4.34  &  0.10 & -0.04 & 0.11 &  90 &  13 &    83 &   14  &  S  &  EL \\ 
    RasTyc2227+2649  & 22 27 27.08 & +26 49 05.3   &	   F8V  &   6348   &    83  &  4.03  &  0.11 & -0.27 & 0.12 &  31 &  14 &    32 &   24  &  S  &  EL \\ 
    RasTyc2233+1040  & 22 33 00.37 & +10 40 34.3   &	 G1.5V  &   5544   &   350  &  4.39  &  0.30 &  0.03 & 0.15 & 282 & 102 &  2974 &  445  &  S  &  EL \\ 
    RasTyc2236+0010  & 22 36 11.99 & +00 10 07.6   &   G8IV-V	&   5252   &	99  &  4.09  &  0.24 &  0.07 & 0.12 &  26 &  34 &   667 &   85  & SB1 &  SA \\ 
    RasTyc2236+7032  & 22 36 15.90 & +70 32 04.1   &	 K0IV	&   5080   &	87  &  3.78  &  0.31 & -0.04 & 0.11 &  75 &  23 &  1551 &  128  &  S  &  SA \\ 
    RasTyc2238+0217  & 22 38 29.22 & +02 17 56.4   &	  G1V  &   5839   &    82  &  4.32  &  0.11 & -0.10 & 0.12 &  88 &  19 &   117 &   23  & S  &	EL \\
    RasTyc2239+0406  & 22 39 50.66 & +04 06 57.1   &	  K2V  &   5040   &    83  &  4.59  &  0.11 & -0.06 & 0.11 &  16 &   5 &   208 &   38  &  S  &  EL \\
    RasTyc2240+1432  & 22 40 52.53 & +14 32 56.0   &   G5IV-V  &   5651   &    61  &  4.23  &  0.11 & -0.01 & 0.10 &  13 &   5 &    11 &    6  &  S  &  EL \\
    RasTyc2241+1430  & 22 41 57.40 & +14 30 59.2   &	K0III  &   4678   &    54  &  2.77  &  0.11 & -0.04 & 0.10 & 362 &  13 &     7 &   11  &  S  &  EL \\
    RasTyc2242+1900$^{*}$  & 22 42 04.91 & +19 00 49.8   &	   K2V  &   5112   &   195  &  4.54  &  0.39 &  0.02 & 0.14 &  36 &  50 &   568 &  129  &  SB2?  &  EL \\   
    RasTyc2244+1341  & 22 44 28.34 & +13 41 10.9   &	   K0V  &   5349   &    67  &  4.23  &  0.13 &  0.08 & 0.10 &	1 &   4 &    29 &    9  &  S  &  EL \\  
    RasTyc2244+1754  & 22 44 41.49 & +17 54 19.0   &	  K1V	&   5161   &	58  &  4.41  &  0.16 & -0.02 & 0.11 & 245 &  18 &   601--720 &   64  & SB1 & EL/SA \\     
    RasTyc2244+3029  & 22 44 46.12 & +30 29 33.6   &	 G8IV	&   5471   &   172  &  4.31  &  0.23 & -0.08 & 0.14 &  73 &  62 &   294 &  196  &  S  &  SA \\ 
    RasTyc2246+5749  & 22 46 13.19 & +57 49 58.0   &     G2V    &   5752   &  145 &  4.32 &  0.10 & -0.05 & 0.11 &  207 &  41 &  410  & 110 & S  &  FR \\
    RasTyc2250+4926  & 22 50 10.89 & +49 26 14.4   &	 G5V	&   5552   &   164  &  4.36  &  0.16 & -0.03 & 0.13 &  38 &  42 &   677 &  123  &  S  &  SA \\ 
    RasTyc2251+8525  & 22 51 28.45 & +85 25 21.1   &	 K1V	&   5177   &	71  &  4.46  &  0.14 & -0.04 & 0.10 & 176 &  22 &   304 &   41  &  S  &  SA \\ 
    RasTyc2252+1730  & 22 52 52.75 & +17 30 29.6   &	 K1V    &   5156   &    59  &  4.53  &  0.11 & -0.05 & 0.10 &	4 &  12 &   219 &   28  &  S  &  EL \\ 
    RasTyc2253+0338  & 22 53 41.72 & +03 38 43.6   &	 G2V    &   5575   &    89  &  4.29  &  0.13 &  0.04 & 0.11 &  65 &  15 &   120 &   28  &  S  &  EL \\ 
    RasTyc2256+0235  & 22 56 49.53 & +02 35 39.2   &   G1.5V    &   5665   &   133  &  4.00  &  0.15 &  0.12 & 0.11 & 106 &  23 &   809 &  115  &  SB1$^{\rm d}$  &  EL \\  
    RasTyc2258-0018  & 22 58 52.89 & -00 18 57.5   &	 K1IV   &   4804   &    66  &  2.96  &  0.15 & -0.02 & 0.11 &  23 &   7 &  295--418 &   22  &  SB1  &  EL \\  
RasTyc2259+4016$^{\ddag}$  & 22 59 25.44& +40 16 35.9 & K1IV    &   4958   &   273  &  3.09  &  1.01 & -0.18 & 0.10 &  35 &   7 &   767 &   71 & SB1  &  AU   \\
    RasTyc2301+3528  & 23 01 47.79 & +35 28 48.7   &    K1III   &   4667   &   155  &  2.97  &  1.07 & -0.08 & 0.13 &  84 &  13 &  1100 &   25 & SB1  &  AU   \\
    RasTyc2302+3515  & 23 02 09.27 & +35 15 39.5   &    K0IV    &   5159   &   123  &  3.98  &  0.48 & -0.06 & 0.09 &  85 &  14 &   309 &   19 & SB1  &  AU   \\
    RasTyc2303+1713  & 23 03 23.57 & +17 13 14.6   &    F8V     &   6117   &    99  &  3.98  &  0.12 & -0.14 & 0.12 &  29 &  34 &    20 &   38 &  S   &  EL \\  
RasTyc2305-0149$^{\ddag}$ &  23 05 13.18 & -01 49 25.4 &   G6V  &   5565   &   252  &  4.33  &  0.19 & -0.08 & 0.18 &  50 &  50 &   452 &  152 & SB2? &  EL  \\ 
    RasTyc2307+3150  & 23 07 24.83 & +31 50 14.1   &    K4V     &   4774   &   108  &  4.58  &  0.13 &  0.05 & 0.10 & 257 &  33 &   629 &   75  &  S  &  SA \\  
\hline
\noalign{\medskip}
\end{tabular}
\end{scriptsize} 
\end{center}
\end{table*}

\clearpage

\addtocounter{table}{-1}
\begin{table*}
\caption{continued.}
\begin{center} 
\begin{scriptsize} 
\begin{tabular}{lccccrccrcrrrrcc}
\hline
\hline
\noalign{\medskip}
\textsl{RasTyc} Name & $\alpha$ (2000) & $\delta$ (2000)	       & Sp. Type &  $T_{\rm eff}$ & $\sigma_{T_{\rm eff}}$ & $\log g$ & $\sigma_{\log g}$ & $[Fe/H]$ & $\sigma_{[Fe/H]}$ & $W_{\rm Li}$ & $\sigma_{W_{\rm Li}}$ & $W_{\rm H\alpha}^{em}$ $^{\bullet}$ & $\sigma_{W_{\rm H\alpha}^{em}}$ &  Bin$^{\rm a}$ & Instr.$^{\rm b}$   \\
                     &  h m s	       & $\degr ~\arcmin ~\arcsec$     &	  &   \multicolumn{2}{c}{(K)}		    &	     \multicolumn{2}{c}{(dex)}	   &	    \multicolumn{2}{c}{(dex)}		  &	    \multicolumn{2}{c}{(m\AA)}  		       &       \multicolumn{2}{c}{(m\AA)}   &  &  \\
\noalign{\medskip}
\hline
\noalign{\medskip}
    RasTyc2308+0207  & 23 08 40.84 & +02 07 39.4   &    G8III   &   5273   &   106  &  3.42  &  0.18 & -0.08 & 0.11 &	8 &   5 &    26 &    9  &  S  &  EL \\ 
    RasTyc2308+0000  & 23 08 50.46 & +00 00 52.8   &    G2V     &   5620   &   211  &  4.38  &  0.23 & -0.00 & 0.16 & 216 &  43 &  1009 &  189  &  S  &  EL \\ 
    RasTyc2309-0225  & 23 09 37.09 & -02 25 54.8   &    K4V     &   4557   &    74  &  4.60  &  0.10 &  0.01 & 0.11 & 195 &  26 &  1411 &   94  &  S  &  EL \\ 
    RasTyc2309+1425  & 23 09 57.17 & +14 25 36.3   &    G1V     &   5855   &    63  &  4.29  &  0.11 & -0.10 & 0.11 &	2 &   7 & 96--127 &  23 & SB1 &  EL \\ 
    RasTyc2317+0941  & 23 17 32.23 & +09 41 36.8   &     K4V    &   4738   &   107  &  4.58  &  0.10 &  0.03 & 0.11 &  10 &   7 &    62 &   21  &  S  &  EL \\ 
    RasTyc2318+4458  & 23 18 48.11 & +44 58 15.7   &     K0V    &   5237   &   123  &  4.34  &  0.26 &  0.01 & 0.15 & 189 &  52 &    88 &   46  &  S  &  FR \\
    RasTyc2320+7414  & 23 20 52.07 & +74 14 07.1   &     K0IV	&   5093   &   100  &  3.58  &  0.40 & -0.02 & 0.12 & 365 &  44 &  4523 &  237  &  S  &  SA \\ 
    RasTyc2321+0211  & 23 21 11.28 & +02 11 50.5   &  K0III-IV  &   4974   &    72  &  3.10  &  0.14 & -0.01 & 0.11 &  10 &  8 & 19--25 &   11  &  S  &  EL \\ 
    RasTyc2321+4510  & 23 21 44.29 & +45 10 34.4   &     K3V    &   4855   &   153  &  4.50  &  0.22 &  0.09 & 0.17 &   3 &  45 &    16 &   24  &  S  &  FR \\
    RasTyc2321+0721  & 23 21 56.36 & +07 21 33.0   &     K0V    &   5207   &    81  &  4.24  &  0.16 &  0.01 & 0.11 & 277 &  25 &   464 &   53  &  S  &  EL \\ 
    RasTyc2323-0635  & 23 23 01.18 & -06 35 43.6   &  K1III-IV  &   4755   &   104  &  2.87  &  0.26 & -0.10 & 0.11 &  87 &  15 &  1345--2852 &  157  &  SB1 & EL \\ 
    RasTyc2324-0733  & 23 24 06.25 & -07 33 02.7   &   G5IV-V   &   5677   &    57  &  4.29  &  0.10 & -0.00 & 0.10 & 136 &  10 &    68 &   23  &  S  &  EL \\   
    RasTyc2324+6215  & 23 24 40.37 & +62 15 51.1   &    F9V     &   6011   &   199  &  3.97  &  0.10 & -0.21 & 0.17 & 136 &  20 &   123 &   23  &  S  &  AU   \\
    RasTyc2328+4522  & 23 28 27.48 & +45 22 40.6   &    K0V	&   5428   &   136  &  4.21  &  0.17 &  0.05 & 0.11 &  50 &  41 &   684 &  149  &  S  &  SA \\     
    RasTyc2331+8124  & 23 31 29.96 & +81 24 42.6   &    K1V	&   5239   &	80  &  4.08  &  0.20 &  0.00 & 0.11 &  18 &  22 &   245 &   38  &  S  &  SA \\ 	
    RasTyc2340-0402  & 23 40 06.09 & -04 02 55.1   &    G1V     &   5853   &   108  &  4.26  &  0.14 & -0.15 & 0.14 & 128 &  29 &   288 &   42  &  SB1$^{\rm d}$  &  EL \\ 
    RasTyc2340-0228  & 23 40 08.35 & -02 28 49.6   &    G3V     &   5762   &    76  &  4.22  &  0.12 &  0.00 & 0.11 &	0 &  22 &   413 &   79  &  SB1$^{\rm d}$  &  EL \\ 
    RasTyc2341-0837  & 23 41 23.80 & -08 37 47.1   &    K1V     &   5202   &   103  &  4.43  &  0.13 & -0.05 & 0.11 &  22 &  15 &   100 &   21  &  S  &  EL \\  
    RasTyc2343+5038  & 23 43 58.22 & +50 38 01.3   &    G5V     &   5731   &   106  &  4.34  &  0.08 & -0.15 & 0.19 & 106 &  11 & \dots & \dots &  S  &  AU   \\
    RasTyc2348+4615  & 23 48 33.85 & +46 15 10.5   &   G8III    &   4961   &   259  &  2.92  &  0.37 & -0.03 & 0.18 & 58  & 111 & \dots & \dots &  S  &  FR \\
    RasTyc2348+4614  & 23 48 35.53 & +46 14 51.1   &   G0IV     &   5937   &   218  &  3.92  &  0.27 & -0.00 & 0.15 & \dots & \dots & 43 &   96 &  S  &  FR \\
    RasTyc2351+7739  & 23 51 17.29 & +77 39 35.3   &    K1V	&   5146   &	78  &  4.52  &  0.17 & -0.07 & 0.11 & 330 &  76 &  1197 &  156  &  S  &  SA \\  	
    RasTyc2352-1143  & 23 52 10.24 & -11 43 14.5   &    G1V     &   5822   &   109  &  4.30  &  0.13 & -0.03 & 0.12 &  87 &  27 &   198 &   44  &  SB1$^{\rm d}$  &  EL \\ 
    RasTyc2353+4413  & 23 53 10.34 & +44 13 58.0   &    G5V     &   5508   &   244  &  4.23  &  0.18 & -0.04 & 0.09 & 184 &   9 &   165 &   14  &  S  &  AU   \\
    RasTyc2358+5140  & 23 58 16.09 & +51 40 39.2   &    K1V	&   5158   &	75  &  4.52  &  0.18 & -0.00 & 0.11 &  90 &  42 &   867 &  108  &  S  &  SA \\ 
 \noalign{\medskip}									
\hline		
\end{tabular}
\begin{list}{}{}									
\item[$^{\bullet}$] The full range of $W_{\rm H\alpha}^{em}$ values is reported for the stars observed in different epochs. 
\item[$^{\rm a}$] S=single or only one RV measure; SB1=single-lined spectroscopic binary; SB2=double-lined. 
\item[$^{\rm b}$] Spectrograph: SA = SARG, EL = ELODIE, AU = AURELIE, FR = FRESCO. 
\item[$^{\rm c}$] Small-amplitude secondary peak in the CCF. Likely SB2 with a faint component. See Table\,\ref{Tab:RV_SB2}. 
\item[$^{\rm d}$] RV variation with respect to SACY \citep{Torres2006}. 
\item[$^{\mathrm{*}}$]  Double-peaked CCF. Possible SB2 with blended lines. See Table\,\ref{Tab:RV_SB2}. 
\item[$^{\mathrm{**}}$] Possible SB2. Broad- + narrow-line components at the same wavelength.  Close visual pair with $\rho=0\farcs4$ \citep{HIPPA97}. 
\item[$^{\dag}$] Asymmetric CCF. Possible SB2 with blended lines. See Table\,\ref{Tab:RV_SB2}. 
\item[$^{\ddag}$] Close visual binary. Composite spectrum. Unreliable parameters.  
\end{list}
\end{scriptsize} 
\end{center}
\end{table*}

\section{Notes on the very young stars and {\it PMS-like} candidates}
\label{Sec:notes}

We briefly summarise the known properties of the very young stars and {\it PMS-like} candidates listed in Table\,\ref{Tab:LiRich} from the Simbad and Vizier databases 
and the literature. 

\begin{itemize}

\item[\#1]  RasTyc~0000+7940 (=BD+78 853)  was  discovered by \citet{Guillout2010} as a WTTS in the surroundings of the Cepheus flare 
 region \citep[e.g.,][and references therein]{Tachihara2005}. \citet{Guillout2010}, on the basis of the different \vsini\  values measured in the two AURELIE 
 spectra, suspected that it was an SB2, but they warned that the different \vsini\  may also be the result of a different CCF shape in the two spectral domains. 
 In several spectra subsequently acquired with FRESCO we measured always the same RV, within the errors, and no significant \vsini\ variation (see Table\,\ref{Tab:RV}). 
 Therefore, we considered this star as a single one.  The position on the HR diagram with the \citet{Siess00} isochrones suggests an age of $\approx$\,20\,Myr 
 (Fig.\,\ref{Fig:HR_PMS}). 

\item[\#2]  RasTyc~0013+7702  (=TYC 4496-780-1) was discovered by \citet{Guillout2010} as a CTTS star, with a broad double-peaked emission, in the surroundings 
 of the Cepheus flare region. They noticed two peaks with a very different height in the CCF that are centered at about 5 and 45\,\kms\, which suggests an SB2 
 system with components of very different luminosity. Other spectra subsequently taken with FRESCO confirm the binary nature of this source (see Table\,\ref{Tab:RV_SB2}). 
 We have assumed that the atmospheric parameters derived with ROTFIT are basically those of the primary component and the effect of the secondary star is negligible, due 
 to its faintness.  RasTyc~0013+7702 is a also a close visual binary \citep[$\rho=1\farcs$41, $\Delta V\simeq 2.2$\,mag,][]{Fabricius2002}. The companion was also detected in the 
 $H$ band by the SEEDS survey \citep{Uyama2017}. 

\item[\#3]  RasTyc~0038+7903  (=TYC 4500-1478-1) was discovered as a PMS star by \citet{Tachihara2005}, who reported a K1 spectral type. It is one of the four co-moving PMS 
stars in the Cepheus flare region investigated by \citet{Guillout2010}. It is a single star, based on the spectrum apperance and the constant RV. Its position on the 
HR diagram suggests an age of $\approx$\,10\,Myr (Fig.\,\ref{Fig:HR_PMS}). 

\item[\#4]  RasTyc~0039+7905  (=BD+78 19) is the fourth PMS discovered by  \citet{Guillout2010} in the surroundings  of the Cepheus flare region. They suggested that
BD+78\,19 may be a binary system of mass ratio $\approx$\,1. 
Indeed, further observations made by us revealed two CCF peaks of the same intensity with a separation of about 25 \kms\ in two epochs (see Table\,\ref{Tab:RV_SB2}). 
The atmospheric parameters listed in Tables\,\ref{Tab:LiRich} and \ref{Tab:APs} were derived from the analysis of the spectra taken at the conjunctions and can be considered 
as typical for both components. The position on the HR diagram suggests and age of $\approx$\,10\,Myr (Fig.\,\ref{Fig:HR_PMS}). However, considering the system composed
of two nearly equal stars, the luminosity of each component should be a factor of two less, bringing them closer to the 20\,Myr isochrone.

\item[\#5]  RasTyc~0046+4808  (=TYC 3266-1767-1) is one the stars with the highest $W_{\rm Li}$ in our sample. No specific reference is found in the Simbad database. This object is 
listed in the catalog of the International Deep Planet Survey \citep{Galicher2016} as a K3.5-type star with an age between 5 and 30~Myr. The classification is in fairly good agreement 
with our own (K2). No distance is reported in the Gaia DR1 catalog, which prevented us from estimating its age from the position on the HR diagram and studying its kinematical properties. 	 

\item[\#6]  RasTyc~0222+5033  (=BD+49 646) was proposed as the optical counterpart of the X-ray source RX J0222.5+5033 by \citet{Motch1998} and classified as an active-corona source on 
the basis of the ROSAT hardness ratios. The star is located just below the Pleaides upper envelope in the $(B-V)_0$--$W_{\rm Li}$ diagram (see Fig.\,\ref{Fig:Li}). 
The space velocity components are fully compatible with the AB~Dor moving group.

\item[\#7]  RasTyc~0230+5656  (=TYC 3695-2260-1). From its sky location \citet{Kharchenko2004} give a possible membership to the h and $\chi$\,Per open clusters. However its magnitude 
and colors are totally inconsistent with the color-magnitude of these clusters, as well as the Gaia distance of 165\,pc, which is much smaller than that of these clusters 
($\sim$2.3 kpc). The position on the HR diagram suggests an age of $\approx$\,20\,Myr (Fig.\,\ref{Fig:HR_PMS}) and its space velocity components are fairly compatible with the Pleiades 
SKG or the TW Hya association.

\item[\#8]  RasTyc~0252+3728  (=TYC 2338-35-1)  is given as a G5\,IV T Tau-type star in Simbad.  
It is listed in the Washington Visual Double Star (WDS) catalog \citep{Mason2001} as a close physical visual binary with a separation of $0\farcs6$ and components of  
9.34 and 10.77 mag in the $K$ band; 
a third component of 13.4\,mag separated by $4\farcs9$ is reported in the same catalog, but it is not given as a physical component. \citet{LiHu1998} 
report an $W_{\rm H\alpha}^{em}=0.50$\,\AA\ and $W_{\rm Li}$\,=\,220\,m\AA. \citet{Li2004} classifies the system as a WTTS candidate 
with proper motions consistent with the Pleiades. From their survey with adaptive optics of young solar analogs \citet{MetchevHillenbrand2009} estimated an age of about 200\,Myr 
and a mass $M=1.1\,M_{\sun}$ by adopting a distance of 170\,pc.
From high resolution spectroscopy \citet{White2007AJ133} found an  $W_{\rm H\alpha}^{em}$\,=\,2.29\AA\ and $W_{\rm Li}$=166\,m\AA.
Moreover they report RV\,=\,$0.78\pm$2.20\,km\,s$^{-1}$ and $v\sin i$=29.59$\pm$1.55\,km\,s$^{-1}$,	
in very good agreement with our determinations. We found a larger lithium EW ($W_{\rm Li}=293\pm 187$\,m\AA) but with a large error, due to the 
low S/N of our spectrum. 
We confirm that the space velocity components are compatible with the Pleiades SKG.	  

\item[\#9]  RasTyc~0300+7225  (=TYC 4321-507-1) 
This star is located just below the Pleaides upper envelope in the $(B-V)_0$--$W_{\rm Li}$ diagram (see Fig.\,\ref{Fig:Li}). The position in the HR diagram is close 
to the ZAMS. The space velocity components are compatible with the Pleiades SKG.
	  
\item[\#10] RasTyc~0311+4810  (=Cl Melotte 20 94) is given as a rotationally variable F9V star in Simbad. 
\citet{Prosser1992} classifies it as a photometric member of the $\alpha$~Per cluster ($d$= 185\,pc, age=\,72\,Myr). 
From kinematic parameters and a distance of 190\,pc assumed for the center of the $\alpha$~Per cluster, \citet{Makarov2006} estimates for this stars an RV=2.65\,km\,s$^{-1}$ 
and a distance $d$=156\,pc, which are in reasonable agreement with our RV and the TGAS distance of 133$\pm$5\,pc. 

\item[\#11] RasTyc~0316+5638  (=TYC 3710-406-1) is listed in the TYCHO reference catalog \citet{Hoogerwerf2000}, which reports this star as a member of the $\alpha$~Per OB 
association. The position on the sky along with the strong lithium line, the strongly filled in H$\alpha$ line, and its kinematics suggest that this star is related to 
the $\alpha$~Per association.

\item[\#12] RasTyc~0323+5843  (=TYC 3715-195-1). 
In the WISE catalog \citep{WISE} a weak probable variability in the W1~(3.35\,$\mu$m) band is reported. The H$\alpha$ line, in our SARG spectrum, is totally filled in 
by core emission that reaches the local continuum. Its position on the HR diagram suggests 
an age of $\approx$\,15\,Myr (Fig.\,\ref{Fig:HR_PMS}). However, its space velocity components are far from the known young SKGs. It also lies
outside of the locus of the young-disc (YD) population (see Fig.\,\ref{Fig:kinematics}). With only one spectrum, we cannot exclude that this source is an SB1 system
observed far from the conjunction.

\item[\#13] RasTyc~0331+4859  (=Cl Melotte 20 935) is a F9.5V star, in the $\alpha$~Per cluster. 
From high-resolution spectroscopy \citet{Prosser1992} found RV$=-2.7$\,km\,s$^{-1}$ and $v\sin i=78$\,km\,s$^{-1}$ and defined the star as a member of $\alpha$~Per 
cluster. In their analysis of rotation in young OCs \citet{Marilli1997} found $V=$10.05\,mag, $B$--$V$=0.62\,mag, a rotation period $P_{\rm rot}=0.275$\,d and a variation 
amplitude of 0.05\,mag in the $V$ band. From high-dispersion spectra of young nearby stars \citet{White2007AJ133} report an absorption H$\alpha$ feature with an
equivalent width of 3.02\,\AA, a lithium absorption with $W_{\rm Li}=$\,110\,m\AA, RV$=-1.8$\,km\,s$^{-1}$, and $v\sin i$=64.44\,km\,s$^{-1}$. 
On our SARG spectrum we measured nearly the same \vsini\ (66.2\,\kms), a slightly higher radial velocity (RV$=3.35\pm 2.17$\kms), and revealed an H$\alpha$ strongly filled in 
by core emission.
 
\item[\#14] RasTyc~0359+4404  (=TYC 2876-1944-1) is given as a rotationally variable star in Simbad. The variability with a period $P_{\rm rot}\simeq0.456$\,d and a variation 
amplitude in the $V$ band of $0.18$\,mag has been detected by ASAS \citep{Kiraga2013}. We found an RV variation of about 11 \kms\ in the two AURELIE spectra, which is, however, 
just smaller than the sum of the two uncertainties.  Therefore, we have classified this star as single, but further observations are needed to exclude binarity.  

\item[\#15] RasTyc~0616+4516  (=TYC 3375-720-1) 
displays a H$\alpha$ line that is strongly filled in by core emission in our SARG spectrum. Its kinematics is not far from the average of the IC~2391 supercluster.  

\item[\#16] RasTyc~0621+5415  (=TYC 3764-338-1)  
has an age of $\approx$\,20\,Myr based on its position on the HR diagram (Fig.\,\ref{Fig:HR_PMS}). However, its space velocity components are far from the 
known young associations and moving groups. It is also outside the locus of the young-disc population (YD, see Fig.\,\ref{Fig:kinematics}).
	 
\item[\#17] RasTyc~1908+5018  (=HD 234808) was already included in Paper\,I (the only reference present in the Simbad database), but the data were not analyzed in that paper. 
The position of this star in the HR diagram is well above the region occupied by the PMS stars and is more consistent with a giant star (see Figs.\,\ref{Fig:HR} and \ref{Fig:HR_PMS}).
The spectral classification (G8III-IV), the \logg\ value of 3.39 dex and the kinematics strengthen the hypothesis of a lithium-rich giant. 
This star shows far-infrared excess (see Sect.\,\ref{subsec:sed} and Fig.\,\ref{Fig:SEDs}). The high lithium content of this target could have been synthesized through the 
\citet{CameronFowler1971} mechanism. This seems to be also supported by the presence of IR excess, which could be due to dust shell possibly generated via the
same mechanism \citep{deLaReza1996}. A conclusive explanation is out of the scope of this work. 

\item[\#18] RasTyc~1925+4429  (=KIC 8429280) was identified as the optical counterpart of the X-ray source RX\,J1925.0+4429 by \citet{Motch1998}. It is an active star observed by
the {\it Kepler} space telescope whose light curves were analyzed by \citet{Savanov2011} and \citet{Frasca2011} reconstructing the starspot pattern and evolution.
In addition to measuring the differential rotation, \citet{Frasca2011} have also measured the atmospheric parameters, RV, \vsini, $W_{\rm Li}$ and chromospheric fluxes from the 
analysis of SARG and FRESCO spectra. These data have been also used in the present work. Unfortunately, the Gaia DR1 TGAS parallax and proper motions are not available for this source.
 
\item[\#19] RasTyc~2004$-$0239  (=BD$-$03 4778) was identified as the optical counterpart of the X-ray source 1RXS~J200449.5-023915 by \citet{Haakonsen2009}. 
A lower-mass stellar companion ($\Delta K\simeq2.2$\,mag at about 2$\farcs$5) was discovered by \citet{Elliott2015}.

From high resolution spectroscopy, \citet{DaSilva2009} found $T_{\rm eff}$=5083\,K, $W_{\rm Li}=290$\,m\AA, and $v\sin i$=8\,km\,s$^{-1}$. \citet{Desidera2015} 
reported a K1\,V spectral type and values of \teff$=5160\pm$30\,K, \vsini=8\,\kms, and $W_{\rm Li}=280$\,m\AA. \citet{Folsom2016}, with high-resolution spectropolarimetry,
determined \teff$=5130\pm$161\,K, \logg$=4.45\pm0.27$, \vsini$=9.6$\,\kms, and measured the magnetic field. 
These values are close to our determinations (\teff$=5183\pm$63\,K, $W_{\rm Li}=298$\,m\AA, \vsini=9\,\kms).
\citet{Messina2010} found it as a member of AB~Dor association and measured a rotation period $P_{\rm rot}=4.68$\,d and a full amplitude of the variation in the $V$ band of about 
0.10\,mag. This star is located just above the Pleiades upper envelope in our $(B-V)_0$--$W_{\rm Li}$ diagram (Fig.\,\ref{Fig:Li}), which indicates an age of $\approx$\,100\,Myr, 
which is in line with the age adopted by \citet{Desidera2015} and with that of AB~Dor \citep[100--125\,Myr,][]{Luhman2005}. Its position on the HR diagram is, however, closer to 
300\,Myr isochrone, but the isochrones at ages larger than 50\,Myr are very close to each other for \teff$>5000$\,K, so that they are not very useful to infer ages in this case.
The space velocity components derived by us (Fig.\,\ref{Fig:kinematics}) confirm the membership to the AB~Dor association.	  

\item[\#20] RasTyc~2016+3106  (=HD 332091) was identified as the optical counterpart of the X-ray source 1RXS~J200449.5-023915 by \citet{Haakonsen2009}. A temperature about 
350\,K cooler than our value was found by \citet{MunozBermejo2013}, who analyzed the same ELODIE spectra acquired by us. We note that our temperature value, \teff=5082\,K
 is closer to the value of 5500\,K derived with the \citet{PecautMamajek2013} calibrations from  the dereddened color index $(B-V)_0=0.715$\,mag than the
 value \teff=4704\,K reported by \citet{MunozBermejo2013}. The H$\alpha$ line is a faint emission feature in all the 
three ELODIE spectra with a variation of $\approx$\,30\%.  
The kinematics of this star suggest membership to the Pleiades SKG.

\item[\#21] RasTyc~2036+3456  (=TYC 2694-1627-1). A temperature about 160\,K cooler than our value was found by \citet{MunozBermejo2013} from the analysis of our ELODIE spectrum. 
The H$\alpha$ line is filled in by core emission. The kinematics of this star suggest membership to the Pleiades SKG.
 
\item[\#22] RasTyc~2039+2644  (=TYC 2178-1225-1) is given as a rotationally variable star in Simbad. The variability with a period $P_{\rm rot}\simeq3.485$\,d and a variation 
amplitude in the $V$ band of $0.06$\,mag has been detected by ASAS \citep{Kiraga2013}. The components of the space velocity indicate a possible association to the
Pleiades supercluster.   

\item[\#23] RasTyc~2106+6906  (=BD+68 1182) 
is a chromospherically active star with the H$\alpha$ line filled in by core emission and 
\vsini\,$\simeq$\,40\,\kms. Both the position on the HR diagram (Fig.\,\ref{Fig:HR_PMS}) and on the $(B-V)_0$--$W_{\rm Li}$ diagram (Fig.\,\ref{Fig:Li}) indicate a PMS
object with an age of $\approx$20--30\,Myr. Its space-velocity components are close to those of the other stars in the Cepheus flare region discovered by \citet{Guillout2010}.
Despite the large apparent distance on the sky, both the kinematics and age suggest that RasTyc~2106+6906 could be related to this stellar group.

\item[\#24] RasTyc~2120+4636  (=TYC 3589-3858-1) was identified as the optical counterpart of the X-ray source RX\,J2120.9+4636  by \citet{Motch1997}, who derived a spectral type 
F9V on the basis of medium-resolution spectra. Our spectral classification (G1.5\,V) is in fairly good agreement with the previous work. 
It is classified as a rotationally variable star in Simbad. The variability with a period $P_{\rm rot}\simeq0.266$\,d and a variation 
amplitude in the $V$ band of $0.08$\,mag has been detected by ASAS \citep{Kiraga2013}.  RasTyc~2120+4636 is a very fast rotating 
star (\vsini\,$\simeq 110$\,\kms) with a filled H$\alpha$ line. Its space velocity components are far from the known young SKGs. 
It is also outside the locus of the young-disc population (YD, see Fig.\,\ref{Fig:kinematics}). With only one spectrum, we cannot exclude that this source is an SB1 system
observed far from the conjunction.

\item[\#25] RasTyc~2223+7741  (=BD+76 857a).
 Both the position on the HR diagram (Fig.\,\ref{Fig:HR_PMS}) and on the $(B-V)_0$--$W_{\rm Li}$ diagram (Fig.\,\ref{Fig:Li}) indicate a PMS object with an age of $\approx$20\,Myr. 
Its space-velocity components are close to the Castor SKG and the Cepheus flare group. 

\item[\#26] RasTyc~2233+1040  (=TYC 1154-1546-1) was identified as the optical counterpart of the X-ray source RX\,J2232.9+1040 by \citet{Appenzeller1998}.
\citet{Zickgraf2005} give an uncertain spectral type K2V:, \teff\,=\,4836\,K, \vsini\,=\,97\,\kms, and $W_{\rm Li}=285$\,m\AA, which lead to
an abundance $A$(Li)\,=\,2.76. With these parameters, they classify the source as a ZAMS star with a lithium abundance between the lower and upper envelope of the Pleiades.  
We find a very similar $W_{\rm Li}=282$\,m\AA, but a very different temperature (\teff\,=\,5544\,K) and projected rotation velocity (\vsini\,=\,234\,\kms). As a consequence of
the higher \teff, we determine a much higher lithium abundance, $A$(Li)\,=\,3.65 and the star lies over the Pleiades upper envelope, both in Figs.\,\ref{Fig:Li} and 
\ref{Fig:NLi}. We note that the TYCHO $B-V=0.558$, that becomes $(B-V)_0=0.521$ considering the reddening, is inconsistent with both the \citet{Zickgraf2005} and our \teff\ 
determinations. 
The color listed in the APASS catalog \citep{Henden2016}, $B-V=0.723$ or $(B-V)_0=0.686$, agrees much better with our \teff\ determination and that of 5361\,K by \citet{Ammons2006}. 
This star, due the aforementioned inconsistencies, deserves future investigations.

\item[\#27] RasTyc~2241+1430  (=HD 214995) was identified as the optical counterpart of the X-ray source RX\,J2241.9+1431 by \citet{Appenzeller1998}.
\citet{Zickgraf2005} report a K0III spectral type, \teff\,=\,5152\,K, $W_{\rm Li}=307$\,m\AA, $A$(Li)\,=\,3.36 and classify this star as an evolved lithium-rich one.
The strong lithium line ($W_{\rm Li}=389$\,m\AA) was confirmed by \citet{Luck2007}.
It was originally included in our list of ELODIE targets, but, based on the literature following our observations and on our results, we discard RasTyc~2241+1430 as PMS 
candidate and confirm its nature of a lithium-rich giant. The values of \teff\ and \logg\ listed in the PASTEL catalog \citep{PASTEL} are in the ranges 4560--4709\,K and 
2.61--2.70\,dex, respectively. Our determinations (Table\,\ref{Tab:APs}) are in very good agreement with them.

\item[\#28] RasTyc~2244+1754  (=BD+17 4799) is given as a K2 star in Simbad. This source was selected by \citet{Jeffries1995} as optical counterpart of an EUV source. 
From intermediate-resolution spectroscopy, he estimated a spectral type K0V/IV, 
RV spanning from 0.7 to 5.9\,km\,s$^{-1}$, $W_{\rm Li}\simeq300$\,m\AA, and was also able to obtain a spectrum of a secondary visual component of spectral 
type M2Ve, $V=12.7$\,mag, and RV\,=\,$-2.5$\,km\,s$^{-1}$, with a separation of about 3\arcsec; from kinematical analysis the system came out to belong to the Local Association. 
From high resolution spectroscopy, \citet{Fekel1997} found for the primary visual component RV\,=\,$-20.2$\,km\,s$^{-1}$, suggesting a SB1 behavior, and $v\sin i$=11.4\,km\,s$^{-1}$. 
\citet{Christian2002} give $v\sin i$=8\,km\,s$^{-1}$ and $W_{\rm Li}$\,=\,244\,m\AA\  ($A$(Li)\,=\,2.8) in excellent agreement with our values (see Table\,\ref{Tab:APs}). 
\citet{Lopez-Santiago2010} found from high-resolution spectra an average RV\,=\,$-16.4$\,km\,s$^{-1}$, 
$v\sin i=11.03$\,km\,s$^{-1}$ and $W_{\rm Li}$\,=\,248\,m\AA, confirming the membership to the Local Association. Their values of H$\alpha$ equivalent width in the subtracted 
spectrum, $W_{\rm H\alpha}^{em}$, are in the range 0.51--0.76\,\AA, in good agreement with our values of 0.60--0.72\,\AA. The spectrum of this source is shown in Fig.\,\ref{Fig:HaLisub}.
The two RV values measured in our spectra are $-16.58$ and $-15.28$ whose difference is larger than the sum of errors. For this reason and from the literature data, we 
consider this star as SB1. The kinematical properties are still consistent with the Local Association, but the velocity components are closer to the AB~Dor association
(see Fig.~\ref{Fig:kinematics}).   

\item[\#29] RasTyc~2246+5749  (=TYC 3992-349-1) has only one reference in Simbad. It has been used as a local photometric standard star by \citet{Henden1999}, who provide 
$BVR_{\rm C}I_{\rm C}$ photometry, in a study of two eclipsing variables in Cepheus. Its space velocity components are far from the known young associations. 
It is also outside the locus of the young-disc population (YD, see Fig.\,\ref{Fig:kinematics}). With only one spectrum, we cannot exclude that this source is an SB1 system
observed far from the conjunction.

\item[\#30] RasTyc~2307+3150  (=TYC 2751-9-1) is given as a K-type variable star in Simbad.  In the General Catalogue of Variable Stars \citep[GCVS,][]{GCVS} 
a period $P_{\rm rot}=7.7129$\,d with an amplitude of $0.04$\,mag is given. The space-velocity components are fairly consistent with Octans association or Castor group.

\item[\#31] RasTyc~2320+7414  (=V395 Cep) is given as a variable star of Orion type in Simbad. 
\citet{Herbig1977} reported a radial velocity of $-10$\,\kms\ with a strong \ion{Li}{i} line and a 
spectral type  G5-8Ve. In the Catalog of Emission-Line Stars of the Orion Population, \citet{HerbigBell1988} reported emission in the H$\alpha$ line with an equivalent
width of about 8\,\AA\ and located the star in the L1259 nebulosity. 
In a study of PMS stars in the Cepheus flare region, \citet{Kun2009} classified this object as a CTTS belonging to the L1261 cloud.
With the exception of the two lithium-rich giants, this is the star with the highest position on the HR diagram, close to the 5-Myr isochrone of \citet{Siess00}.
Our SARG spectrum displays a broad double-peaked H$\alpha$ emission profile with a central absorption, which is typical of CTT stars.

\item[\#32] RasTyc~2321+0721  (=TYC 584-343-1) is given as a rotationally variable star in Simbad.  A rotation period $P_{\rm rot}\simeq2.5$\,d
and an amplitude $\Delta V\simeq0.076$\,mag are reported by \citet{aavso}. \cite{Torres2006} quote a K0V spectral type, \vsini\,=\,14.6\,\kms\ and
$W_{\rm Li}\simeq$\,300\,m\AA. 
\citet{MunozBermejo2013}, from the analysis of our ELODIE spectrum, found \teff\,=\,5221\,K. Our determinations (K0V, \teff\,=\,5207\,K, \vsini\,=\,14.3\,\kms, 
$W_{\rm Li}=$\,277$\pm$25\,m\AA) are in excellent agreement with the aforementioned values from the literature.
A visual companion at $5\farcs1$, which is 2\,mag fainter in the $J$ band, has been discovered by \citet{Elliott2015}.
Both the position on the HR diagram (Fig.\,\ref{Fig:HR_PMS}) and on the $(B-V)_0$--$W_{\rm Li}$ diagram (Fig.\,\ref{Fig:Li}) indicate a young
object with an age of $\approx$30\,Myr. Its space-velocity components are just outside the YD locus but they are not far from those of Octans association.

\item[\#33] RasTyc~2351+7739  (=TYC 4606-740-1). 
In the TYCHO Double Star Catalogue \citep{Fabricius2002} it is quoted as a close visual binary with a separation of $0\farcs82$ and components of $V$ magnitudes 
11.34 and 11.49\,mag. The position on the $(B-V)_0$--$W_{\rm Li}$ diagram (Fig.\,\ref{Fig:Li}) indicates a {\it PMS-like} object on the upper envelope of the IC~2602 cluster
($age\approx$ \,30\,Myr). Unfortunately, the parallax is not available, preventing us to place this object on the HR diagram and to investigate its kinematical properties. 

\item[{\it a}] RasTyc~0106+3306 (=TYC 2282-1396-1) is quoted as a close visual pair \citep[$\rho=0\farcs$85, $\Delta V\simeq 0.29$\,mag,][]{Fabricius2002}.   
The $B-V$ color reported in TYCHO catalog ($B-V=1.092$\,mag) is probably unreliable, as well as the individual colors of both components \citep[$B-V\approx 1.0-1.1$\,mag,][]{Fabricius2002}, 
which are at odds with our spectral classification.  
This could be due to binary nature. However, the value of $B-V=0.646$\,mag reported by \citet{Henden2016} agrees much better with our G4V type and \teff\,=\,5606\,K.
With this color RasTyc~0106+3306 would move on the left side in the $(B-V)_0$--$W_{\rm Li}$ diagram (Fig.\,\ref{Fig:Li}), overcoming the Pleiades upper envelope.
The parallax is not available, preventing us to place this object on the HR diagram and to investigate its kinematical properties. 
Anyway, \citet{Dias2014} consider this star as a member of the cluster Platais~2 ($age \sim$\,350\,Myr). 

\item[{\it b}] RasTyc~0249+4255 (=BD+42 636) is given as a G3 star of the NGC~1039 OC ($d$= 499\,pc, age=\,180\,My) in Simbad.  
\citet{Kharchenko2004} found that the star is not a member of NGC~1039. Indeed, the Gaia DR1 TGAS distance, $d=118.3\pm13.9$\,pc, is inconsistent with NGC~1039.
The space-velocity components are close to those of the TW~Hya association.

\item[{\it c}] RasTyc~0344+5043 (=TYC 3325-98-1) is reported as a close pair in the WDS catalog \citep{Mason2001}, with a separation of $0\farcs7$ and components of magnitudes 
10.80 and 12.80\,mag. This star is well inside the inner radius of the $\alpha$~Per cluster, but its photometry and kinematics exclude membership to the cluster \citep{Kharchenko2004}. 
The Gaia DR1 parallax is not available, preventing us to place this object on the HR diagram and to investigate its kinematical properties.

\item[{\it d}] RasTyc~0646+4147 (=TYC 2949-780-1) has no specific reference in Simbad. 

\item[{\it e}] RasTyc~1731+2815 (=TYC 2087-1742-1) is quoted as a rotationally variable star in Simbad. The rotation period of about 1.26 d was measured both by
SuperWASP \citep{Norton2007} and ASAS \citep{Kiraga2012} observations. \citet{Dragomir2007} associated the X-ray source  1RXS\,J173103.4+281510
to this optical counterpart and classified it as a G/K star from low-resolution spectra. This star was observed spectroscopically by \citet{Binks2015} who found 
RV\,=\,-18.2$\pm$0.5\,\kms, $W_{\rm Li}=$\,265$\pm$36\,m\AA, and a faint H$\alpha$ absorption line with equivalent width of 0.19\,\AA. They estimated the age in 
the range 30--150\,Myr, based on $W_{\rm Li}$, and a lower limit of 37\,Myr, on the basis of the H$\alpha$ and $(V-I)$ color.
Our values of RV\,=\,$-19.22\pm$0.62\,\kms\ and $W_{\rm Li}=243\pm22$\,m\AA\ are not significantly different from that of \citep{Binks2015}. The H$\alpha$ line 
is totally filled in with emission.

\item[{\it f}] RasTyc~1956+4345 (=KIC 7985370) was discovered as a periodic variable in the field of the {\it Kepler} space telescope by \citet{Pigulski2009}, who
found a period $P_{\rm rot}=0.4249$\,d and amplitudes of 0.04\,mag and 0.02\,mag in the $V$ and $I$ light curves, respectively. The {\it Kepler} light curves were 
analyzed by \citet{Froehlich2012} who redetermined the period as 2.86\,d and reconstructed the starspot distribution and evolution over 229 days with a Bayesian spot model.
In addition to measuring the differential rotation, \citet{Froehlich2012} have also measured the atmospheric parameters, RV, \vsini, $W_{\rm Li}$, and chromospheric 
fluxes from the analysis of SARG and FRESCO spectra. These data have been also used in the present work. As already discussed by \citet{Froehlich2012}, the kinematical
properties of KIC~7985370 indicate it is a member of the Local Association (Pleiades SKG).
 
\item[{\it g}] RasTyc~2038+3546 (=BD+35 4198) was included in the bright catalog (Paper\,I) but the ELODIE spectrum was not analyzed with ROTFIT. In the present paper, the 
analysis of the ELODIE and a new SARG spectrum provided us with revised values of atmospheric parameters, $W_{\rm Li}$, and $W_{\rm H\alpha}^{em}$.
The value of \teff\ we find (5895\,K) is slightly larger than in Paper\,I and agrees better with the determination of 5954\,K by \citet{MunozBermejo2013}.
Based on its kinematical properties, it could be a member of the Pleiades SKG.

\item[{\it h}] RasTyc~2114+3941 (=BD+39 4490) was included in the RasTyc bright catalog (Paper\,I) but the ELODIE spectrum was not analyzed with ROTFIT. In that paper the 
parameters were derived from two AURELIE spectra which provided \teff\,=\,5731\,K, $W_{\rm Li}=101$\,m\AA, and $W_{\rm H\alpha}^{em}=64$\,m\AA. The ELODIE spectrum, with its
higher quality (especially in terms of the wider spectral range), provides us with a higher value of \teff\ (5930\,K), which agrees better with the value of 5955\,K derived
by \citet{MunozBermejo2013}. The slightly larger $W_{\rm Li}=131\pm27$\,m\AA\ measured in the ELODIE spectrum leads to a lithium abundance $A$(Li)=3.01, which brings the star
just above the Pleiades upper envelope in Fig.\,\ref{Fig:NLi}. We also find a higher $W_{\rm H\alpha}^{em}=149\pm34$\,m\AA, which is likely due
to variations of the chromospheric activity. The $U_{\sun}$ and $V_{\sun}$ velocity components are close to Castor, but the $W_{\sun}$ is not.

\item[{\it i}] RasTyc~2203+3809 (=TYC 3198-1809-1) has no specific reference in Simbad. 
We did not find any spectroscopic study in the literature. This object lies under the Pleiades upper envelope in the $(B-V)_0$--$W_{\rm Li}$ diagram (Fig.\,\ref{Fig:Li}),
but it is just above it in the diagram \teff--$A$(Li) (Fig.\,\ref{Fig:NLi}). 

\item[{\it j}] RasTyc~2308+0000 (=TYC 576-1220-1) is quoted as a PMS star in Simbad. It was associated to the X-ray source RX\,J2308.8+0001 by \citet{Zickgraf2003}.
\cite{Torres2006} quote a G8V spectral type, RV\,=\,7.4\,\kms, \vsini\,=\,39.0$\pm$9.0\,\kms\ and $W_{\rm Li}$\,=\,250\,m\AA. We find an earlier spectral type (G2V)
and a slightly lower $W_{\rm Li}$\,=\,216$\pm$43\,m\AA. The \vsini\ value of 42.8$\pm$5.2\,\kms\ agrees with that of \citet{Torres2006}, but the RV\,=\,5.30$\pm$0.22\,\kms\ 
can be indicative of an SB1 system, even if \citet{Torres2006} do not provide the error of their RV measure. However, an RV variation is also indicated by the value
of RV\,=\,1.696$\pm$2.642\,\kms\ reported in the RAVE 5th data release \citep{Kunder2017}. In this catalog, values for the atmospheric parameters of
\teff\,=\,6000\,K, \logg\,=\,4.00, and [Fe/H]\,=\,-0.25	are also reported. The effective temperature of \teff\,=\,5533\,K derived by \citet{Kunder2017} with the infrared
flux method is in better agreement with our determination of \teff\,=\,5620$\pm$211\,K. This star lies between the 20- and 30-Myr isochrones in the HR diagram 
(Fig.\,\ref{Fig:HR_PMS}) and close to the UVW position of Octans association. 

\item[{\it k}] RasTyc~2324+6215 (=TYC 4283-219-1) has no specific reference in Simbad, where an F8 spectral type is reported. The Gaia DR1 parallax is not available, preventing 
us to place this object on the HR diagram and to investigate in detail its kinematical properties.

\end{itemize}

\end{appendix}

\end{document}